\documentclass[11pt]{article}
\usepackage{amsmath}
\usepackage{amssymb}
\usepackage{subfig}
\usepackage{epsfig}

\usepackage{graphicx}
\usepackage{cite}
\usepackage{color} 

\makeatletter
\renewcommand{\@biblabel}[1]{\quad#1.}
\makeatother

\date{}

\begin{document}

\begin{flushleft}
{\Large
\textbf{A cell-type based model explaining co-expression patterns of genes in the brain\\ 
\vspace{.2cm}
}
}

Pascal Grange$ ^1$, Jason Bohland$ ^2$, Hemant Bokil$ ^1$, Sacha Nelson$ ^3$, Benjamin Okaty$ ^3$,\\
 Ken Sugino$ ^3$, Lydia Ng$ ^5$, Michael Hawrylycz$ ^5$, Partha P. Mitra$ ^1$\\

\vspace{1cm}
$ ^1$ Cold Spring Harbor Laboratory,\\
   One Bungtown Road, Cold Spring Harbor, New York 11724, United States\\
$ ^2$ Health Sciences Department, Boston University, Boston, Massachussets, United States\\ 
$ ^3$ Department of Biology, Brandeis University, Waltham, Massachusetts, United States\\
$ ^4$ Department of Genetics, Harvard Medical School, Boston, Massachussets, United States\\
$ ^5$ Allen Institute for Brain Science, Seattle, Washington 98103, United States\\
\vspace{0.4cm}
$\ast$ E-mail: pascal.grange@polytechnique.org
\end{flushleft}
\vspace{1cm}
\begin{abstract}

Much of the genome is expressed in the vertebrate brain with
individual genes exhibiting different spatially-varying patterns of
expression. These variations are not independent, with pairs of genes
exhibiting complex patterns of co-expression, such that two genes may
be similarly expressed in one region, but differentially expressed in
other regions. These correlations have been previously studied
quantitatively, particularly for the gene expression atlas of the
mouse brain, but the biological meaning of the
co-expression patterns remains obscure. We propose a simple model of the co-expression patterns in terms
of spatial distributions of underlying cell types. We establish the
plausibility of the model in terms of a test set of cell types for
which both the gene expression profiles and the spatial distributions
are known.
\end{abstract}

\pagebreak

\tableofcontents





\section{The Allen Gene Expression Atlas}
\subsection{Gene-expression energies}

The gene expression energies we analyzed were obtained from ISH images
of thousands of genes in the Allen
Gene Expression Atlas \cite{AllenGenome,AllenAtlasMol}. For each of the genes, an 
eight-week old C57Bl/6J male mouse brain was prepared as unfixed, 
fresh-frozen tissue.
 The following steps 
 were taken in an
automatized pipeline\footnote{For more details on the
processing of the ISH image series, see the NeuroBlast User Guide,
{\ttfamily{http://mouse.brain-map.org/documentation/index.html}}}:\\ 
\begin{itemize}
\item {\bf{Colorimetric {\it{in
        situ}} hybridization}};
\item {\bf{Automatic processing of the
    resulting images.}} Find tissue area eliminating artifacts, look
for cell-shaped objects of size $\simeq 10-30$ microns to minimize
artefacts;\\ 
\item {\bf{Aggregation of the raw pixel data into a grid.}}
The mouse brain is partitioned into $V = 49,742$ cubic voxels of
side 200 microns. For every voxel $v$, the {\it{expression energy}} of
the gene $g$ is defined as a weighted sum of the greyscale-value
intensities of pixels $p$ intersecting the voxel:
$$E(v,g) := \frac{\sum_{p\in v} M( p ) I(p)}{\sum_{p\in v} 1},$$ where
$M( p )$ is a Boolean mask worked out by step 2 with value 1 if the
pixel is expressing and 0 if it is non-expressing.\\ 
\end{itemize}

The present analysis is focused on 4104 genes for which sagittal and coronal 
data are available. We computed the correlation coefficients
between sagittal and coronal data and selected the genes in the top-three 
quartiles of correlation (3041 genes) for further analysis.

\subsection{Classical neuroanatomy: systems of annotation}
Partitions of the
brain (or of the left hemisphere) into regions, at the same resolution (200 microns) as the gene-expression data,
 are available from the Allen Reference Atlas \cite{AllenAtlas}. The Allen reference Atlas
 was obtained using the same mouse strain and methodology as for the the gene-expression data.
 Each voxel in the mouse brain therefore comes
with a label containing the name of the brain region to which it belongs.\\

 As separable parts
of the brain (connected components of the gene-expression data in gene space) are of special interest to us,
we will make use of two non-hierarchical systems of annotation, available for the left hemisphere. 
We will refer to them as:\\
- 'Big 12',consisting of the 12 regions of the left hemisphere (together with a more
patchy group of voxels termed 'basic cell groups of regions') whose names, sizes and shapes are shown in Table \ref{fig:referenceTableBig12};\\
- 'Fine', a refinement of 'Big 12' into 94 regions.\\

\begin{table}
\begin{tabular}{|l|l|p{5cm}|}
\hline
\textbf{Name of region}&\textbf{Percentage of hemisphere}&\textbf{Profile of region (maximal-intensity projection)}\\\hline
Basic cell groups and regions&4.6&\includegraphics[width=1.8in,keepaspectratio]{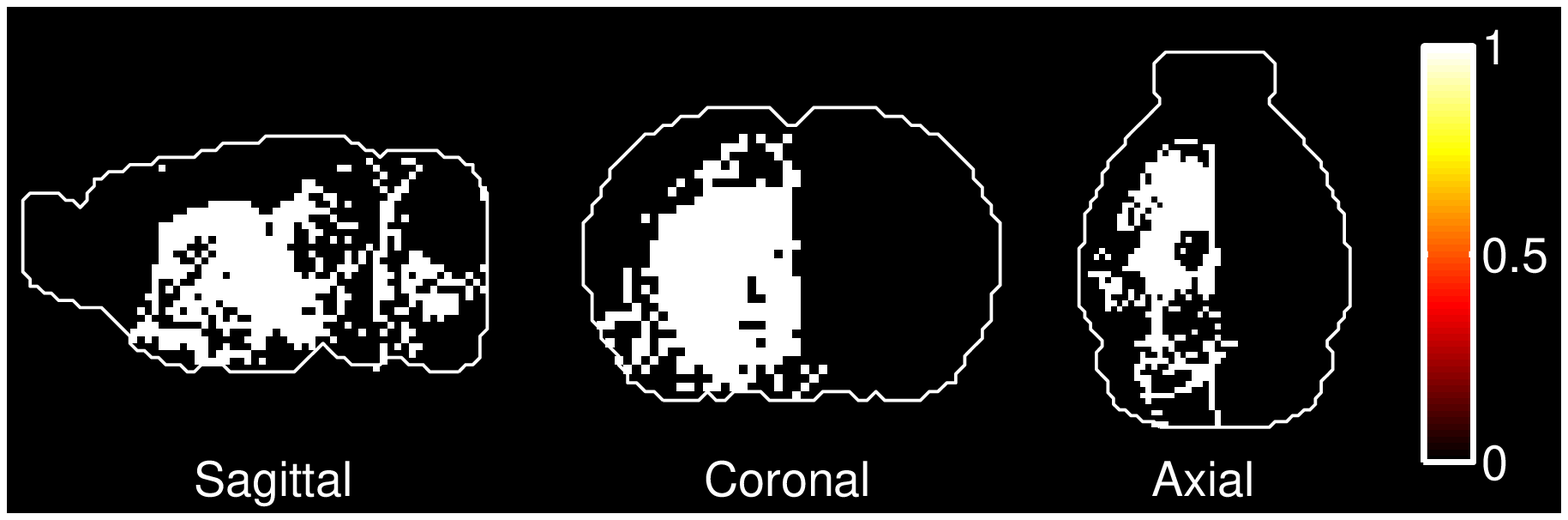}\\\hline
Cerebral cortex&29.5&\includegraphics[width=1.8in,keepaspectratio]{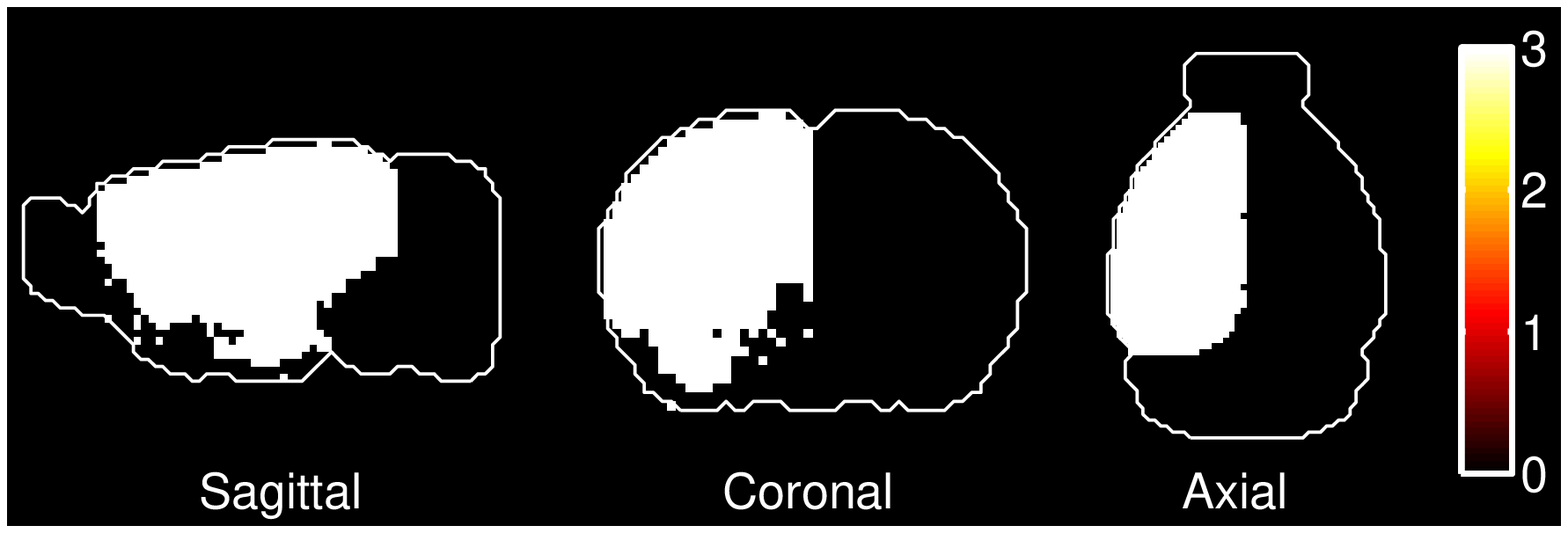}\\\hline
Olfactory areas&9.2&\includegraphics[width=1.8in,keepaspectratio]{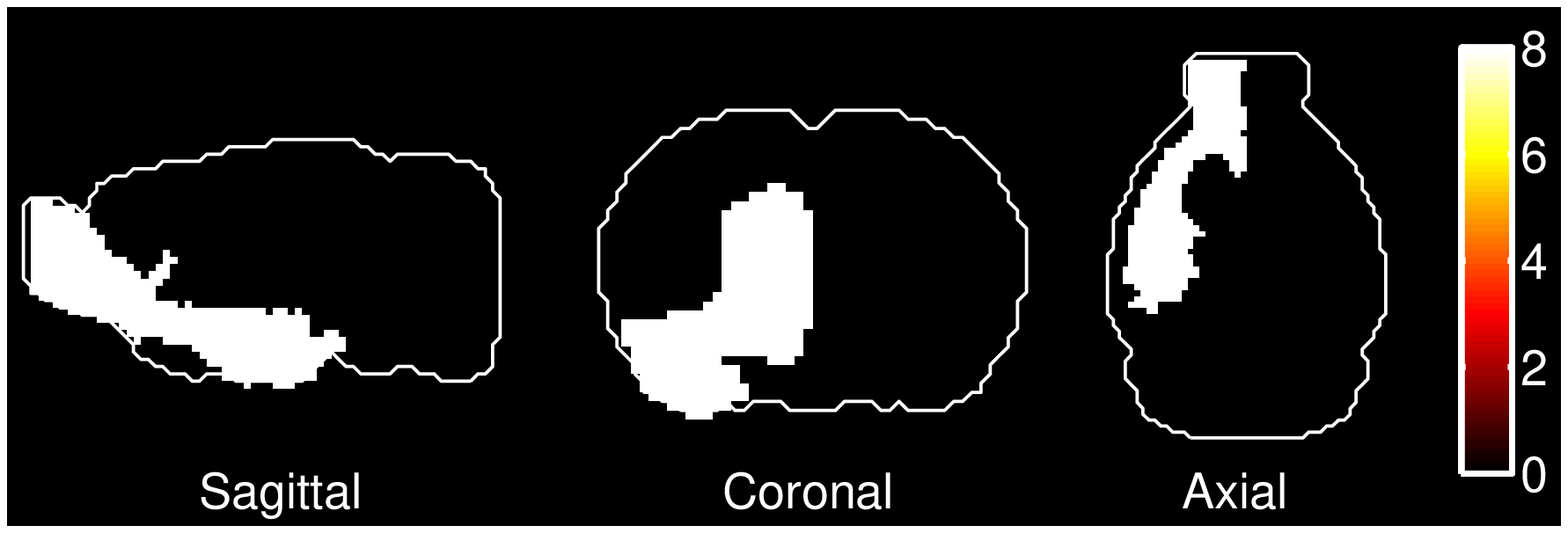}\\\hline
Hippocampal region&4.3&\includegraphics[width=1.8in,keepaspectratio]{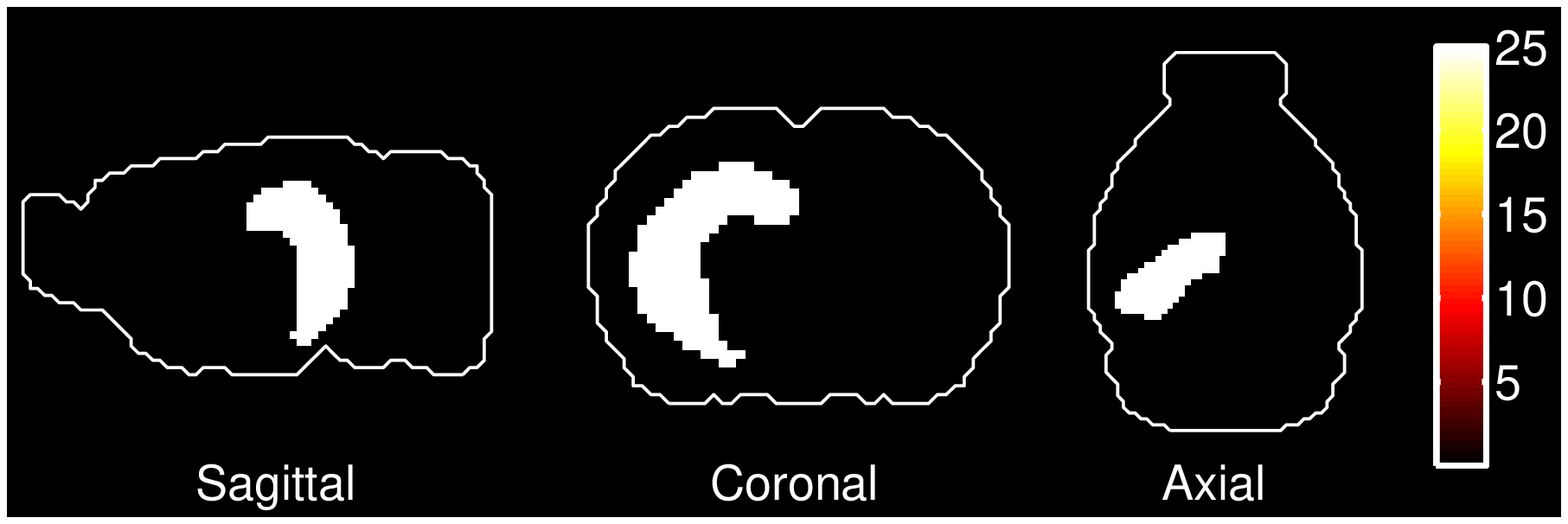}\\\hline
Retrohippocampal region&4&\includegraphics[width=1.8in,keepaspectratio]{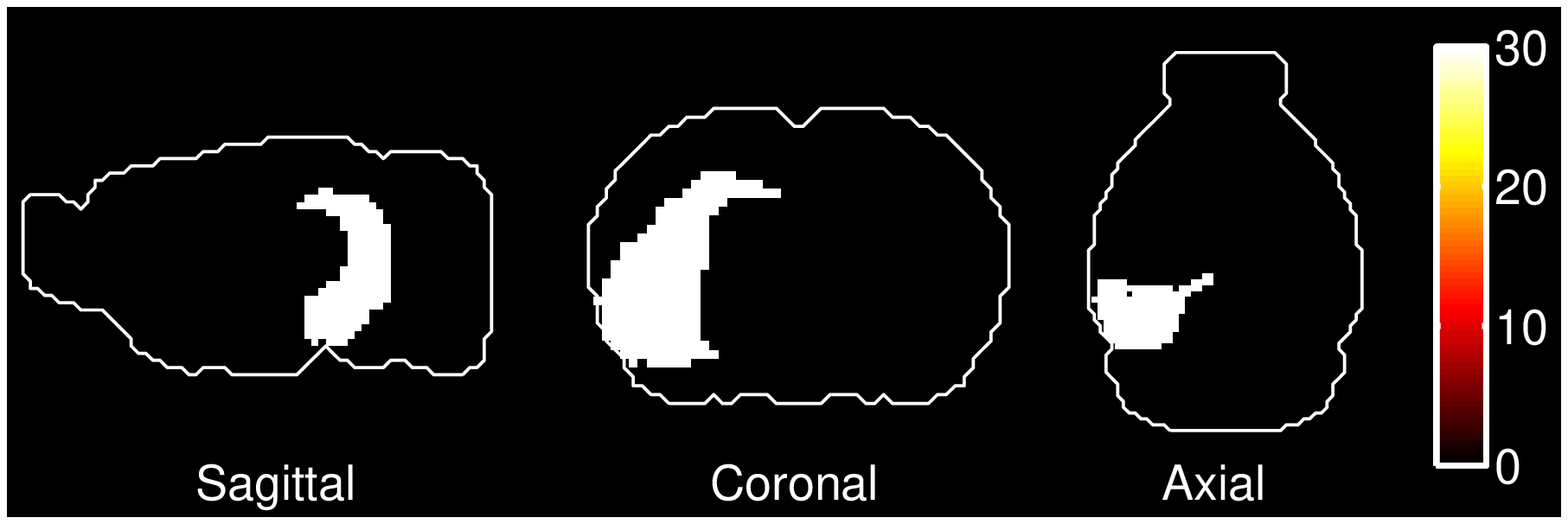}\\\hline
Striatum&8.6&\includegraphics[width=1.8in,keepaspectratio]{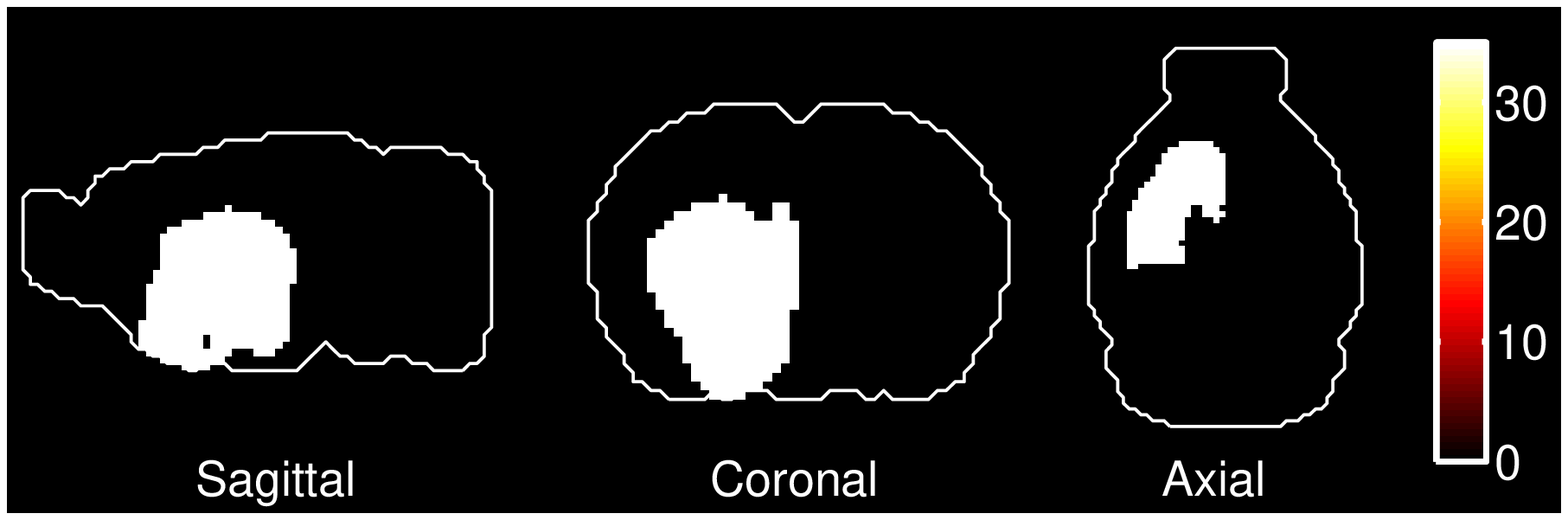}\\\hline
Pallidum&1.9&\includegraphics[width=1.8in,keepaspectratio]{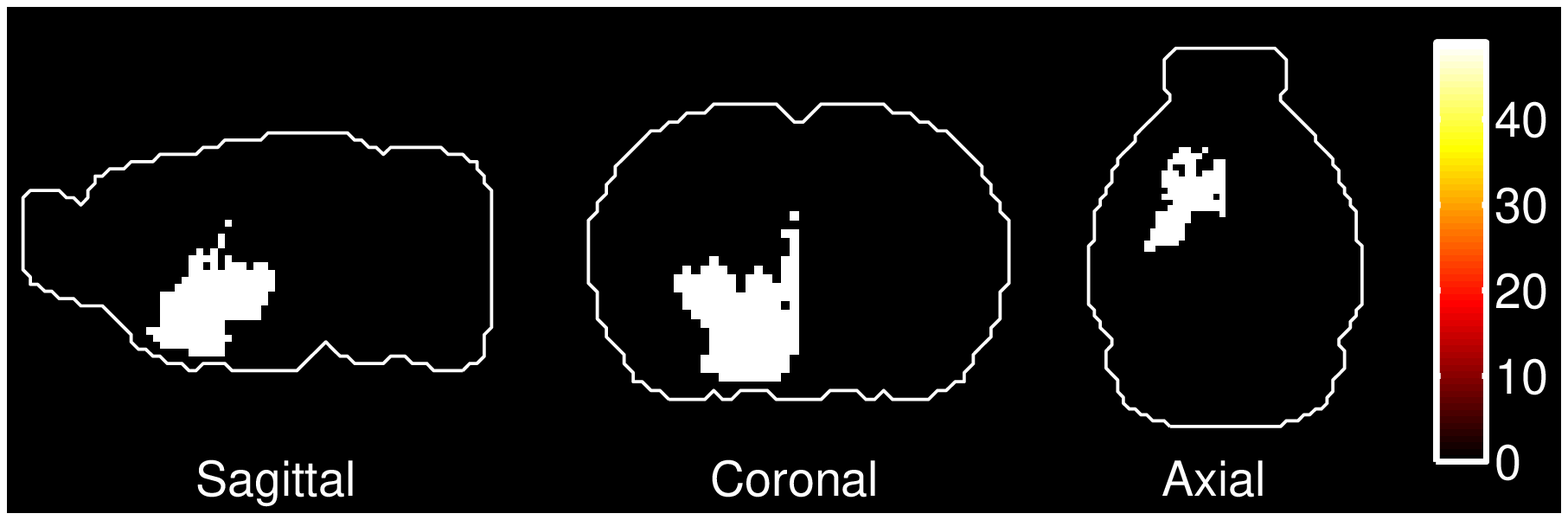}\\\hline
Thalamus&4.3&\includegraphics[width=1.8in,keepaspectratio]{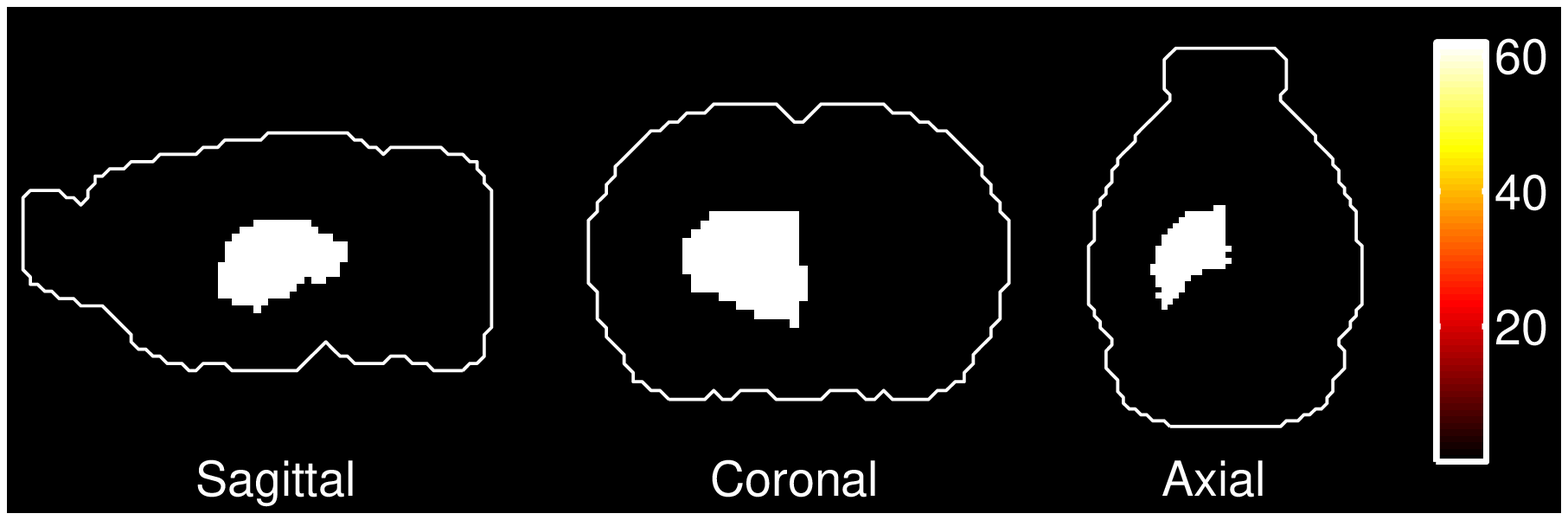}\\\hline
Hypothalamus&3.5&\includegraphics[width=1.8in,keepaspectratio]{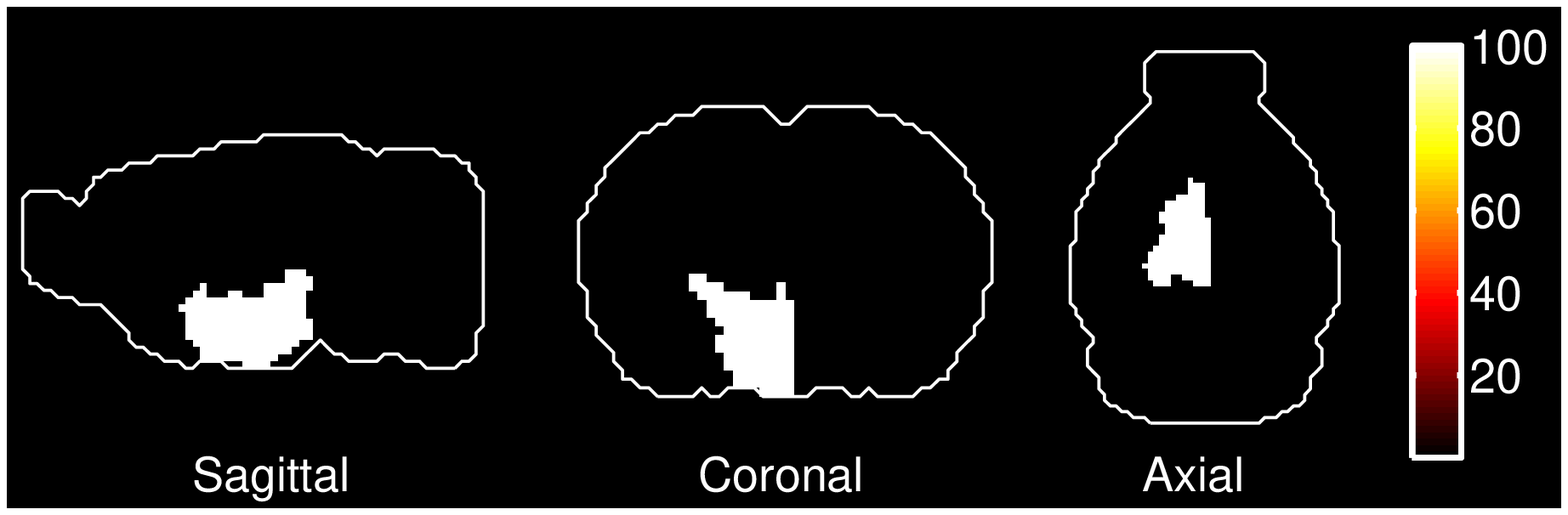}\\\hline
Midbrain&7.8&\includegraphics[width=1.8in,keepaspectratio]{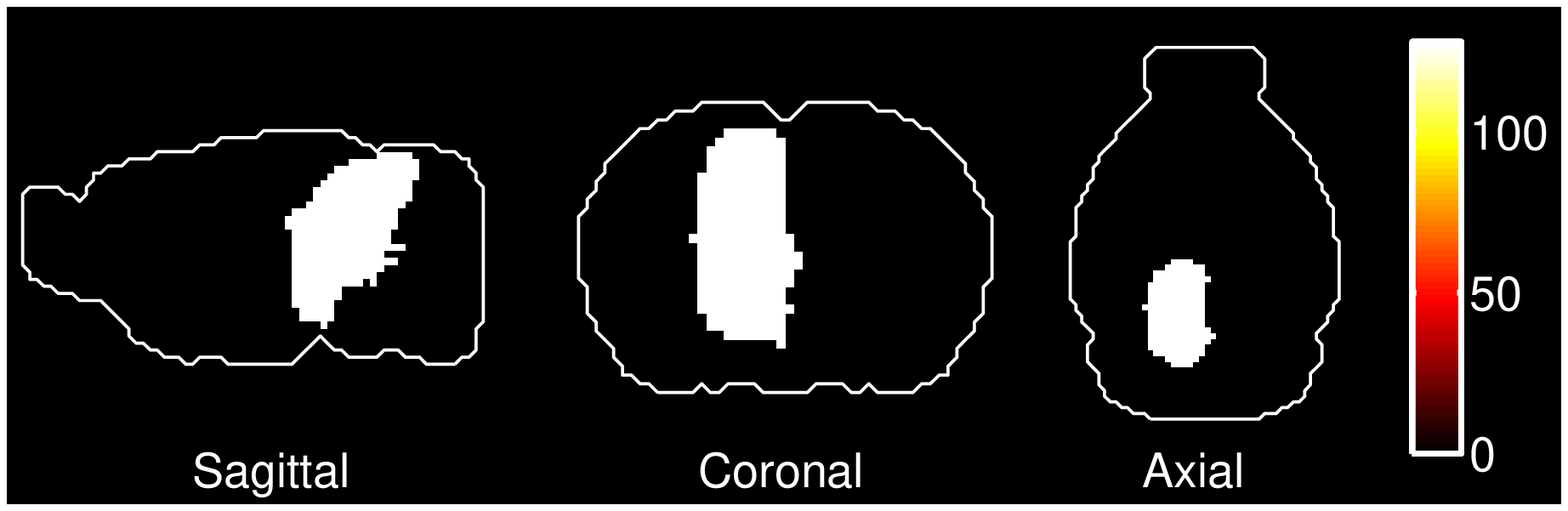}\\\hline
Pons&4.6&\includegraphics[width=1.8in,keepaspectratio]{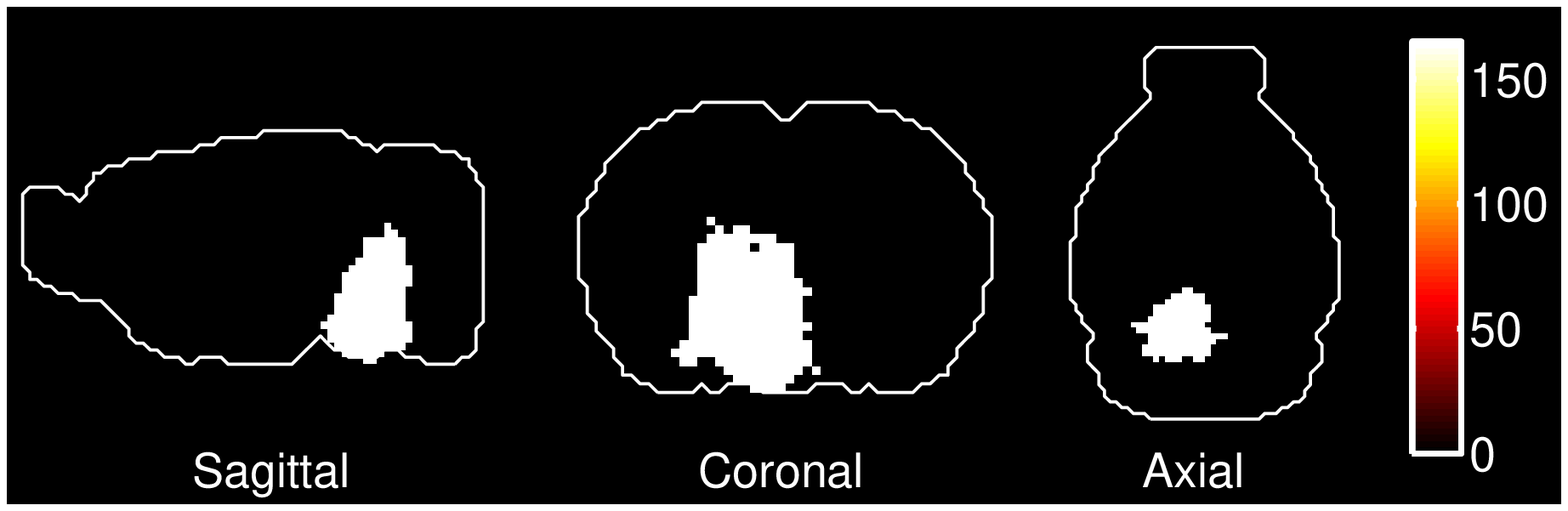}\\\hline
Medulla&6.2&\includegraphics[width=1.8in,keepaspectratio]{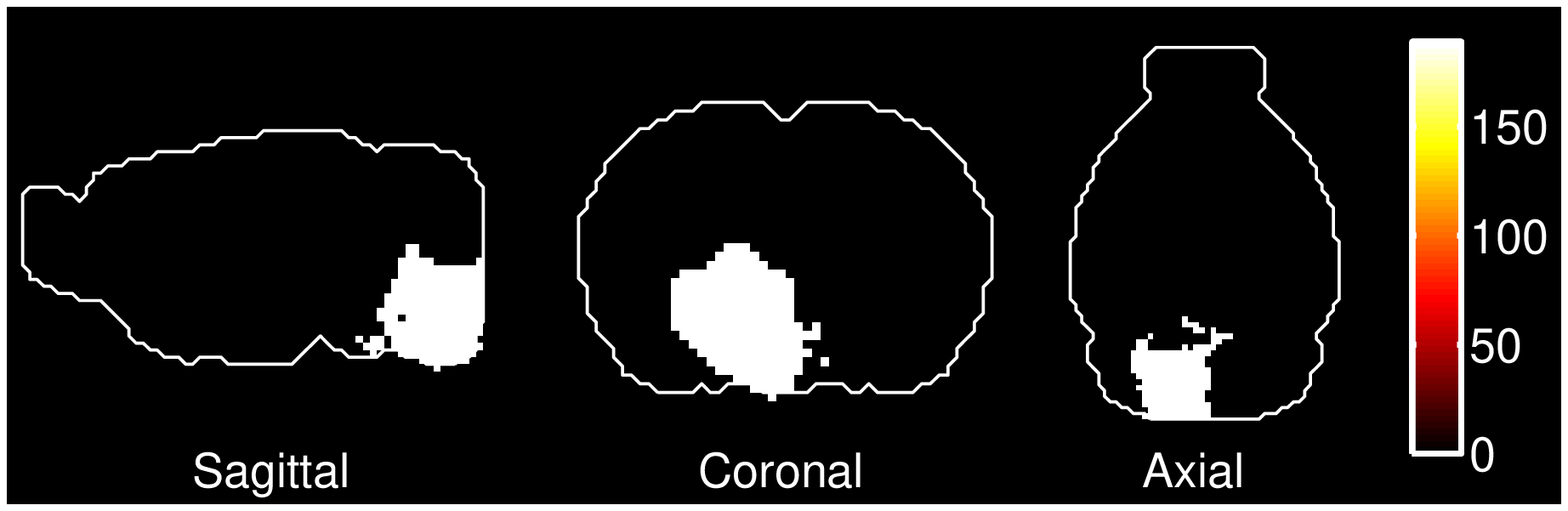}\\\hline
Cerebellum&11.5&\includegraphics[width=1.8in,keepaspectratio]{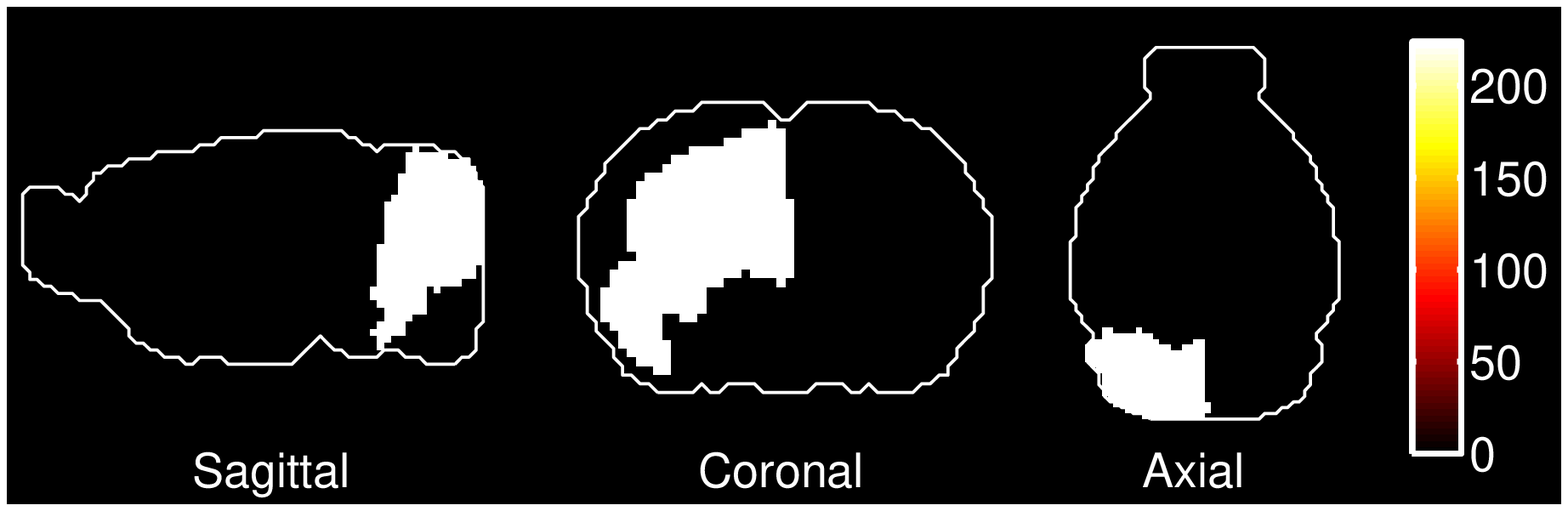}\\\hline
\end{tabular}

\caption{Brain regions in the coarsest annotation of the left hemisphere
in the Allen Reference Atlas, with the relative volume of the hemisphere they occupy, and 
 maximal-intensity projections of their three-dimensional profiles.}
\label{fig:referenceTableBig12}
\end{table}

\section{The Allen Gene Expression Atlas and cell types}

\subsection{Gene expression energies and cell-types}

 A major challenge for neurobiology in this era of complete genomes is to map gene
expression patterns with cellular resolution in the brain.
 The Allen Gene Expression Atlas takes a gene-by-gene approach to this problem, obtaining
a brainwide expression profile for each gene. Co-expression patterns of genes can then be studied by 
comparing those maps. A complementary approach uses microarray experiments
to study co-expression patterns in a small set of neurons of the same type. We studied
two datasets using these two approaches, each of which boils down to a matrix of positive numbers
estimating numbers of mRNAs:
 \begin{itemize}
\item {\bf{Allen Gene Expression Atlas (co-registered ISH data in the whole brain).}}
As explained in the first section, the mouse brain
is partitioned into $V= 49,742$ cubic voxels of side 200
microns. For a given gene $g$ and a given voxel $v$, the AGEA 
 provides the expression energy of $g$
at voxel $v$, or $E(v,g)$, defined as the average intensity of pixels
in the set of ISH images that intersect $v$. It is an estimator of the number
of mRNAs for gene $g$ in voxel $v$. We focus on about $G_A=3041$ genes that have
 the best correlation of signals between sagittal
and coronal sections. 
\item {\bf{Cell-type specific microarray data.}} We analyzed microarray  data
  for $T=64$ different cell types \cite{foreBrainTaxonomy,RossnerCells,CahoyCells,DoyleCells,OkatyCells}.
 Each of these cell types is
characterized by expression  of $G_T = 14580$ genes. These data can be arranged in a $T$ by
$G_T$ matrix $C$, where $C( t, g ) $ is the expression level of gene $g$
in type $t$.
\end{itemize}
Studying these two datasets together can help us gain some insight into brainwide co-expression patterns in terms of neuronal types.\\

We worked out the intersection of the two sets of
genes for corresponding to the two datasets. We picked the columns of
both matrices corresponding to genes that are in both datasets (let
$G=2131$ denote the number of such genes), and rearranged them in the
same order. From now on, $E$ is assumed to be a $V$ times $G$ matrix,
and $C$ to be a $T$ times $G$ matrix, with columns of both matrices corresponding 
to the same set of genes, ordered in the same way. The quantitative techniques 
used in this section do not depend on the precise value of $G$, $V$ and $T$, and the computations
can be repeated with richer datasets involving more genes (larger $G$), higher resolution (larger $V$), 
and/or larger numbers of cell types (larger $T$).\\

\subsection{Correlation between gene-expression data for cell-types and the AGEA}
We first studied the correlations between the AGEA and cell-type data.
For each cell-type $t$, we have a vector in gene-space
with components $C_t( g ) = C(t,g)$, $1\leq g \leq G$ from microarray data. 
We also have a vector in gene space at every voxel $v$ from the AGEA, with 
components $E(g,v )$, $1\leq g \leq G$. We can compute the 
correlation coefficient between cell-type $t$ and voxel $v$ as the 
cosine of the angle between the deviation
from the average cell-type data and from the average voxel, in gene space:

\begin{equation}
\mathrm{Corr}(t,v) = \frac{\sum_{g=1}^G( C(t,g) - \bar{C}(g))( E( v,g ) - \bar{E}(g) )}{\sqrt{\sum_{g=1}^G( C(t,g) - \bar{C}(g))^2}\sqrt{\sum_{g=1}^G ( E( v,g ) - \bar{E}(g) )^2}},
\end{equation}
\begin{equation}
\bar{C}(g) = \frac{1}{T}\sum_{t=1}^T C(t,g),
\end{equation}
\begin{equation}
\bar{E}(g) = \frac{1}{V}\sum_{v=1}^V E(v,g).
\end{equation}
For a fixed type $t$, the correlations with all the voxels $v$ can be rearranged into the 
volume of the brain, and plotted (see the second column of the tables in section 5).\\
It is quite encouraging to observe that correlation patterns 
for some cell types exhibit striking resemblance with neuroanatomical patterns: for instance 
correlations between the AGEA and the microarray data for medium spiny neurons
are highest in the striatum (see Figure \ref{fig:cellType16}).

\subsection{Fitting a linear combination of cell types to the AGEA}

The maps from genes to the whole brain given by the expression energies
 obtained gene by gene in the AGEA must come from the gene-expression
activity of the brain cells. Their co-expression properties must come from the
co-expression of genes in given cell types, and from the spatial variation 
of the abundance of cell types.\\

 We would like to decompose the signal in the AGEA into its cell-type specific components.
Let us introduce the quantity $\rho_t( v )$ denoting the contribution of cell type $t$ at voxel $v$, 
and propose the following linear model:
\begin{equation}
E( v,g) = \sum_{ t = 1 }^T \rho_t( v )C_t( g ) + {\mathrm{Err(v,g)}}.
\label{linearModel}
\end{equation}
Both sides are estimators of the number of mRNAs for gene $g$ at voxel $v$.  
The quantity denoted by ${\mathrm{Err}}$ is an error term (reflecting noise in the measurements, reproducibility
issues, the non-linearity of the relations between numbers of mRNAs, expression energies and microarray data, and the 
 fact that $T=64$ types are not enough to sample the whole diversity of cell types 
 in the mouse brain).\\

In order to find the best fit of the model to the AGEA, we have to solve the
following minimization problem under positivity constraints, that corresponds to the least squares of the error function:
$$\left(\rho_t( v )\right)_{1\leq t \leq T, 1\leq v \leq V} = {\mathrm{argmin}}_{\phi\in {\mathbf{R}}_+^T\times {\mathbf{R}}_+^V}\mathcal{ E}_{E,C}(\phi ),$$
where $T$ is the total number of types available from the microarray data, $V$ is the
total number of voxels in the brain at a resolution of 200 microns, and the function $\mathcal{ E}_{E,C}$ is taken to 
be the sum of squares of the duifferences between the expresison energies and the superposition of
cell types weighted by the positive coefficients of the matrix $\phi$:
$$\mathcal{ E}_{E,C}( \phi ) = \sum_{v=1}^V\left( \sum_{g=1}^G \left( E(v,g) - \sum_{t=1}^T\phi(t,v)C_t(g)\right)^2 \right).$$
As the terms of the sum over voxels that contain coefficients of a fixed line voxel $v$ of the matrix $E$,   involve only terms of  the test function that with the same voxel index $v$, 
the problem can be solved voxel by 
voxel. At each voxel one has to minimize a quadratic function of a vector with $T$ positive components. As the 
function is quadratic and the domain of the unknown is convex, the problem at each voxel is a convex optimization problem, which admits a global minimum:
$$\forall v \in [1..V],  (\rho_t( v ))_{1\leq t \leq T}= {\mathrm{argmin}}_{\nu\in {\mathbf{R}}_+^T}\sum_{g=1}^G \left( E(v,g) - \sum_{t=1}^T\nu(t) C_t(g)\right)^2.$$ 
For each cell-type index  $t$,  a heat map of the maximal-intensity projections 
of $(\rho_t( v ))_{1\leq v \leq V}$ is shown in the rightmost column of the tables of section 5 (N/A indicates
that the optimization returned a zero at all voxels for the corresponding type).

\subsection{Results}
The maximal-intensity projections of correlations and fittings
are presented in the tables of section 5 for all cell types in our study.
 Some striking neuroanatomical patterns appear in the results,
 both in correlations and in fittings. We only made use of the microarray data and of the AGEA
in the above analysis. We compared the results of the computations to the Allen Reference Atlas,
and found out that some major cell-types have strongly inhomogeneous correlations 
and fittings across the brain. So it is interesting to compare the observed neuroanatomical
 patterns to the metadata available from 
\cite{foreBrainTaxonomy,RossnerCells,CahoyCells,DoyleCells,OkatyCells}, specifying from which regions 
of the brain the cell samples were taken.

\subsubsection{Consistent neuroanatonical patterns}

The inspection of the tables of results in section 5 yields several classes
of patterns, both in correlations and fittings, that exhibit similarity with neuroanatomical structures,
such as cerebral cortex (and subregions thereof), cerebellum, hippocampus, striatum
 thalamus, and white matter (including {\emph{arbor vitae}}). We 
studied a few examples in more details, using sections in addition to maximal-intensity projections.

\begin{itemize}

\item{\bf{Cortical patterns.}} See Figure \ref{fig:cellType47} for a layer-specific 
class of pyramidal neurons in the cortex \cite{foreBrainTaxonomy}, and Figure \ref{fig:cellType48} for a class of pyramidal neurons
taken from the amygdala and whose fitting coefficients indeed exhibit a pattern similar to the amygdala.
 The patterns of fitting coefficients are much sparser 
than those of correlations.

\begin{figure}
  \centering 
 \subfloat[Maximal-intensity projection of correlations.]{\label{fig:cellTypeProj47}\includegraphics[width=0.5\textwidth]{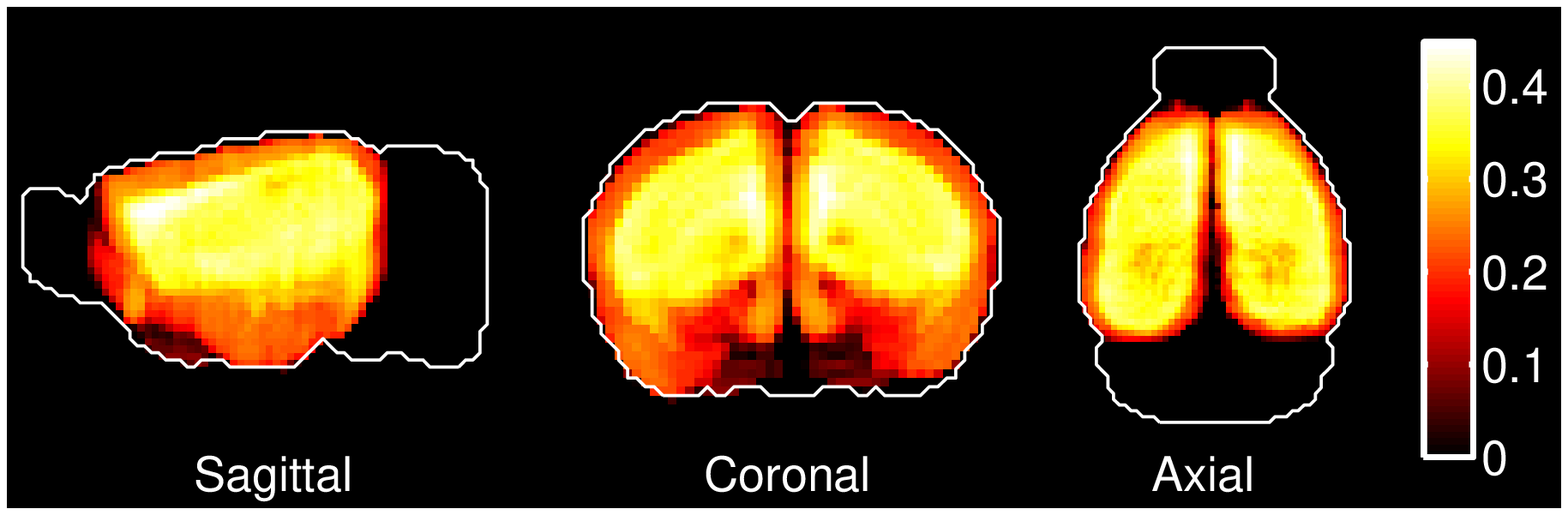}}
  \subfloat[Maximal-intensity projection of the linear fitting.]{\label{fig:cellTypeModelFit47}\includegraphics[width=0.5\textwidth]{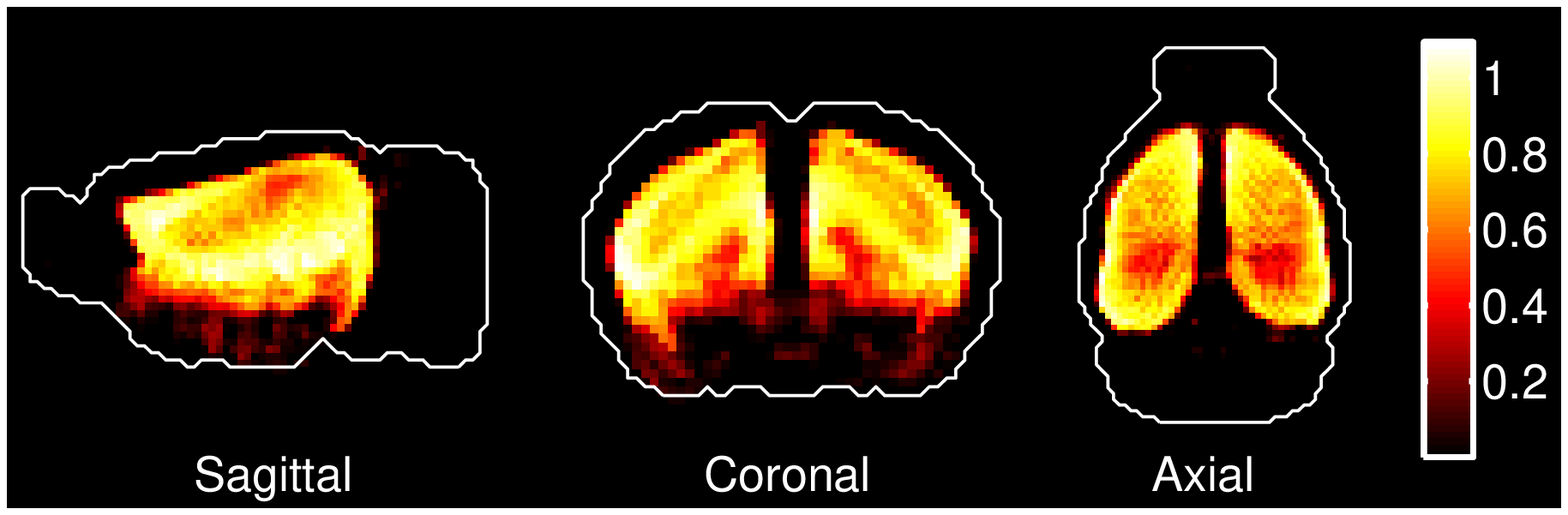}}\\
  \subfloat[A section of correlations, with atlas boundaries]{\label{fig:correlationsIntenseSection47}\includegraphics[width=\textwidth]{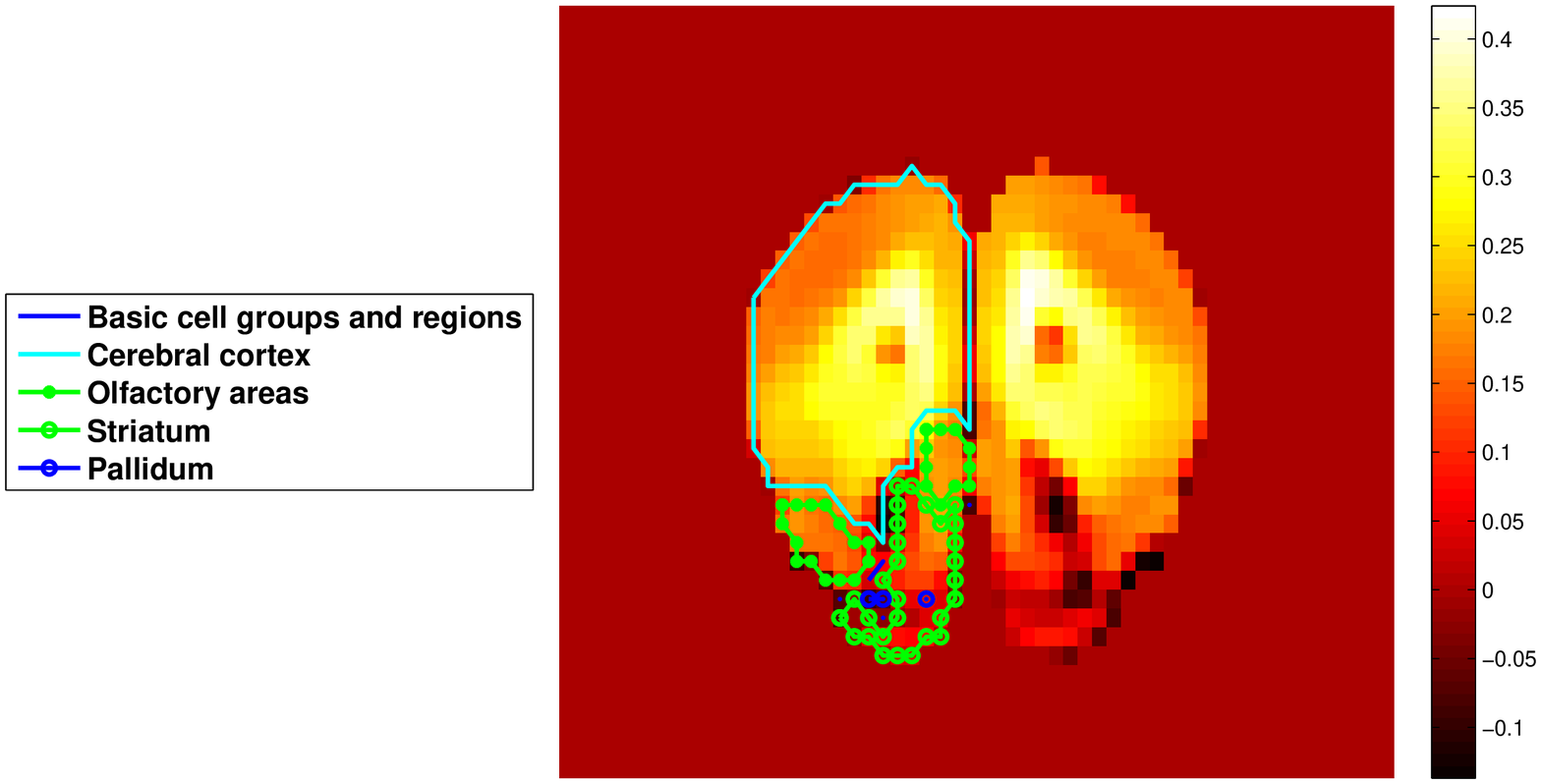}}\\
  \subfloat[A section of fitting coefficients, with atlas boundaries]
                      {\label{fig:fittingsIntenseSection47}\includegraphics[width=\textwidth]{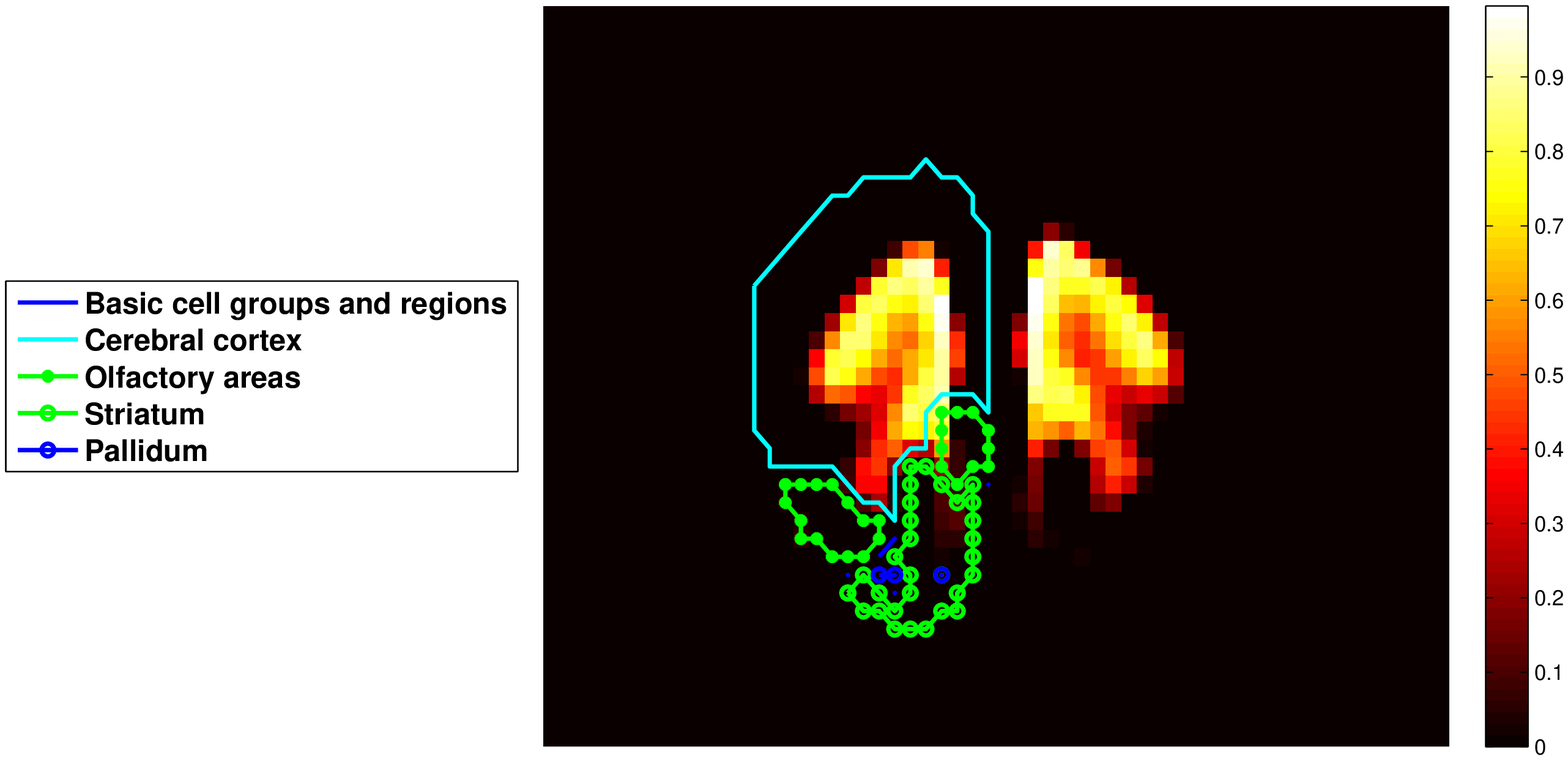}}\\
 \caption{Correlations and fitting coefficients between the AGEA and pyramidal neurons taken from primary somatosensory area, layer 5 \cite{foreBrainTaxonomy}.}
  \label{fig:cellType47}
\end{figure}

\begin{figure}
  \centering 
 \subfloat[Maximal-intensity projection of correlations.]{\label{fig:cellTypeProj48}\includegraphics[width=0.5\textwidth]{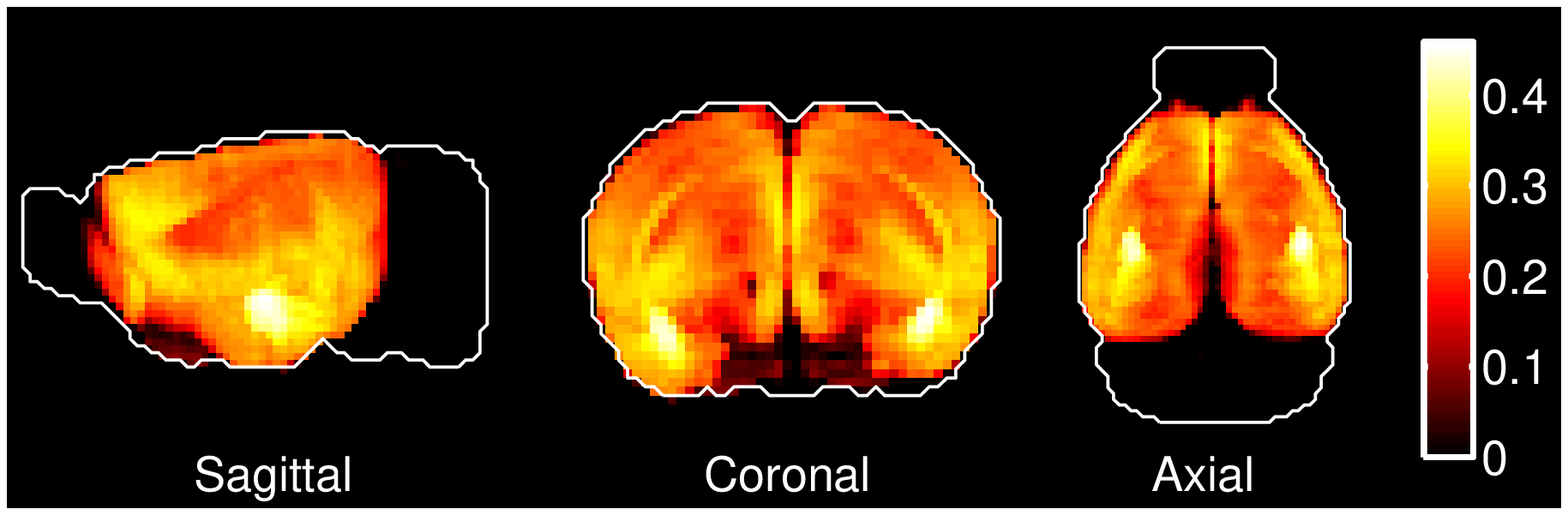}}
  \subfloat[Maximal-intensity projection of the linear fitting.]{\label{fig:cellTypeModelFit48}\includegraphics[width=0.5\textwidth]{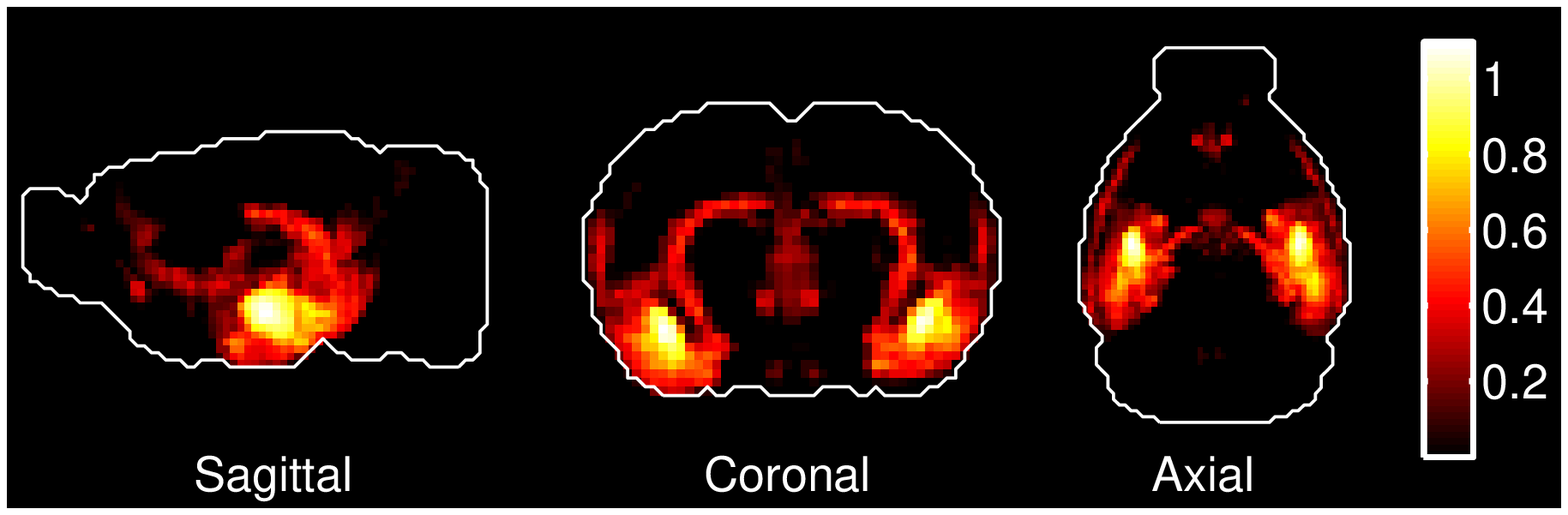}}\\
  \subfloat[A section of correlations, with atlas boundaries]{\label{fig:correlationsIntenseSection48}\includegraphics[width=\textwidth]{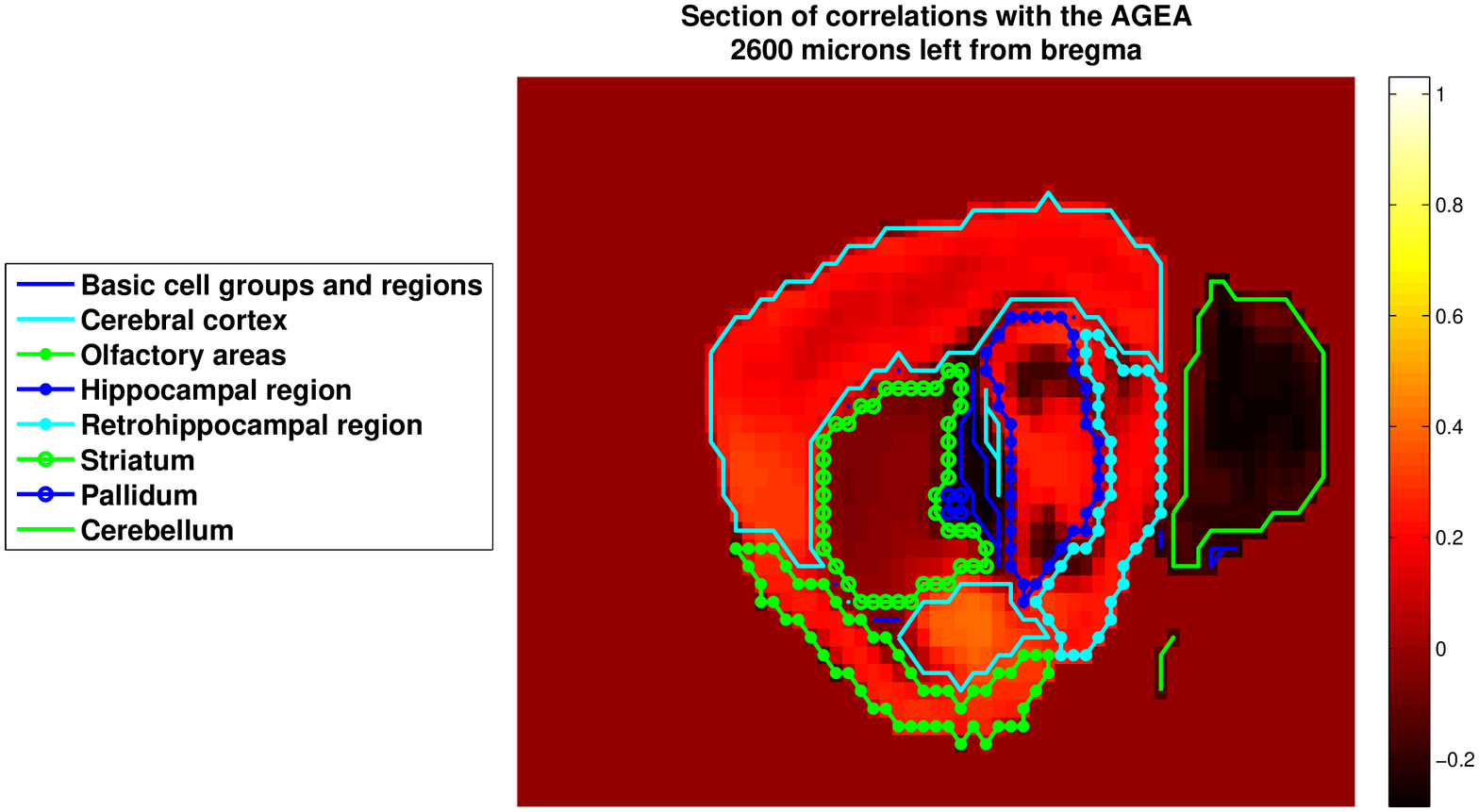}}\\
  \subfloat[A section of fitting coefficients, with atlas boundaries]
                      {\label{fig:fittingsIntenseSection48}\includegraphics[width=\textwidth]{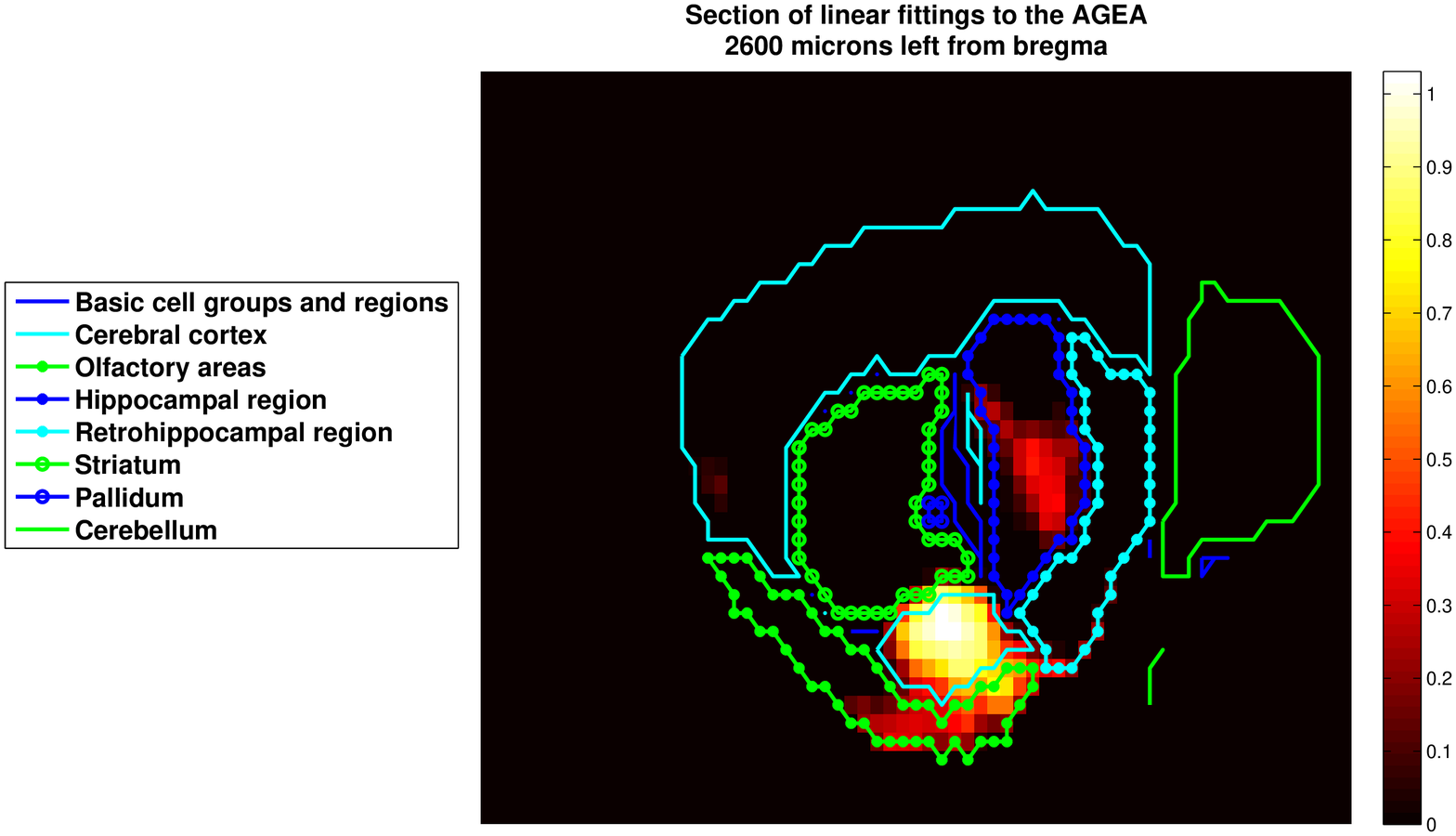}}\\
 \caption{Correlations and fitting coefficients between the AGEA and pyramidal neurons taken from the amygdala (which falls into the 
cerebral cortex, layer 6B in the Allen Reference Atlas) \cite{foreBrainTaxonomy}.}
  \label{fig:cellType48}
\end{figure}

\item{\bf{Cerebellar patterns.}} See Figure \ref{fig:cellType1} for a class of Purkinje cells \cite{foreBrainTaxonomy}  
and Figure \ref{fig:cellType21} for a class of mature oligodendrocytes \cite{DoyleCells}, both extracted from the cerebellum. Their correlation and
fitting coefficients are indeed mostly localized in cerebellum.

\begin{figure}
  \centering 
 \subfloat[Maximal-intensity projection of correlations.]{\label{fig:cellTypeProj1}\includegraphics[width=0.5\textwidth]{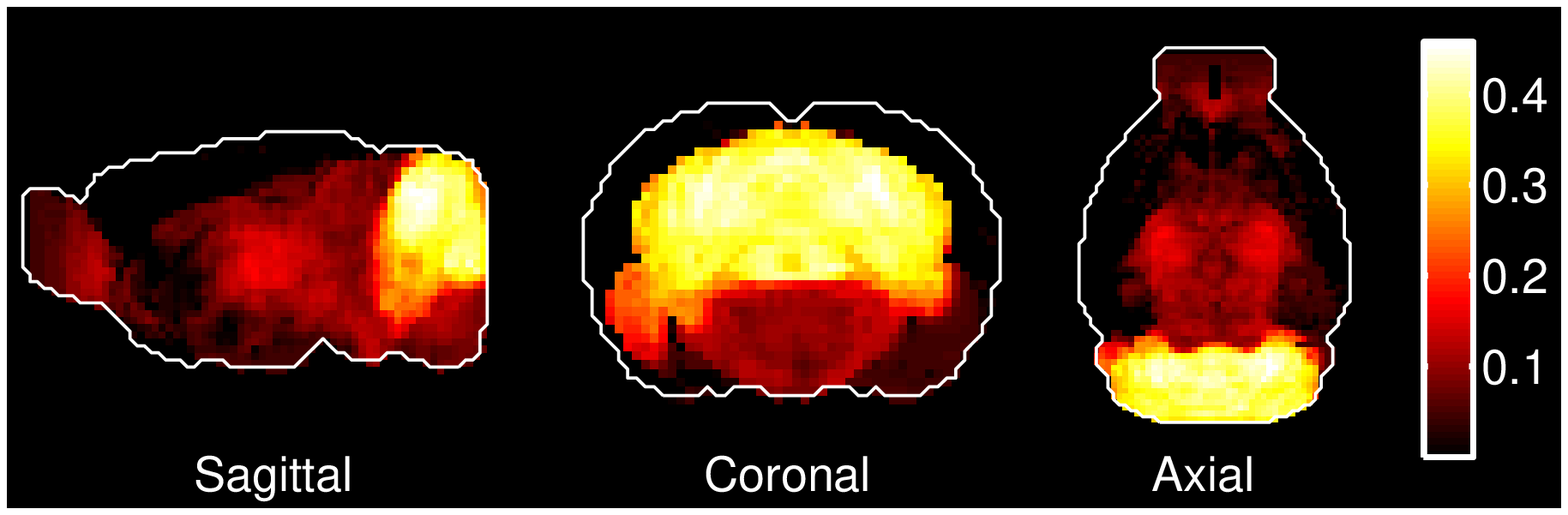}}
  \subfloat[Maximal-intensity projection of the linear fitting.]{\label{fig:cellTypeModelFit1}\includegraphics[width=0.5\textwidth]{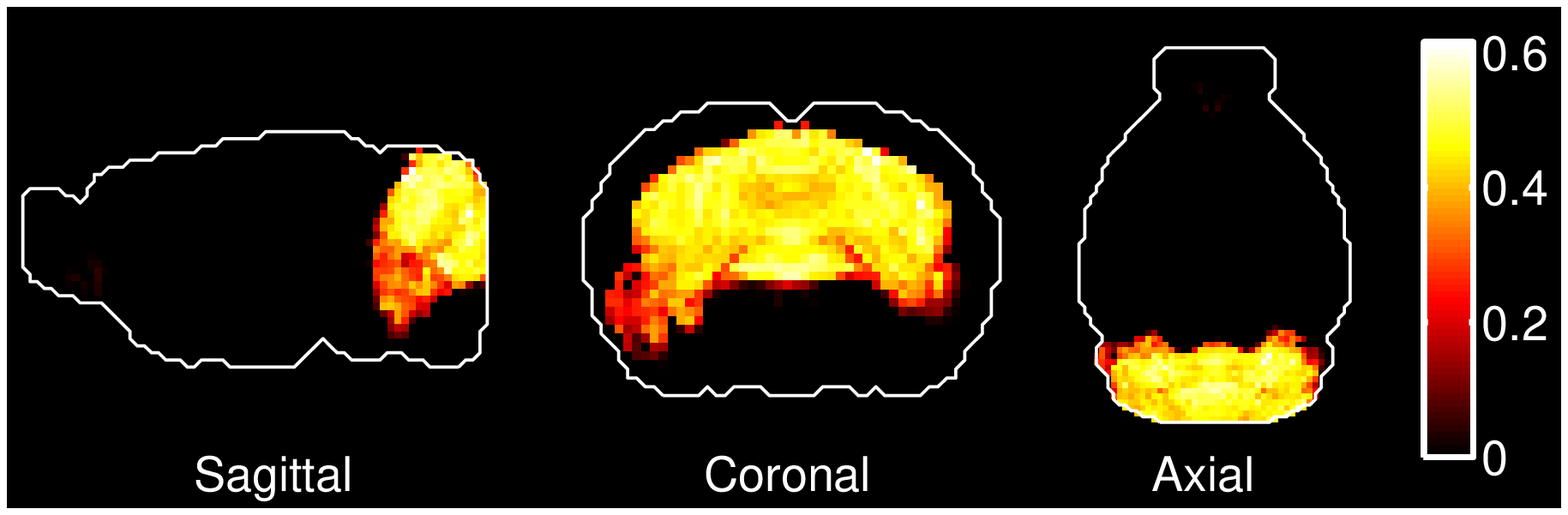}}\\
  \subfloat[A section of correlations, with atlas boundaries]{\label{fig:correlationsIntenseSection1}\includegraphics[width=\textwidth]{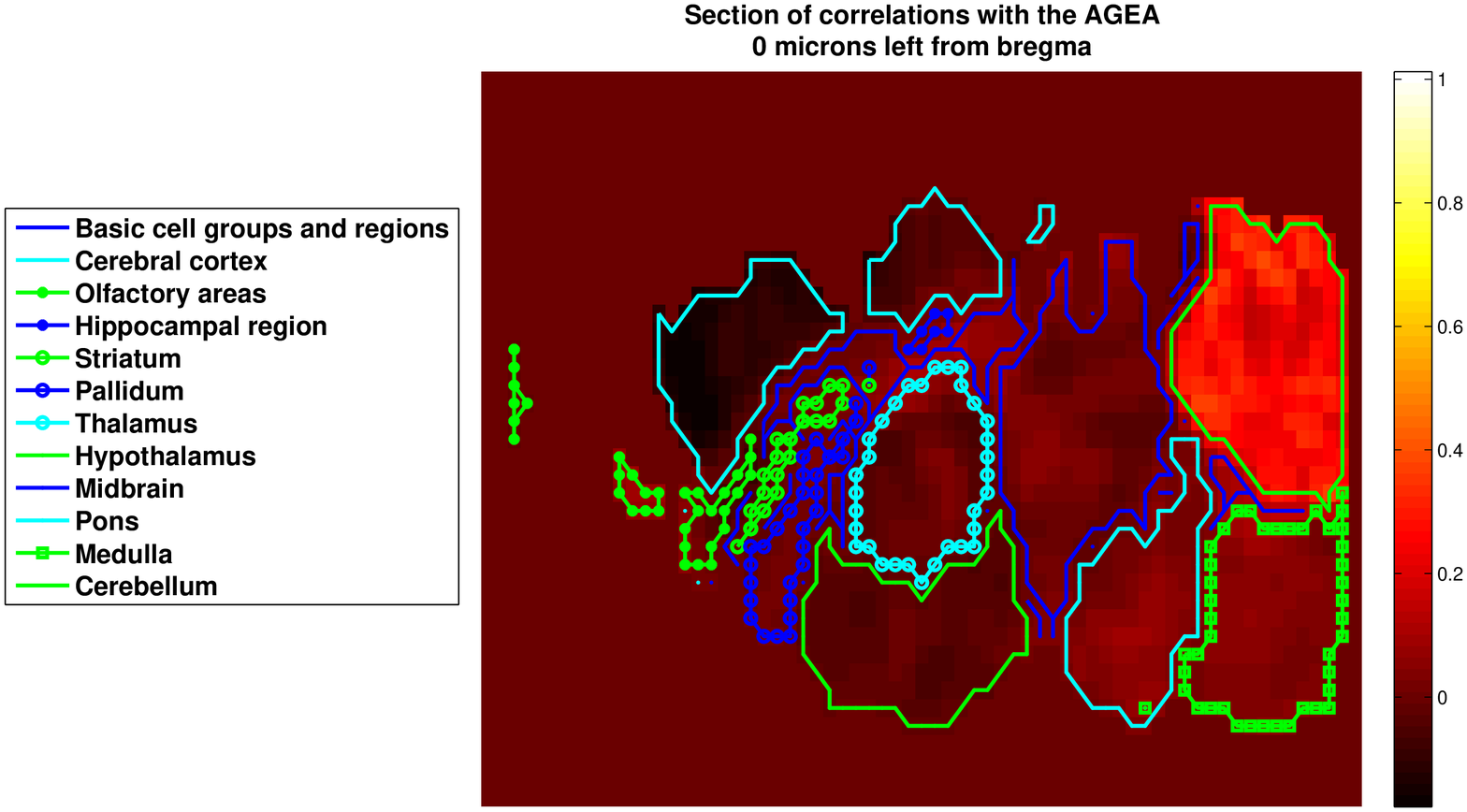}}\\
  \subfloat[A section of fitting coefficients, with atlas boundaries]
                      {\label{fig:fittingsIntenseSection1}\includegraphics[width=\textwidth]{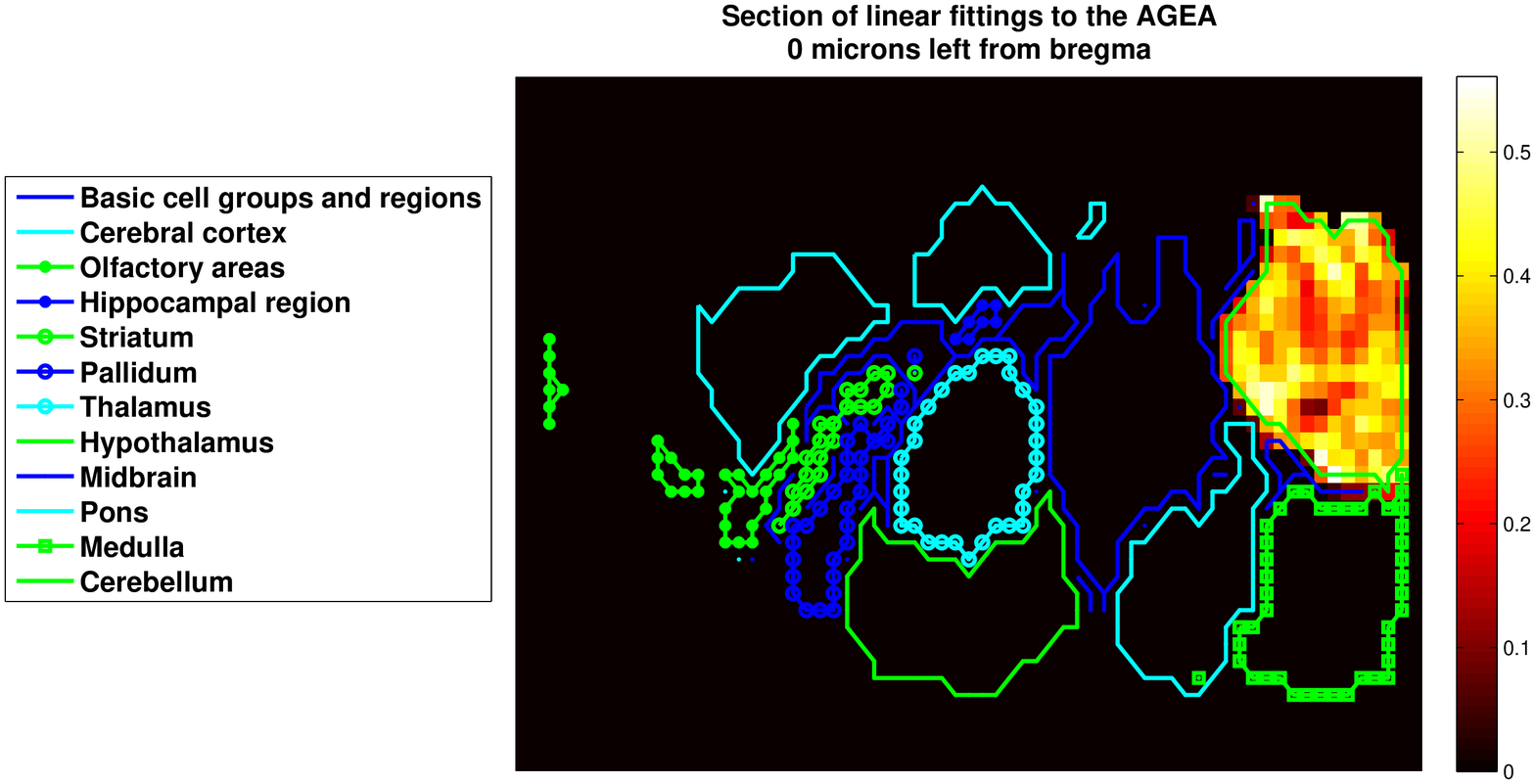}}\\
 \caption{Correlations and fitting coefficients between the AGEA and Purkinje cells \cite{RossnerCells} (index 1 in the table of cell-type results).}
  \label{fig:cellType1}
\end{figure}

\begin{figure}
  \centering 
 \subfloat[Maximal-intensity projection of correlations.]{\label{fig:cellTypeProj21}\includegraphics[width=0.5\textwidth]{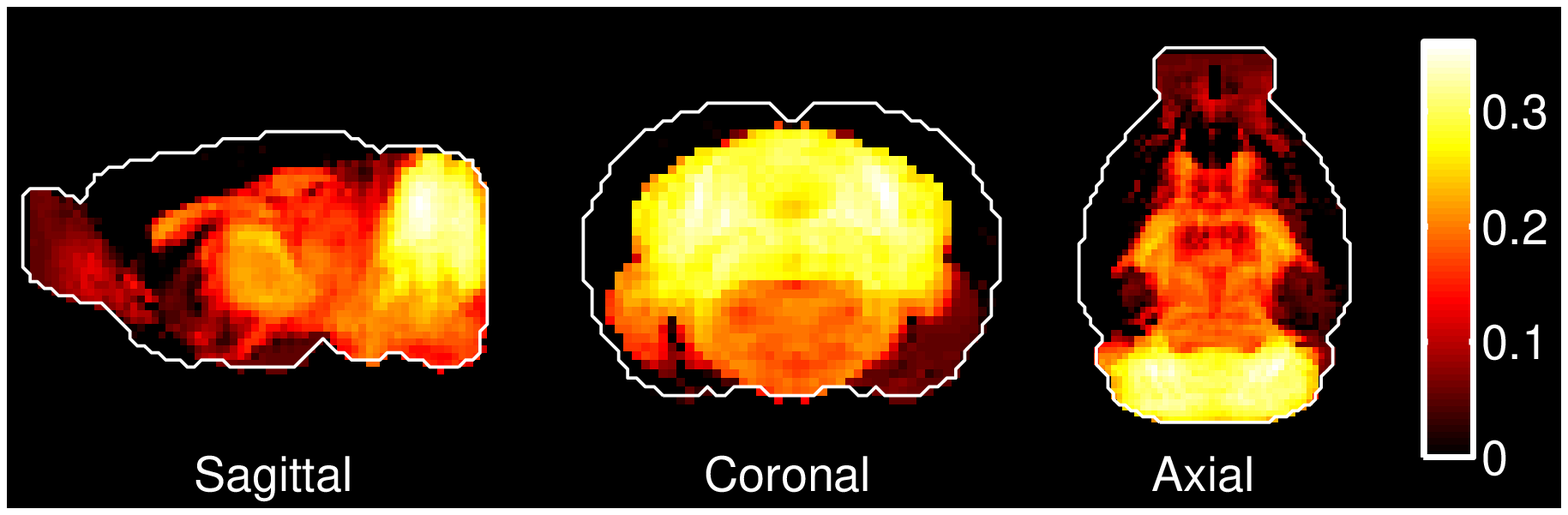}}
  \subfloat[Maximal-intensity projection of the linear fitting.]{\label{fig:cellTypeModelFit21}\includegraphics[width=0.5\textwidth]{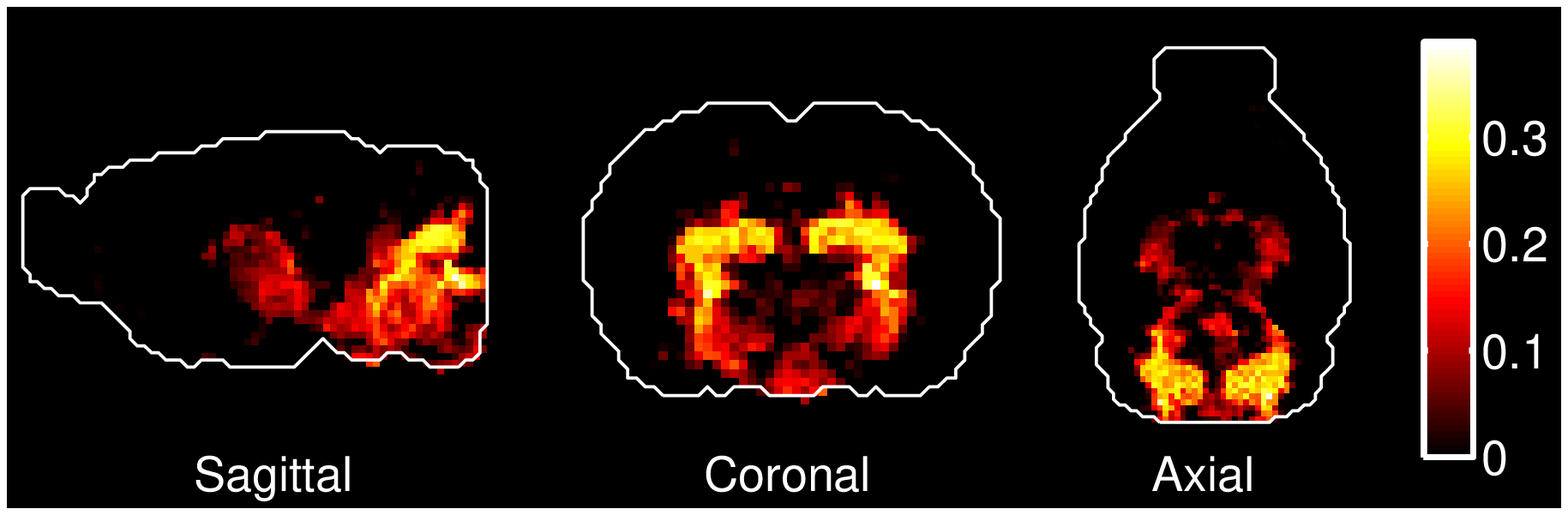}}\\
  \subfloat[A section of correlations, with atlas boundaries]{\label{fig:correlationsIntenseSection21}\includegraphics[width=\textwidth]{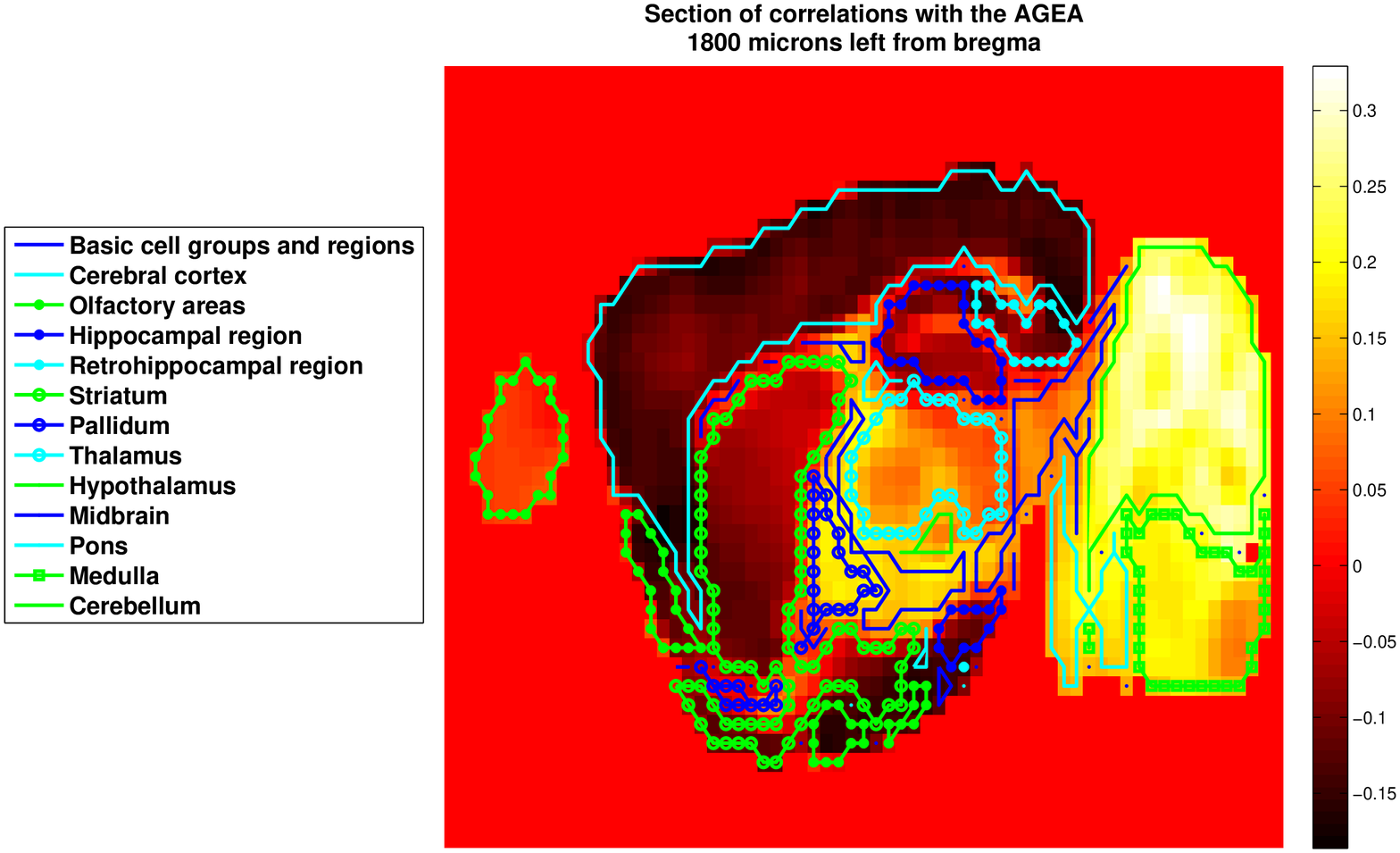}}\\
  \subfloat[A section of fitting coefficients, with atlas boundaries]
                      {\label{fig:fittingsIntenseSection1}\includegraphics[width=\textwidth]{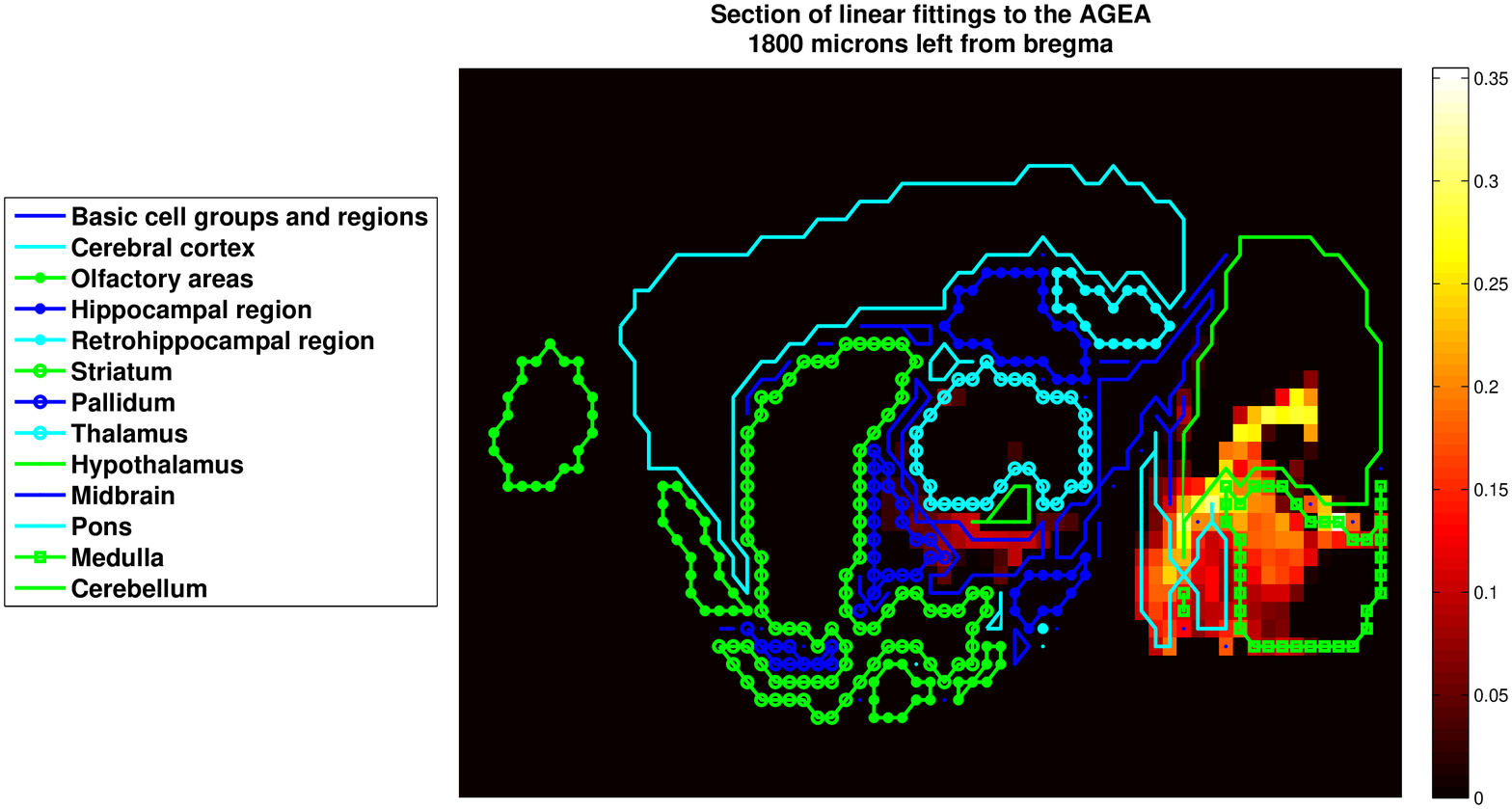}}\\
 \caption{Correlations and fitting coefficients between the AGEA and mature oligodendrocytes
extracted from the cerebellum \cite{DoyleCells} (index 21 in the table of cell-type results).}
  \label{fig:cellType21}
\end{figure}

\item{\bf{Hippocampal pattern.}} See  Figure \ref{fig:cellType49} for a class of pyramidal neurons \cite{foreBrainTaxonomy} extracted from the
hippocampus. Its correlation pattern is highest in the hippocampal region and
fitting coefficients are mostly localized in the hippocampal region.

\begin{figure}
  \centering 
 \subfloat[Maximal-intensity projection of correlations.]{\label{fig:cellTypeProj49}\includegraphics[width=0.5\textwidth]{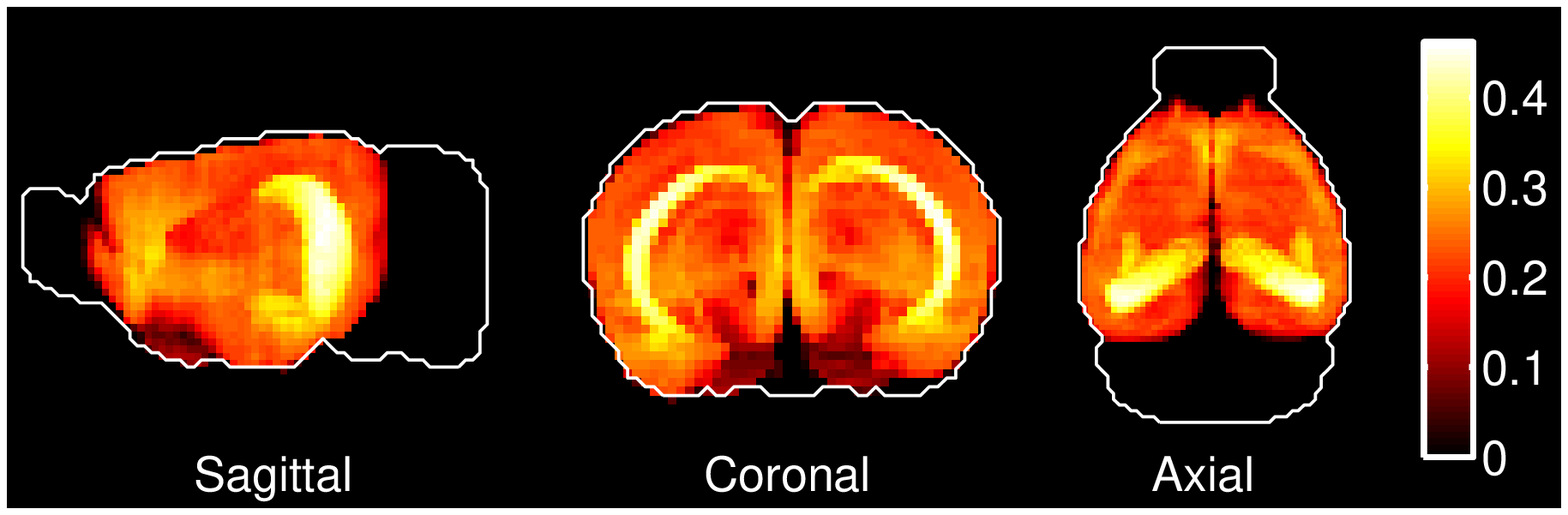}}
  \subfloat[Maximal-intensity projection of the linear fitting.]{\label{fig:cellTypeModelFit49}\includegraphics[width=0.5\textwidth]{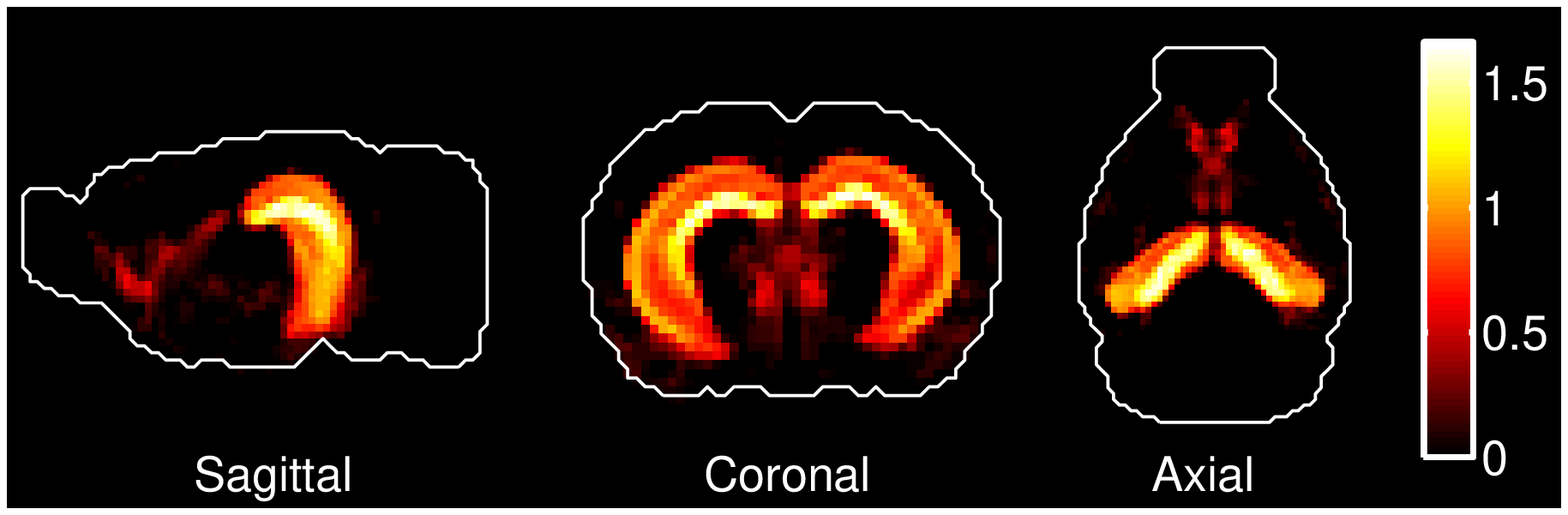}}\\
  \subfloat[A section of correlations, with atlas boundaries]{\label{fig:correlationsIntenseSection49}\includegraphics[width=\textwidth]{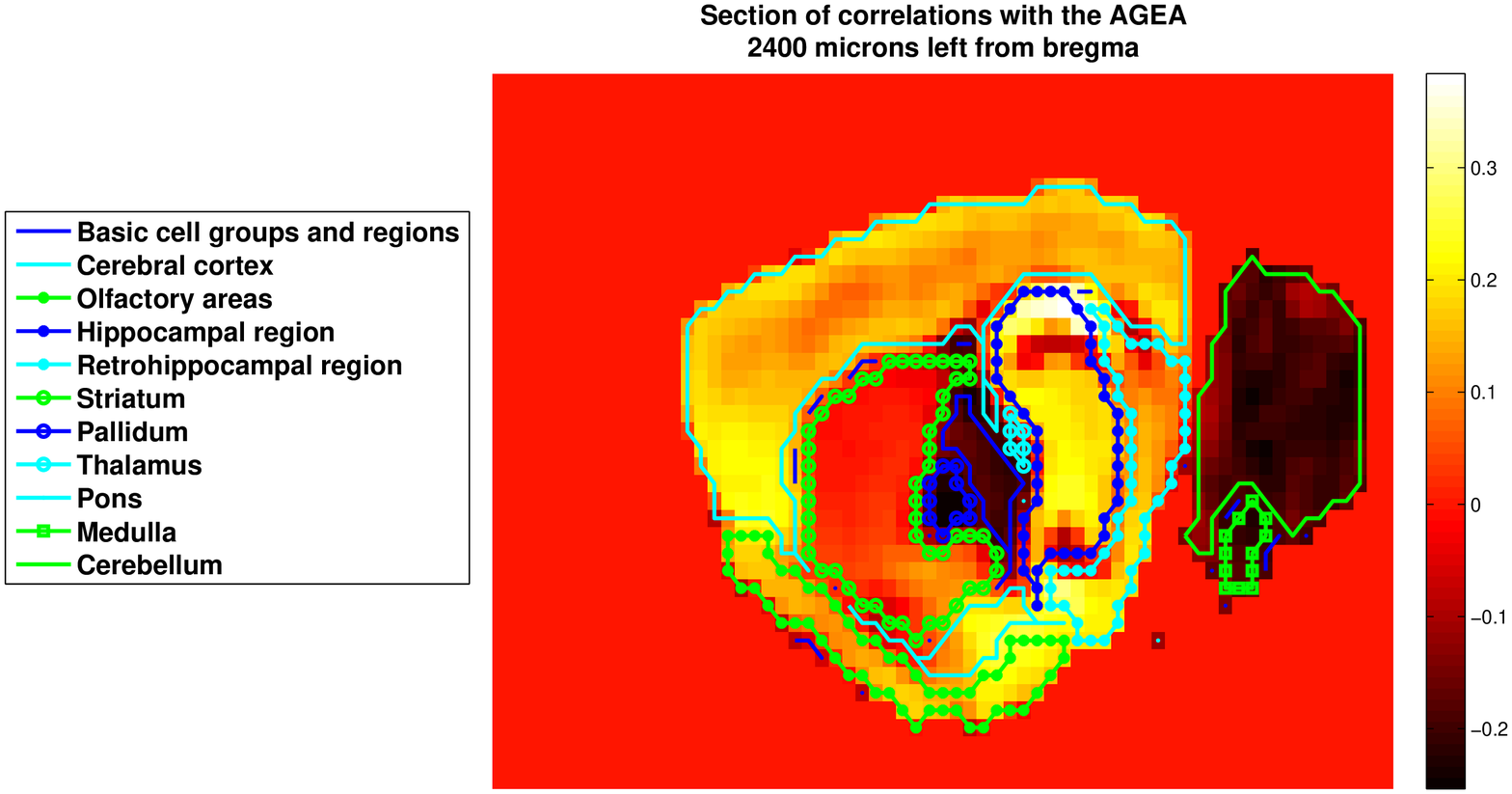}}\\
  \subfloat[A section of fitting coefficients, with atlas boundaries]
                      {\label{fig:fittingsIntenseSection49}\includegraphics[width=\textwidth]{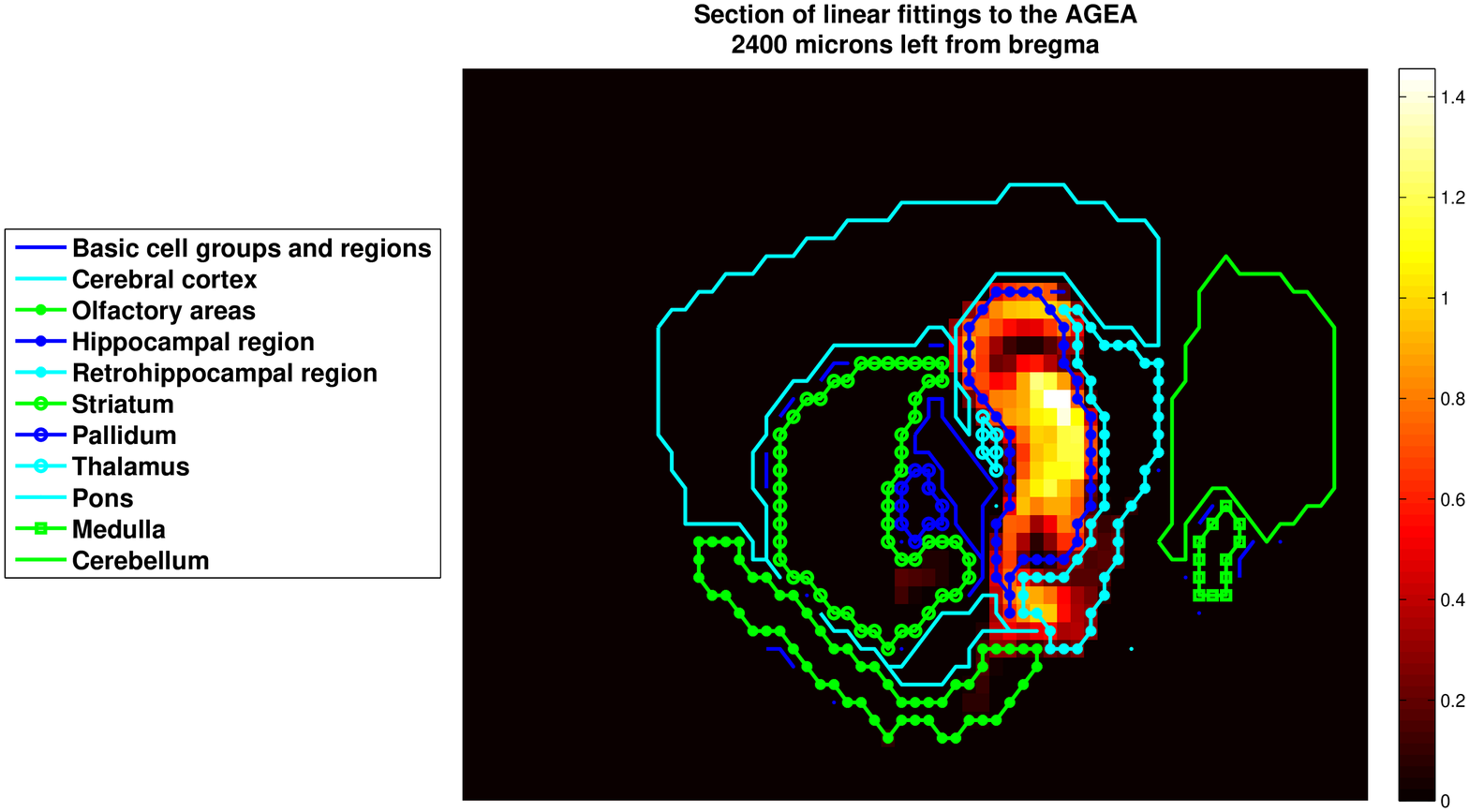}}\\
 \caption{Correlations and fitting coefficients between the AGEA and pyramidal neurons taken from the hippocampus (Ammon's horn) \cite{foreBrainTaxonomy} (index 49 in the table of cell-type results).}
  \label{fig:cellType49}
\end{figure}

\item{\bf{Striatal pattern.}} See Figure \ref{fig:cellType16} for a class of medium spiny neurons \cite{DoyleCells} extracted from the striatum.
Its correlation and
fitting coefficients are indeed mostly localized in the striatum.

\begin{figure}
  \centering 
 \subfloat[Maximal-intensity projection of correlations.]{\label{fig:cellTypeProj16}\includegraphics[width=0.5\textwidth]{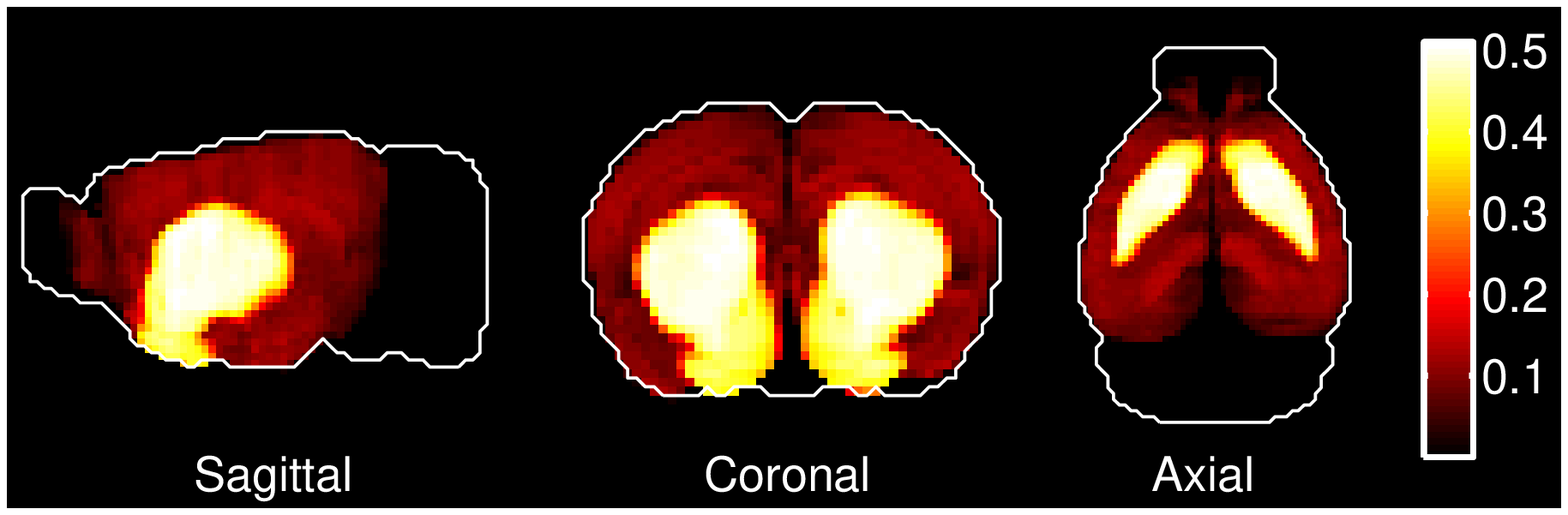}}
  \subfloat[Maximal-intensity projection of the linear fitting.]{\label{fig:cellTypeModelFit16}\includegraphics[width=0.5\textwidth]{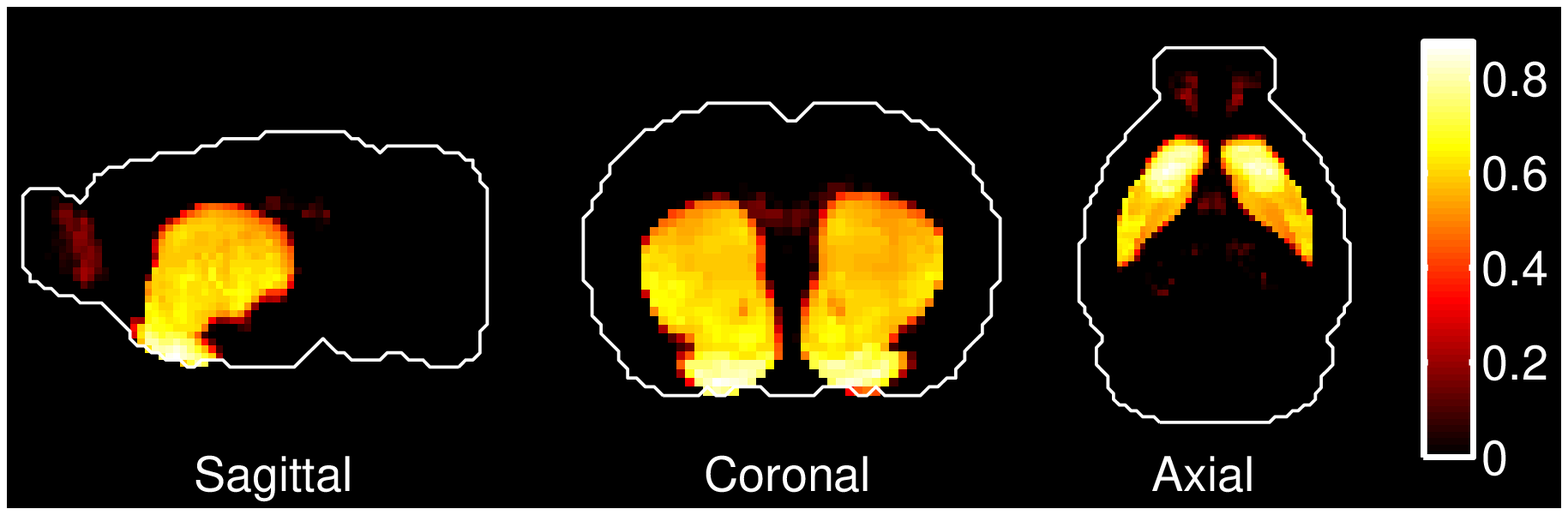}}\\
  \subfloat[A section of correlations, with atlas boundaries]{\label{fig:correlationsIntenseSection16}\includegraphics[width=\textwidth]{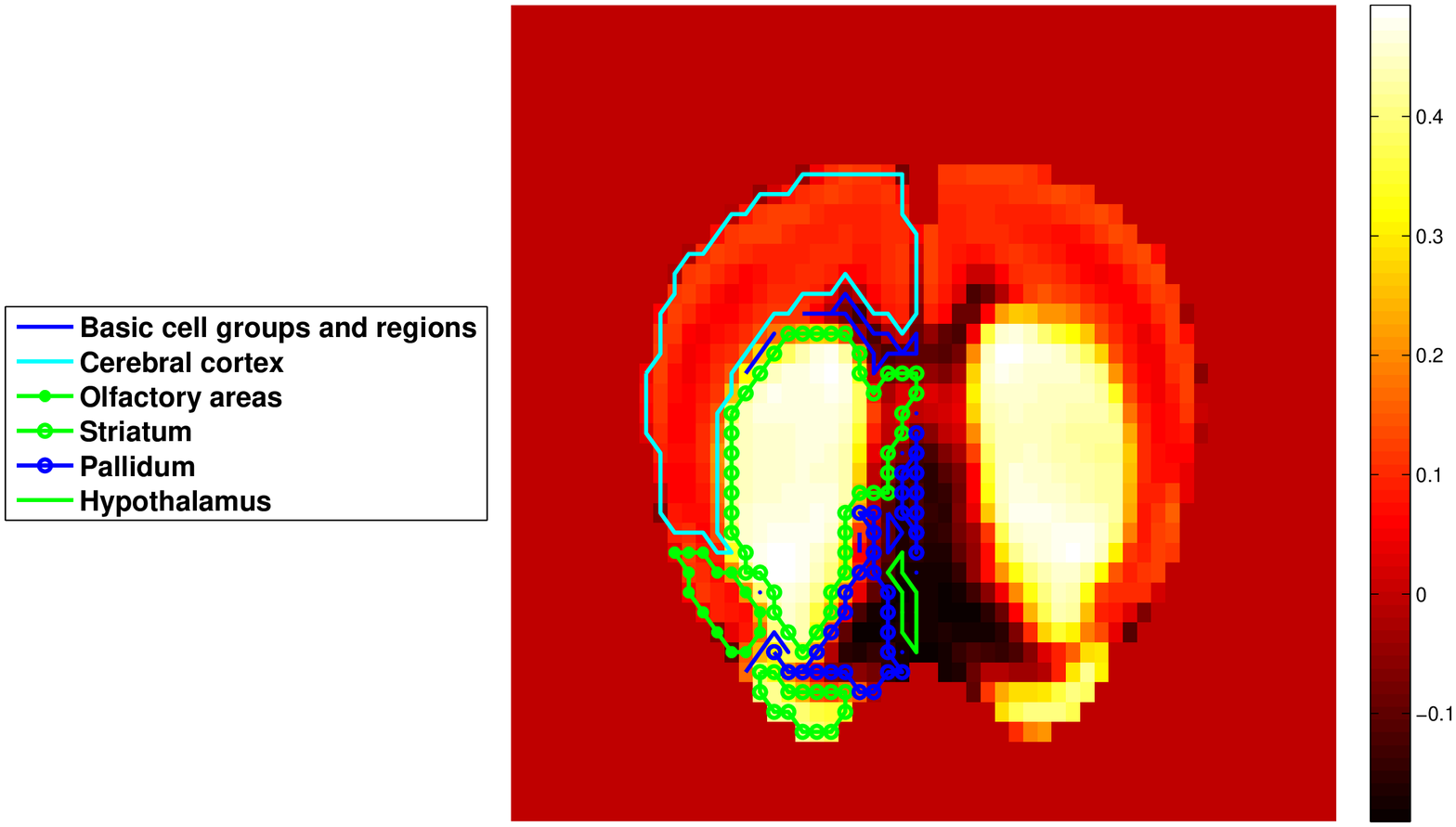}}\\
  \subfloat[A section of fitting coefficients, with atlas boundaries]
                      {\label{fig:fittingsIntenseSection16}\includegraphics[width=\textwidth]{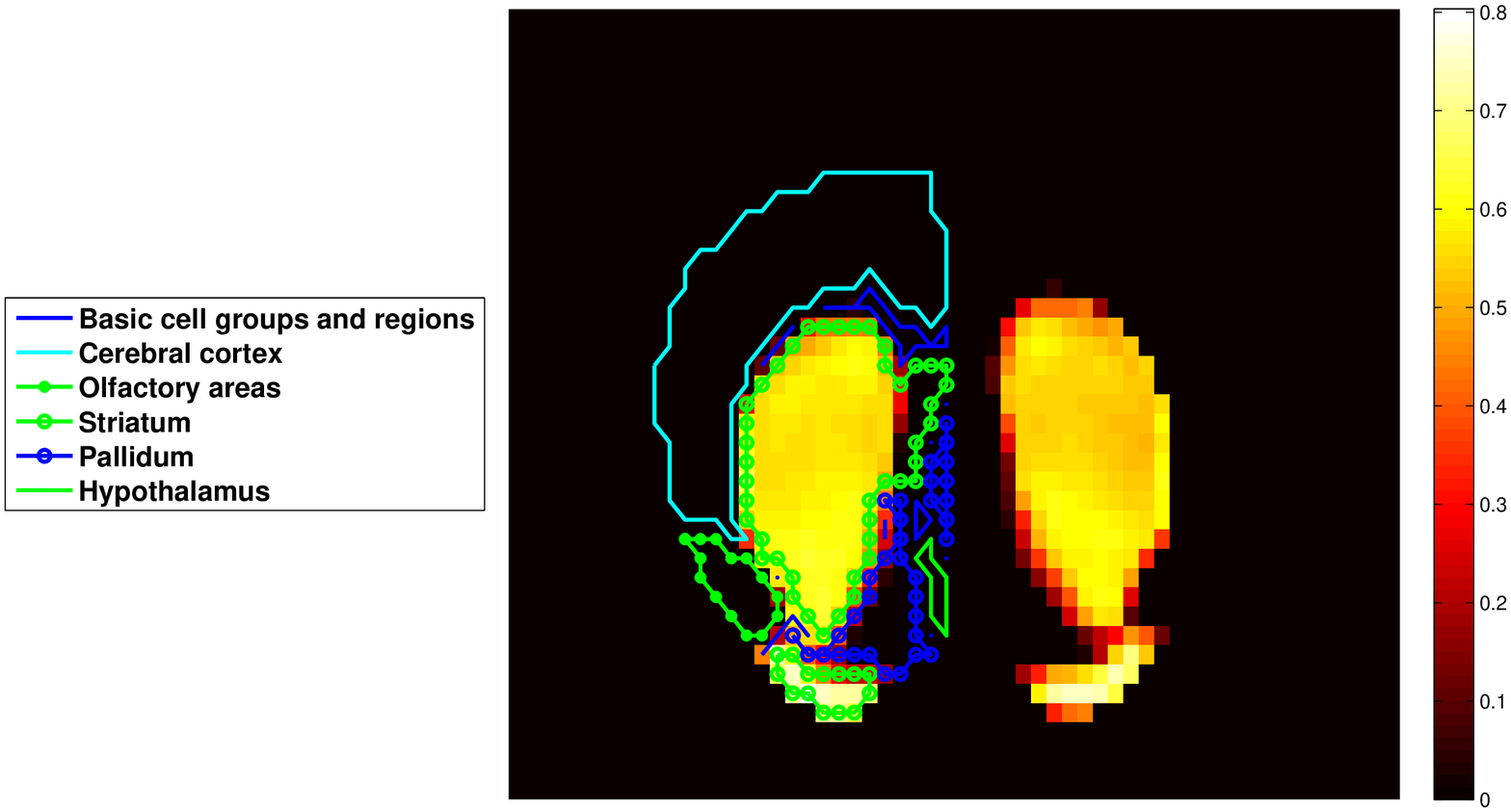}}\\
 \caption{Correlations and fitting coefficients between the AGEA and {\emph{Drd2}} medium spiny neurons taken from the striatum \cite{DoyleCells}, index 16 in the table of cell types.}
  \label{fig:cellType16}
\end{figure}

\item{\bf{White-matter pattern.}} See Figure \ref{fig:cellType31}  for a class of astrocytes \cite{CahoyCells} extracted from the cortex.
 Its correlation and fitting coefficients look exhibit a singular pattern that looks like white matter, with the most caudal component 
  corresponding to  the {\emph{arbor vitae}}.

\begin{figure}
  \centering 
 \subfloat[Maximal-intensity projection of correlations.]{\label{fig:cellTypeProj31}\includegraphics[width=0.5\textwidth]{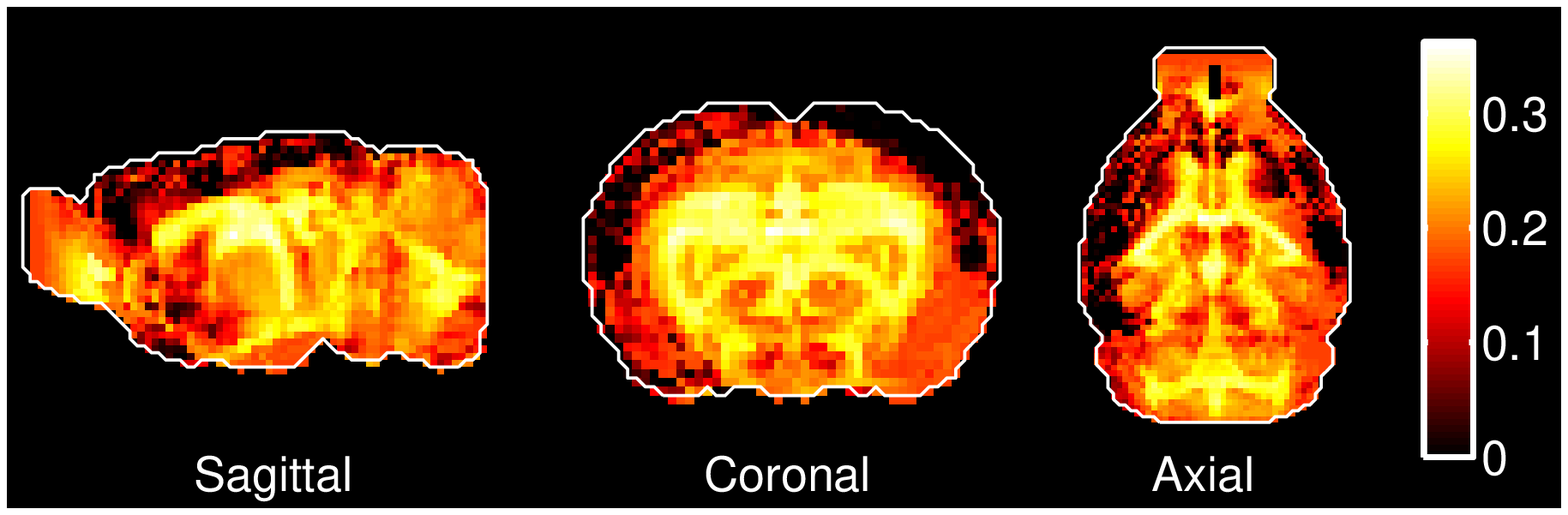}}
  \subfloat[Maximal-intensity projection of the linear fitting.]{\label{fig:cellTypeModelFit31}\includegraphics[width=0.5\textwidth]{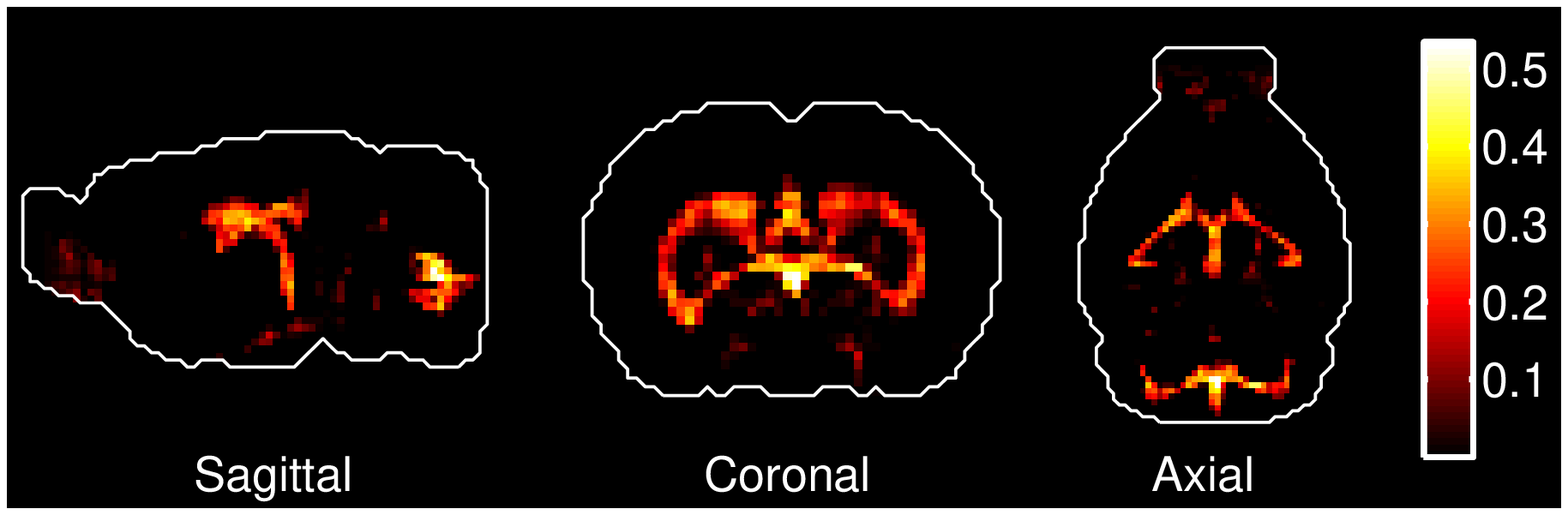}}\\
  \subfloat[A section of correlations, with atlas boundaries]{\label{fig:correlationsIntenseSection31}\includegraphics[width=\textwidth]{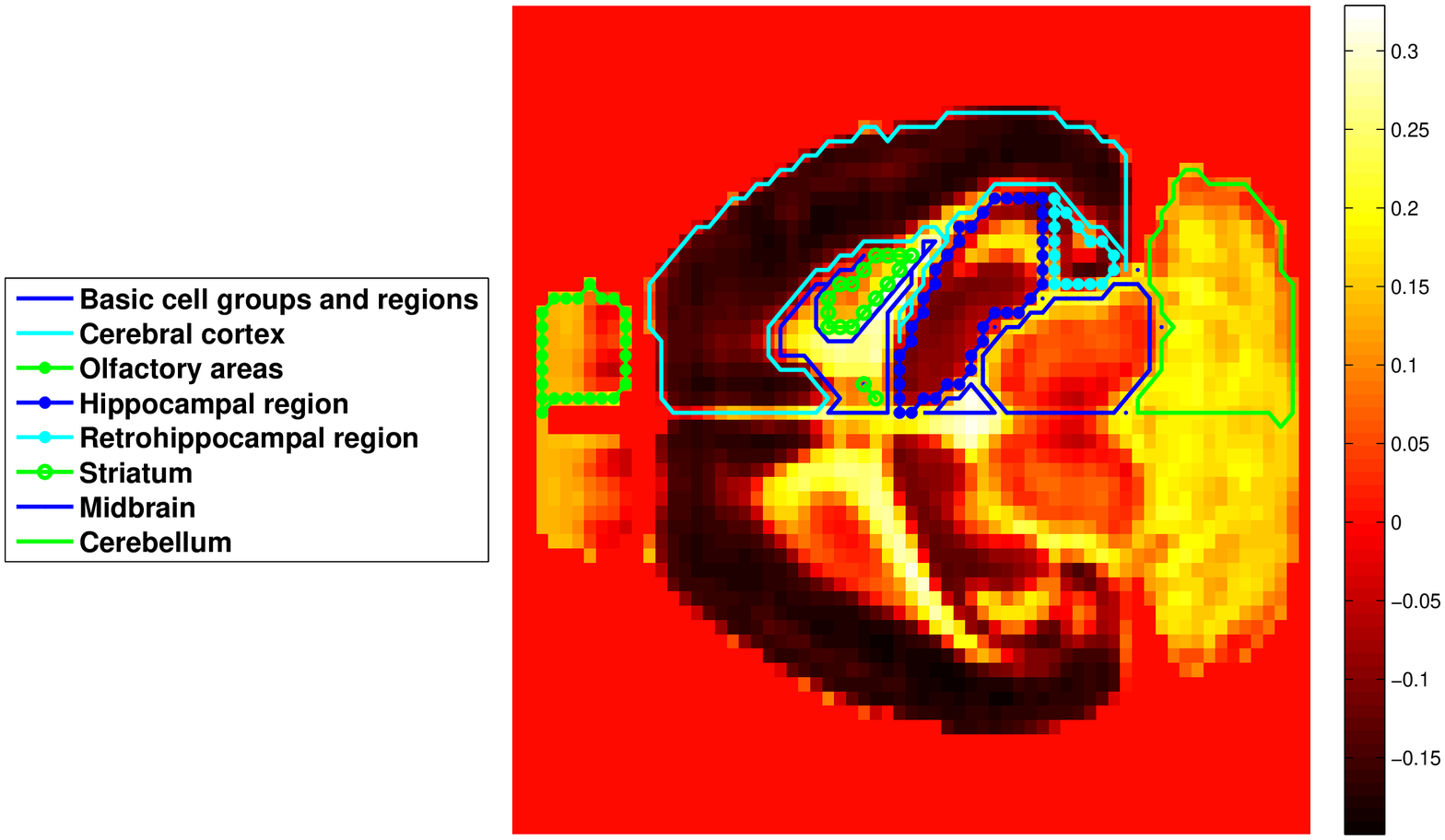}}\\
  \subfloat[A section of fitting coefficients, with atlas boundaries]
                      {\label{fig:fittingsIntenseSection31}\includegraphics[width=\textwidth]{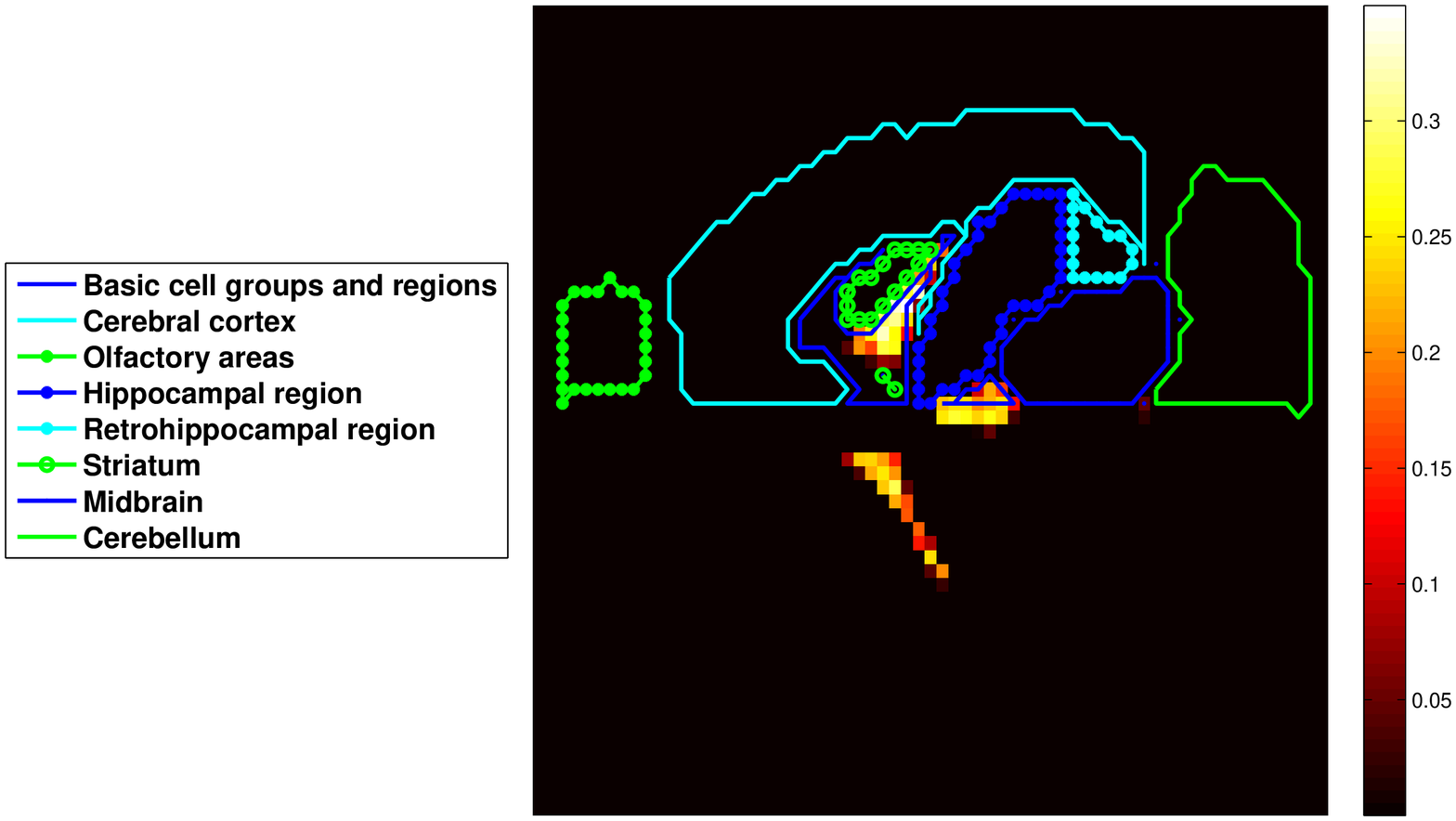}}\\
 \caption{Correlations and fitting coefficients between the AGEA and astrocytes taken from the cerebral cortex (Ammon's horn) \cite{CahoyCells} (index 31 in the table of cell-types).}
  \label{fig:cellType31}
\end{figure}

\end{itemize}

These examples of agreement between the support of the fitting 
coefficients and the metadata establish the plausibility of the 
linear model we proposed in Equation \ref{linearModel}. Some of the fitting 
profiles indeed have a considerably cleaner profile than the correlation,
and are able to reveal fine neuroanatomical details The microarray dataset was rich enough to distinguish
each of the regions described in the patterns above from the rest of the brain. 
It is interesting to note that four of the non-cortical patterns described above emerge from bi-clustering
of the data at seven biclusters (see section 3 on clustering analysis), if the list of genes is restricted to
 the most localized genes in the 
sense of the Kullback--Leibler distance. So the diversity of cell-types found in the
microarray dataset reveals partitions of the brains by sets of genes that can be 
obtained from (a special subset of ) the AGEA without using extra data. The major patterns listed above
are therefore not a mere accident of the microarray dataset, but reflect a reasonable 
first list of data-driven neuroanatomical structures.\\

\subsubsection{Unexpected fitting patterns}
 The results of the model contain some surprises that can be linked to the relative 
paucity of cell types in the study, compared to the whole diversity of cells in the mouse brain, and 
to the fact that the cell samples are not distributed uniformly across the brain.  
Two of the anatomical patterns (thalamus and olfactory areas) that emerge from biclustering of localized genes (see
 section 3 on clustering analysis) are not 
as convincingly illustrated by the results of our analysis. A class of Gabaergic Interneurons (index 55, \cite{foreBrainTaxonomy})
has fitting coefficients well-localized in the olfactory areas, but the cells were extracted from the somatosensory cortex (primary
somatosensory area in the Allen reference Atlas). As the olfactory areas are not represented in our microarray dataset,
this gives an indication of the cell types in the dataset that are closest in terms of gene expression to cell-types represented in the 
 olfactory areas. As for the thalamus, it is represented by only one cell type in 
our microarray dataset (GABAergic interneurons, PV+, index 60, \cite{foreBrainTaxonomy}). The cells were extracted 
from the lateral geniculate complex, and the fitting coefficients are positive in the olfactory areas and in midbrain
as well as in the thalamus. Again clustering suggests that some cell types are specific to thalamus,
and a richer microarray dataset with better sampling of the thalamus will move some of the fitting coefficients 
to these cell types.\\

A few cell-types have surprising fitting coefficients, some of which may be traced
to developmental properties.
\begin{itemize}

\item{\bf{A class of Purkinje cells does not fit to the cerebellum.}} See 
Figure \ref{fig:PurkinkjeThalamusPattern} for a class of Purkinje cells (index 52, unpublished) that correlate 
best with the AGEA both in the thalamus and in the cerebellum, but 
that fits only in the thalamus. This indicates that thalamus must contain cell types 
whose gene expression profile is closest to Purkinje cells in the present dataset, but that
thalamus is not sampled in enough detail by our microarray dataset for these cell types
to be distinguished from this class of Purkinje cells.

\begin{figure}
  \centering 
 \subfloat[Maximal-intensity projection of correlations.]{\label{fig:cellTypeProj52}\includegraphics[width=0.5\textwidth]{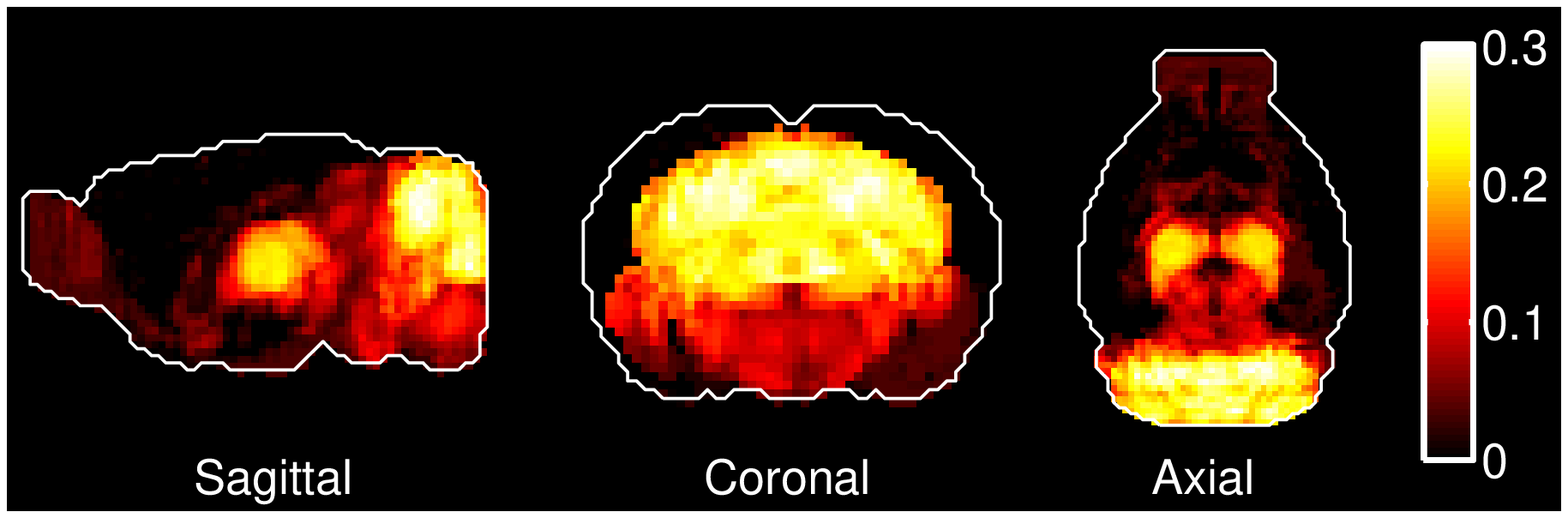}}
  \subfloat[Maximal-intensity projection of the linear fitting.]{\label{fig:cellTypeModelFit52}\includegraphics[width=0.5\textwidth]{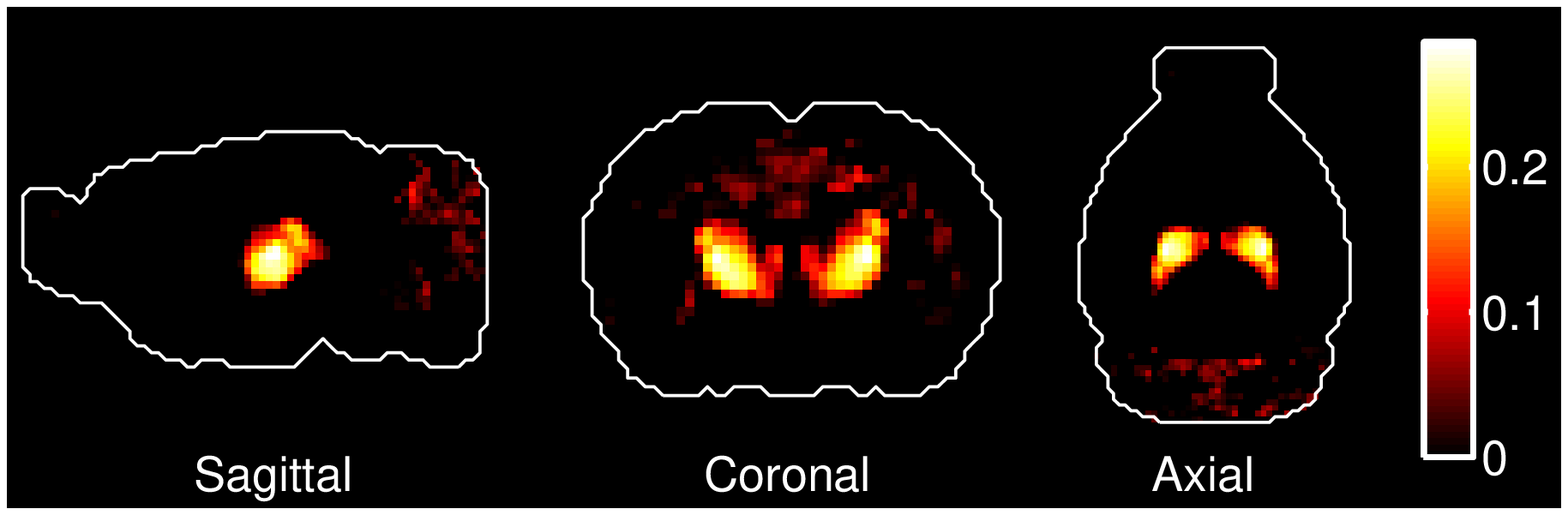}}\\
  \subfloat[A section of correlations, with atlas boundaries]{\label{fig:correlationsIntenseSection52}\includegraphics[width=\textwidth]{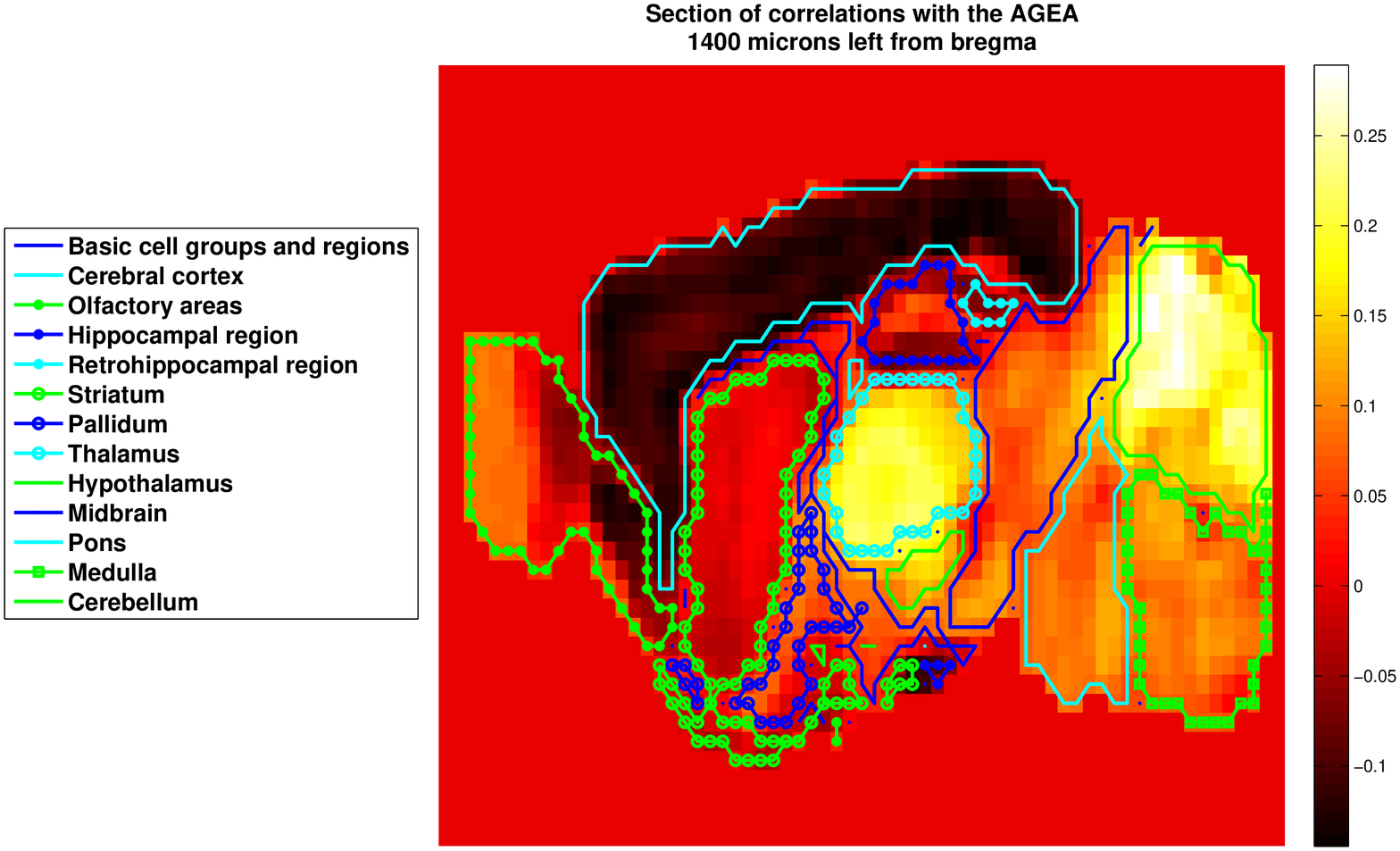}}\\
  \subfloat[A section of fitting coefficients, with atlas boundaries]
                      {\label{fig:fittingsIntenseSection52}\includegraphics[width=\textwidth]{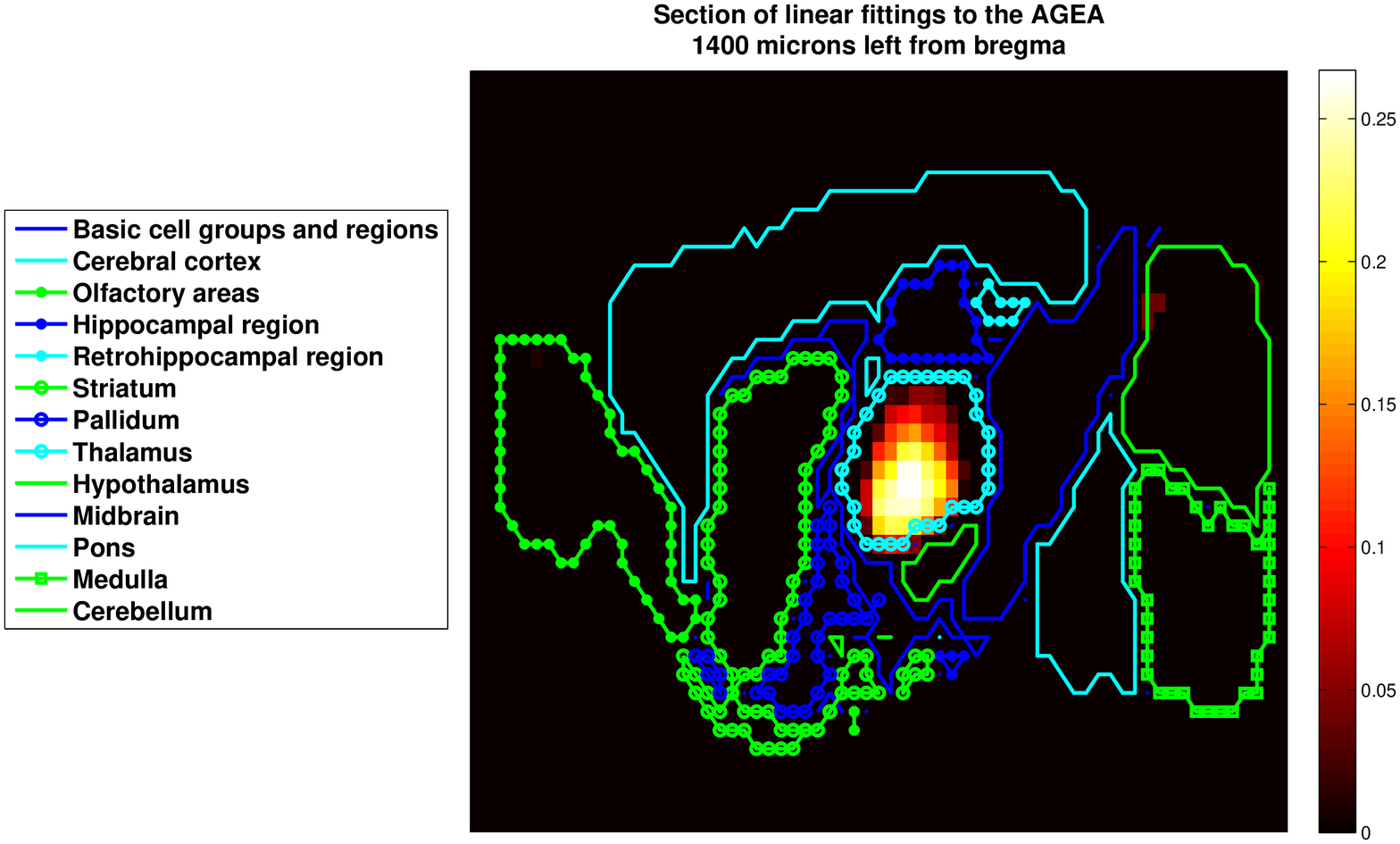}}\\
 \caption{Correlations and fitting coefficients between the AGEA and a class of Purkinje cells extracted from the cerebellum (index 52 in the table of cell-type results). The correlation profile is high both in the cerebellum and in the thalamus, but the fitting coefficients are localized in thalamus}
  \label{fig:PurkinkjeThalamusPattern}
\end{figure}

\item{\bf{Some pyramidal neurons, including corticospinal, fit poorly
    to the AGEA.}} Some pyramidal neurons, extracted from the cortex,
  have very sparse fitting coefficients, even though their correlation
  profiles with the AGEA have a maximal-intensity projection that
  looks like the cerebral cortex (see indices 38, 39, 41, 42, 43).  It
  could be that the different non-linearities between numbers of
  mRNAs, ISH data and microarray data, prevent these cell types to be
  correctly detected by our linear model. However, one can notice that
  the ages of the mice are P3, P6, P14 for these classes of pyramidal
  neurons. The poor fitting to the AGEA could therefore come from a
  developmental effect, with corticospinal pyramidal neurons and
  callosally-projecting pyramidal neurons maturing late, while the 
  AGEA corresponds to the adult mouse brain.

\end{itemize}

\section{Clustering analysis}

\subsection{Biclustering of genes and voxels}

Some of the cell types for which we have microarray data are predicted 
by the linear model \ref{linearModel} to be very localized in the brain. However, the validity model 
is limited 
by the number of cell types in the study. The $T$ different cell types only span a $T$-dimensional 
subspace of gene space, while the AGEA contains data for thousands of 
genes. However, in the absence of microarray data for more cell types,
we can still ask if the Allen Gene Expression Atlas has some sets of genes whose expression is strongly localized in
some sets of voxels. 
The sets of voxels would be the brain regions in which some cell types are expected to be localized,
and the sets of genes would be the genes whose expression is highest in those cell types.\\

 Mathematically, this problem is a biclustering problem: we need to partition 
the set of voxels {\emph{and}} the set of genes in the AGEA in an optimal way given the gene expression energies.
Gene expression energies can be used to turn voxels and genes into a bipartite graph, and this graph can be  partitioned 
  using a biclustering algorithm.\\

 \subsubsection{From the AGEA to a bipartite graph}

The AGEA can be mapped to a \emph{weighted bipartite graph} in the following way:
\begin{itemize}
\item the first set of vertices consists of voxels, numbered from $1$ to $V$,
\item the second set of vertices consists of genes, numbered from $1$ to $G$,
\item each of the  edges connect one voxel to one gene, and has a weight given by the expression energy of the gene at the voxel. If the
expression energy $E(v,g)$  is zero, there is no edge between voxel $v$  and gene $g$.\\
\end{itemize}
We looked for partitions of this weighted bipartite graph into subgraphs such that the 
weights of the internal edges of the subgraphs are strong compared to the
 weights of the edges between the subgraphs. This is the isoperimetric problem 
addressed by the algorithm of \cite{biclustering} (the graph need not be bipartite 
to apply this algorithm).\\

Given a weighted graph, the algorithm cuts some of the links, thus partitioning the graph 
into a subset $S$ and its complementary $\bar{S}$,  
such that the sum of weights (in this case expression energies) in the set of cut edges is minimized relative to the 
total weight of internal edges in $S$. The sum of weights in the set of cut edges is analogous to a boundary term,
while the total weight of  internal edges is analogous to a volume term. In that sense the problem
is an isoperimetric optimization problem, and the optimal set $S$ minimizes the {\emph{isoperimetric ratio}} $\rho$ over
 all the possible subgraphs:

\begin{equation}
S = {\mathrm{argmin}}_{{\mathrm{Vol}}(s) \leq {\mathrm{Vol}}(\bar{s}) } \,\rho( s ),
\end{equation}

\begin{equation}
\rho( s ) := \frac{|\partial s|}{\mathrm{Vol}(s)},
\end{equation}

\begin{equation}
|\partial s| = \sum_{i\in s, j \in \bar{ s}} W_{ij},
\end{equation}

\begin{equation}
{\mathrm{Vol}}(s) = \sum_{ i\in s, j\in s} W_{ij},
\end{equation}
 where the quantity $W_{ij}$ is the weight of the link between vertex $i$ and vertex $j$. 
Once $S$ has been worked out, the algorithm can be applied to $S$ and its complementary $\bar{S}$. This 
recursive application goes on until the isoperimetric ratio reaches a stopping ratio, representing the 
highest allowed isoperimetric ratio. This value is a parameter of the 
algorithm. Rising it results in a higher number of clusters, as it rises the number of acceptable cuts. 

\subsubsection{Restriction of the Allen Atlas to the most localized genes}

We studied the localization properties of the genes in the Allen Gene Expression Atlas 
by comparing their expression energies to a uniform function over the brain. We computed the Kullback-Leibler 
divergence of each gene from a uniform distribution:
\begin{equation}
KL(g ) = -\sum_{v = 1}^{V}\frac{1}{V}{\mathrm{log}}\left( V E_{\mathrm{prob}}(v,g)\right),
\end{equation}
\begin{equation}
E_{\mathrm{prob}}(v,g) = \frac{E(v,g)}{\sum_{v = 1 }^V E(v,g).}
\end{equation}
where $V$ is the total number of voxels in the brain and $E_{\mathrm{prob}}$ is a normalized version of  $E$  such that the expression energy of each gene is 
a probability density over the volume of the brain.\\

We took the highest 5 percentiles of genes by KL-divergence,
 and
thresholded the expression energies, thus reducing the
number of voxels.
 These are the most localized genes by that measure,
so solving the ispoerimetric problem on the graph constructed with
those genes only will be more likely to produce the localized
biclusters we are looking for. Thresholding the expression value gets
rid of many voxels with low expression that would be (uninteresting)
optimal subgraphs of their own because they are hardly connected to
any genes.

\begin{figure}
\includegraphics{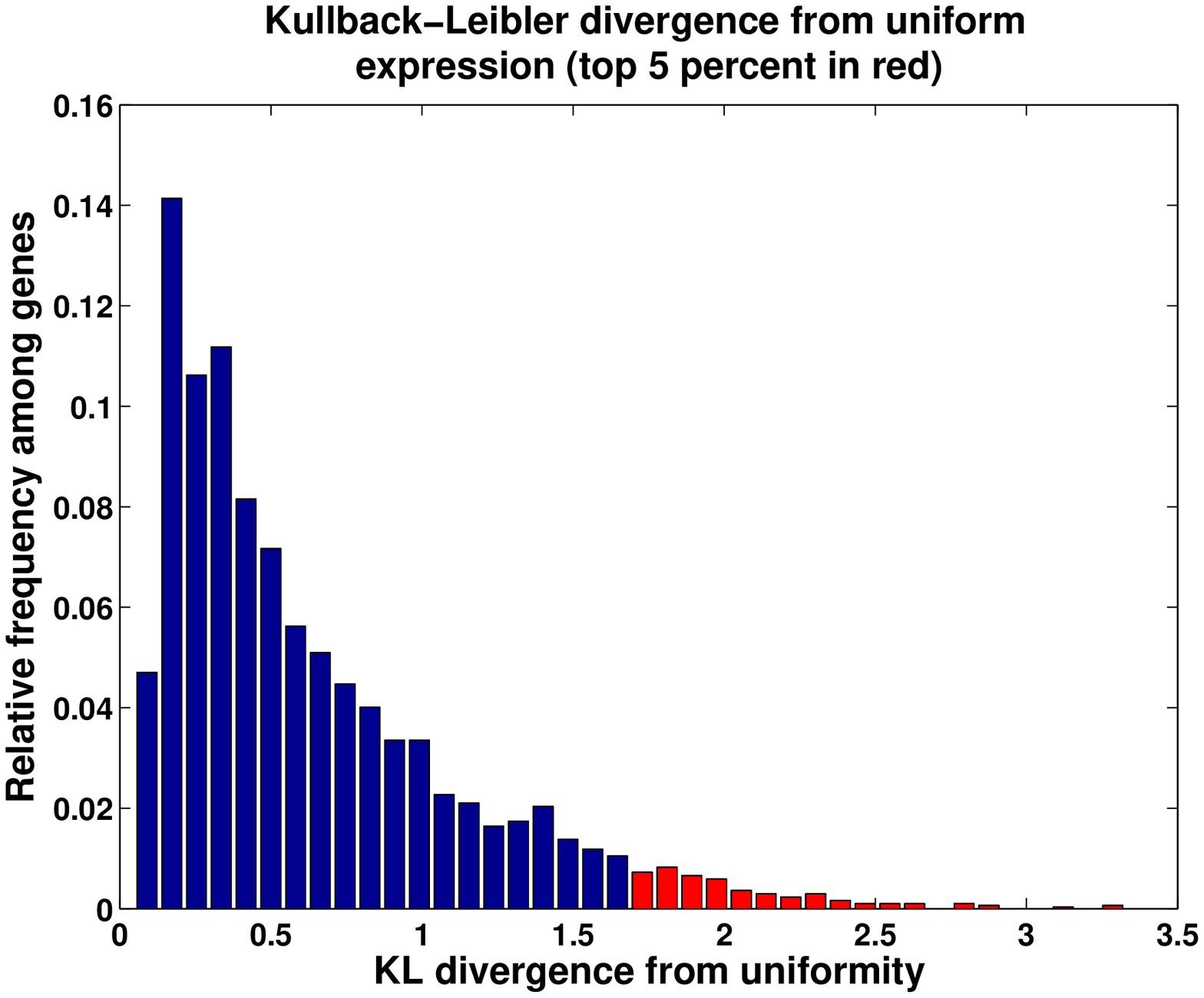}
\caption{A histogram of the occurence of Kullback--Leibler divergences from
a uniform expression across 3041 genes. The top 5 percentiles of KL divergences are highlighted in red. They consist
 of the genes selected for biclustering analysis. The names of those genes appear in the tables describing each of 
the biclusters.}
\end{figure}

\subsection{Results}

We illustrate biclusters corresponding to the plateau at $B=7$ biclusters in the 
number of bicluster as a function of the stopping ratio.\\


For each of the biclusters, the algorithm returns a set of  voxels and a set of genes.
Let is call  the set of voxels ${\mathcal{V}}_b$ and the set of genes ${\mathcal{G}}_b$.
 For each of the biclusters, we plotted:
\begin{itemize}
\item the voxel profile of the bicluster: the maximal-intensity projections of the set of voxels ${\mathcal{V}}_b$  in the bicluster;
 \item the expression profile of the bicluster: the maximal-intensity projection of the the sum of the expression energies of the 
set of genes ${\mathcal{G}}_b$ in the bicluster.
\end{itemize}
The expression profiles follow the shapes of the sets of voxels quite nicely. For each gene, we  
computed the fraction of the total expression of the bicluster that it represents.\\
 Moreover, we noticed many of the biclusters look similar to brain regions defined in the {\hbox{'Big 12'}} partition 
of the left hemisphere 
in the Allen Reference Atlas \cite{AllenAtlas}, 
illustrated on Table \ref{fig:referenceTableBig12}. In order to assess this resemblance quantitatively,
we took the Allen Reference Atlas of the left hemisphere and for each region $R$, we computed the fraction of the 
voxels in  ${\mathcal{V}}_b$ , and the fraction of the expression energy that is contained in the region.
These quantities are expressed as:
\begin{equation}
{\mathrm{Fraction\;of \;voxels}}(R,b) = \frac{|{\mathcal{V}}_b\cap R|}{| R |},\;{\mathrm{for\;brain\; region}}\; R\;
 {\mathrm{and \;bicluster}} \; b,
\end{equation}
\begin{equation}
{\mathrm{Fraction\;of \;expression}}(R,b) = \frac{\sum_{ v\in {\mathcal{V}}_b\cap R}\sum_{g\in{\mathcal{G}}_b}E(v,g)}{ \sum_{ v\in {\mathcal{V}}_b}\sum_{g\in{\mathcal{G}}_b}E(v,g)},\;{\mathrm{for\;brain\;  region}}\; R\; {\mathrm{and \;bicluster}} \; b.
\end{equation}
Only regions that have non-zero overlap with the biclusters are shown in tables. These regions
are ordered by decreasing value of the fraction of voxels. The biclusters are described by the following 
figures:
 \begin{itemize}
\item {\bf{Bicluster 1: 'cerebellum-like'}}, see Figure \ref{fig:biCluster1}.

\begin{figure}
  \centering 
 \subfloat[Sum of gene-expression energies in the
    cerebellum-like
    bicluster]{\label{fig:biClusterExpression1}\includegraphics[width=0.5\textwidth]{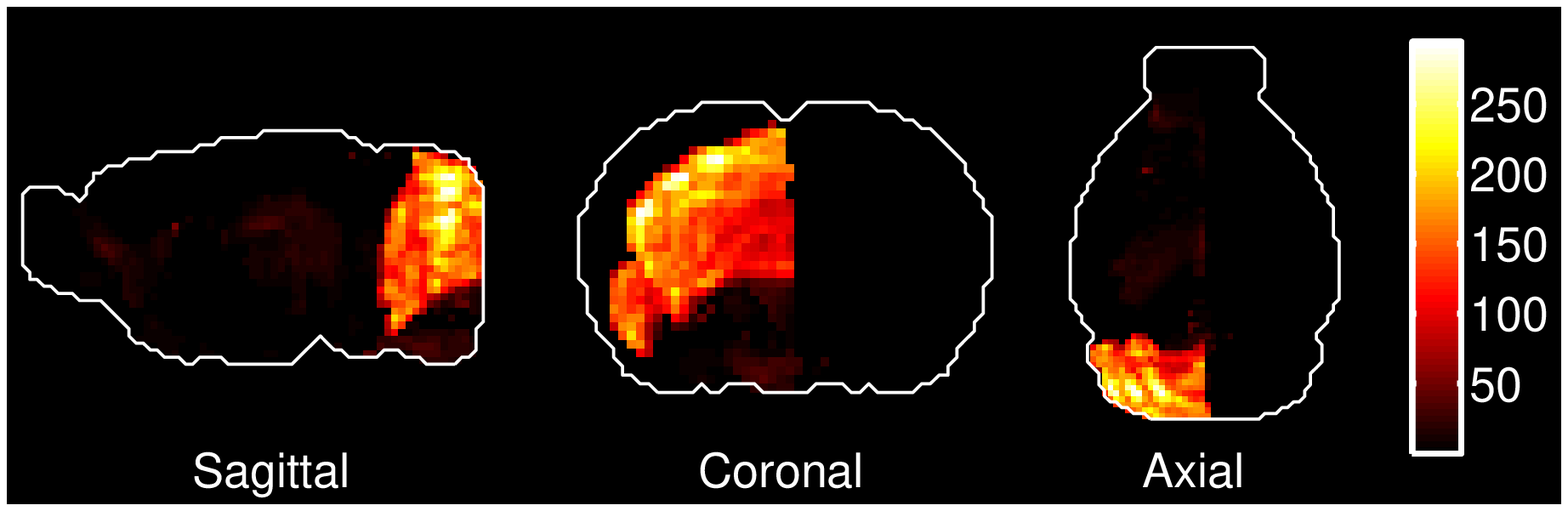}}
  \subfloat[The set of voxels in the cerebellum-like
    bicluster]{\label{fig:biClusterMask1}\includegraphics[width=0.5\textwidth]{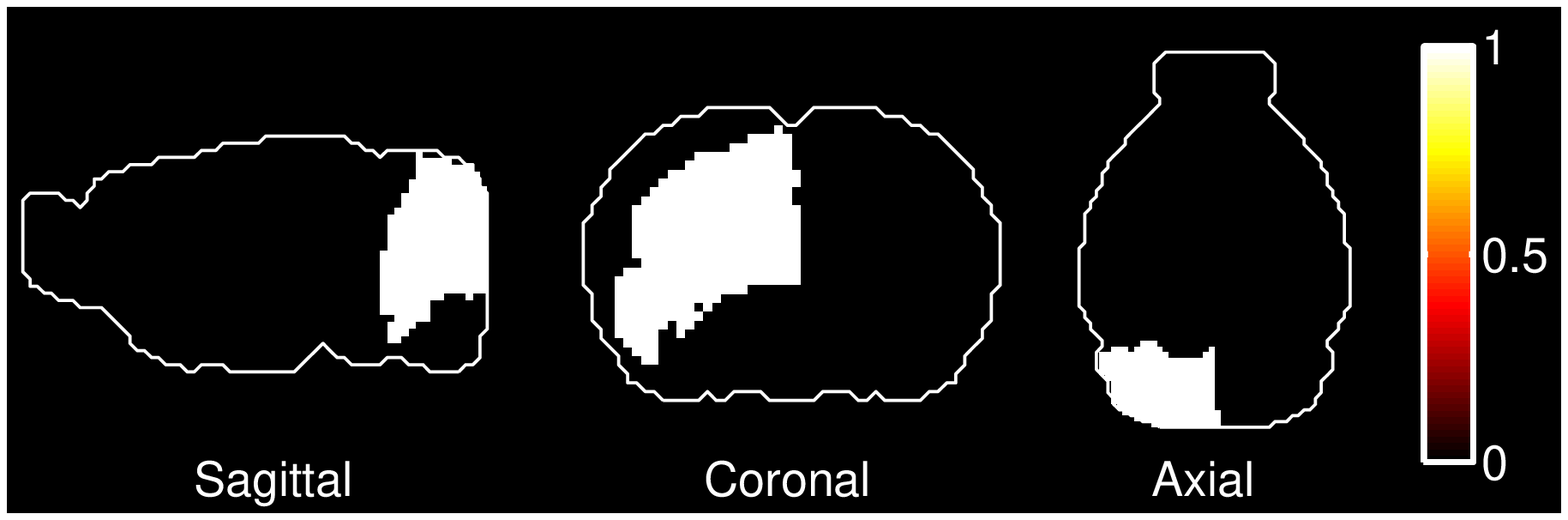}}\\
  \subfloat[The list of genes in the cerebellum-like
    bicluster, ordered by decreasing expression]{\label{fig:biClusterTable1}\begin{tabular}{|p{3cm}|p{3cm}|}
\hline
\textbf{GeneNames}&\textbf{Fraction of cluster energy (pct)}\\\hline
Il16&19.04\\\hline
Neurod1&17.74\\\hline
Cnksr3&14.48\\\hline
Slc6a5&9.35\\\hline
Gng13&6.55\\\hline
3110001A13Rik&6.44\\\hline
Gabra6&6.19\\\hline
Col18a1&3\\\hline
Plxdc1&2.45\\\hline
En2&2.14\\\hline
AW049765&2.12\\\hline
B3gnt5&2.07\\\hline
2010106G01Rik&1.96\\\hline
Lhx5&1.64\\\hline
Htr5b&1.44\\\hline
Ptprm&1.16\\\hline
Slc22a3&0.72\\\hline
Zfp423&0.63\\\hline
Gbx2&0.33\\\hline
Ghrh&0.27\\\hline
Nr5a1&0.26\\\hline
\end{tabular}
}
  \subfloat[The set of voxels in the cerebellum-like
    bicluster]{\label{fig:biClusterTableAnatomy1}\begin{tabular}{|l|l|}
\hline
\textbf{Brain region}&\textbf{Percentage of cluster}\\\hline
Cerebellum&96.58\\\hline
Medulla&2.87\\\hline
Basic cell groups and regions&0.56\\\hline
\end{tabular}
}\\
  \caption{Bicluster 1: 'cerebellum-like'.}
  \label{fig:biCluster1}
\end{figure}

\item{\bf{Bicluster 2}: 'white-matter like'}, see Figure \ref{fig:biCluster2}.

\begin{figure}
  \centering 
 \subfloat[Sum of gene-expression energies in 
    bicluster 2]{\label{fig:biClusterExpression2}\includegraphics[width=0.5\textwidth]{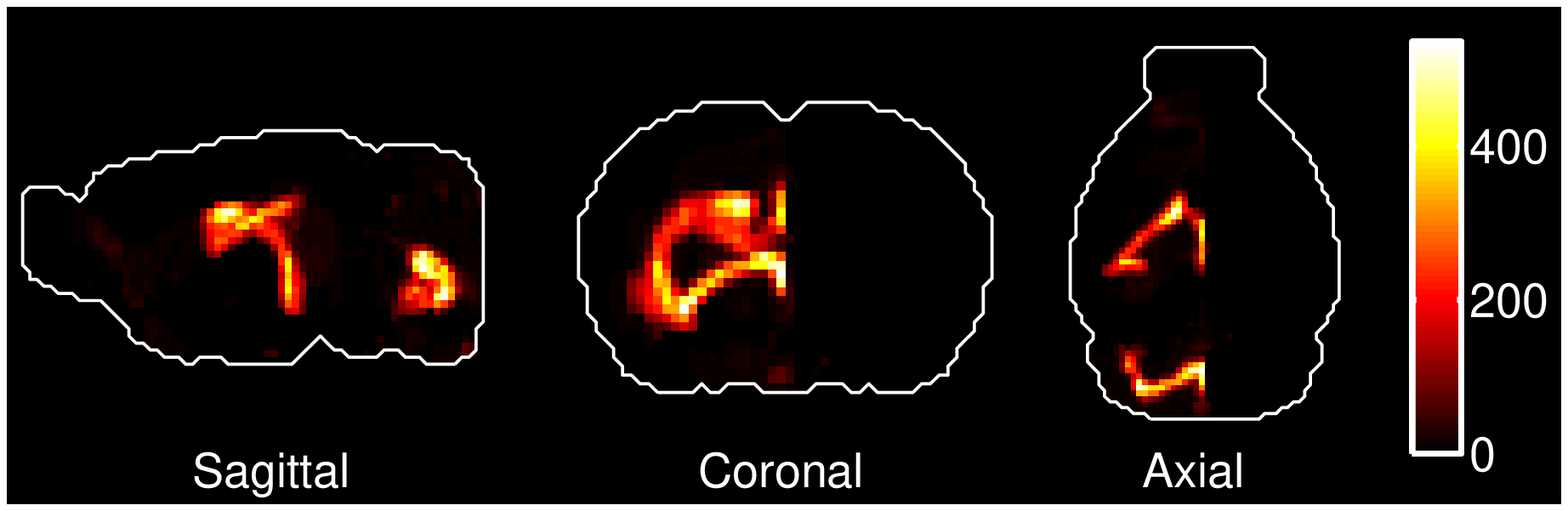}}
  \subfloat[The set of voxels in bicluster 2]{\label{fig:biClusterMask2}\includegraphics[width=0.5\textwidth]{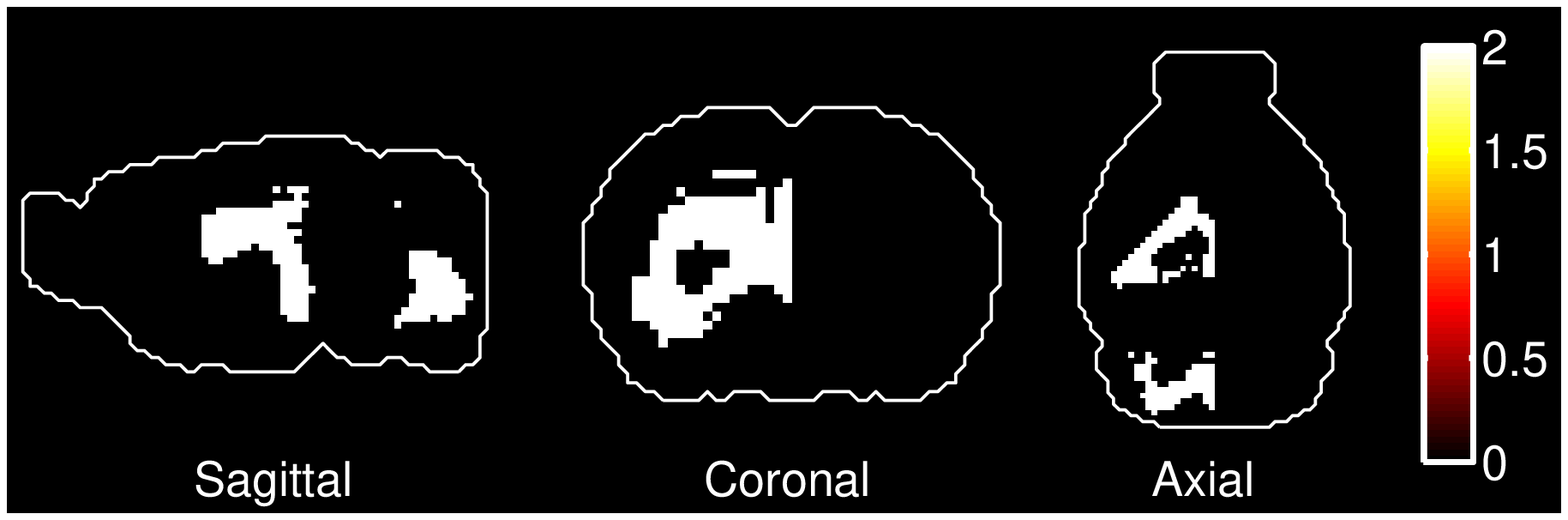}}\\
  \subfloat[The list of genes in bicluster 2 ordered by decreasing expression]{\label{fig:biClusterTable2}\begin{tabular}{|p{3cm}|p{3cm}|}
\hline
\textbf{GeneNames}&\textbf{Fraction of cluster energy (pct)}\\\hline
Kl&11.22\\\hline
Clic6&9.24\\\hline
Ace&8.33\\\hline
Cab39l&6.34\\\hline
F5&5.99\\\hline
Tgfb2&5.88\\\hline
Rdh5&5.4\\\hline
Prlr&4.69\\\hline
AI987712&4.58\\\hline
Acaa2&4.37\\\hline
E030013G06Rik&4.31\\\hline
Tcn2&3.96\\\hline
Gm967&2.44\\\hline
Frmpd2&2.36\\\hline
Trpv4&2.34\\\hline
Cd59a&2.2\\\hline
LOC436099&2.14\\\hline
Vil2&2.12\\\hline
Sntb1&2.07\\\hline
Fzd4&2.06\\\hline
6820408C15Rik&1.92\\\hline
D9Ertd280e&1.71\\\hline
Thbs4&1.68\\\hline
B3gat2&1.45\\\hline
Itpkb&1.2\\\hline
\end{tabular}
}
  \subfloat[The set of voxels in bicluster 2]{\label{fig:biClusterTableAnatomy2}\begin{tabular}{|l|l|}
\hline
\textbf{Brain region}&\textbf{Percentage of cluster}\\\hline
Basic cell groups and regions&36.85\\\hline
Medulla&15.61\\\hline
Hippocampal region&12.86\\\hline
Cerebral cortex&12.43\\\hline
Striatum&9.39\\\hline
Cerebellum&6.79\\\hline
Thalamus&5.49\\\hline
Midbrain&0.29\\\hline
Pons&0.29\\\hline
\end{tabular}
}\\
  \caption{Bicluster 2: 'white-matter like'. The pattern is disconnected and looks like white matter, with the most caudal component
coinciding with the {\emph{arbor vitae}}. Some of the white matter is annotated by the Allen reference Atlas as 'Basic cell groups
and regions'.}
  \label{fig:biCluster2}
\end{figure}

\item{\bf{Bicluster 3: 'hippocampus-like'}}, see Figure \ref{fig:biCluster3}.

\begin{figure}
  \centering 
 \subfloat[Sum of gene-expression energies in 
    bicluster 3]{\label{fig:biClusterExpression3}\includegraphics[width=0.5\textwidth]{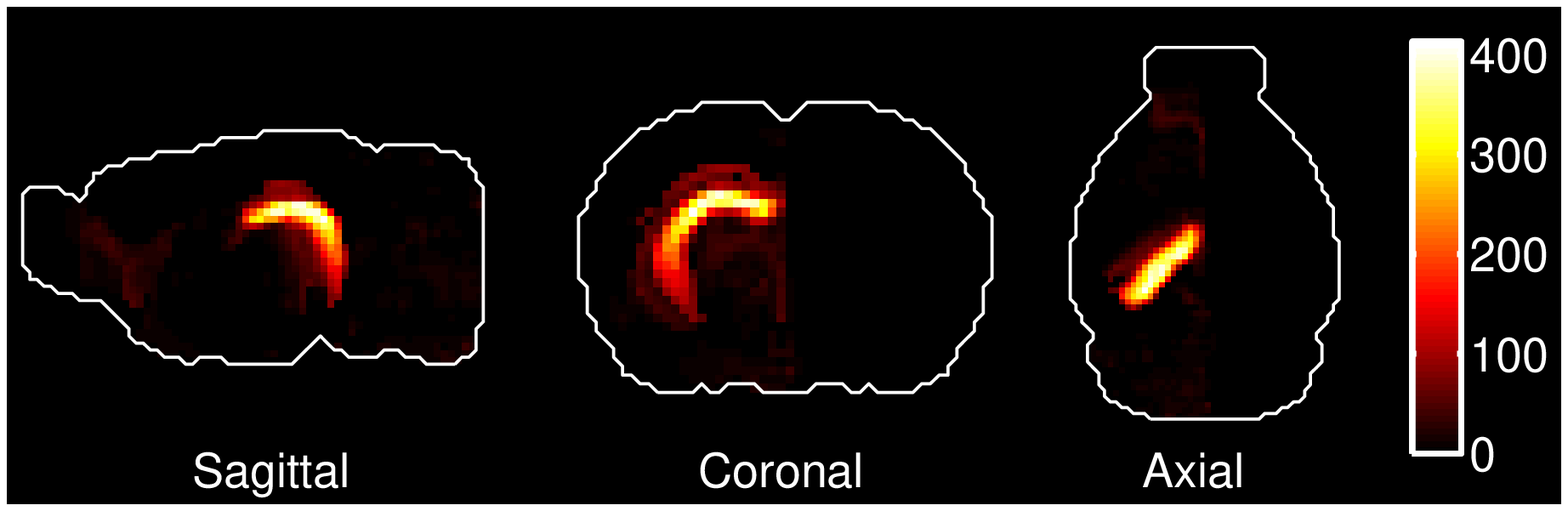}}
  \subfloat[The set of voxels in bicluster 3]{\label{fig:biClusterMask3}\includegraphics[width=0.5\textwidth]{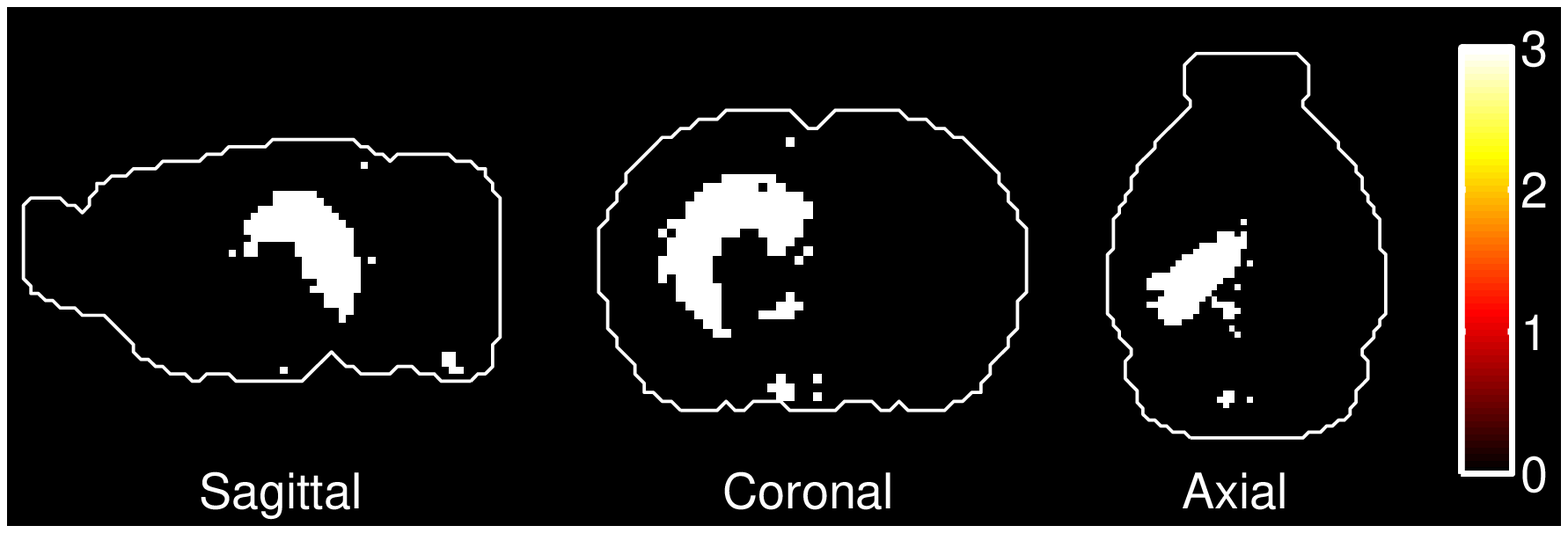}}\\
  \subfloat[The list of genes in bicluster 3 ordered by decreasing expression]{\label{fig:biClusterTable3}\begin{tabular}{|p{3cm}|p{3cm}|}
\hline
\textbf{GeneNames}&\textbf{Fraction of cluster energy (pct)}\\\hline
C1ql2&11.64\\\hline
TC1412430&7.35\\\hline
Zbtb20&6.48\\\hline
Sipa1l2&6.27\\\hline
C630041L24Rik&5.79\\\hline
Dsp&5.48\\\hline
Crlf1&5.46\\\hline
A330019N05Rik&5.25\\\hline
Slc39a6&5.12\\\hline
C78409&5.07\\\hline
Tnfrsf25&4.71\\\hline
Cyp7b1&4.36\\\hline
Lct&3.71\\\hline
Klk8&3.31\\\hline
Gpc4&2.94\\\hline
Prox1&2.66\\\hline
Nr3c2&2.47\\\hline
Pkp2&2.47\\\hline
Arl15&2.16\\\hline
Itga7&1.82\\\hline
Fat4&1.49\\\hline
Sema5a&1.47\\\hline
Csf2rb2&1.37\\\hline
Rnf19&1.14\\\hline
\end{tabular}
}
  \subfloat[The set of voxels in bicluster 3]{\label{fig:biClusterTableAnatomy3}\begin{tabular}{|l|l|}
\hline
\textbf{Brain region}&\textbf{Percentage of cluster}\\\hline
Hippocampal region&80.39\\\hline
Thalamus&9.48\\\hline
Midbrain&3.59\\\hline
Basic cell groups and regions&2.94\\\hline
Medulla&1.8\\\hline
Cerebral cortex&1.47\\\hline
Retrohippocampal region&0.16\\\hline
Hypothalamus&0.16\\\hline
\end{tabular}
}\\
  \caption{Bicluster 3: 'hippocampus-like'.}
  \label{fig:biCluster3}
\end{figure}

\item{\bf{Bicluster 4: 'olfactory-like'}}, see Figure \ref{fig:biCluster4}.
\begin{figure}
  \centering 
 \subfloat[Sum of gene-expression energies in 
    bicluster 4]{\label{fig:biClusterExpression4}\includegraphics[width=0.5\textwidth]{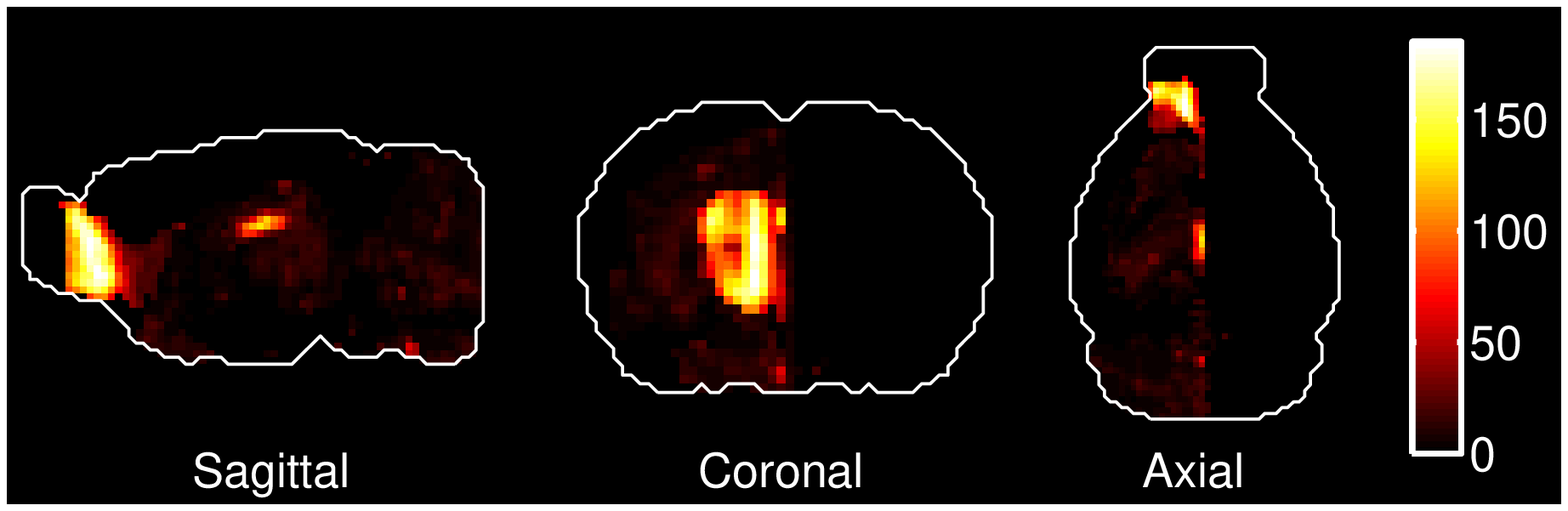}}
  \subfloat[The set of voxels in bicluster 4]{\label{fig:biClusterMask4}\includegraphics[width=0.5\textwidth]{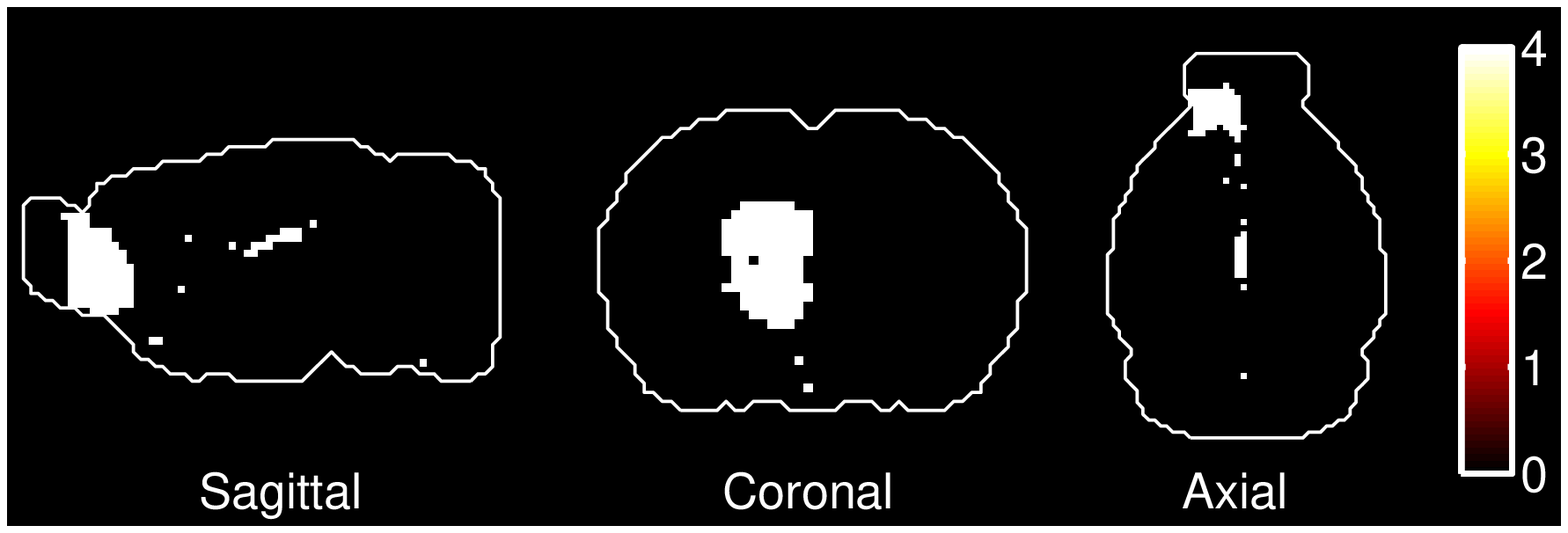}}\\
  \subfloat[The list of genes in bicluster 4 ordered by decreasing expression]{\label{fig:biClusterTable4}\begin{tabular}{|p{3cm}|p{3cm}|}
\hline
\textbf{GeneNames}&\textbf{Fraction of cluster energy (pct)}\\\hline
Lrrc55&10.29\\\hline
Stard8&8.21\\\hline
9830123M21Rik&6.79\\\hline
Nmb&6.6\\\hline
C230040D10Rik&5.89\\\hline
Kcnb2&5.84\\\hline
4930589M24Rik&5.45\\\hline
Adamts19&5.26\\\hline
Raver2&4.46\\\hline
Syt6&4.35\\\hline
Chrna3&3.45\\\hline
Cdh23&3.01\\\hline
Sp8&3\\\hline
B930011P16Rik&2.75\\\hline
Nmbr&2.62\\\hline
A230065H16Rik&2.31\\\hline
Vipr2&2.17\\\hline
Jam2&2.12\\\hline
Eomes&1.95\\\hline
B430201A12Rik&1.89\\\hline
mCG1049722.1&1.66\\\hline
Atp6v1c2&1.61\\\hline
Rgnef&1.42\\\hline
Atp10a&1.4\\\hline
Dnahc11&1.19\\\hline
Scube2&1.17\\\hline
AW456874&1.08\\\hline
Tmem16a&1.05\\\hline
Lhx8&1.02\\\hline
\end{tabular}
}
  \subfloat[The set of voxels in bicluster 4]{\label{fig:biClusterTableAnatomy4}\begin{tabular}{|l|l|}
\hline
\textbf{Brain region}&\textbf{Percentage of cluster}\\\hline
Olfactory areas&86.68\\\hline
Cerebral cortex&7.38\\\hline
Thalamus&3.48\\\hline
Basic cell groups and regions&1.64\\\hline
Striatum&0.61\\\hline
Medulla&0.2\\\hline
\end{tabular}
}\\
 \caption{Bicluster 4: 'olfactory-like'}
  \label{fig:biCluster4}
\end{figure}

\item{\bf{Bicluster 5:}} a disconnected bunch with medulla and other regions, see Figure \ref{fig:biCluster5}.
\begin{figure}
  \centering 
 \subfloat[Sum of gene-expression energies in 
    bicluster 5]{\label{fig:biClusterExpression5}\includegraphics[width=0.5\textwidth]{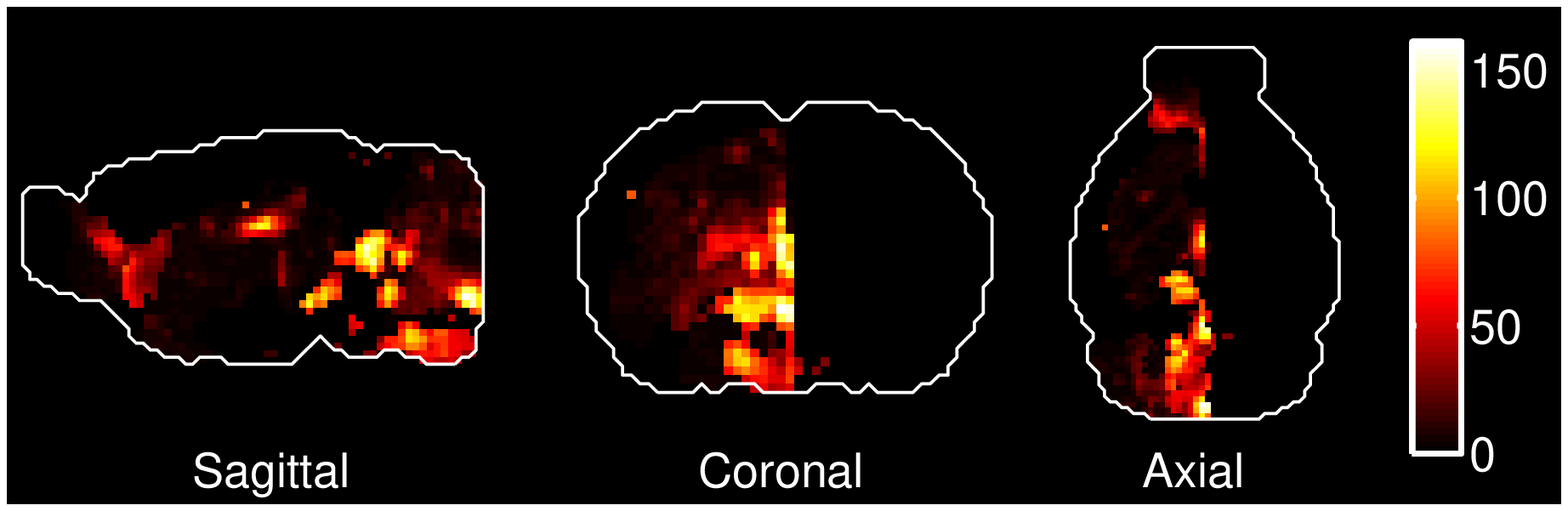}}
  \subfloat[The set of voxels in bicluster 5]{\label{fig:biClusterMask5}\includegraphics[width=0.5\textwidth]{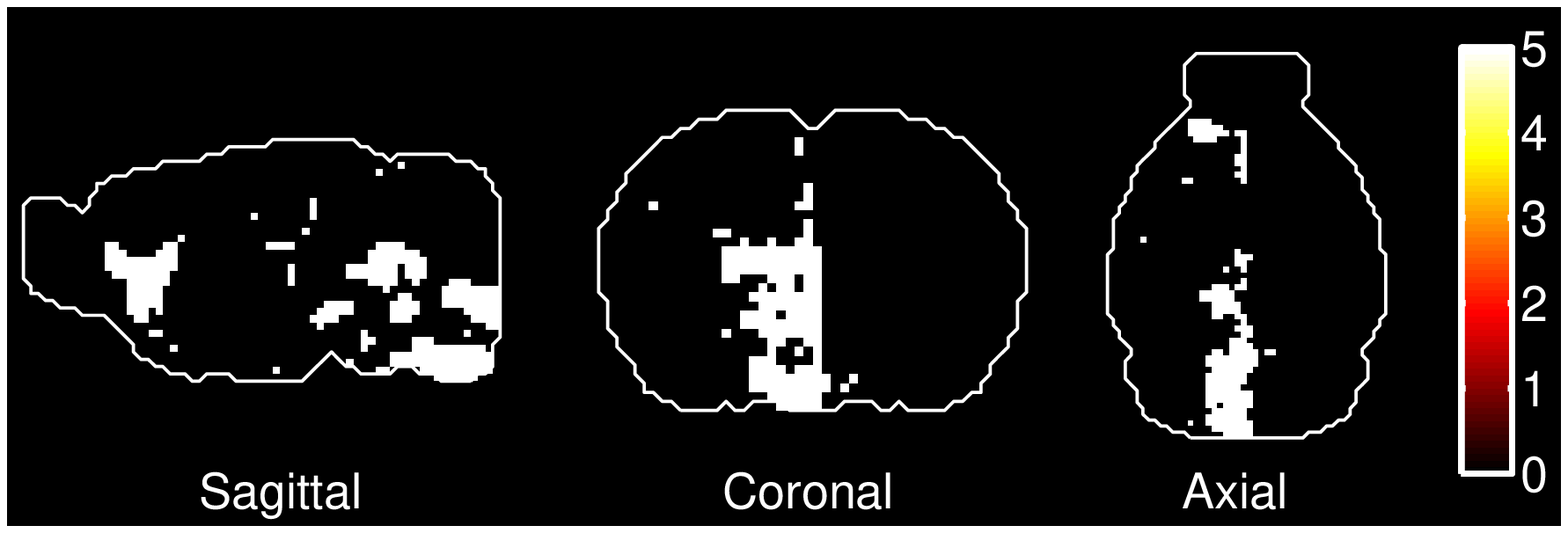}}\\
  \subfloat[The list of genes in bicluster 5 ordered by decreasing expression]{\label{fig:biClusterTable5}
{\small{
\begin{tabular}{|p{3cm}|p{3cm}|}
\hline
\textbf{GeneNames}&\textbf{Fraction of cluster energy (pct)}\\\hline
Glra1&22.11\\\hline
Slc18a3&5.75\\\hline
Slc18a2&4.86\\\hline
Slc5a7&4.78\\\hline
Ddc&4.39\\\hline
Slc6a3&3.79\\\hline
1810023C24Rik*&3.48\\\hline
Chrna6&3.24\\\hline
Pou4f1&3.09\\\hline
LOC244958&2.36\\\hline
Chrnb4&2.26\\\hline
Gpr151&2.24\\\hline
Tph2&2.24\\\hline
Avp&2.23\\\hline
Calca&2.19\\\hline
Anxa2&2.07\\\hline
Frmd6&2.05\\\hline
C130034I18Rik&1.98\\\hline
Slc6a4&1.85\\\hline
Epha1&1.79\\\hline
Postn&1.69\\\hline
Stk24&1.66\\\hline
Il13ra1&1.64\\\hline
C130021I20Rik&1.63\\\hline
Calcb&1.61\\\hline
A330102H22Rik&1.51\\\hline
Dbh&1.27\\\hline
Ppap2a&1.26\\\hline
Gpr3&1.22\\\hline
Hspb1&1.15\\\hline
Tal1&0.98\\\hline
Layn&0.97\\\hline
Cdh6&0.95\\\hline
Tspan12&0.93\\\hline
Wif1&0.86\\\hline
Scml2&0.72\\\hline
Pscdbp&0.48\\\hline
Gucy2c&0.41\\\hline
Ntsr1&0.3\\\hline
\end{tabular}
}}
}
  \subfloat[The set of voxels in bicluster 5]{\label{fig:biClusterTableAnatomy5}\begin{tabular}{|l|l|}
\hline
\textbf{Brain region}&\textbf{Percentage of cluster}\\\hline
Medulla&47.11\\\hline
Midbrain&16.33\\\hline
Pons&14.46\\\hline
Cerebral cortex&8.5\\\hline
Olfactory areas&6.8\\\hline
Basic cell groups and regions&4.08\\\hline
Thalamus&1.53\\\hline
Striatum&0.51\\\hline
Cerebellum&0.51\\\hline
Hypothalamus&0.17\\\hline
\end{tabular}
}\\
 \caption{Bicluster 5: a disconnected bunch with medulla and other regions.}
  \label{fig:biCluster5}
\end{figure}

\item{\bf{Bicluster 6: 'thalamus-like'}}, see Figure \ref{fig:biCluster6}.
\begin{figure}
  \centering 
 \subfloat[Sum of gene-expression energies in 
    bicluster 6]{\label{fig:biClusterExpression6}\includegraphics[width=0.5\textwidth]{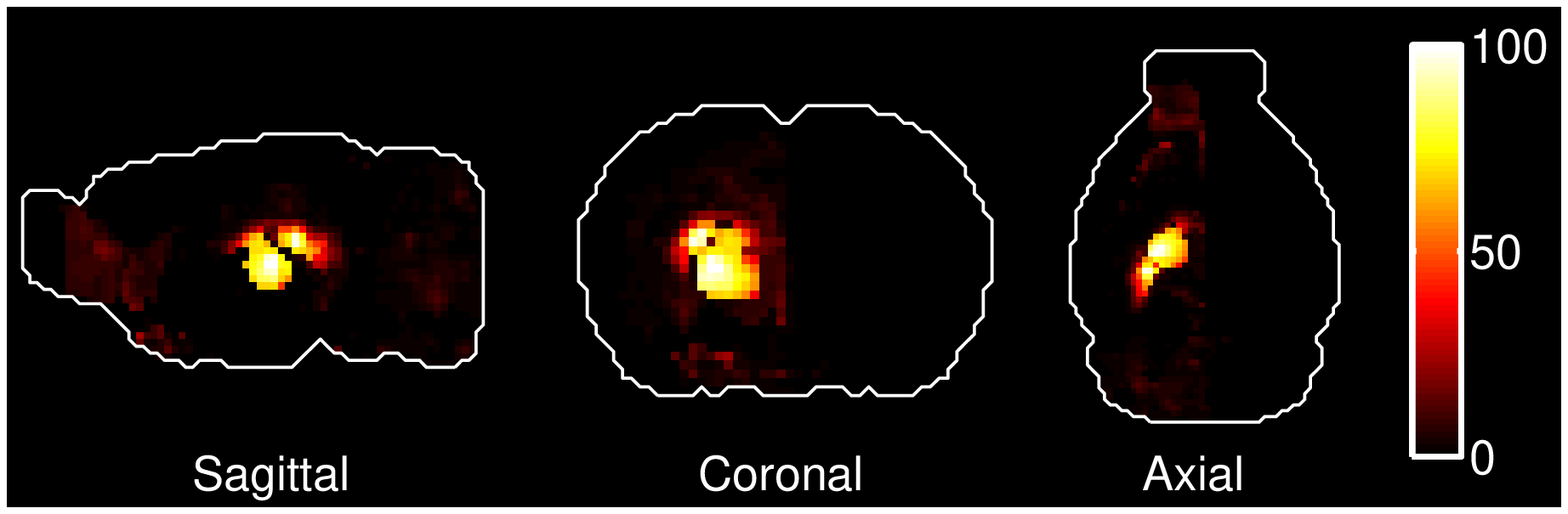}}
  \subfloat[The set of voxels in bicluster 6]{\label{fig:biClusterMask6}\includegraphics[width=0.5\textwidth]{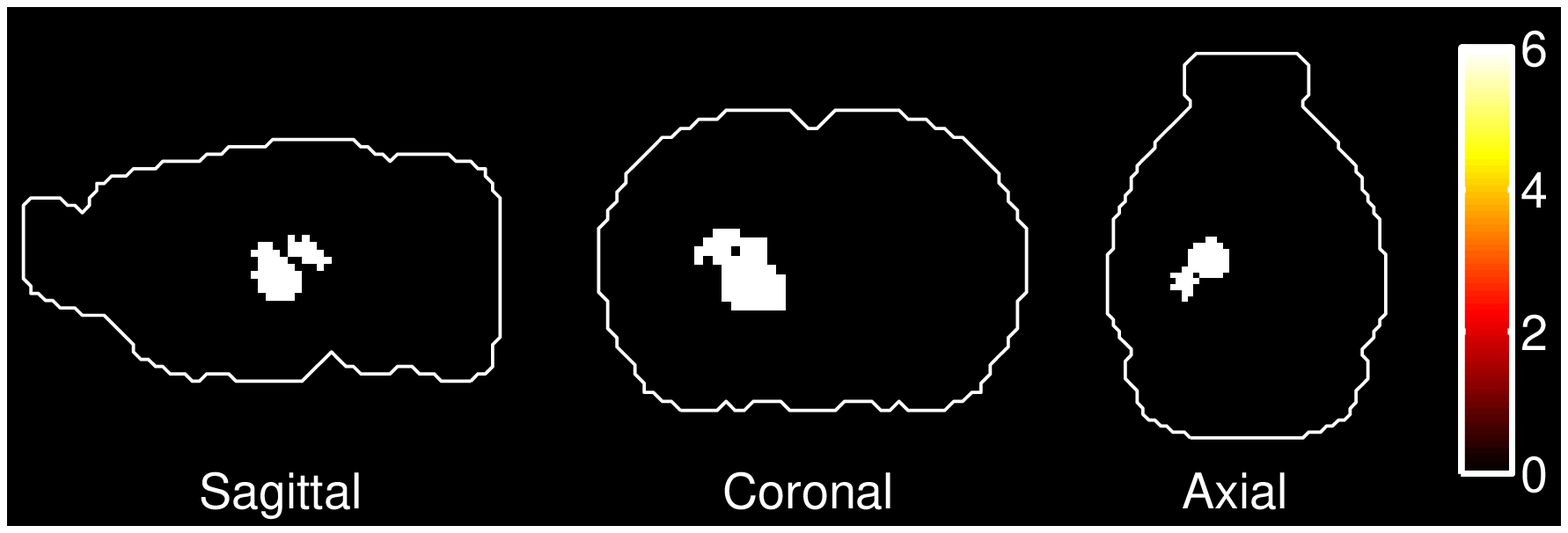}}\\
  \subfloat[The list of genes in bicluster 6 ordered by decreasing expression]{\label{fig:biClusterTable6}\begin{tabular}{|p{3cm}|p{3cm}|}
\hline
\textbf{GeneNames}&\textbf{Fraction of cluster energy (pct)}\\\hline
Plekhg1&26.95\\\hline
1110069I04Rik*&22.71\\\hline
Tnnt1&16.72\\\hline
Lef1&14.68\\\hline
Rab37&14.08\\\hline
Slitrk6&4.86\\\hline
\end{tabular}
}
  \subfloat[The set of voxels in bicluster 6]{\label{fig:biClusterTableAnatomy6}\begin{tabular}{|l|l|}
\hline
\textbf{Brain region}&\textbf{Percentage of cluster}\\\hline
Thalamus&98.4\\\hline
Basic cell groups and regions&0.53\\\hline
Hippocampal region&0.53\\\hline
Hypothalamus&0.53\\\hline
\end{tabular}
}\\
 \caption{Bicluster 6: 'thalamus-like'.}
  \label{fig:biCluster6}
\end{figure}

\item{\bf{Bicluster 7: 'striatum-like'}}, see Figure \ref{fig:biCluster7}.
\begin{figure}
  \centering 
 \subfloat[Sum of gene-expression energies in 
    bicluster 7]{\label{fig:biClusterExpression7}\includegraphics[width=0.5\textwidth]{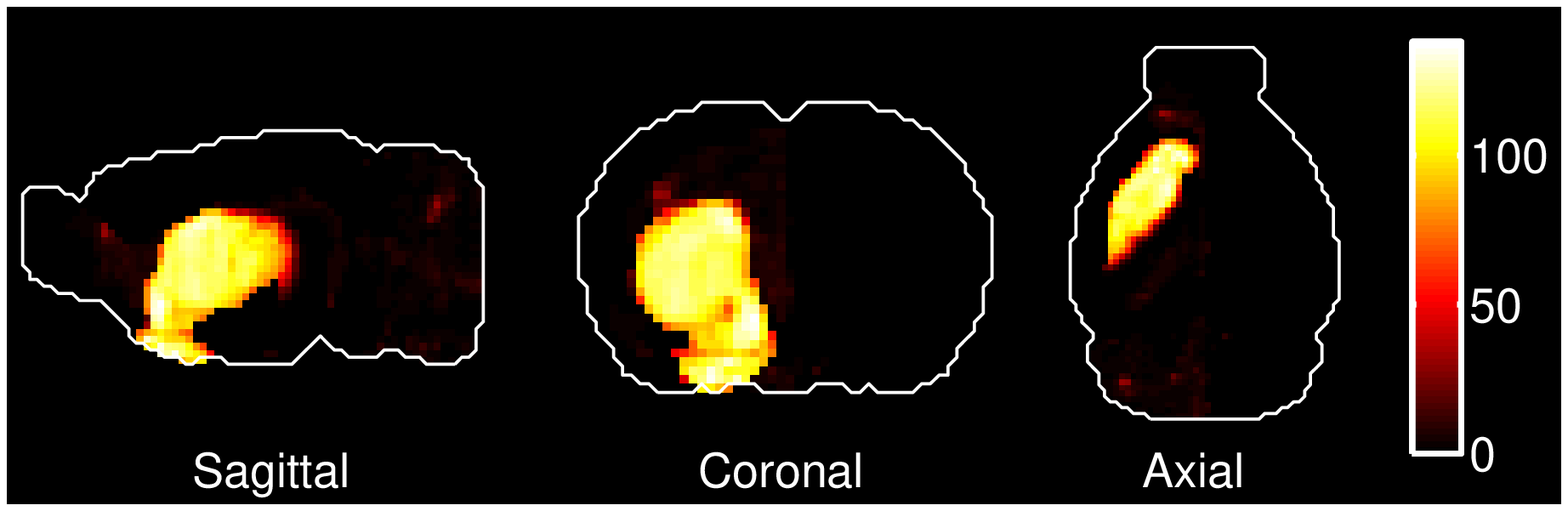}}
  \subfloat[The set of voxels in bicluster 7]{\label{fig:biClusterMask7}\includegraphics[width=0.5\textwidth]{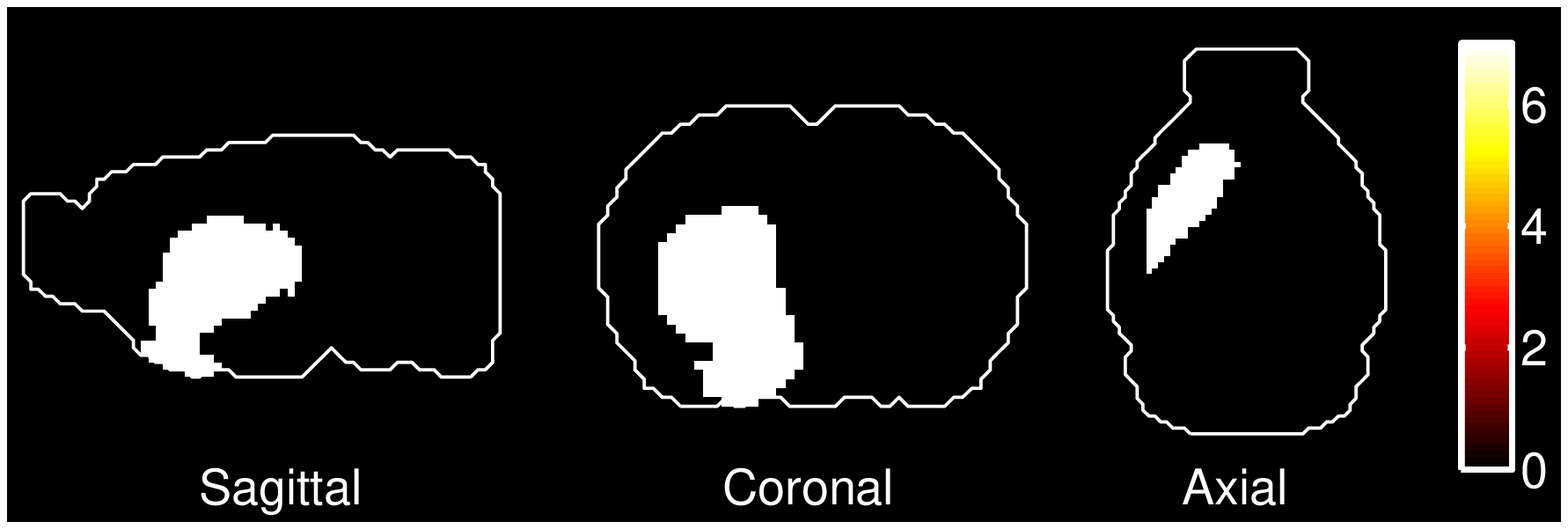}}\\
  \subfloat[The list of genes in bicluster 7 ordered by decreasing expression]{\label{fig:biClusterTable7}\begin{tabular}{|p{3cm}|p{3cm}|}
\hline
\textbf{GeneNames}&\textbf{Fraction of cluster energy (pct)}\\\hline
Rgs9&25.68\\\hline
Pde1b&20.98\\\hline
Adora2a&15.7\\\hline
Rarb&12.92\\\hline
Gprin3&8.32\\\hline
Gpr6&8\\\hline
Serpina9&5.72\\\hline
Cd4&2.69\\\hline
\end{tabular}
}
  \subfloat[The set of voxels in bicluster 7]{\label{fig:biClusterTableAnatomy7}\begin{tabular}{|l|l|}
\hline
\textbf{Brain region}&\textbf{Percentage of cluster}\\\hline
Striatum&95.23\\\hline
Basic cell groups and regions&1.87\\\hline
Cerebral cortex&1.73\\\hline
Pallidum&1.18\\\hline
\end{tabular}
}\\
 \caption{Bicluster 7: 'striatum-like'.}
  \label{fig:biCluster7}
\end{figure}

\end{itemize}

\pagebreak

\section{Dimensionality estimates of brain regions in gene space}

\subsection{Scaling argument and algorithm}
Due to the large number of genes that are analyzed in the AGEA atlases,
our data live in very high-dimensional space. On the other hand, many genes
have non-zero expression in the gene-expression profiles of cell-types. So the multiplicity 
of genes expressed  may not reflect the complexity of the underlying biological objects.\\

A fixed cell-type is just a one-dimensional subspace of gene space. From the inspection of the 
projection of the SVD of gene-expression energies onto the subspace spanned by the first three 
singular vector in gene space, there seems to be  
clouds of voxels that are localized on low-dimensional subspaces. How can we 
detect such localization properties? The following scaling argument 
has been used to discover the intrinsic dimension of datasets that are only known 
by their pairwise dissimilarities \cite{conical}.\\
\begin{figure}
\centering
\noindent
\includegraphics[width=5in,keepaspectratio]{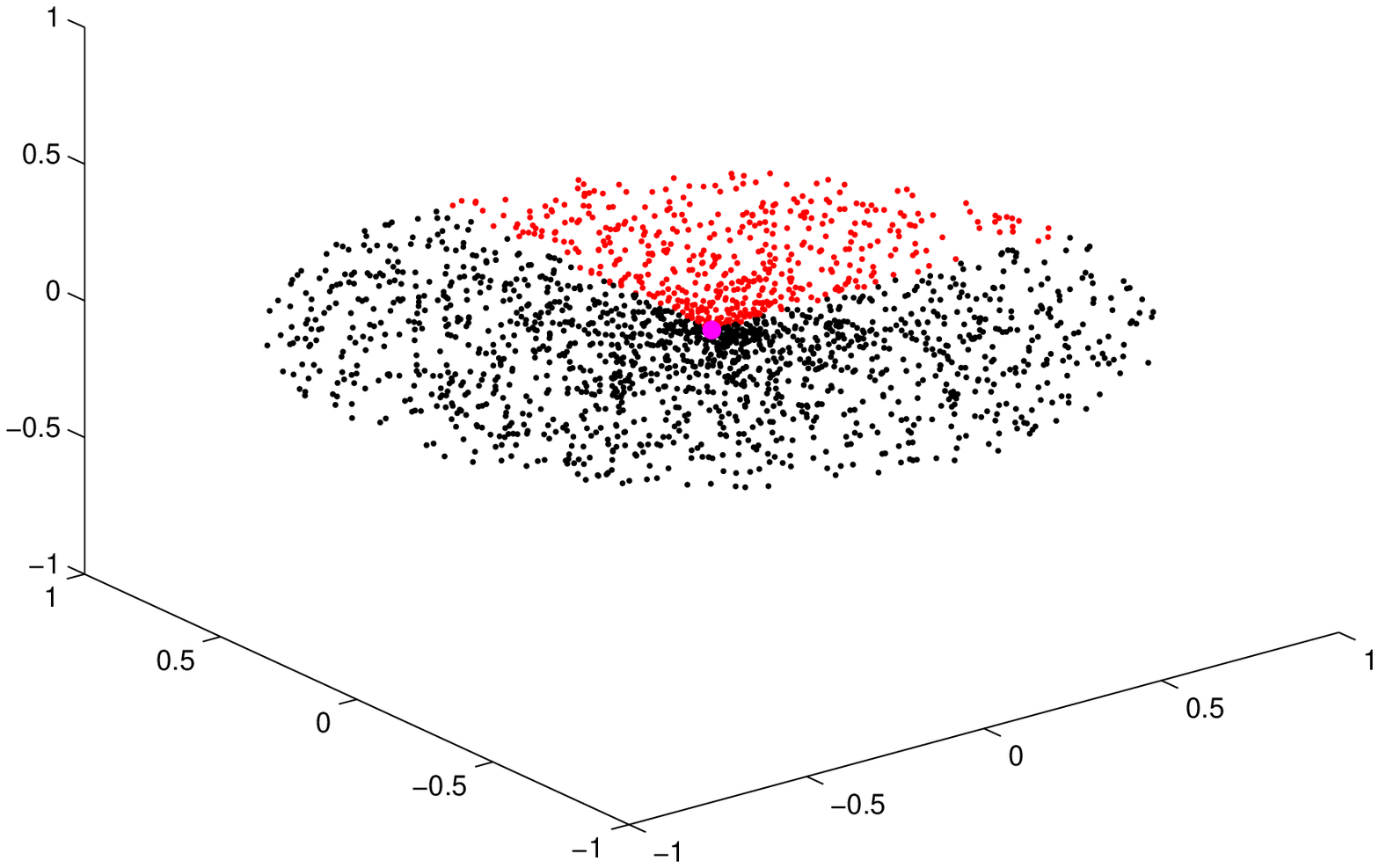}\\ 
\caption{{\bf{A disc sampled by 2000 random points with uniform probability, embedded into a three-dimensional space.}} A cone with tip at the center
of the disc (shown in magenta), with an opening of 90 degrees, intercepts roughly a quarter of the points. This fraction would be the same
if the disc was embedded into a higher-dimensional space.}
\centering
\label{fig:embeddedDisc}
\end{figure}
If the points are drawn from an underlying submanifold of the ambient space, of 
dimension $d$, then a cone with tip at the center of the cloud of points and
an opening of 90 degrees should intercept a fraction of the points
 that decreases exponentially with the dimension $d$. This is illustrated on Figure (\ref{fig:embeddedDisc}) 
for a disc embedded into three dimensions.\\

For a given region $R_r$, the dimension of the underlying subspace of
gene space is estimated by the following procedure:\\

{\bf{1. Compute the pairwise distances in the region.}} Compute the
pairwise distances between the voxels $(v_i)_{1,\leq r \leq V_r}$
belonging to region $R_r$ (these are the relevant lines of the data
matrix $D$):
 $$ \forall i \in [ 1..V_r],\;  v_i : = (D_{ig} )_{1\leq g \leq G} \in {\mathbf{R}}^G,$$
 $$ \forall i, j \in [ 1..V_r], \; d_{ij} := || v_i - v_j||_{L^2}.$$

{\bf{2. Find the center of the region.}} Find out the center $C$ of this cloud defined as the point
  {\emph{in the set}}
 which minimizes the dispersion of points around itself. This is as
 close to the average as we get, within the set (this is to
 pathologies with very hollow, non-convex regions, where the ordinary
 mean may be far from all the voxels). Let us call $c$ the index of
 the center, meaning $C = v_c$ with:
 $$ c = {\mathrm{argmin}} \sum_j d_{cj}^2.$$
This is a Fr\'echet mean with uniform weight in the image of the region $R_r$ under $D$.\\

{\bf{3. Define a typical cone in the region.}} Pick a point $P$ (with
index $p$, or $P = v_p$) in the region, which is at one standard
deviation away from the center (or as close to it as you can get
while staying within the region):
   $$ p = {\mathrm{argmin}} \left(d_{cp}^2 - \sigma_d^2\right),$$
  $$ \sigma_d^2 = \sum_j d_{cj}^2.$$ Consider the cone whose tip is at
   $C$, whose axis points towards $P$, and whose opening angle is $\pi /
   2$.

{\bf{3. Compute the number of voxels in the region that fall within the cone.}}
For a point $M$ with index $m$, the property of being inside the
   cone is defined in terms of its distances to $C$ and $P$, as the
   critical distance from $P$ is attained \footnote{This is obtained by Pythagoras' theorem in the
     triangle $CMP$: the angle $\widehat{PCM}$ is smaller than a chosen angle
     $\beta$ (here $\beta = \pi / 4 $), if its cosine is larger than
     $\cos \beta$. Pythagoras' theorem applied to the triangle $PCM$
     expresses this condition as 
 $\frac{d_{pc}^2+d_{mc}^2 - d_{mp}^2}{2 d_{pc}d_{mc}} < \cos \beta $, i.e.
 $d_{mp} < \sqrt{d_{pc}^2 +d_{mc}^2 - 2\cos\beta d_{pc}d_{mc}}$} when the angle $\widehat{PCM}$
   equals $\pi/4$:
$$ d_{mp} < \sqrt{ d_{pc}^2 + d_{mc}^2-\sqrt{ 2 } d_{pc} d_{mc}}.$$

We can compute the fraction of points in the cloud that is inside the cone: 
$$ f_{C,P} = \frac{1}{V}\left|\left\{m \in [ 1..V] , d_{mp} < \sqrt{
   d_{pc}^2 + d_{mc}^2-\sqrt{ 2 } d_{pc} d_{mc}} \right\}\right|.$$

{\bf{4. The fraction of points intercepted by the cone scales exponentially with the dimension of the underlying subspace.}}
 If
 the points are taken from a $\delta$-dimensional manifold which is
 not too anisotropic (see below for a quantitative study of the dispersion of the values that are 
returned by this algorithm for different choices of cones), this fraction should scale as
 $$ f_{C,P} \simeq 2^{-\delta},$$ which gives a global estimate of the
 dimension of the (putative) manifold underlying the cloud of points:
 $$ \delta_P = -\frac{\log f_{C,P} }{ \log 2 }.$$
 The results are quite small (typically below ten) for all the structures in the 'Fine'
 annotation of the Allen Reference Atlas \cite{AllenAtlas},
  but should {\emph{not}} be compared directly to $G$,
 as the output of the calculation described above is constrained by
 the sample size, and many regions have quite small size.\\
 
{\bf{4. Probabilistic interpretation of the dimensions.}}
Repeat the above computation for all points $P$ distinct from the center $C$.
This induces a distribution of estimates of the dimension for the cloud of
points, with a value of $\delta_P$ associated to each point $P$. The more peaked
this distribution is, the better the cloud of point is localized on a manifold.
The value of the peak estimates the dimension of this manifold. 
\begin{equation}
{\mathrm{Prob}}( \delta = \delta_0 )  = {\mathrm{Prob}}( f_{C,P} = 2^{-\delta_0}V ) 
= \frac{1}{V-1}\left| \left\{ P, f_{C,P} = 2^{-\delta_0}V  \right\}\right|.
\end{equation}
The values of the dimension are binned into integer values. If the brain structure 
 considered is 1) well-localized and 2) reasonably homogeneous, the plot of the probability distribution 
should be 1) well-separated from a Dirac mass placed at the critical dimension of the structure, 
and 2) be peaked. Results for caudoputamen are shown on Figure 18.

\subsection{Results and limitations}

What is the maximal value of the dimension that can be found using the
conical trick described above?  This value, which we will call
{\emph{critical dimension}} is attained in situations where the cloud
of points is so dilute that the cone is essentially empty: it
intercepts no point in the region apart from its tip $C$ and the point
$P$ that defines the orientation of the cone. This critical dimension,
or $\delta_{\mathrm{crit}}$, is therefore defined by
$$ 2^{-\delta_{\mathrm{crit}}} = \frac{ 2 }{ V},$$
hence
$$ \delta_{\mathrm{crit}} = \frac{ \log V }{ \log 2 } - 1.$$ When a
  cloud of $V$ points has critical dimension, it must therefore be
  considered to be a genuine cloud of points, living in high
  dimension, rather than a sample taken from a lower-dimensional
  space.\\

 Of course, the critical dimension is a decreasing function of the sample size, 
and at the resolution of 200 microns we are working with, the critical dimensions of  
the brain structures are all orders of magnitude below the total number of genes $G=3041$, 
which is a big threat to our method: as the value of $\delta$ fluctuates with the orientation of the cone,
it may well be that cones that are close in terms of Euclidean distances 
 induce fluctuation in the dimension so big that they saturate the critical dimension.
In what follows, the distribution of dimensions across all possible orientations of the cone is studied.

Can we trust this estimate? If the manifold from which the voxels are
taken had dimension larger than the $\log_2$ of the number of voxels
for structure, we would not be able to detect it.  So for the many
small structures in the fine annotation, the value of the dimension is
much limited by the resolution scale. But this is an inherent
characteristic of dimension for physical object defined through
observations of clouds of points: dimension is not a well-defined quantity and depends on
scale.\\
\begin{figure}
\centering
 \includegraphics[width=0.8\textwidth]{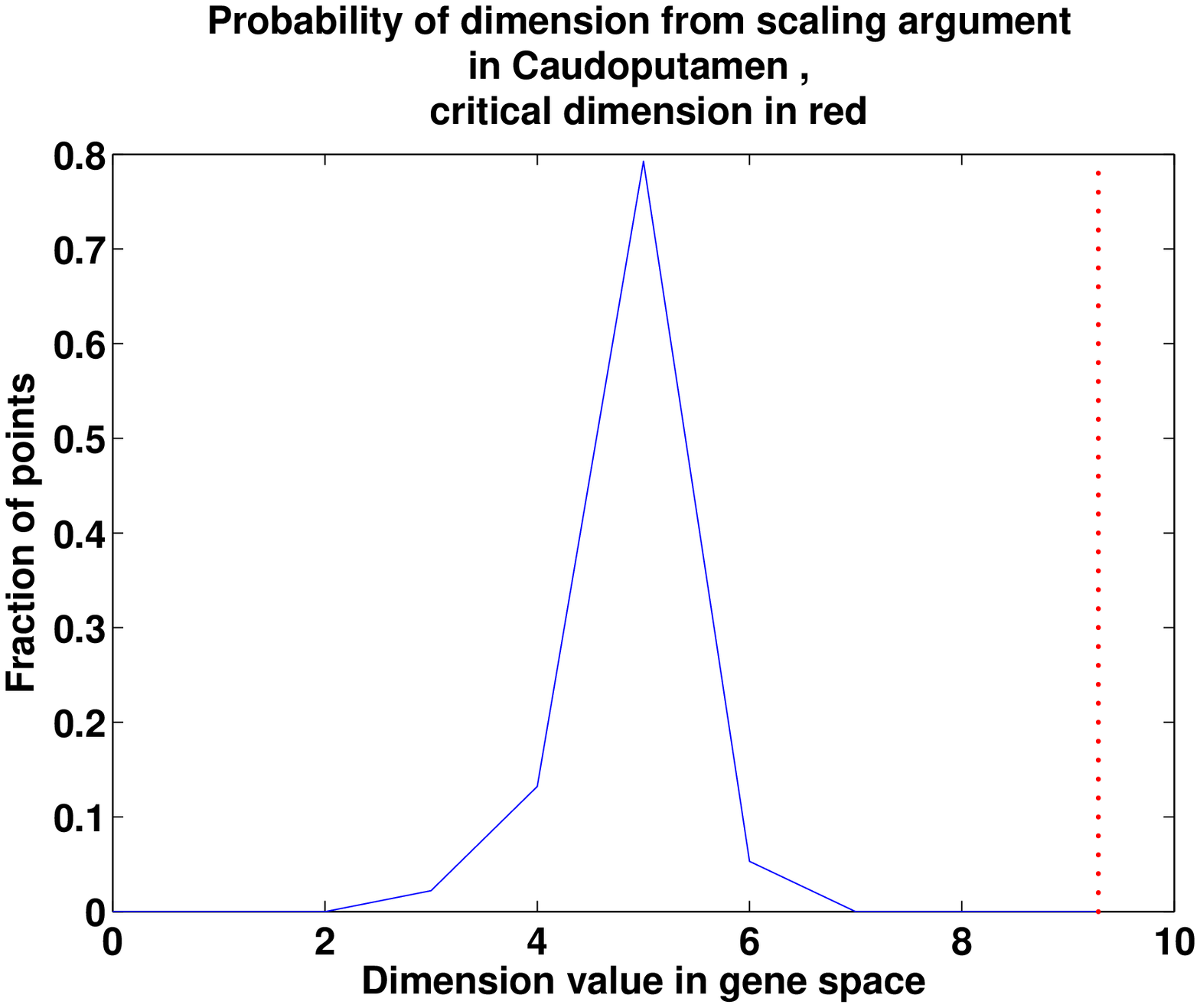}
\caption{Distribution of the dimension estimates for the voxels belonging to caudoputamen. At a resolution of 200 microns,
caudoputamen consists of 1248 voxels, which induces a value of 9.28 for the critical dimension, the maximal dimension
that can be returned by the algorithm. 80 percent of the voxels correspond to an underlying dimension of 5, and the probablity of the critical
dimension is zero.}
\label{fig:CDimensionsGraph}
\end{figure}

\subsection{What distribution of dimensions would be expected by chance?}

 The center and shape of this probability distribution have to be
 compared to the ones that would be obtained by drawing a set of
 random points in $G$ dimensions, with the same sample size. Consider
 a fixed center $C$ and a second point $P$ defining a reference
 quadrant. The $G$-dimensional ambient space is partitioned into $2^G$
 quadrants by completing the axis $[CP)$ into an orthogonal basis.
 When drawing $V-2$ points randomly, we draw each of them from one of those 
quadrant, with equal probability 
$$p = \frac{1}{2^G}.$$  For a given quadrant, the number of drawn points
that falls into it follows a binomial law of parameters $V-2$ (as
two points are used to define the quadrant) and $p$. The dimension
$\delta_{V,G}$ of the set of drawn points is therefore distributed as the
(opposite of) the $\log_2$ of a binomial variable:
$$d_{V,G} = \frac{1}{\log 2}\left(\log V - \log  (N_{V,G}+2)\right),\;\;\; N_{V,G}\sim B( V-2, 2^{-G}),$$
$$P( d_{V,G} = D ) = P( N_{V,G} = 2^{-D}(V - 2) ),$$
$$ {\mathrm{i.e.}}\; P( d_{V,G} = D )= \frac{ V'!}{( 2^{-D}V')!( V' - 2^{-D}V' )!}p^{ 2^{-D}V'}( 1 - p )^{V' - 2^{-D}V' },\;\; V' = V - 2.$$
 Any sample of the
size of those we are dealing with ($V$ up to a few thousands of
points, with fixed $G=3041$) is insufficient to parse a given
dimension: the probability distribution of $ N_{V,G}$ is extremely
peaked at zero: $d_{V,G}$ is therefore extremely peaked at the
critical dimension:
$$P( N_{V,G} = 0 ) = ( 1- p ) ^{V-2} = 1 - (V-2) \times p + o(p).$$
So, in order to have one percent of the probability away from the critical dimension 
put the following lower bound $V^{pc}_{\mathrm{min}}$ on the sample size:
$$P( N_{V,G} = 0 ) < 0.99 \Rightarrow  V > V^{pc}_{\mathrm{min}} = \frac{0.01}{p} + 2 \simeq 2^{3039},$$
which is hopeless, as the whole brain has $V = 49742 \simeq 2^{15.6}$ voxels, so the voxel size 
would have to be $2^{(3039 - 15.6) / 3} \simeq 2^{1008}$ times smaller, way out of bounds of 
the size of known physical objects. Anther way of putting one percent of the probability away from the critical dimension is to lower the 
number of genes:
$$G< G_{\mathrm{max}}\frac{1}{\log( 2 )}\times(\log( V - 2 ) - \log( 0.01 ) ) \simeq 16,$$
which is a very low bound on the dimension of gene space, compared to the size of our data.

 So our null hypothesis of points drawn binomially into a flat gene space is not anywhere
near the actual distribution of gene expression energies, even if we processed the data
 at a resolution of one micron. So even if the distribution of 
dimensions for brain structures in gene space
 is not very sharply peaked because of noise or anisotropy or underlying disconnected components, the very 
fact that the probability is concentrated away from the critical dimension makes the 
gene expression data extremely packed in gene space, with a number of independent degrees of freedom much lower than 
the total number of genes in our data.

\pagebreak
\vspace{-0.5cm}
\section{Tables of correlations and fittings between cell-types and gene-expression energies}
\begin{tabular}{|l|l|l|l|}
\hline
\textbf{index}&\textbf{Cell type}&\textbf{Heat map of correlations}&\textbf{Heat map of weight}\\\hline
1&\tiny{Purkinje Cells}&\includegraphics[width=2in,keepaspectratio]{cellTypeProj1.eps}&\includegraphics[width=2in,keepaspectratio]{cellTypeModelFit1.eps}\\\hline
2&\tiny{Pyramidal Neurons}&\includegraphics[width=2in,keepaspectratio]{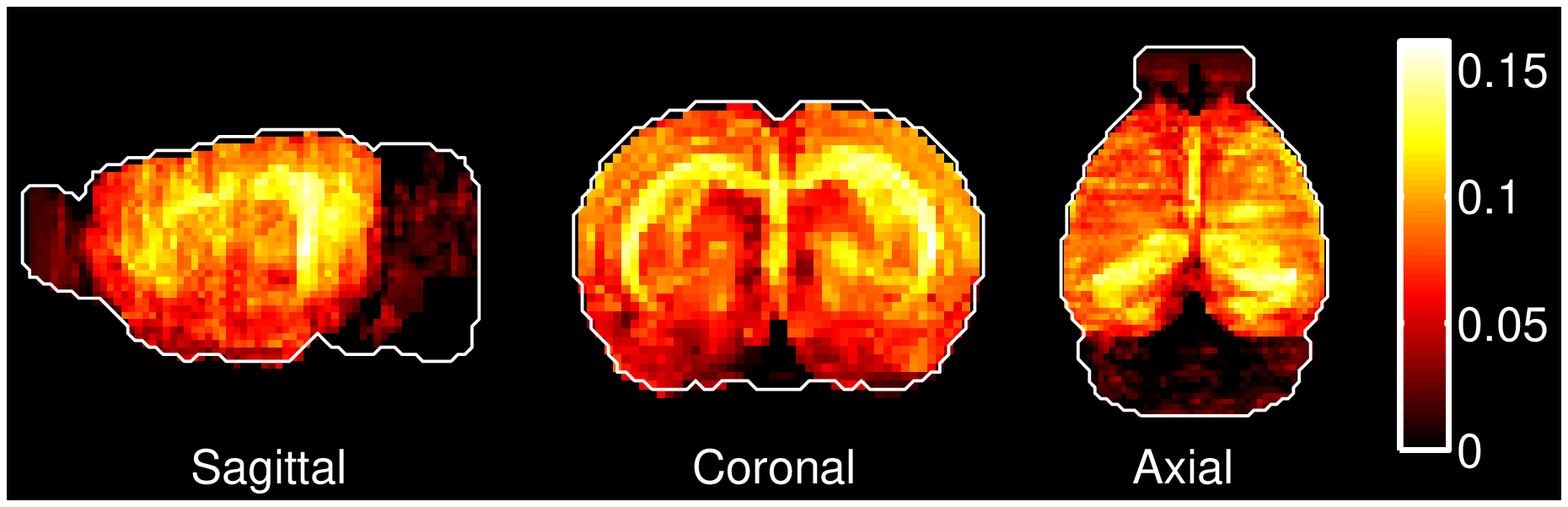}&\includegraphics[width=2in,keepaspectratio]{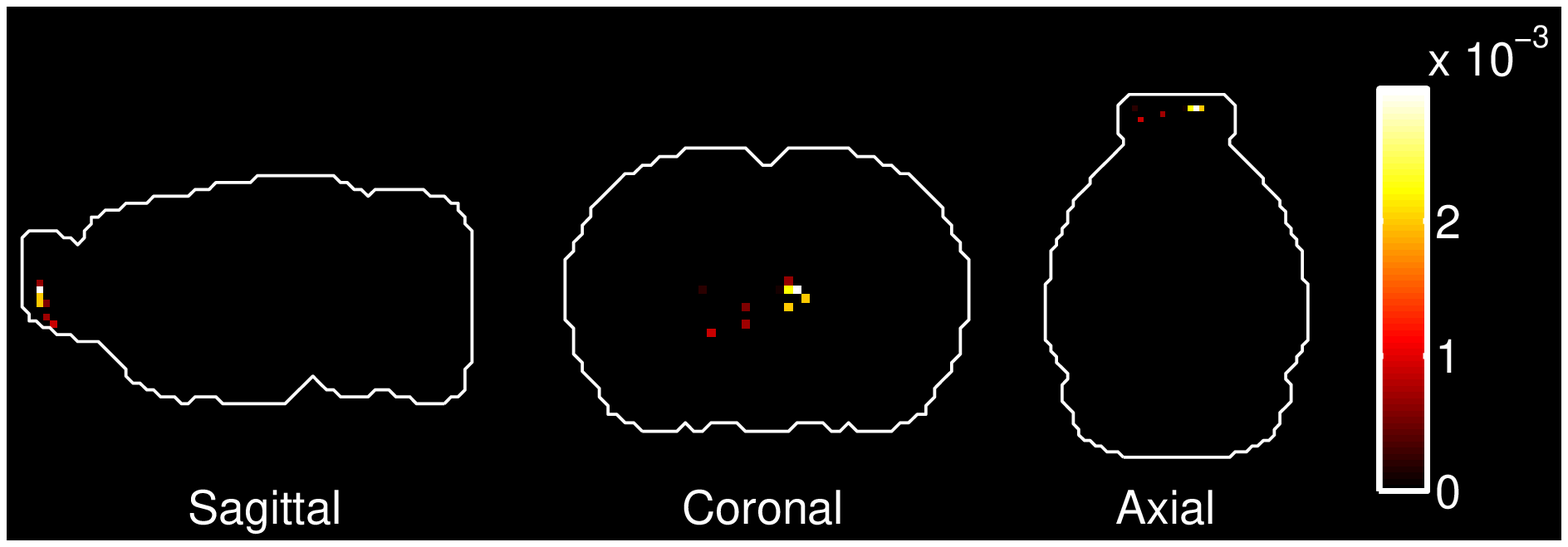}\\\hline
3&\tiny{Pyramidal Neurons}&\includegraphics[width=2in,keepaspectratio]{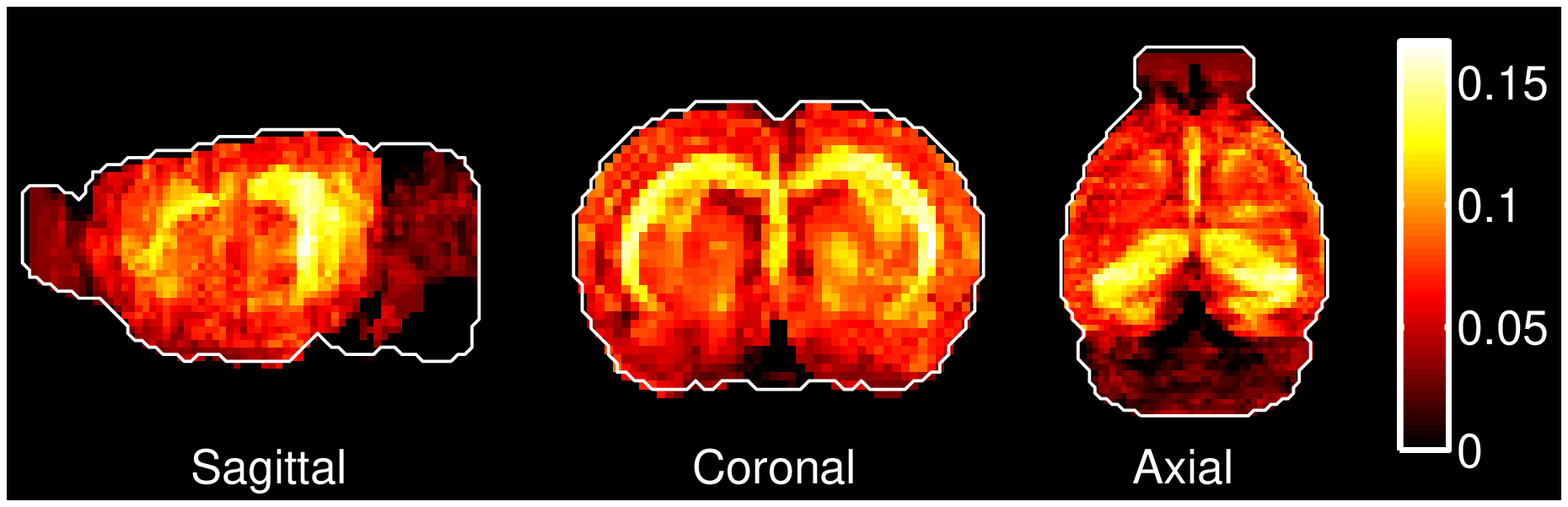}&\includegraphics[width=2in,keepaspectratio]{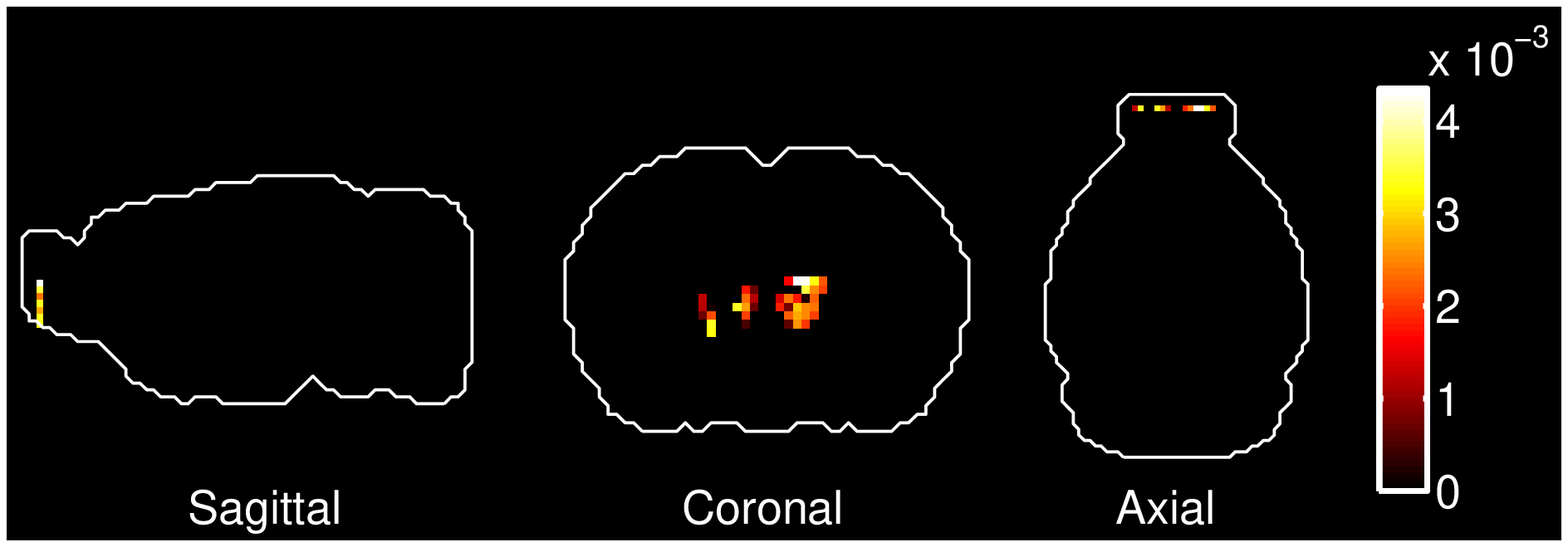}\\\hline
4&\tiny{A9 Dopaminergic Neurons}&\includegraphics[width=2in,keepaspectratio]{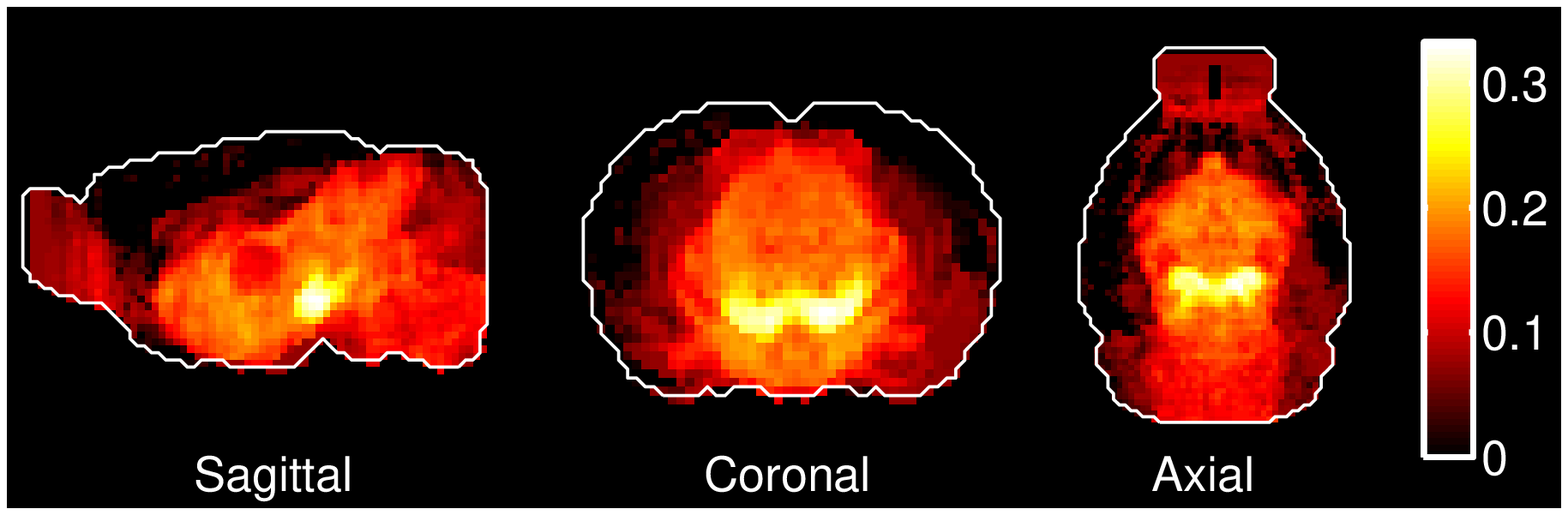}&\includegraphics[width=2in,keepaspectratio]{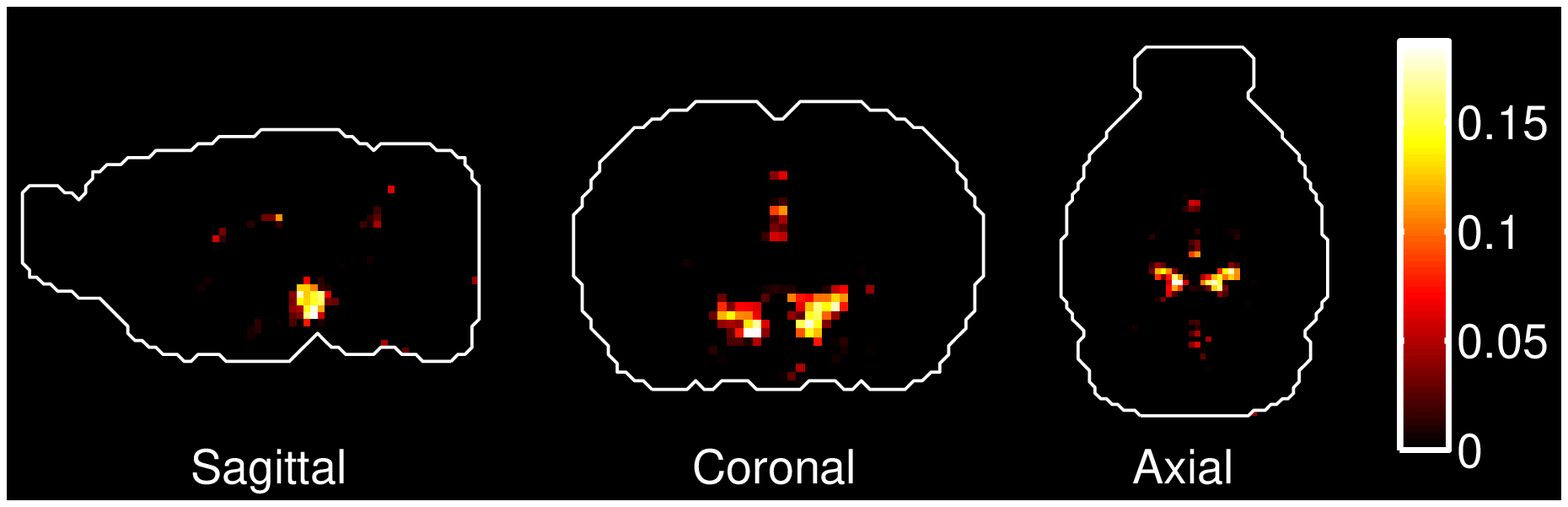}\\\hline
5&\tiny{A10 Dopaminergic Neurons}&\includegraphics[width=2in,keepaspectratio]{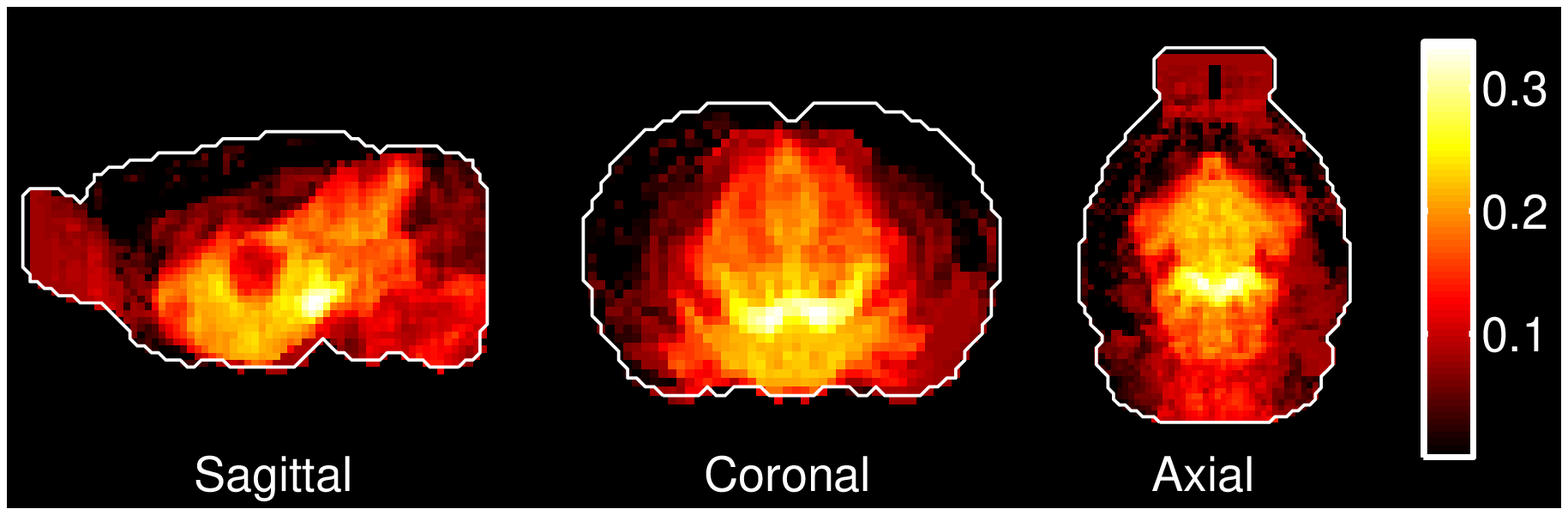}&\includegraphics[width=2in,keepaspectratio]{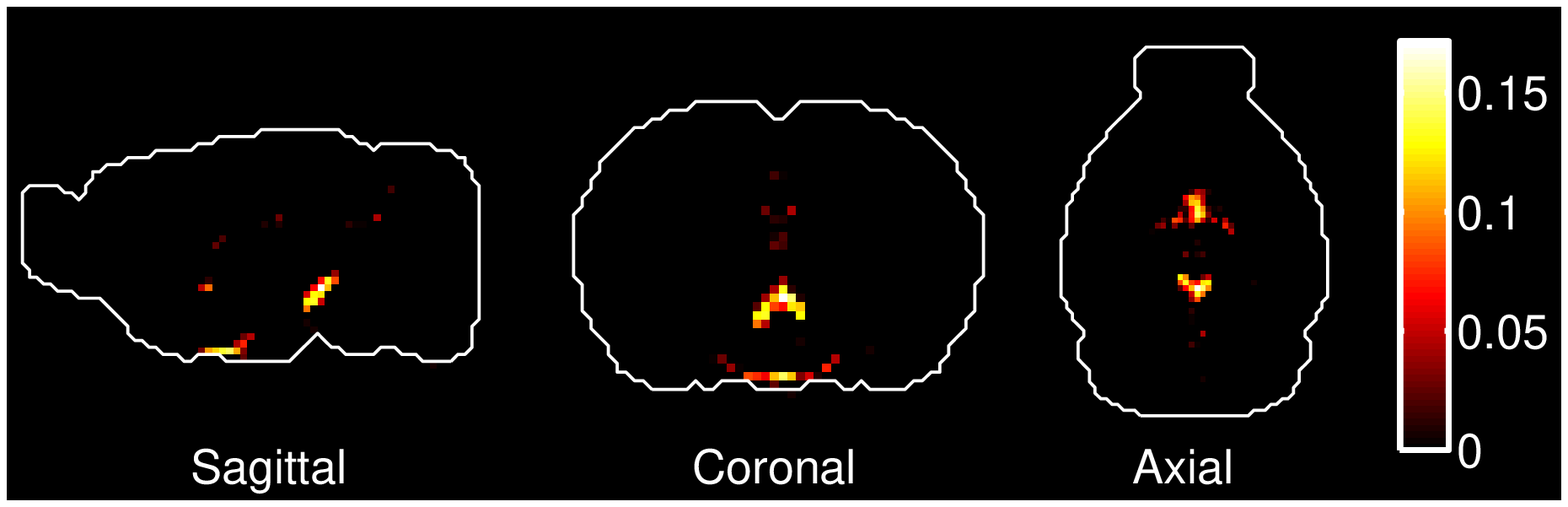}\\\hline
6&\tiny{Pyramidal Neurons}&\includegraphics[width=2in,keepaspectratio]{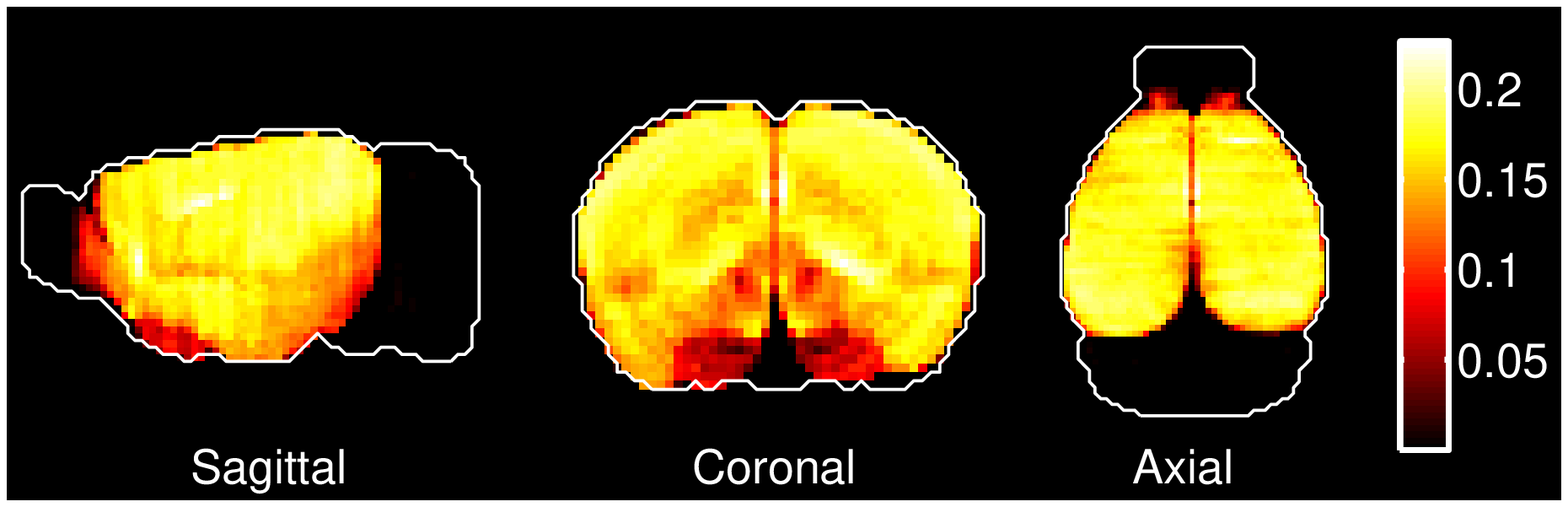}&\includegraphics[width=2in,keepaspectratio]{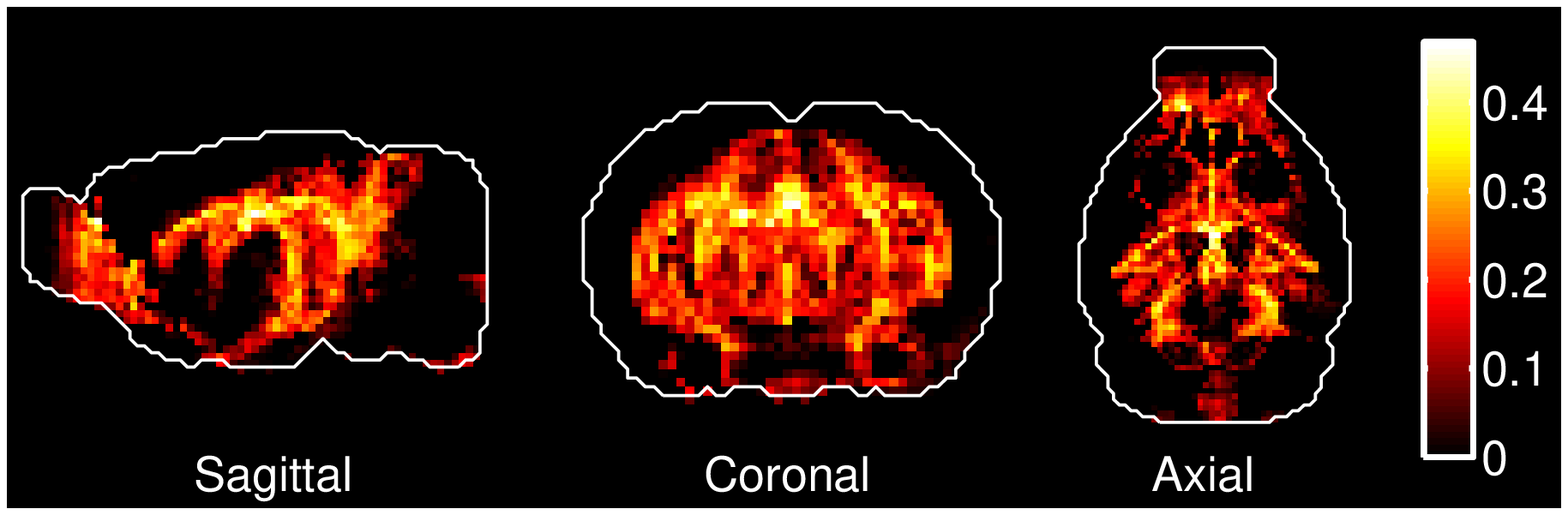}\\\hline
7&\tiny{Pyramidal Neurons}&\includegraphics[width=2in,keepaspectratio]{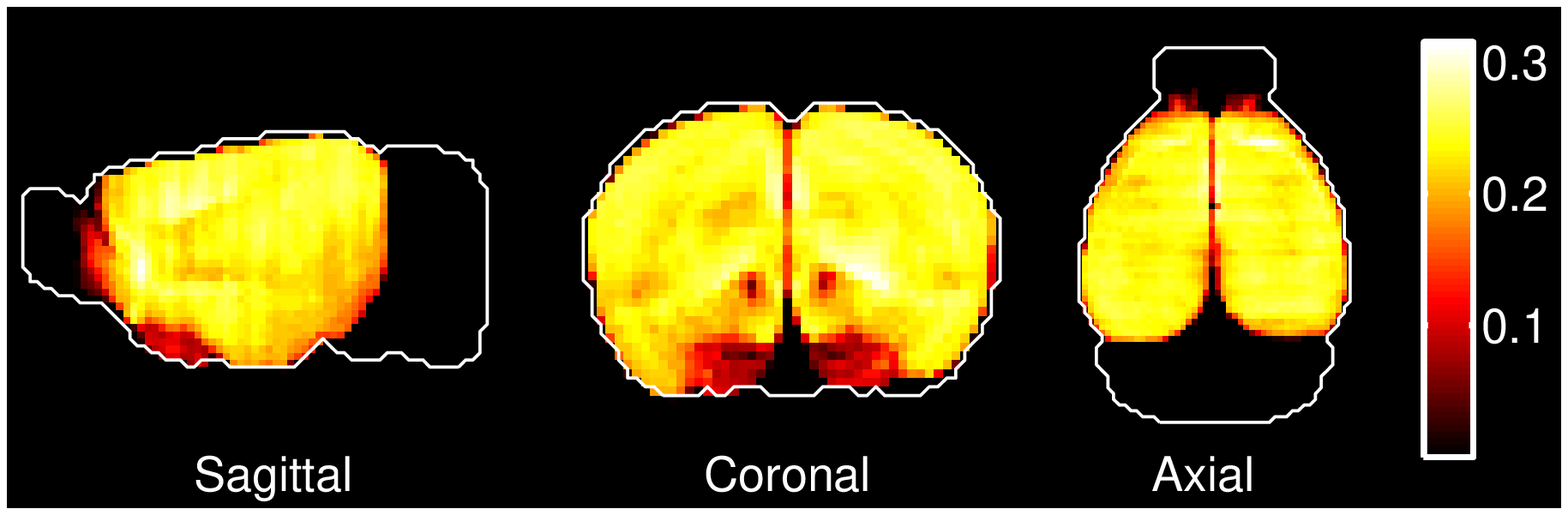}&\includegraphics[width=2in,keepaspectratio]{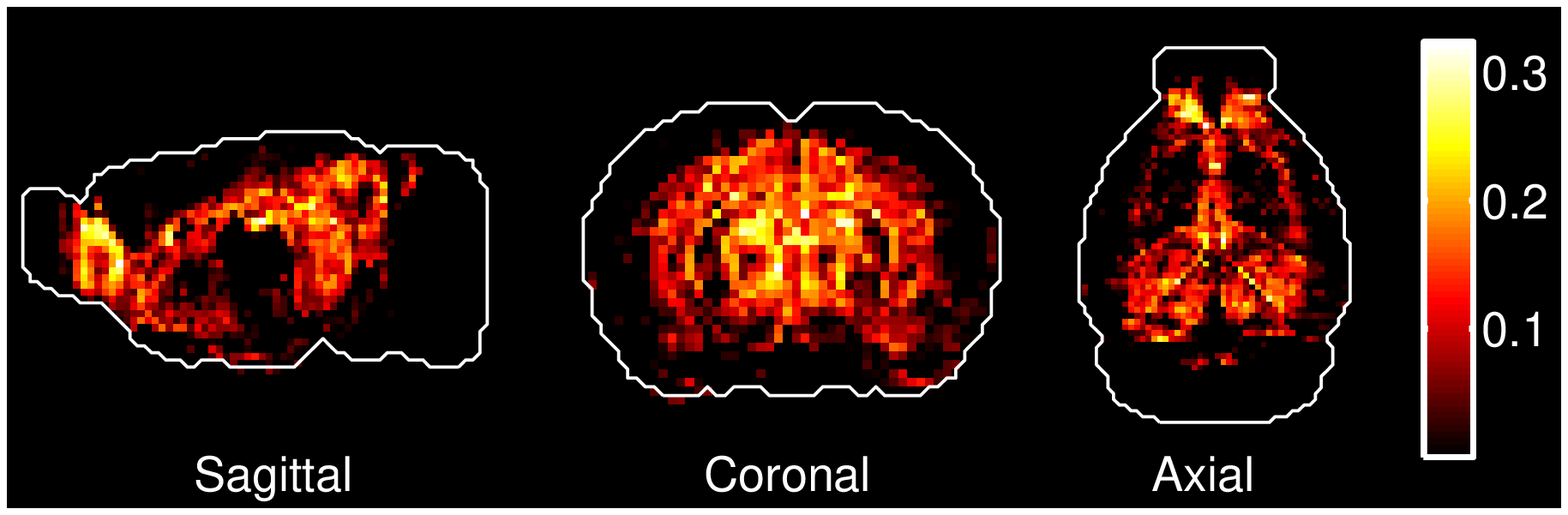}\\\hline
8&\tiny{Pyramidal Neurons}&\includegraphics[width=2in,keepaspectratio]{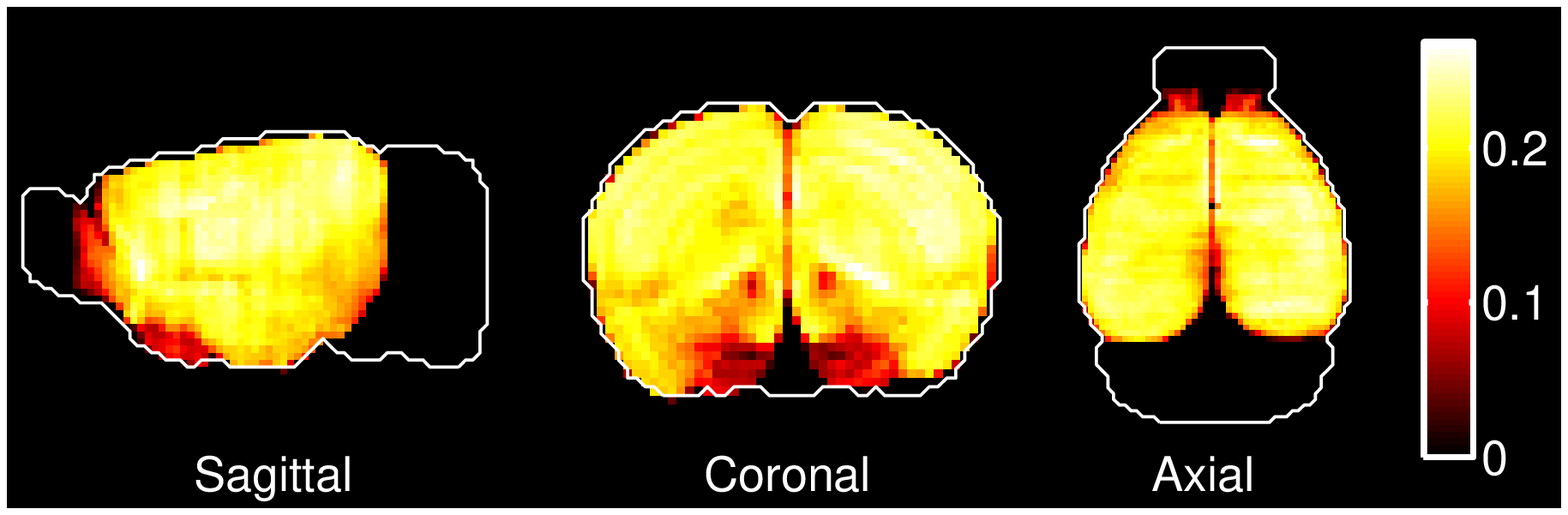}&\includegraphics[width=2in,keepaspectratio]{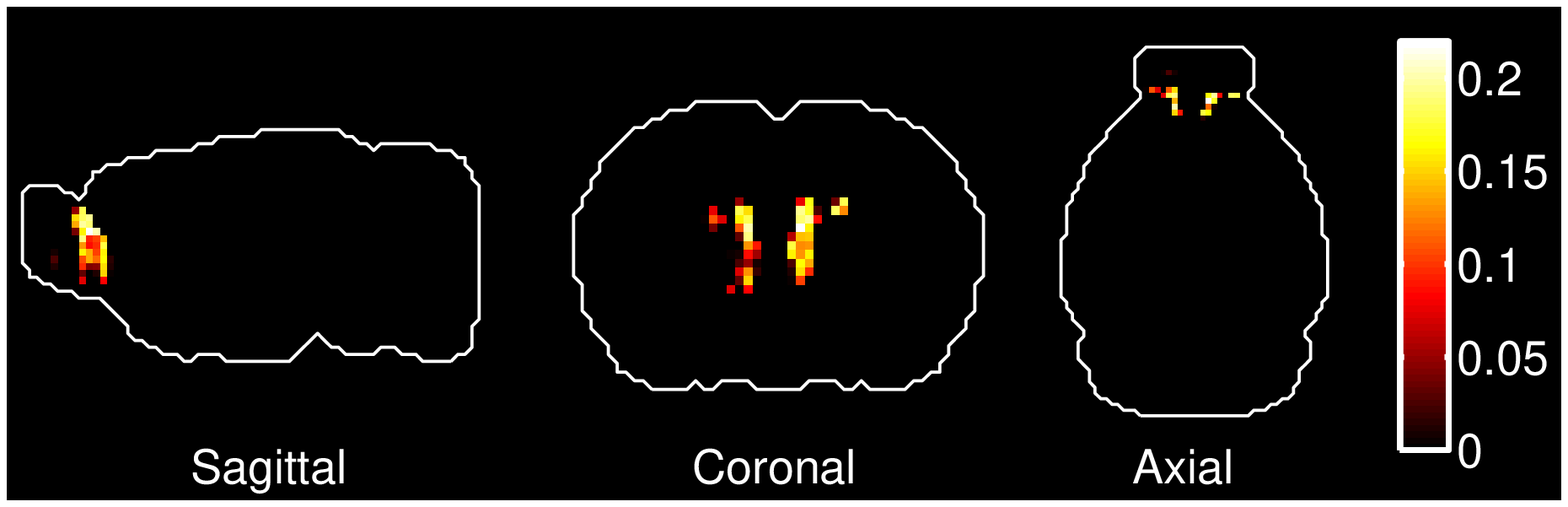}\\\hline
9&\tiny{Mixed Neurons}&\includegraphics[width=2in,keepaspectratio]{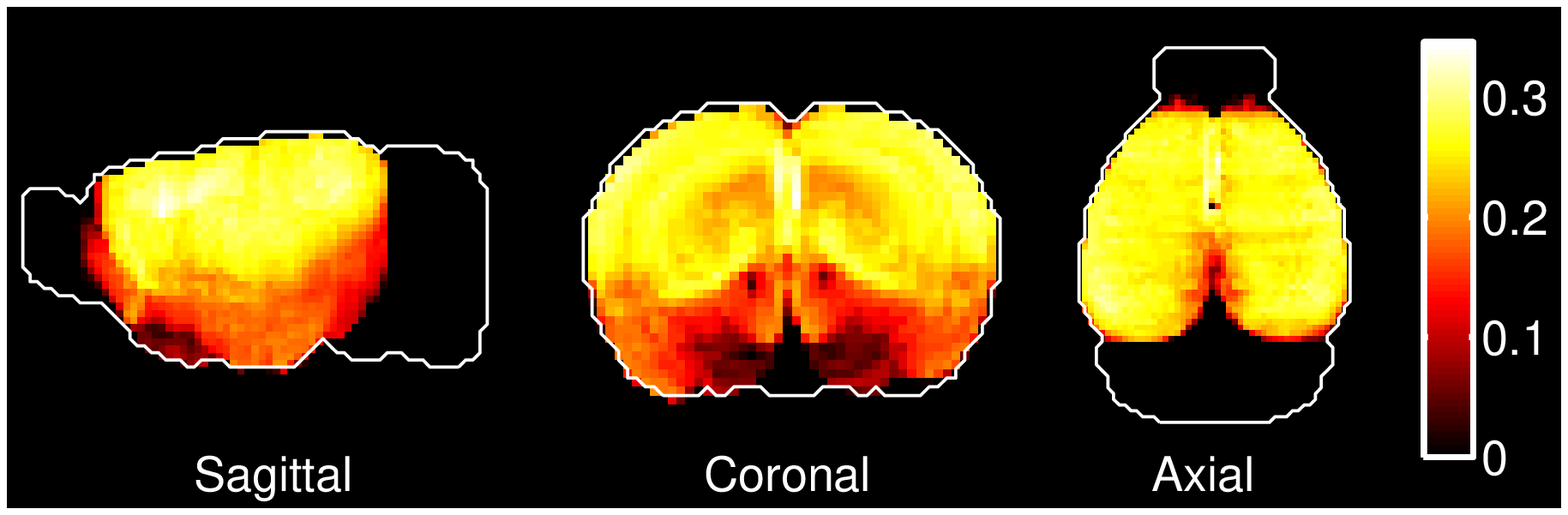}&\includegraphics[width=2in,keepaspectratio]{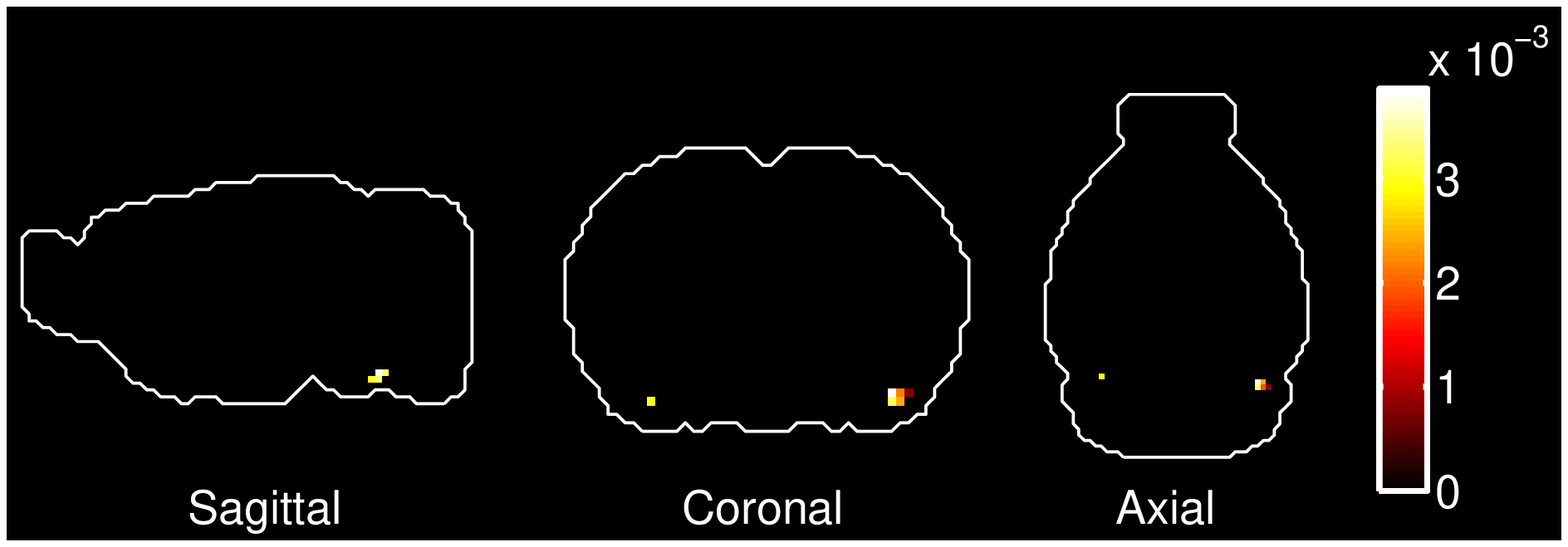}\\\hline
10&\tiny{Motor Neurons, Midbrain Cholinergic Neurons}&\includegraphics[width=2in,keepaspectratio]{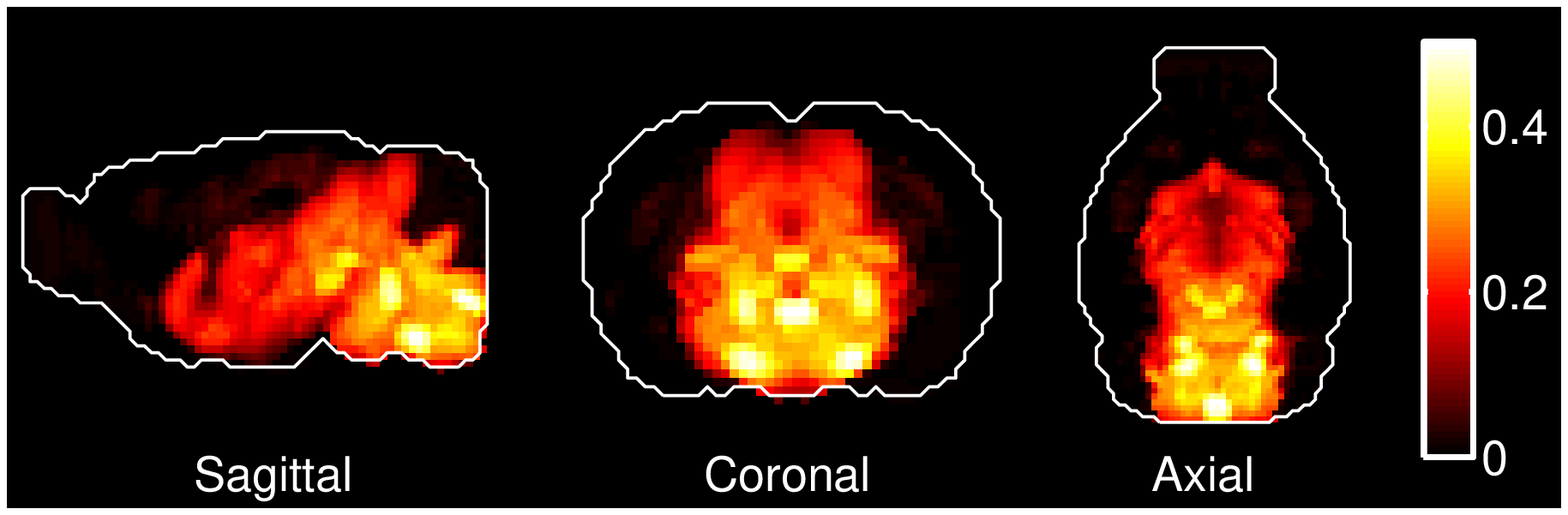}&\includegraphics[width=2in,keepaspectratio]{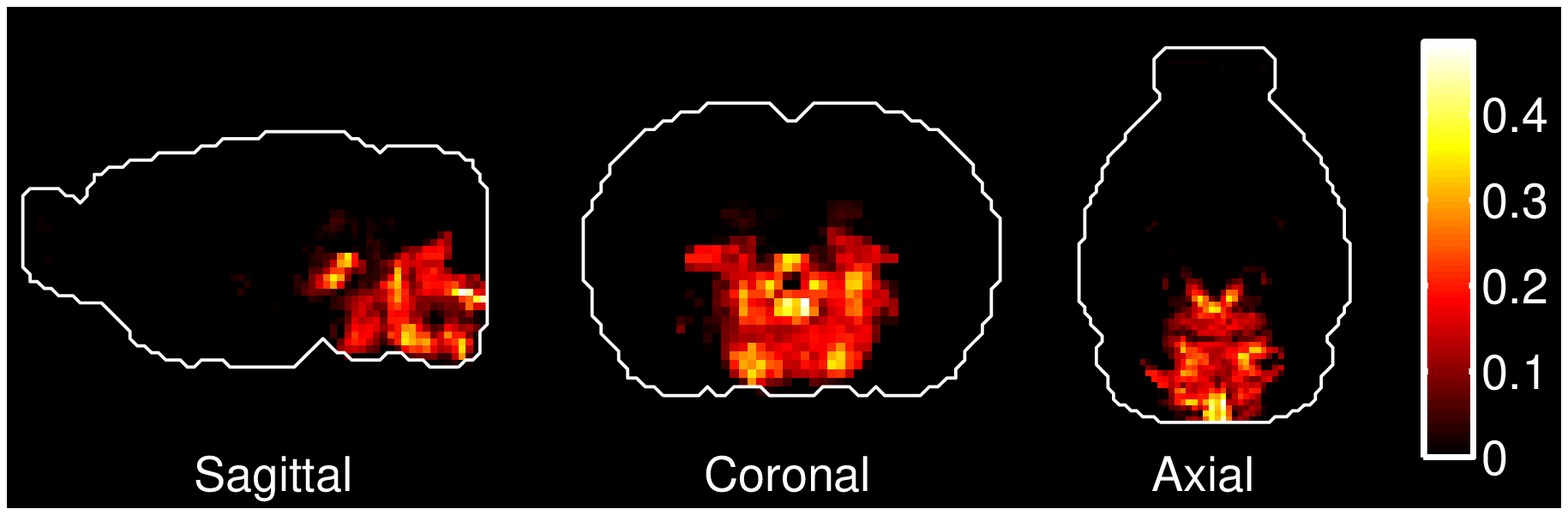}\\\hline
11&\tiny{Cholinergic Projection Neurons}&\includegraphics[width=2in,keepaspectratio]{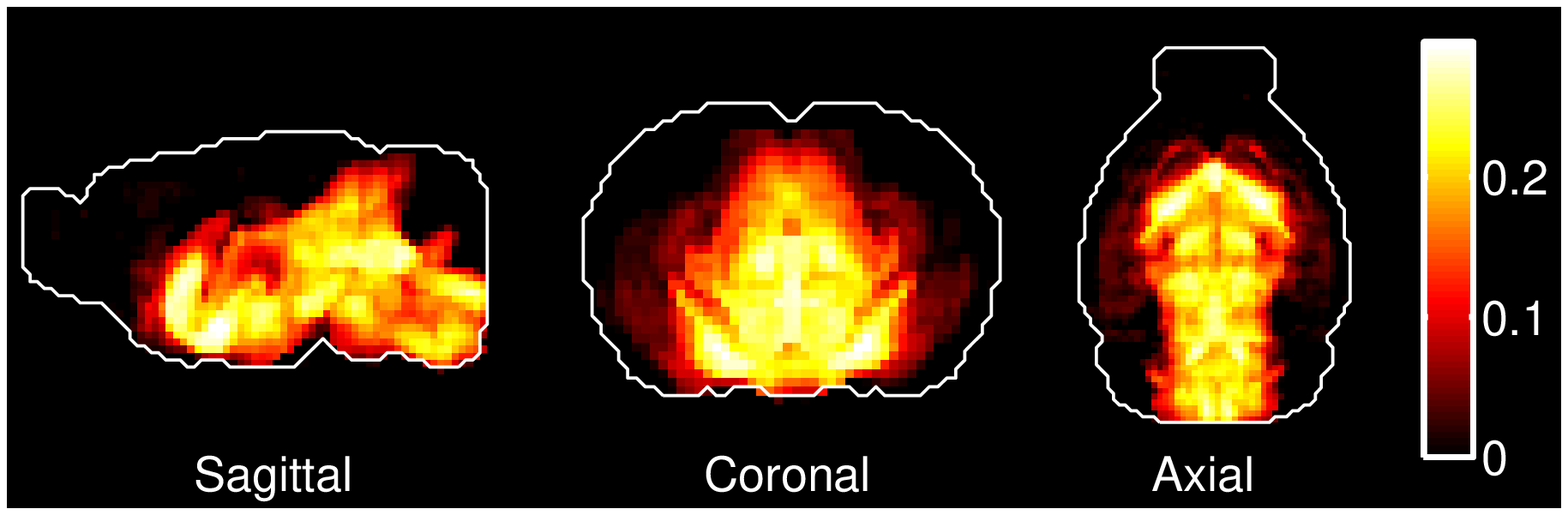}&\includegraphics[width=2in,keepaspectratio]{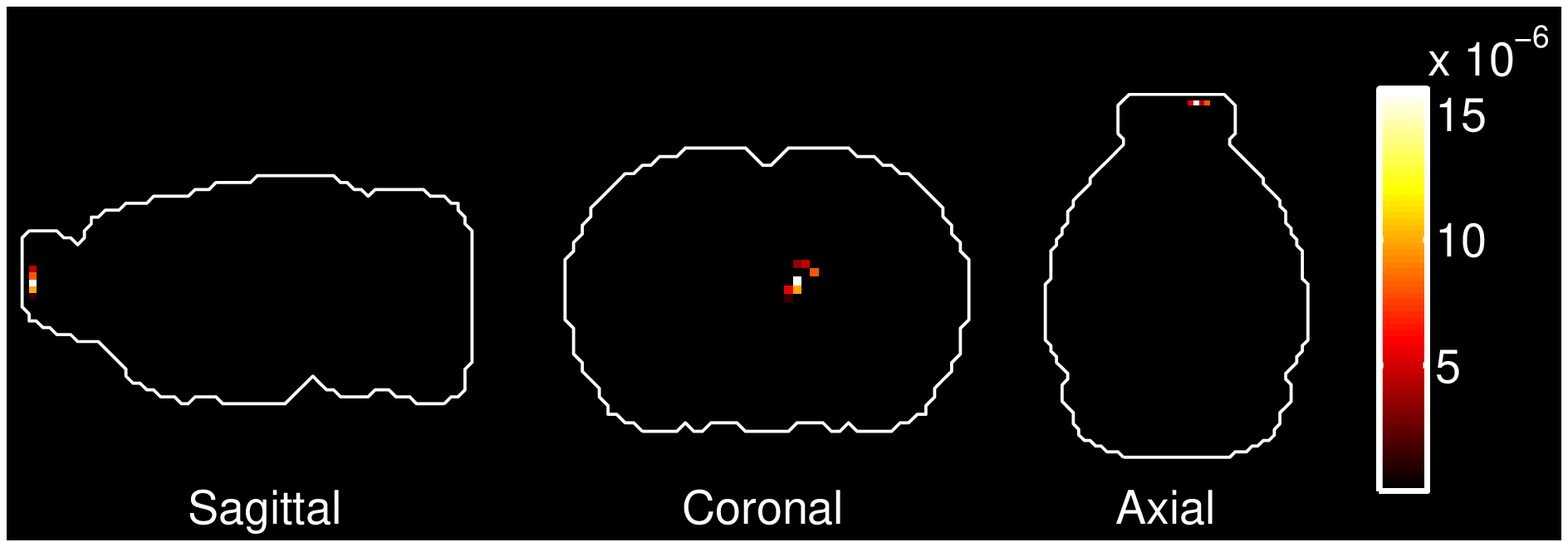}\\\hline
12&\tiny{Motor Neurons, Cholinergic Interneurons}&\includegraphics[width=2in,keepaspectratio]{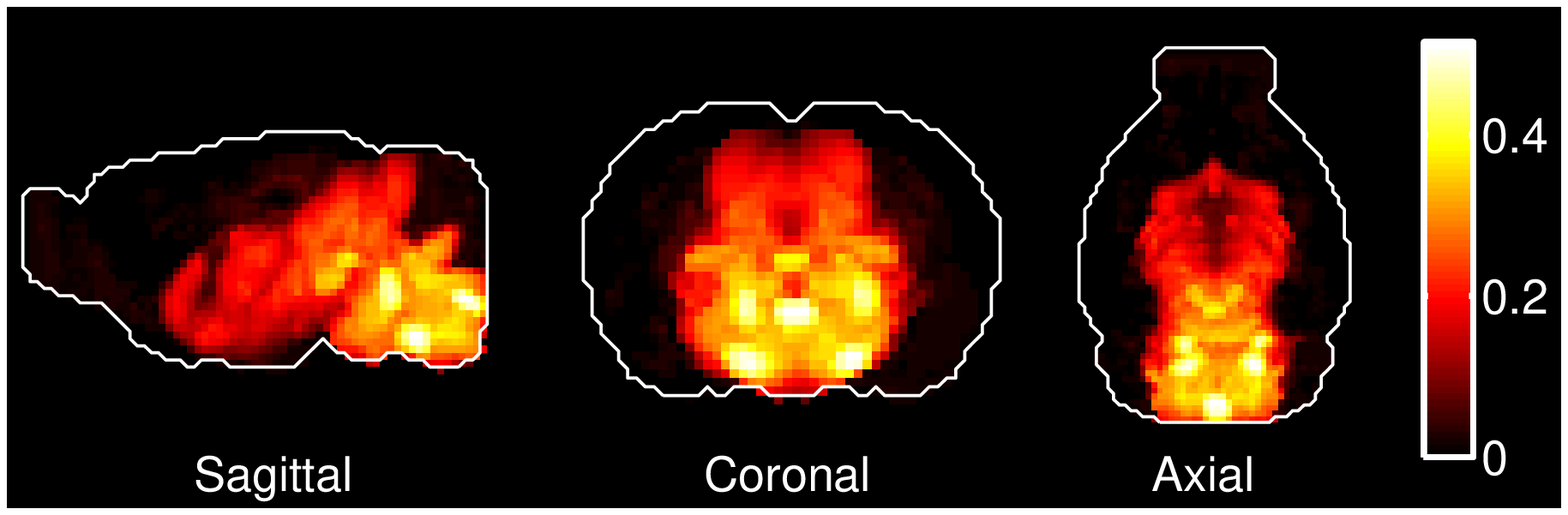}&\includegraphics[width=2in,keepaspectratio]{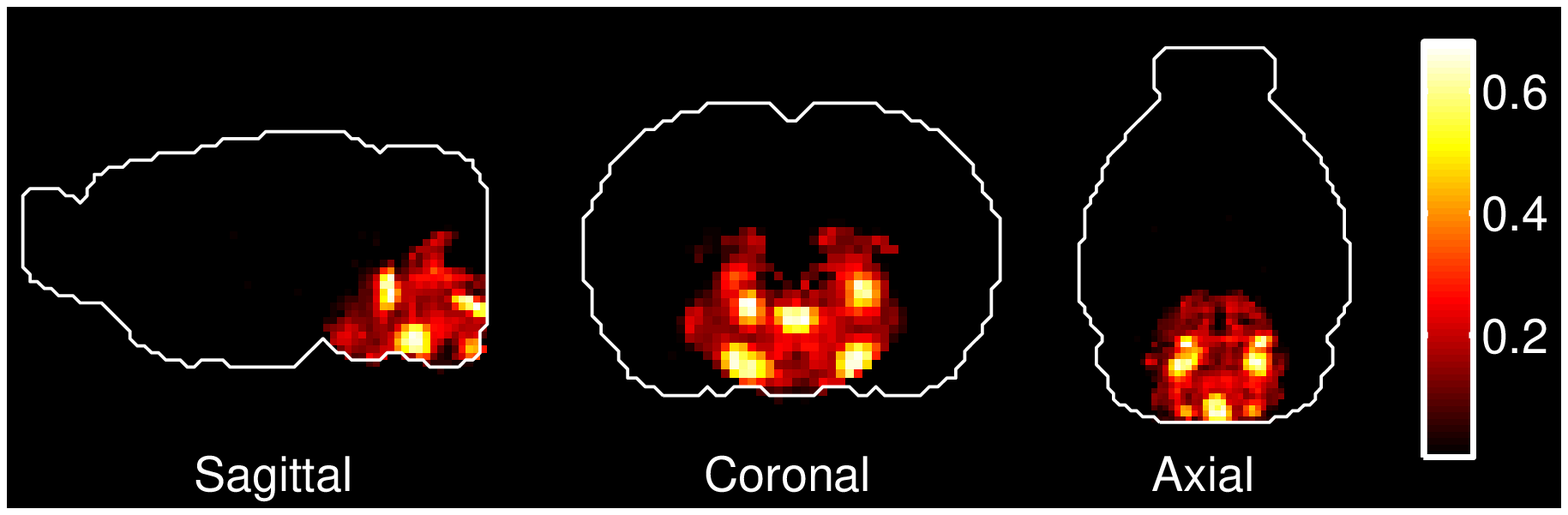}\\\hline
\end{tabular}
\\
\begin{tabular}{|l|l|l|l|}
\hline
\textbf{index}&\textbf{Cell type}&\textbf{Heat map of correlations}&\textbf{Heat map of weight}\\\hline
13&\tiny{Cholinergic Neurons}&\includegraphics[width=2in,keepaspectratio]{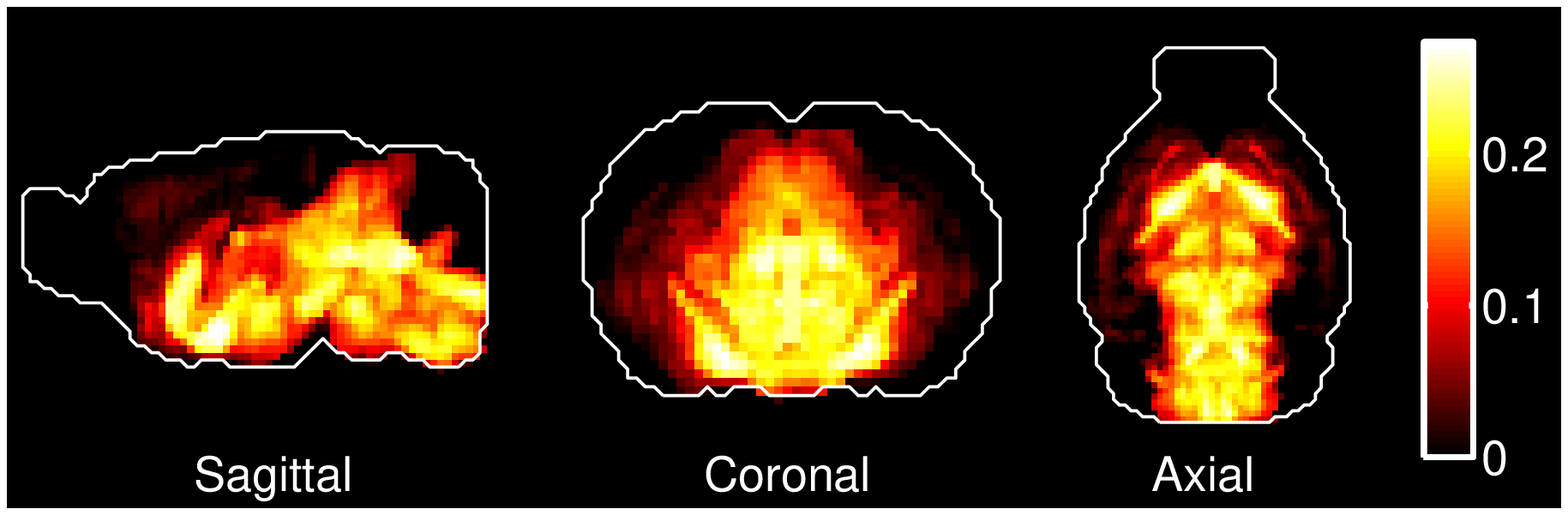}&\includegraphics[width=2in,keepaspectratio]{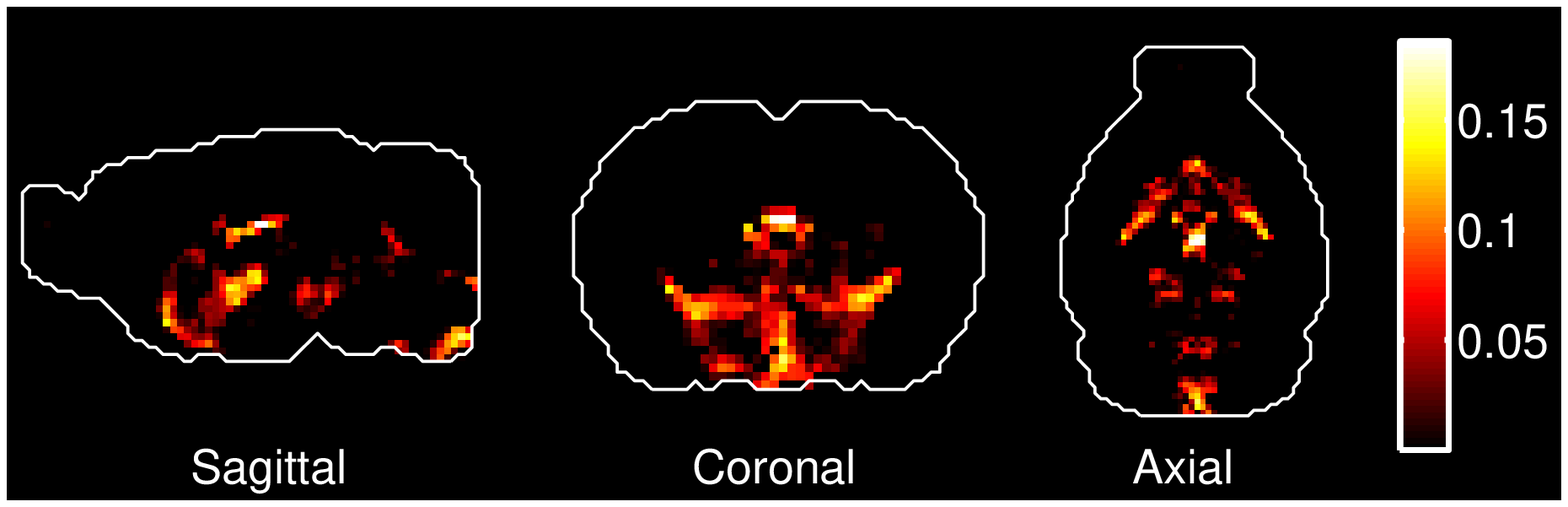}\\\hline
14&\tiny{Interneurons}&\includegraphics[width=2in,keepaspectratio]{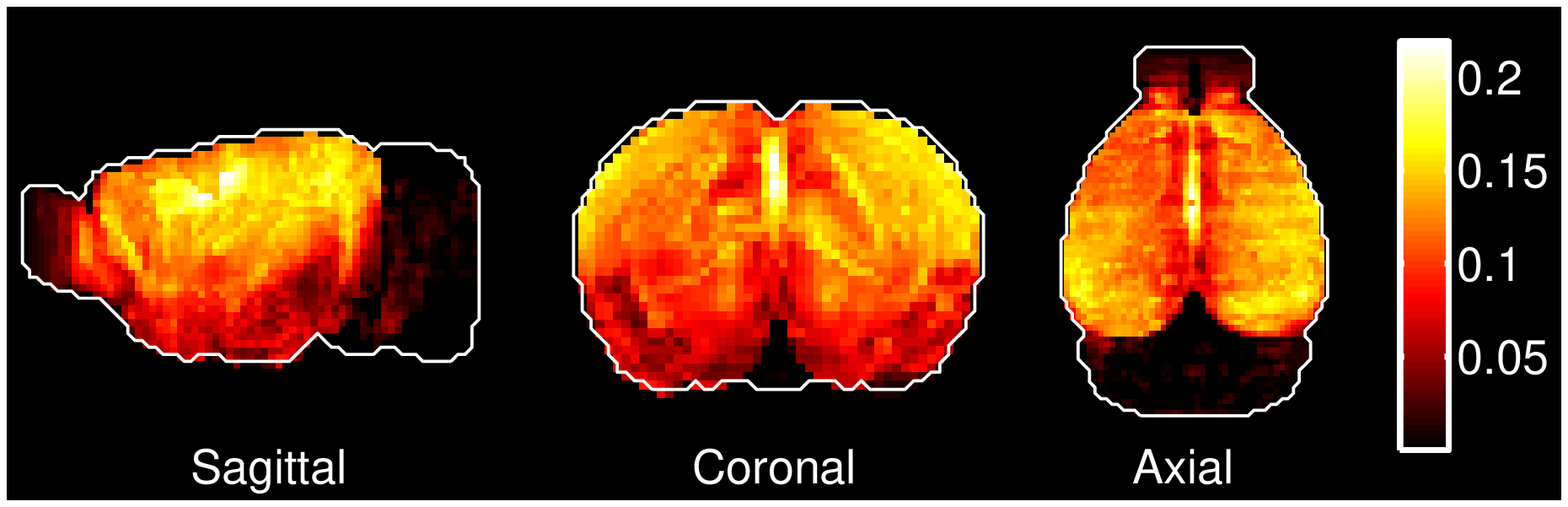}&\includegraphics[width=2in,keepaspectratio]{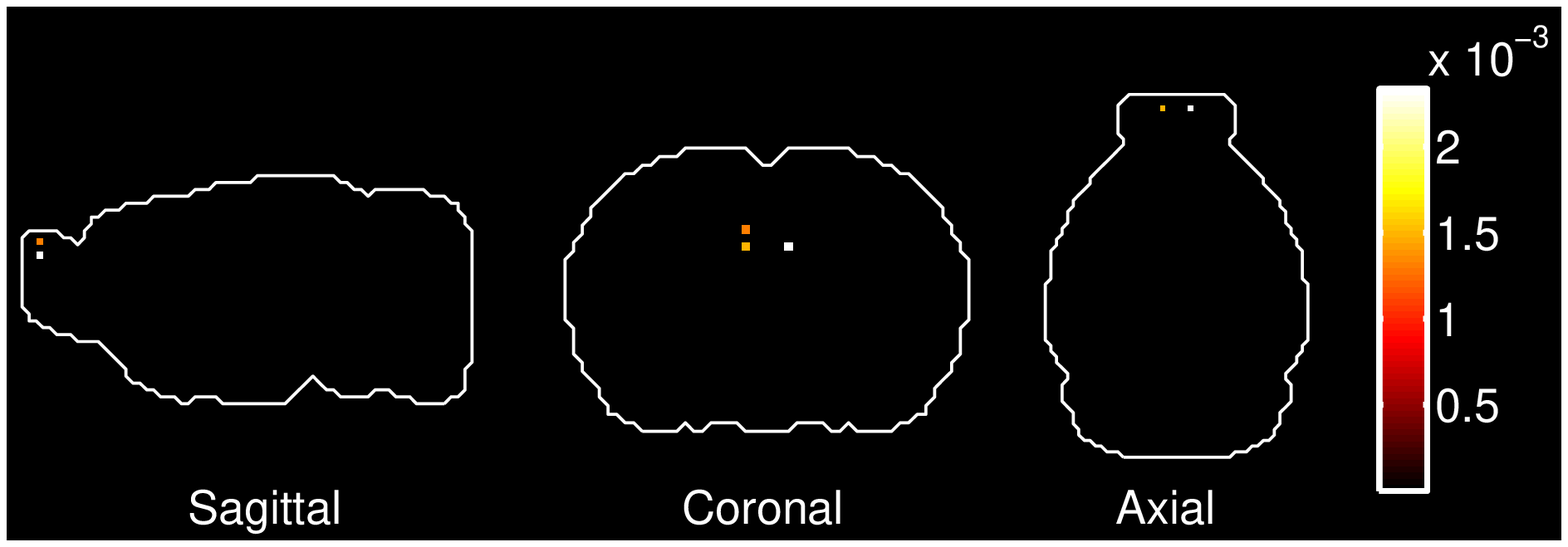}\\\hline
15&\tiny{Drd1+ Medium Spiny Neurons}&\includegraphics[width=2in,keepaspectratio]{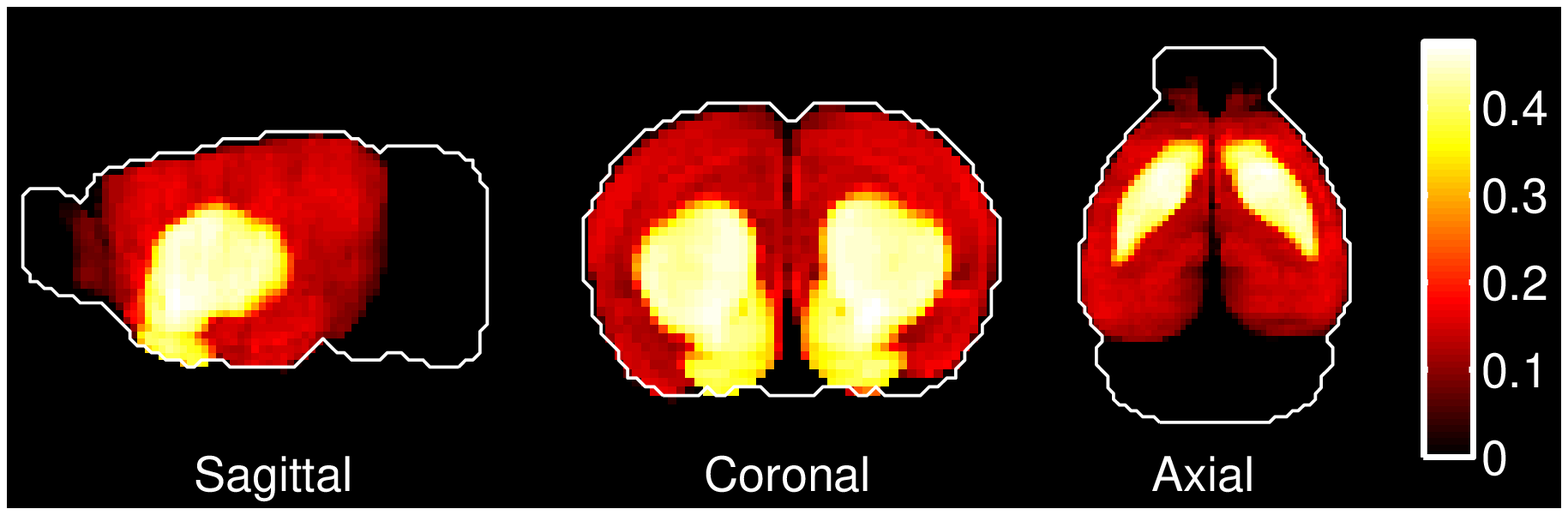}&\includegraphics[width=2in,keepaspectratio]{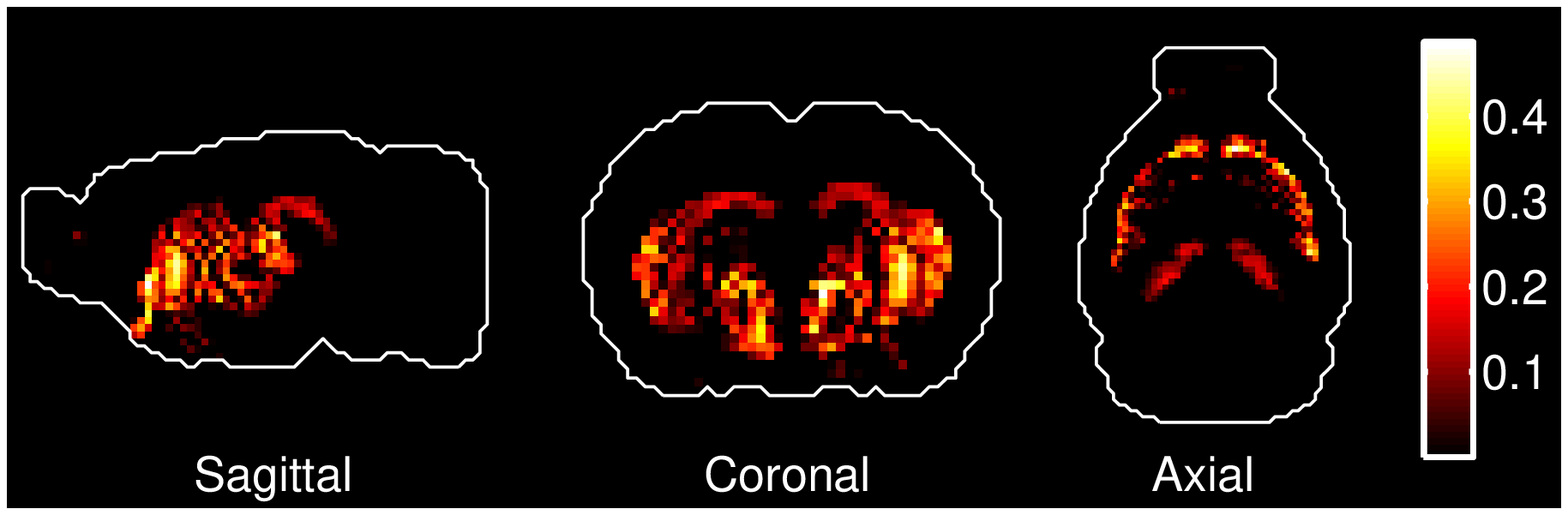}\\\hline
16&\tiny{Drd2+ Medium Spiny Neurons}&\includegraphics[width=2in,keepaspectratio]{cellTypeProj16.eps}&\includegraphics[width=2in,keepaspectratio]{cellTypeModelFit16.eps}\\\hline
17&\tiny{Golgi Cells}&\includegraphics[width=2in,keepaspectratio]{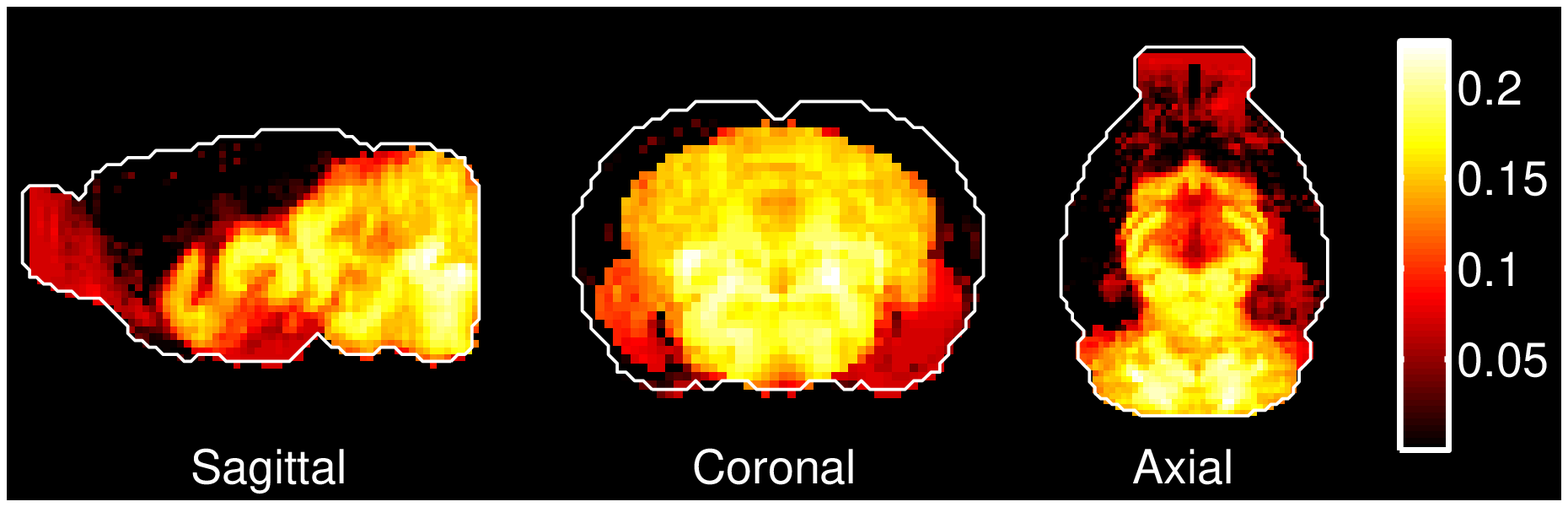}&\includegraphics[width=2in,keepaspectratio]{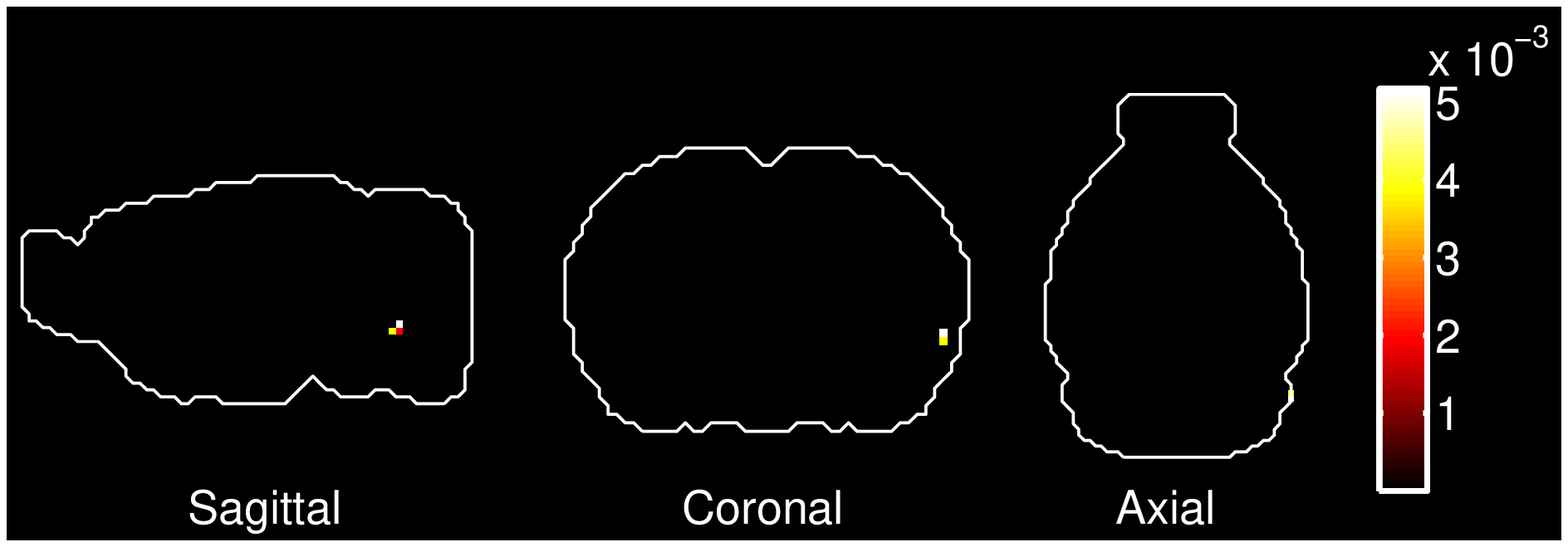}\\\hline
18&\tiny{Unipolar Brush cells (some Bergman Glia)}&\includegraphics[width=2in,keepaspectratio]{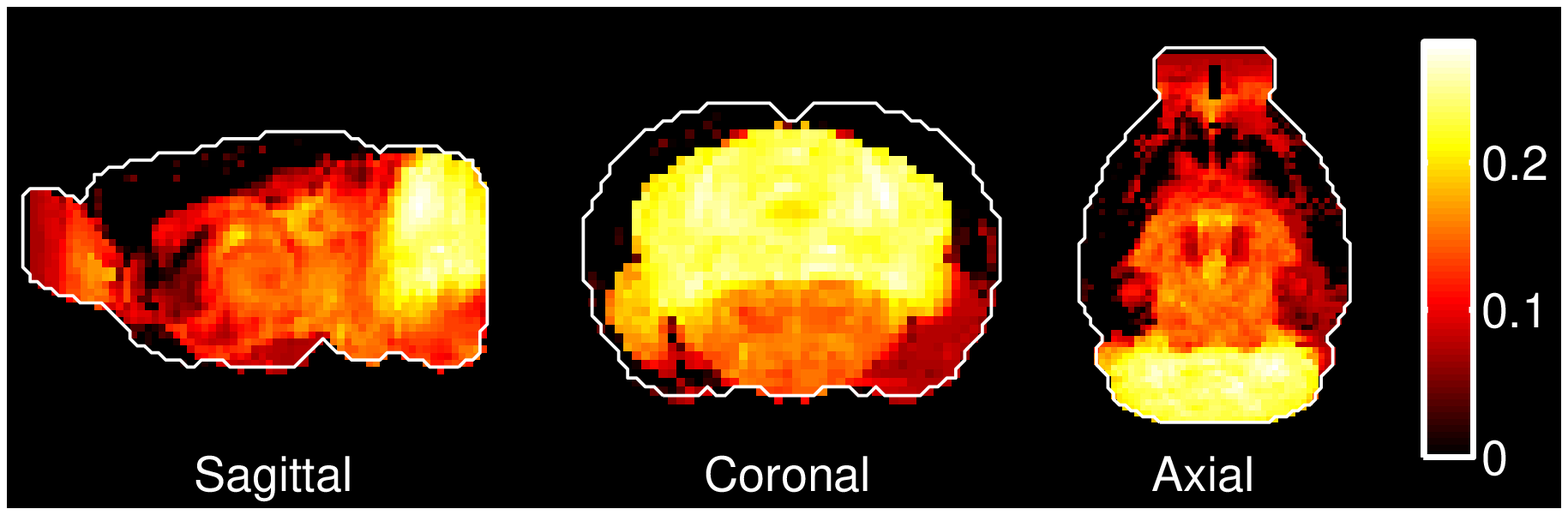}&\includegraphics[width=2in,keepaspectratio]{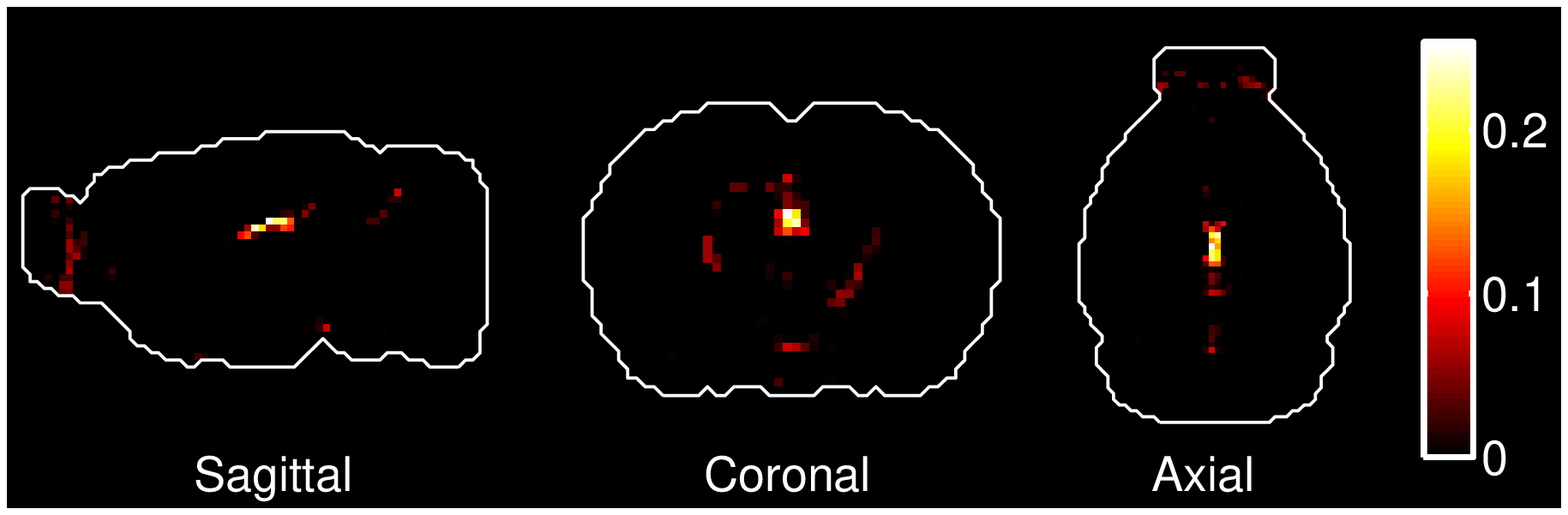}\\\hline
19&\tiny{Stellate Basket Cells}&\includegraphics[width=2in,keepaspectratio]{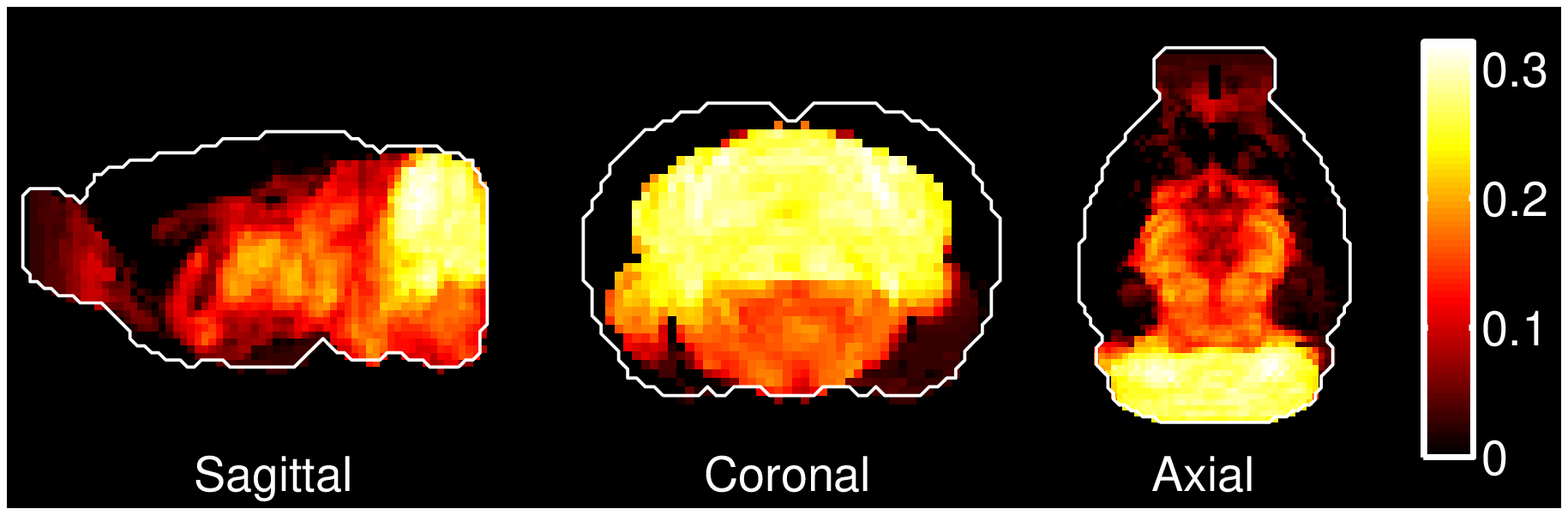}&\includegraphics[width=2in,keepaspectratio]{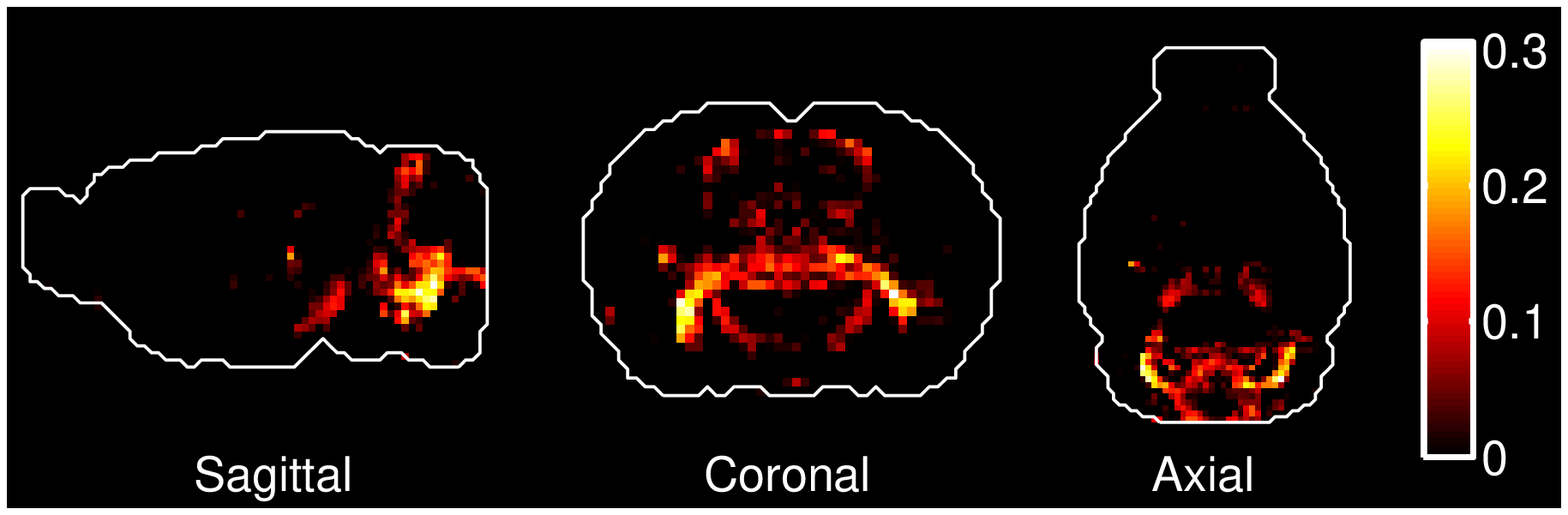}\\\hline
20&\tiny{Granule Cells}&\includegraphics[width=2in,keepaspectratio]{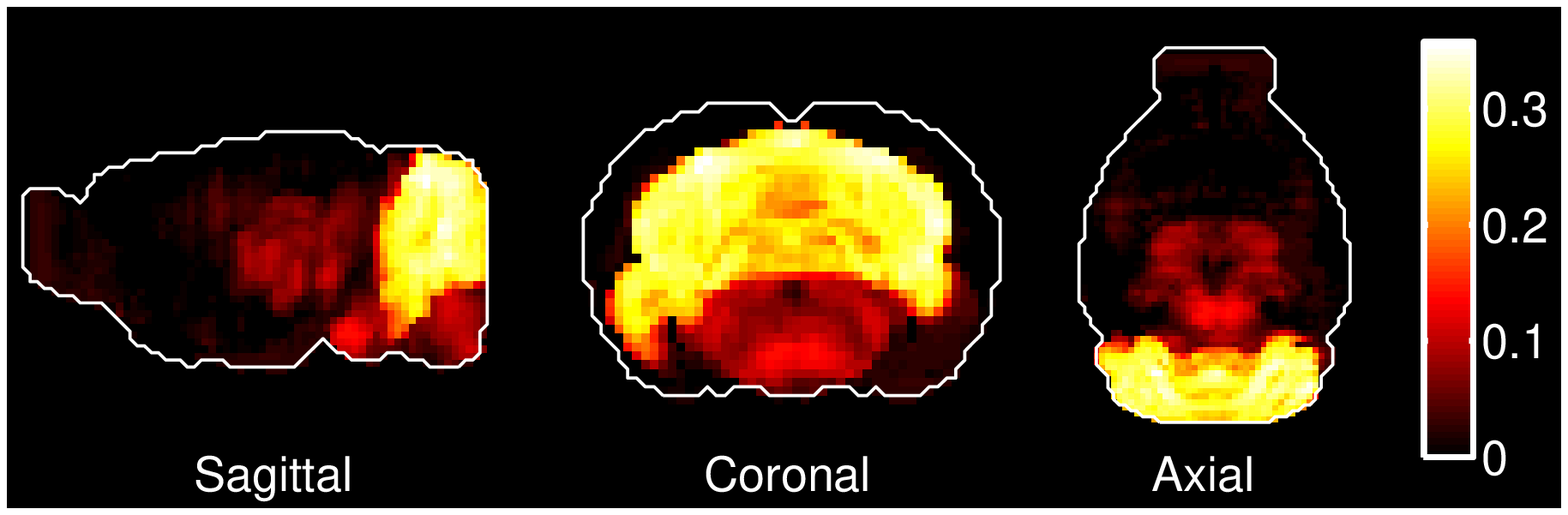}&\includegraphics[width=2in,keepaspectratio]{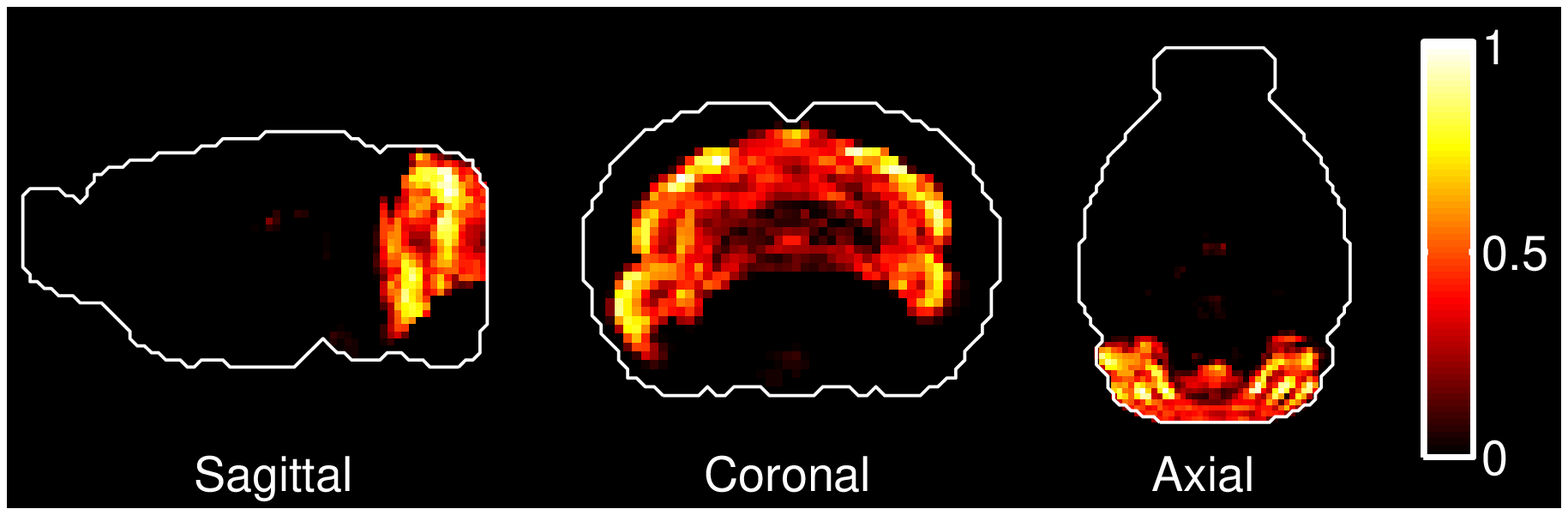}\\\hline
21&\tiny{Mature Oligodendrocytes}&\includegraphics[width=2in,keepaspectratio]{cellTypeProj21.eps}&\includegraphics[width=2in,keepaspectratio]{cellTypeModelFit21.eps}\\\hline
22&\tiny{Mature Oligodendrocytes}&\includegraphics[width=2in,keepaspectratio]{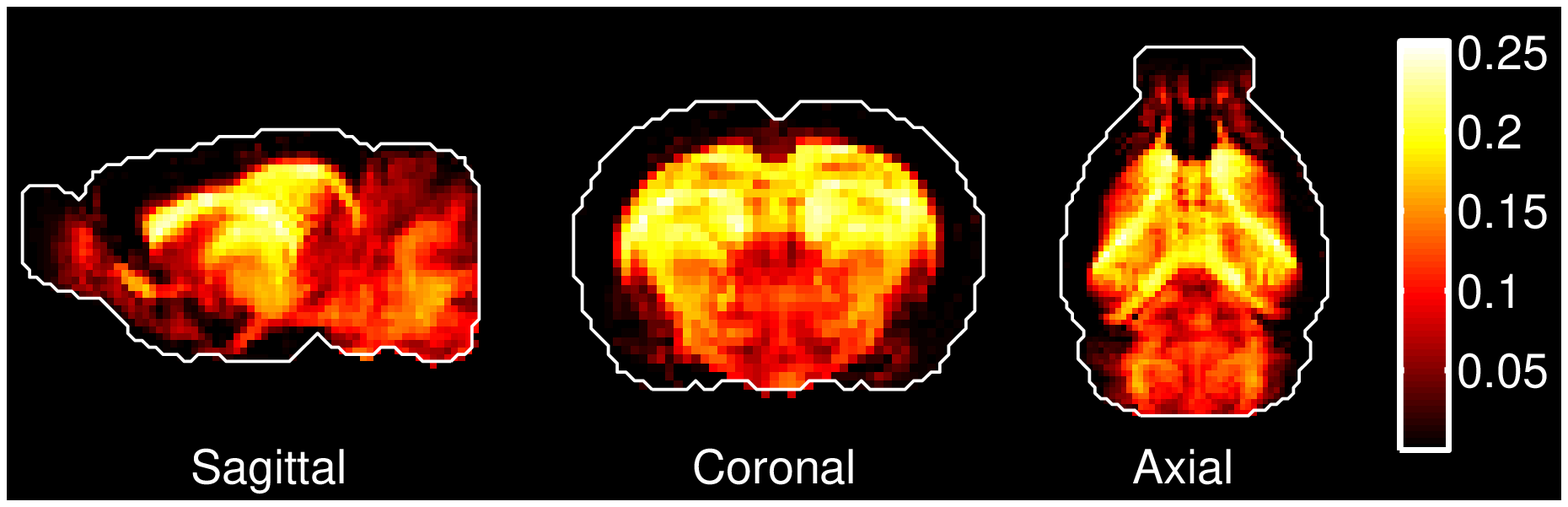}&\includegraphics[width=2in,keepaspectratio]{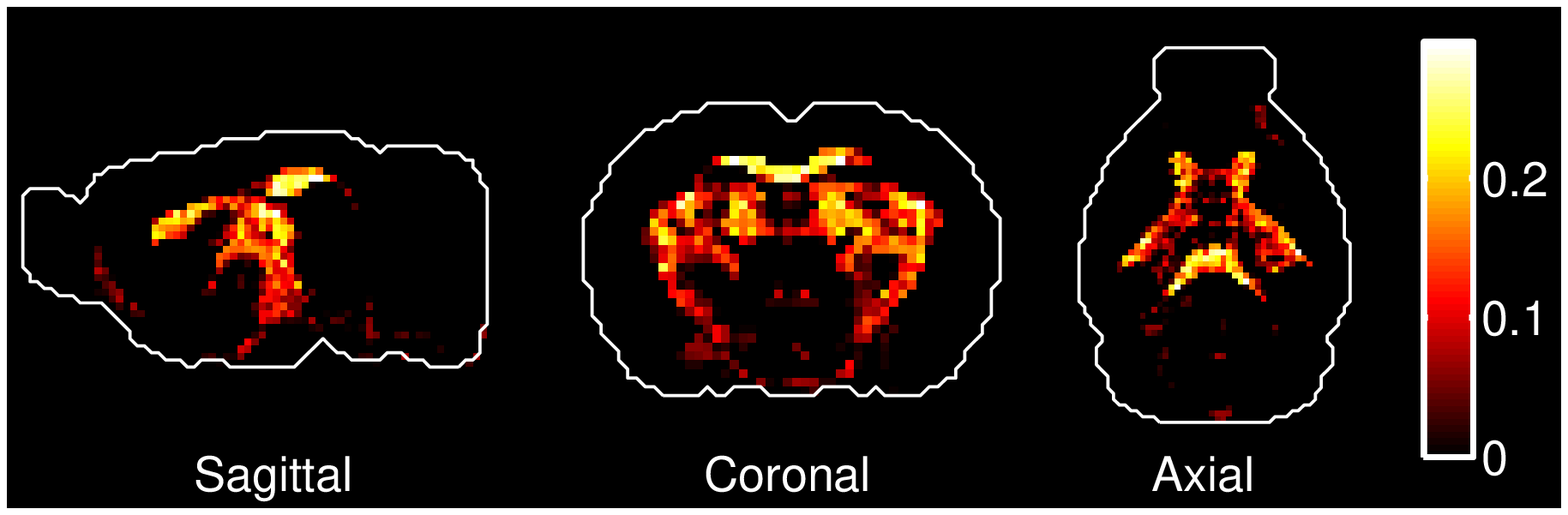}\\\hline
23&\tiny{Mixed Oligodendrocytes}&\includegraphics[width=2in,keepaspectratio]{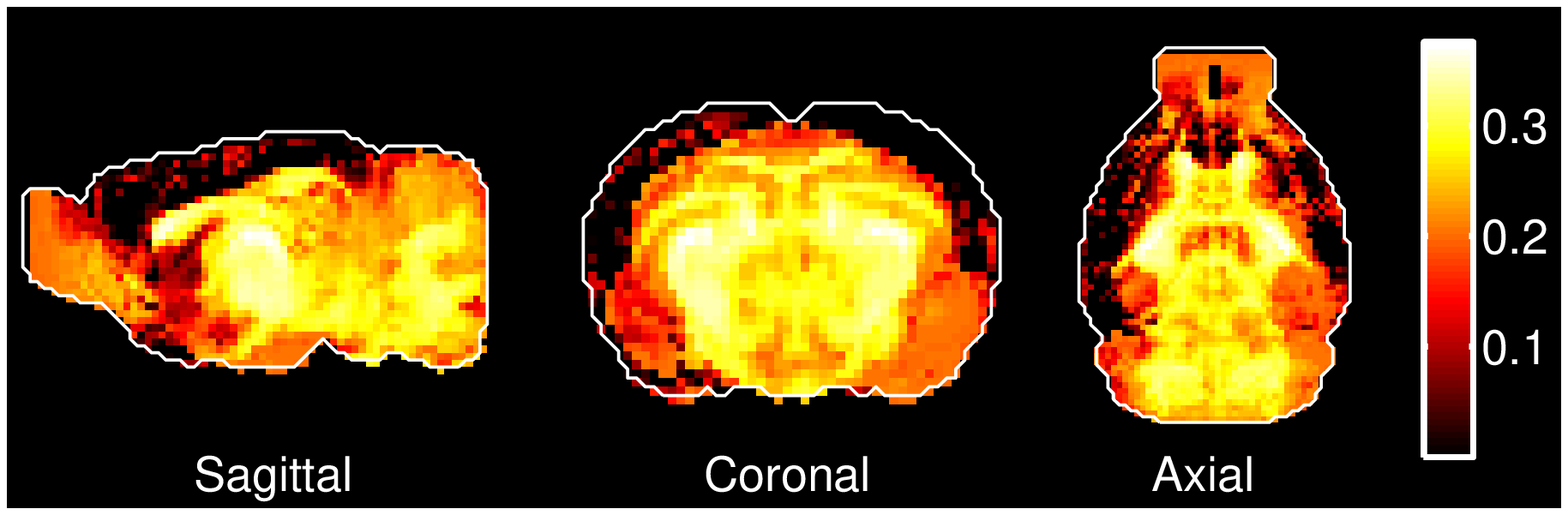}&N/A\\\hline
24&\tiny{Mixed Oligodendrocytes}&\includegraphics[width=2in,keepaspectratio]{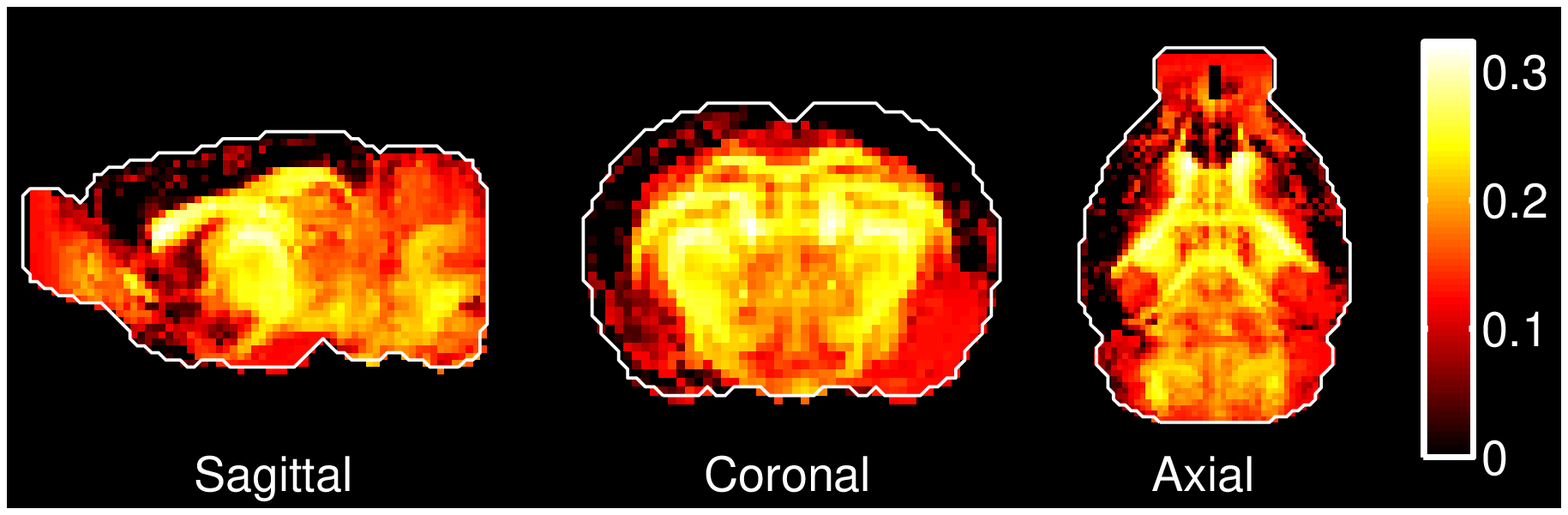}&\includegraphics[width=2in,keepaspectratio]{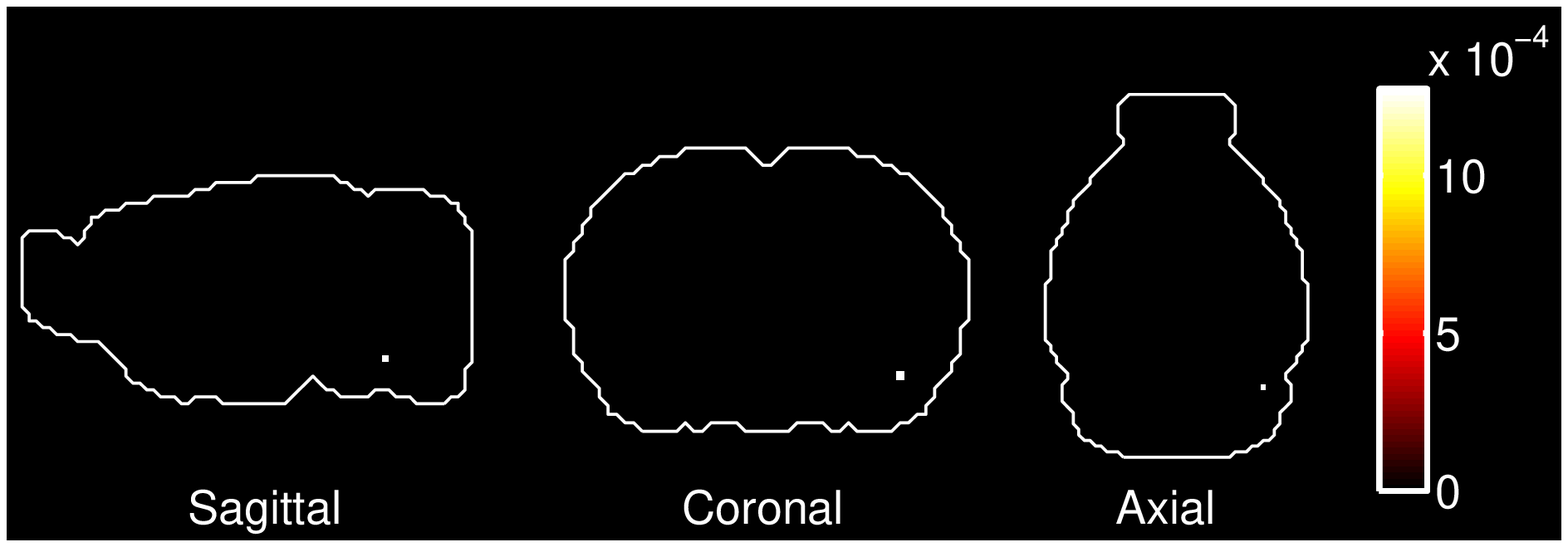}\\\hline
\end{tabular}
\\
\begin{tabular}{|l|l|l|l|}
\hline
\textbf{index}&\textbf{Cell type}&\textbf{Heat map of correlations}&\textbf{Heat map of weight}\\\hline
25&\tiny{Purkinje Cells}&\includegraphics[width=2in,keepaspectratio]{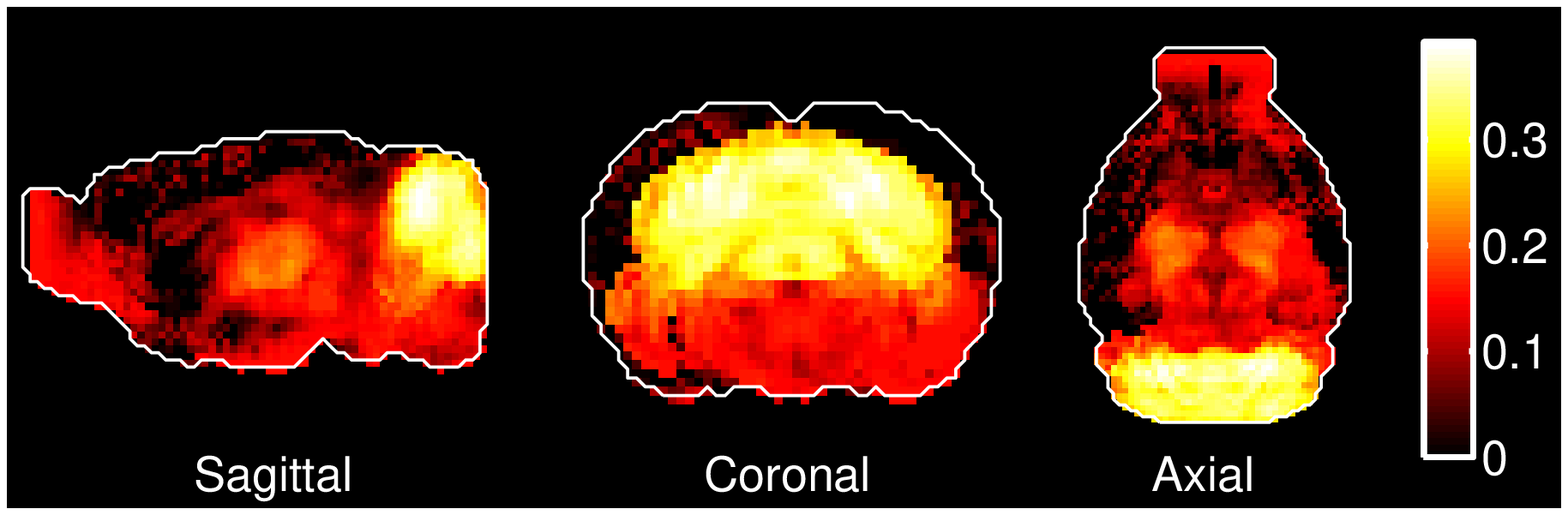}&\includegraphics[width=2in,keepaspectratio]{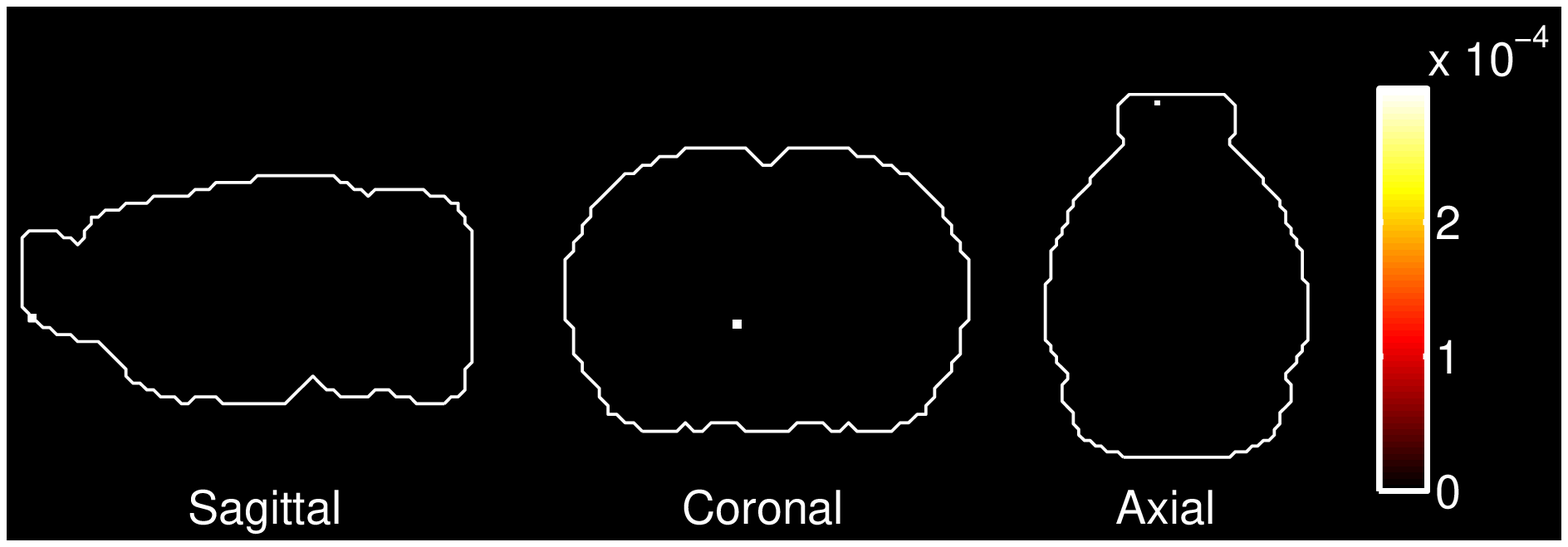}\\\hline
26&\tiny{Neurons}&\includegraphics[width=2in,keepaspectratio]{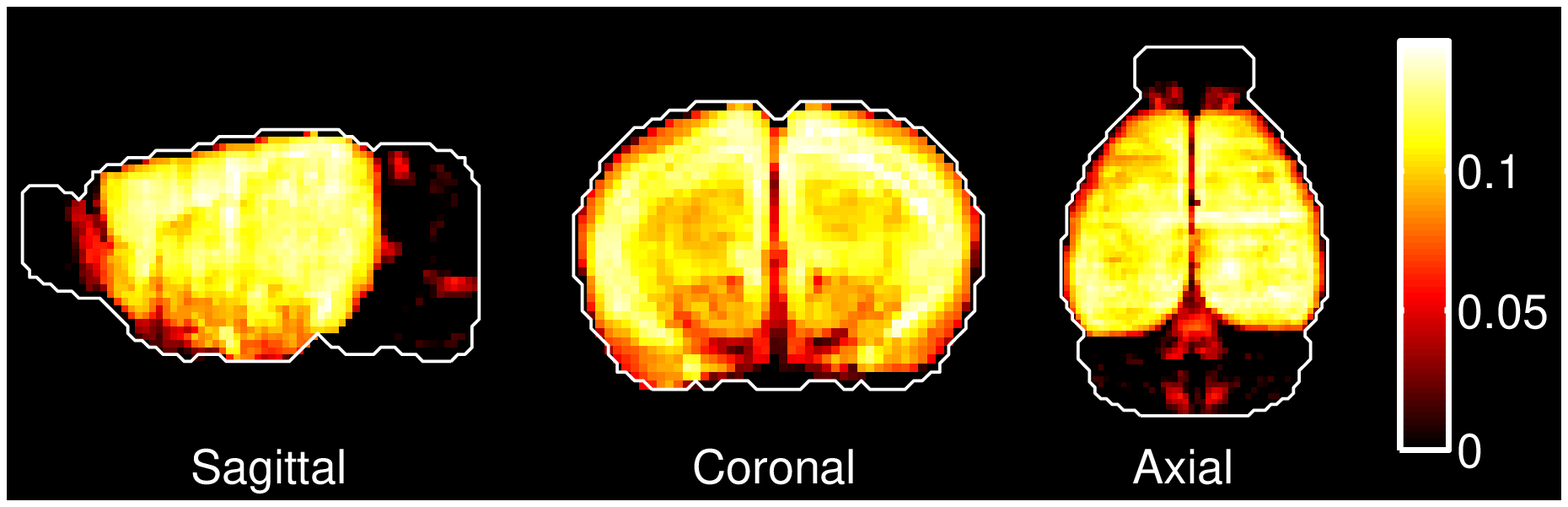}&\includegraphics[width=2in,keepaspectratio]{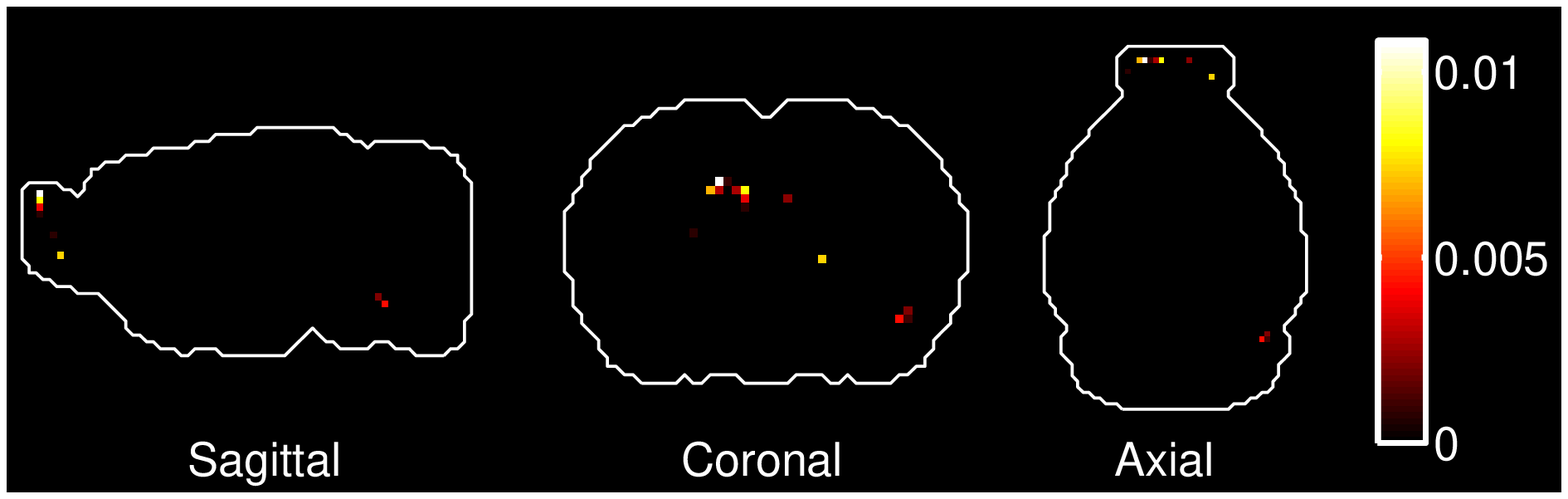}\\\hline
27&\tiny{Bergman Glia}&\includegraphics[width=2in,keepaspectratio]{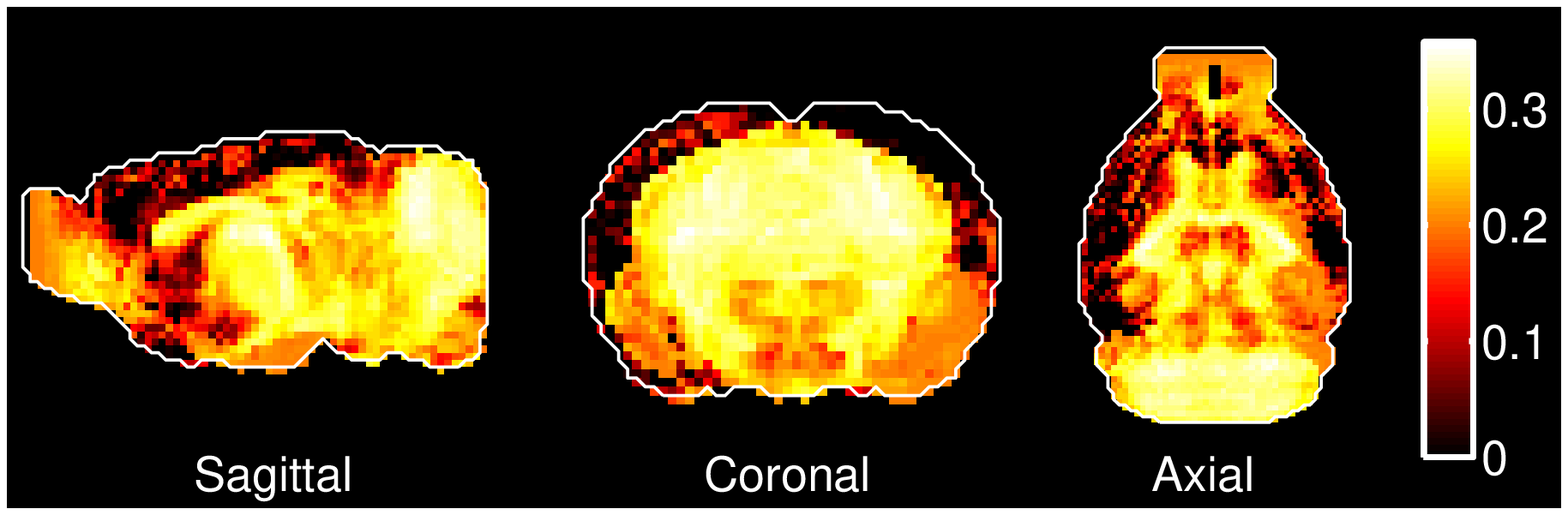}&\includegraphics[width=2in,keepaspectratio]{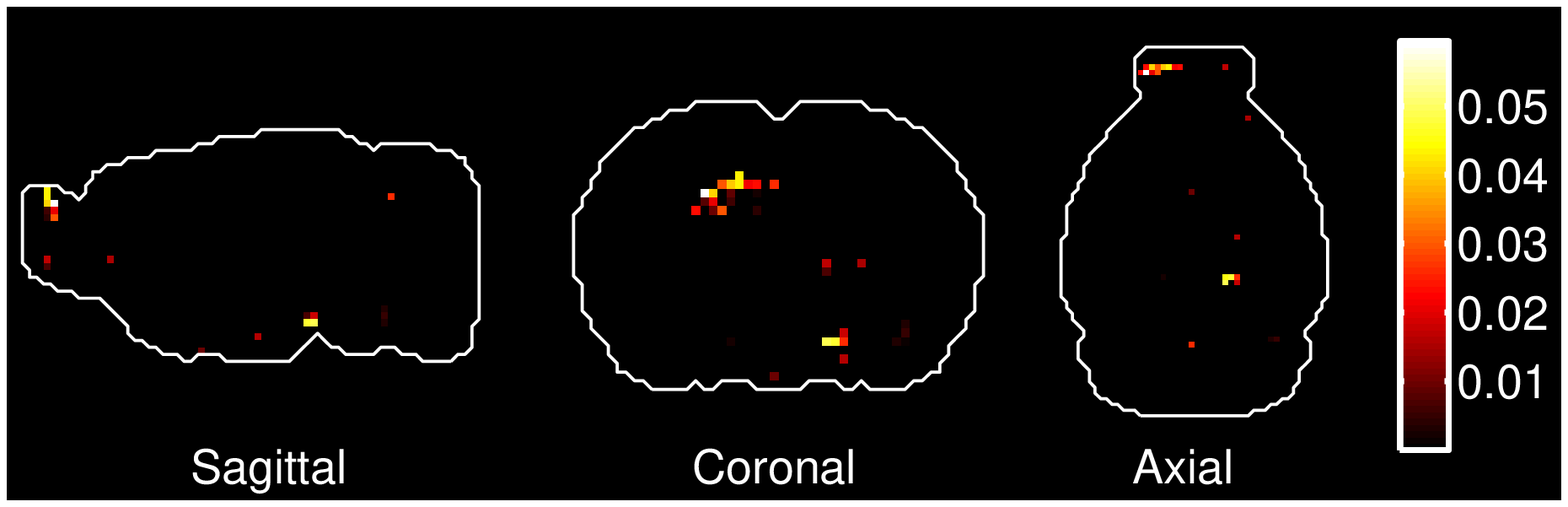}\\\hline
28&\tiny{Astroglia}&\includegraphics[width=2in,keepaspectratio]{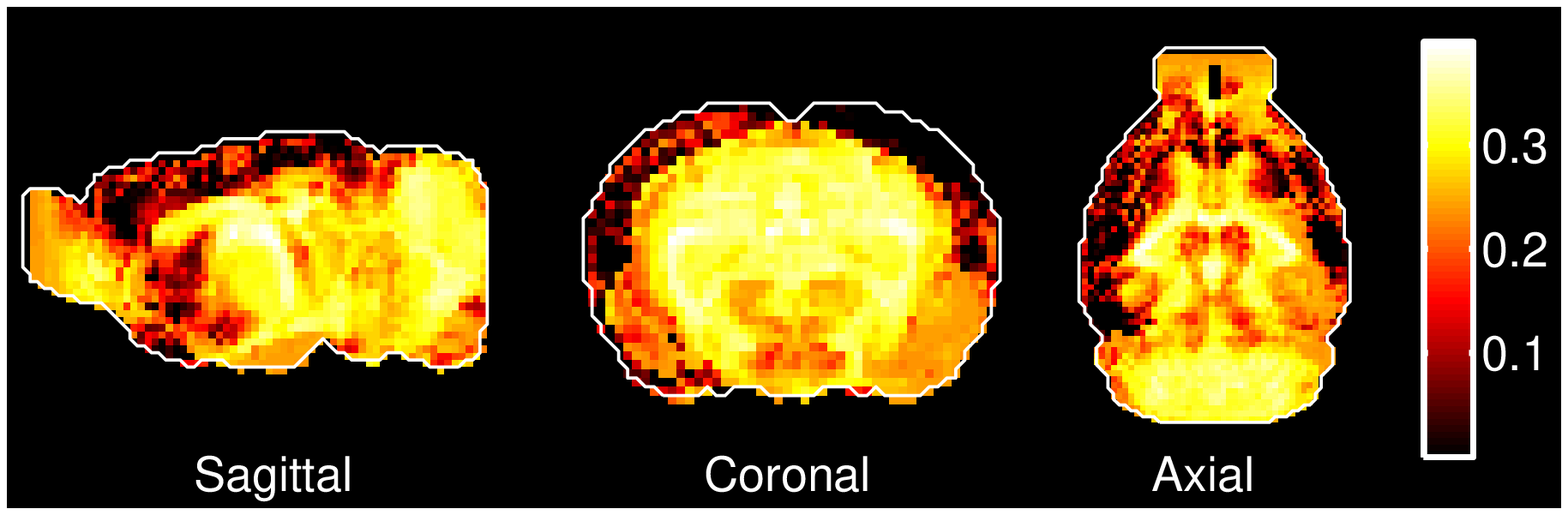}&\includegraphics[width=2in,keepaspectratio]{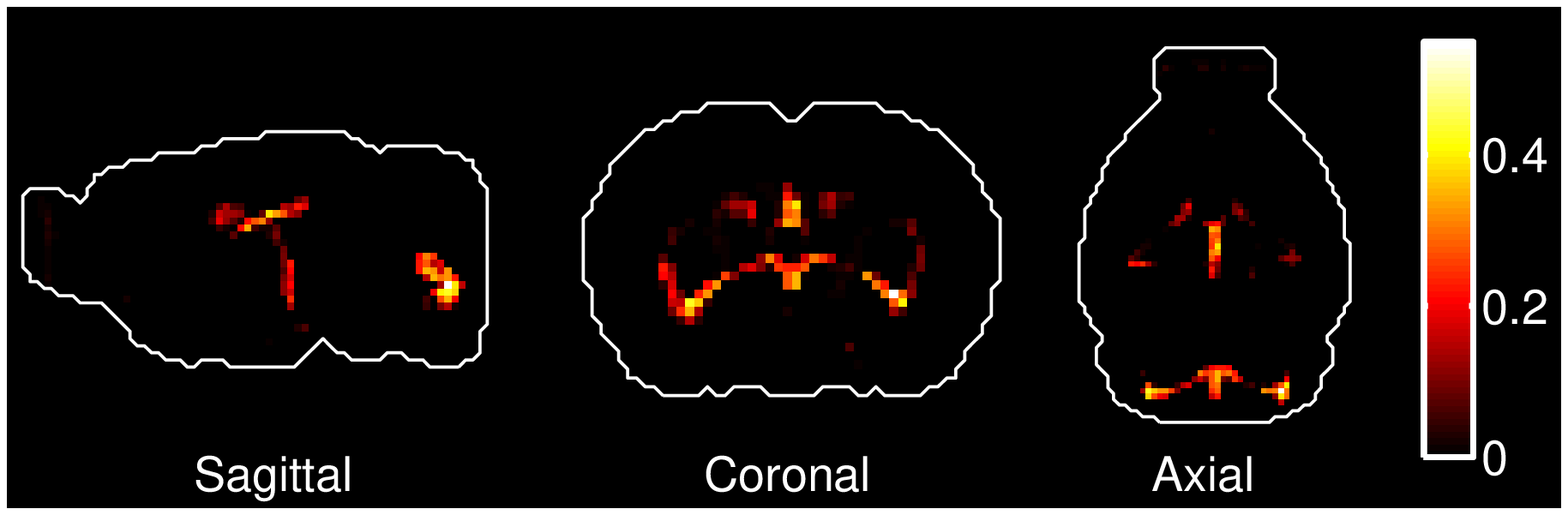}\\\hline
29&\tiny{Astroglia}&\includegraphics[width=2in,keepaspectratio]{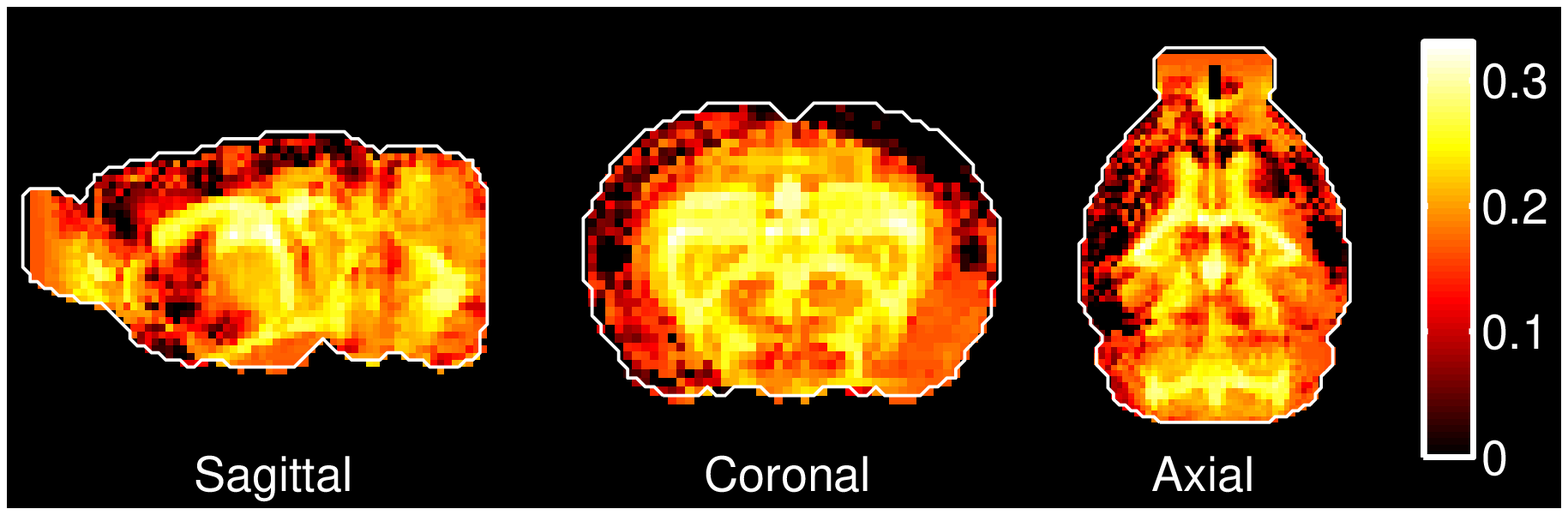}&\includegraphics[width=2in,keepaspectratio]{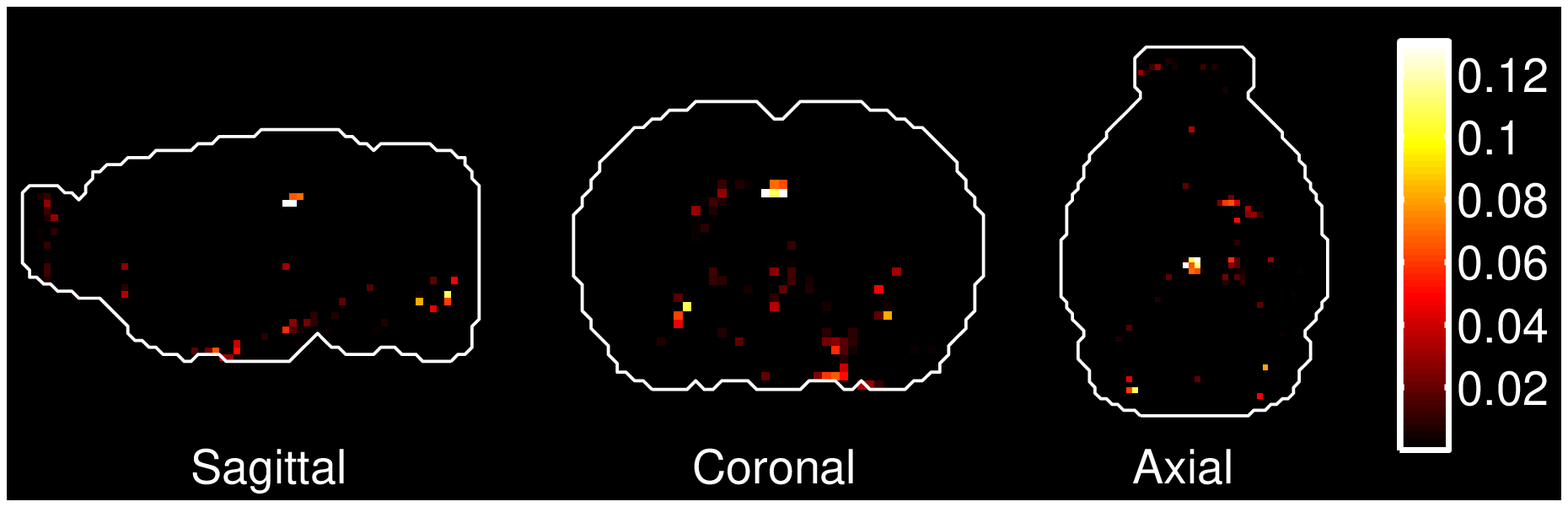}\\\hline
30&\tiny{Astrocytes}&\includegraphics[width=2in,keepaspectratio]{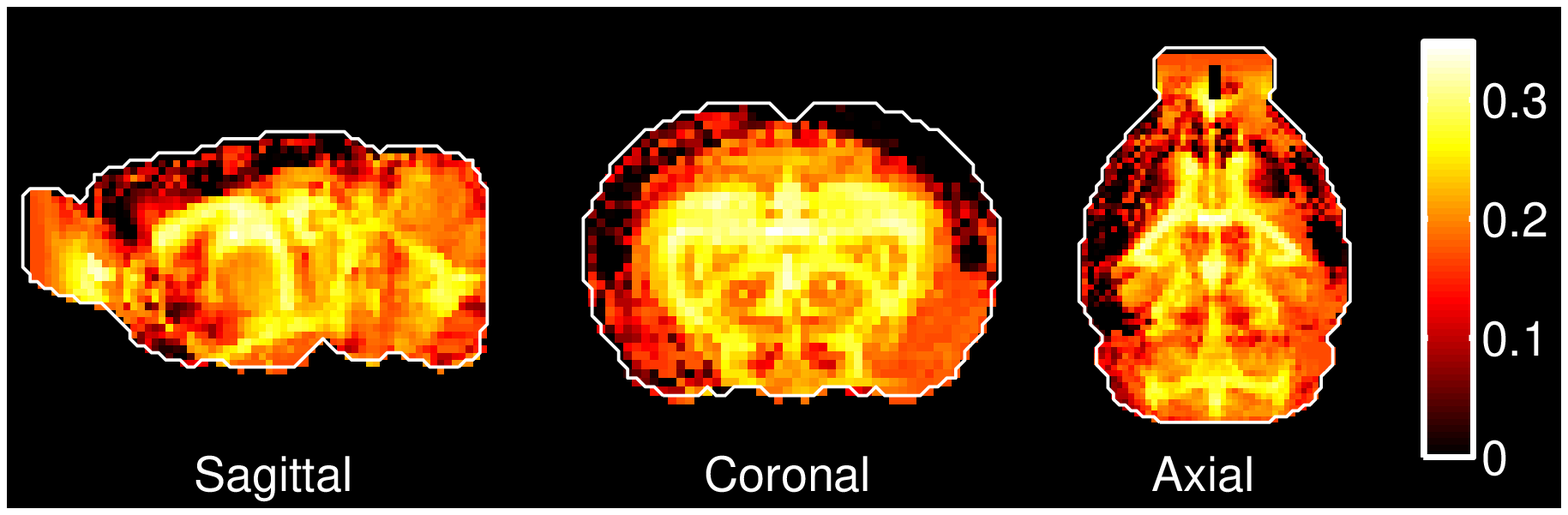}&\includegraphics[width=2in,keepaspectratio]{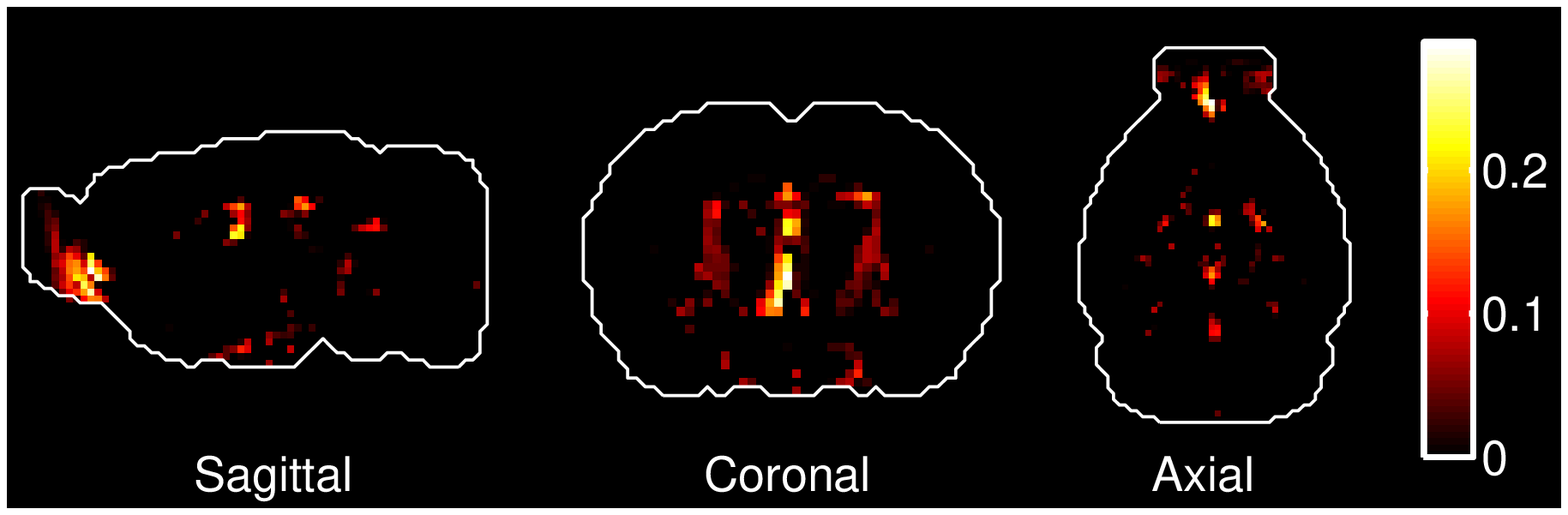}\\\hline
31&\tiny{Astrocytes}&\includegraphics[width=2in,keepaspectratio]{cellTypeProj31.eps}&\includegraphics[width=2in,keepaspectratio]{cellTypeModelFit31.eps}\\\hline
32&\tiny{Astrocytes}&\includegraphics[width=2in,keepaspectratio]{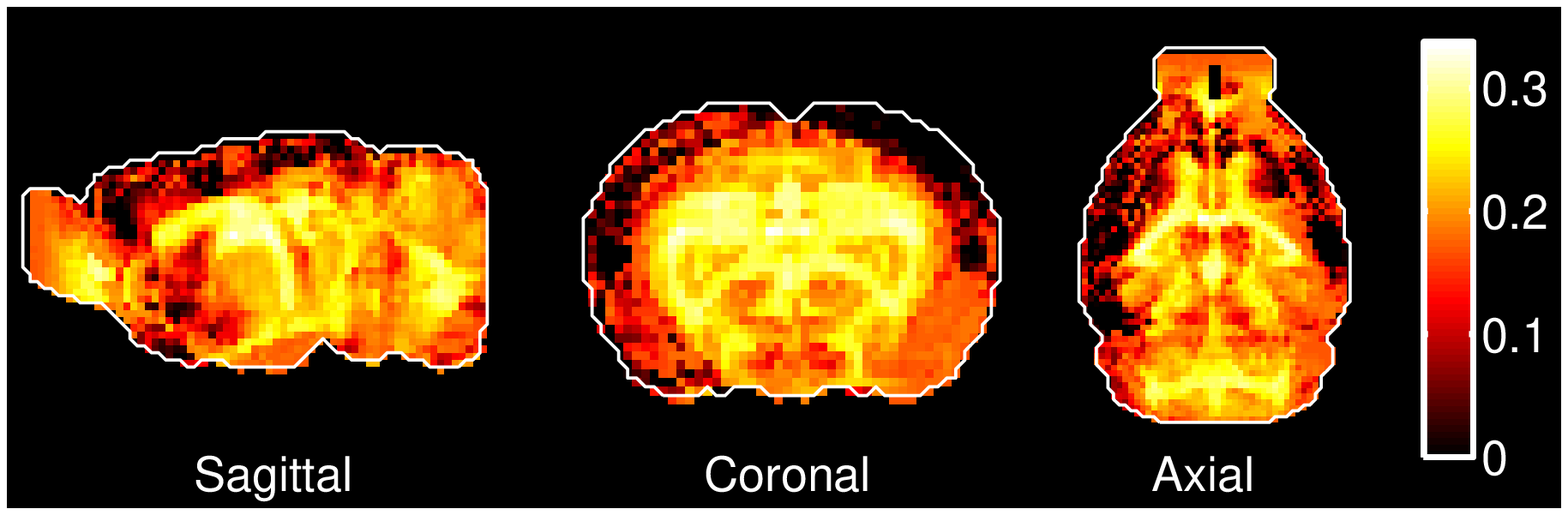}&\includegraphics[width=2in,keepaspectratio]{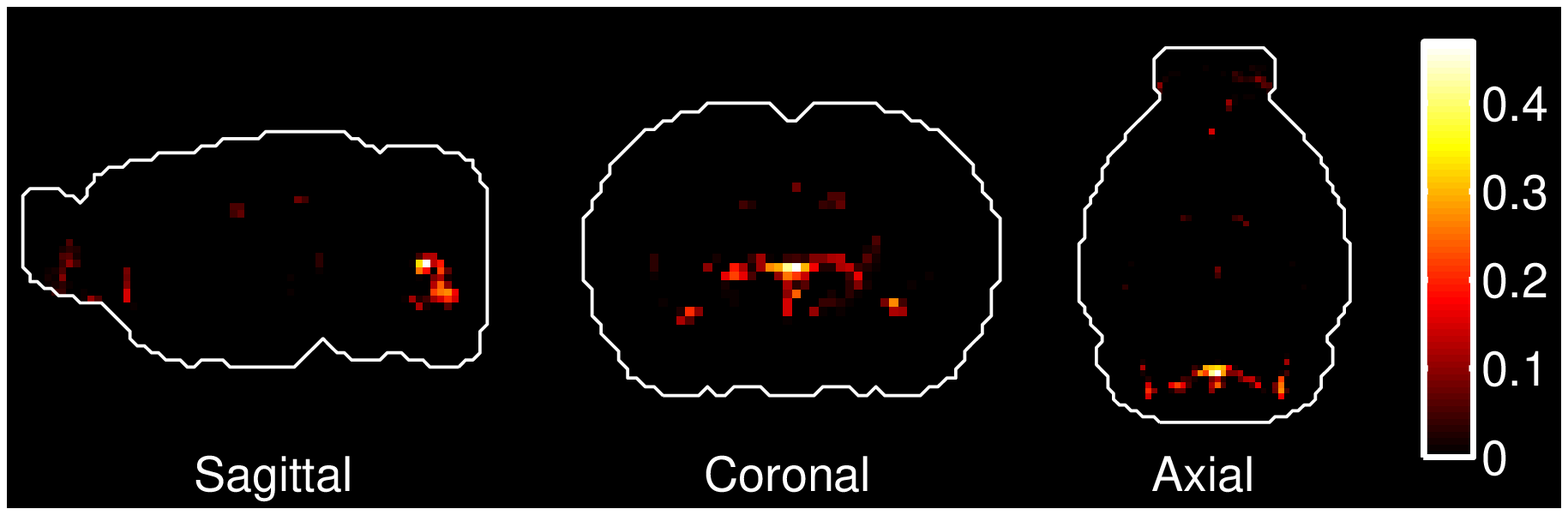}\\\hline
33&\tiny{Mixed Neurons}&\includegraphics[width=2in,keepaspectratio]{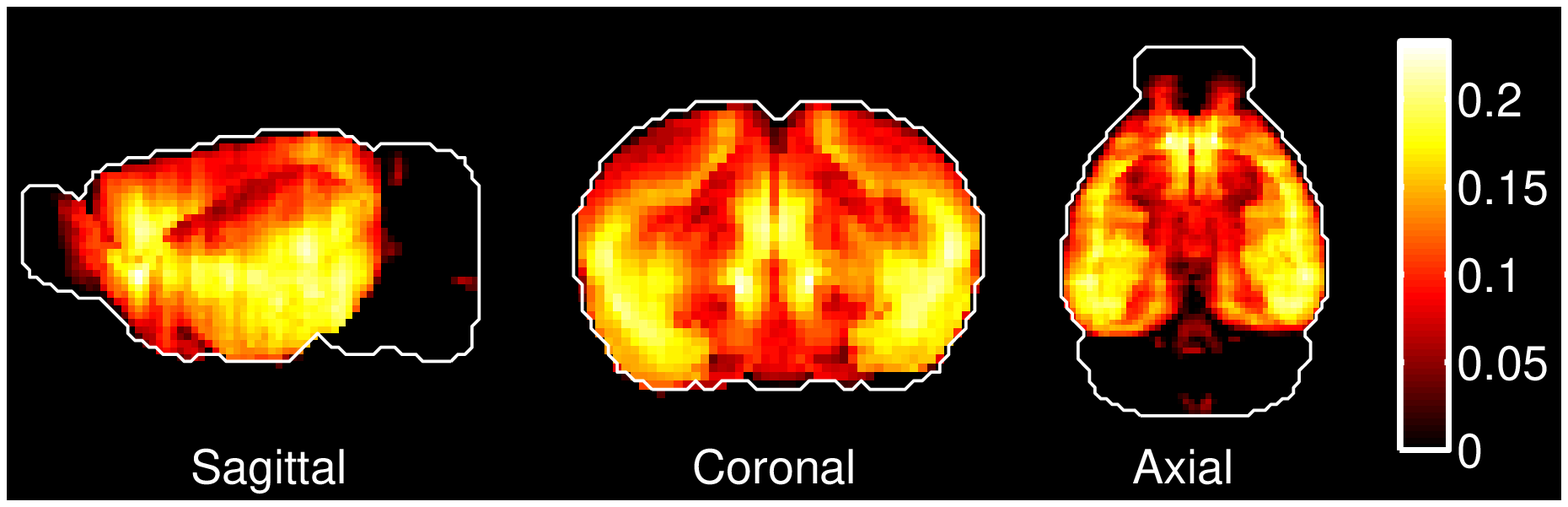}&\includegraphics[width=2in,keepaspectratio]{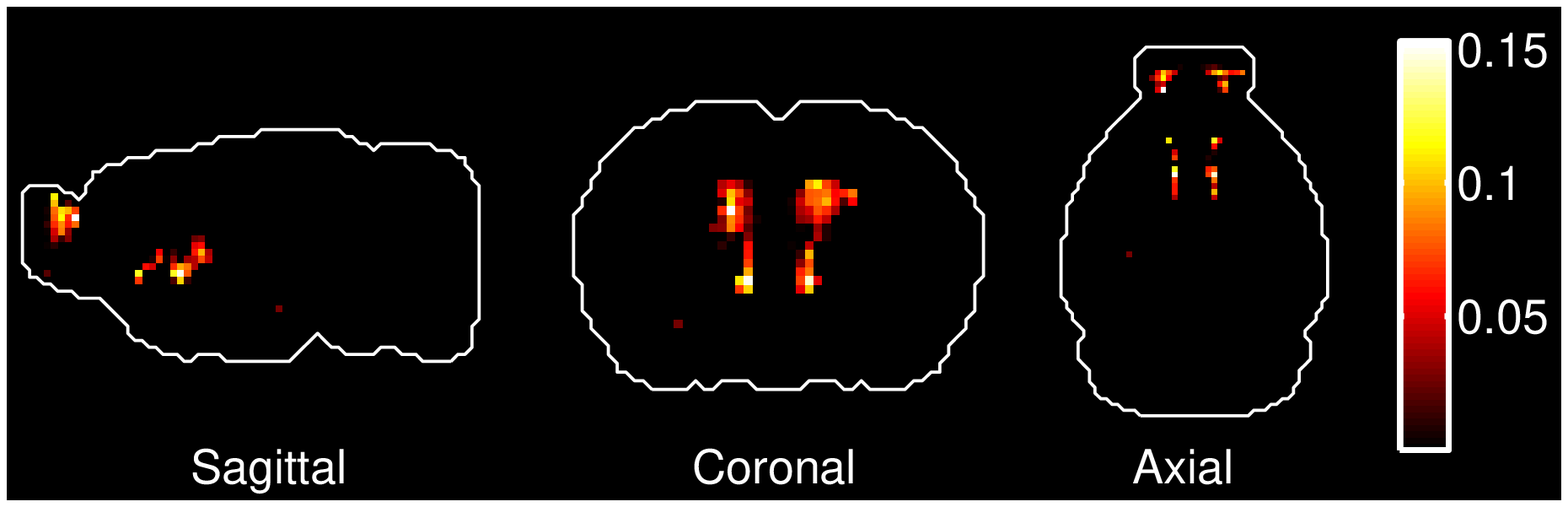}\\\hline
34&\tiny{Mixed Neurons}&\includegraphics[width=2in,keepaspectratio]{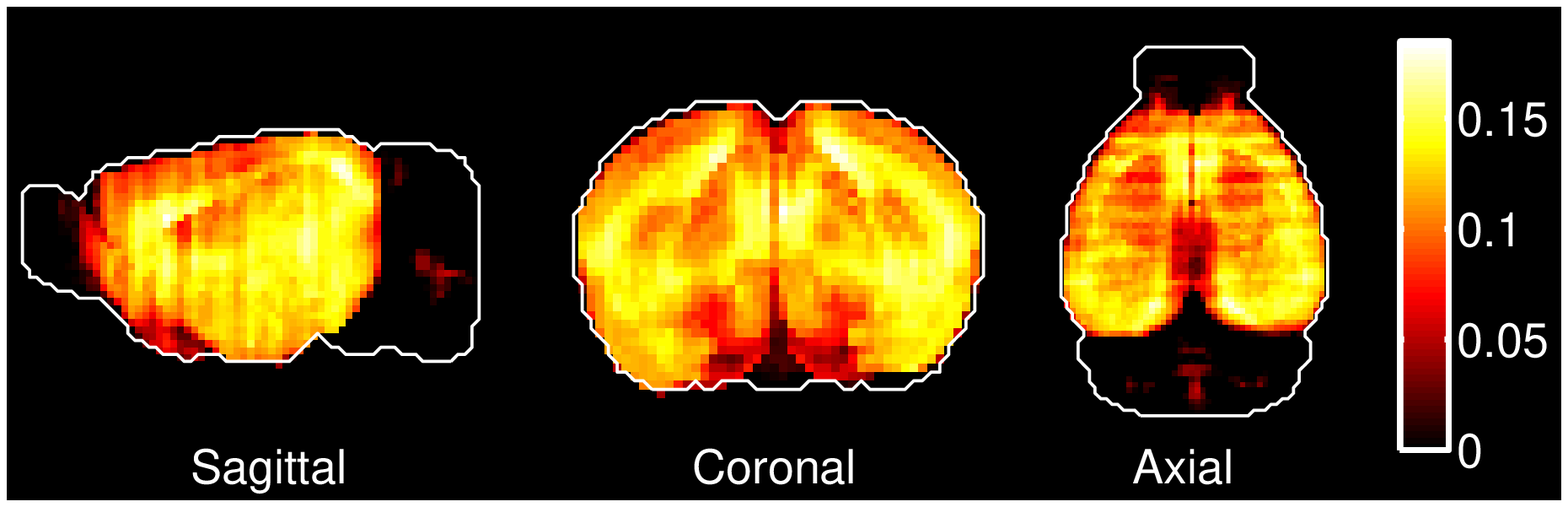}&\includegraphics[width=2in,keepaspectratio]{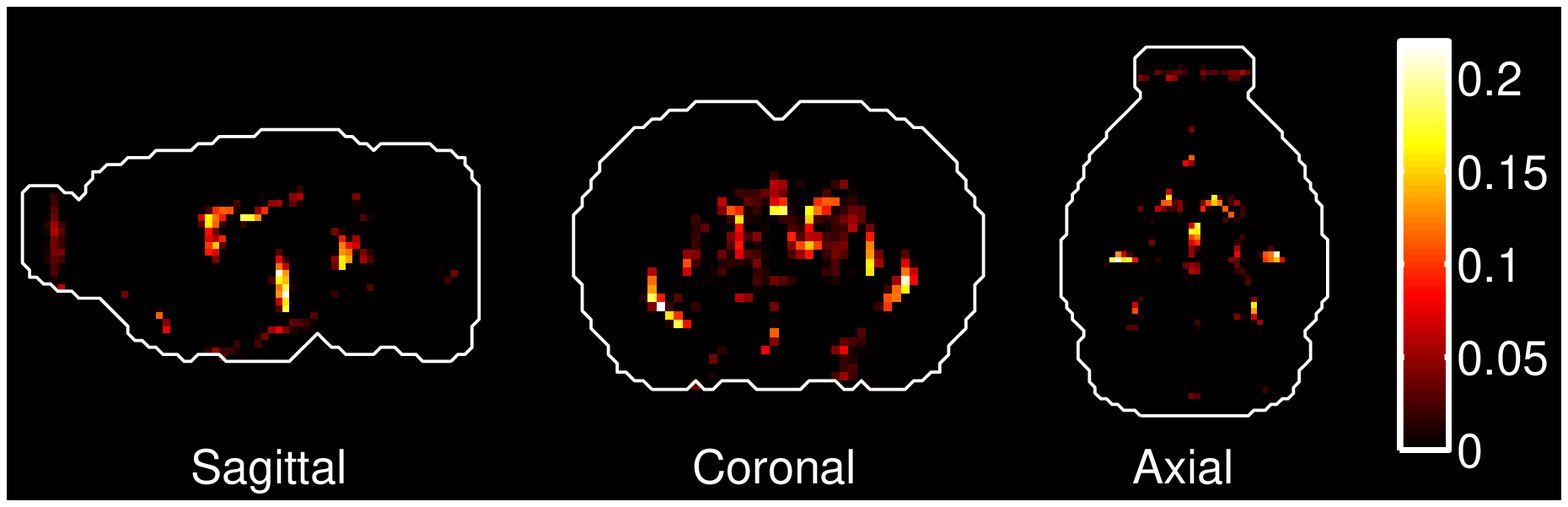}\\\hline
35&\tiny{Mature Oligodendrocytes}&\includegraphics[width=2in,keepaspectratio]{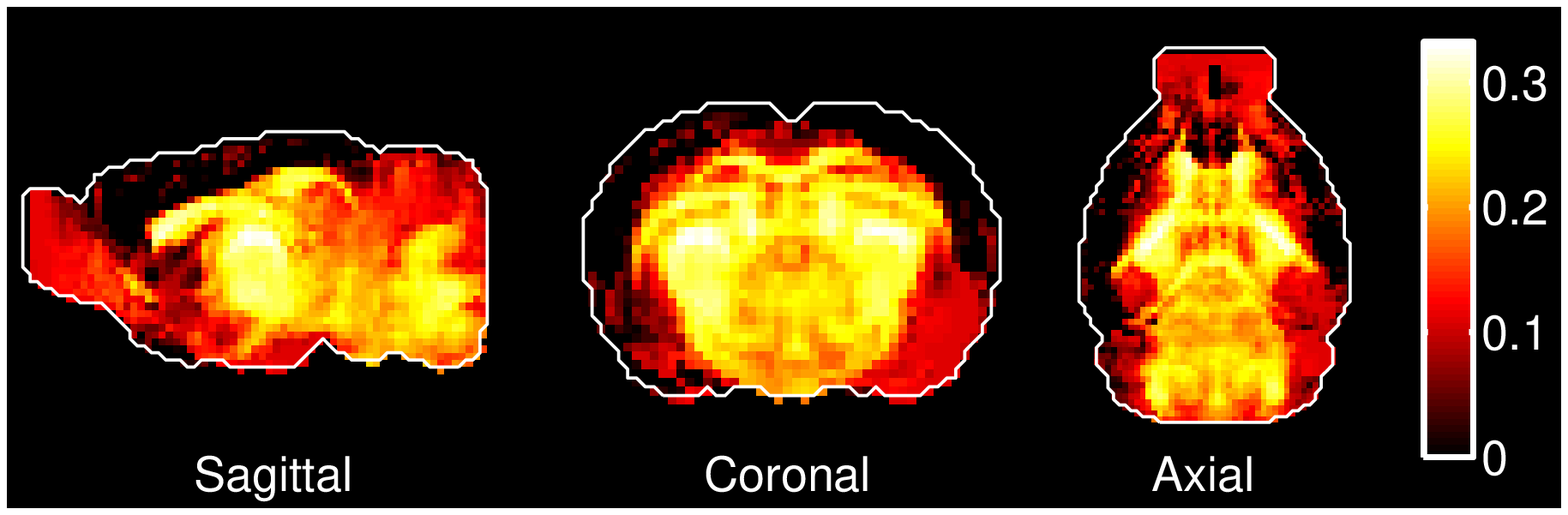}&\includegraphics[width=2in,keepaspectratio]{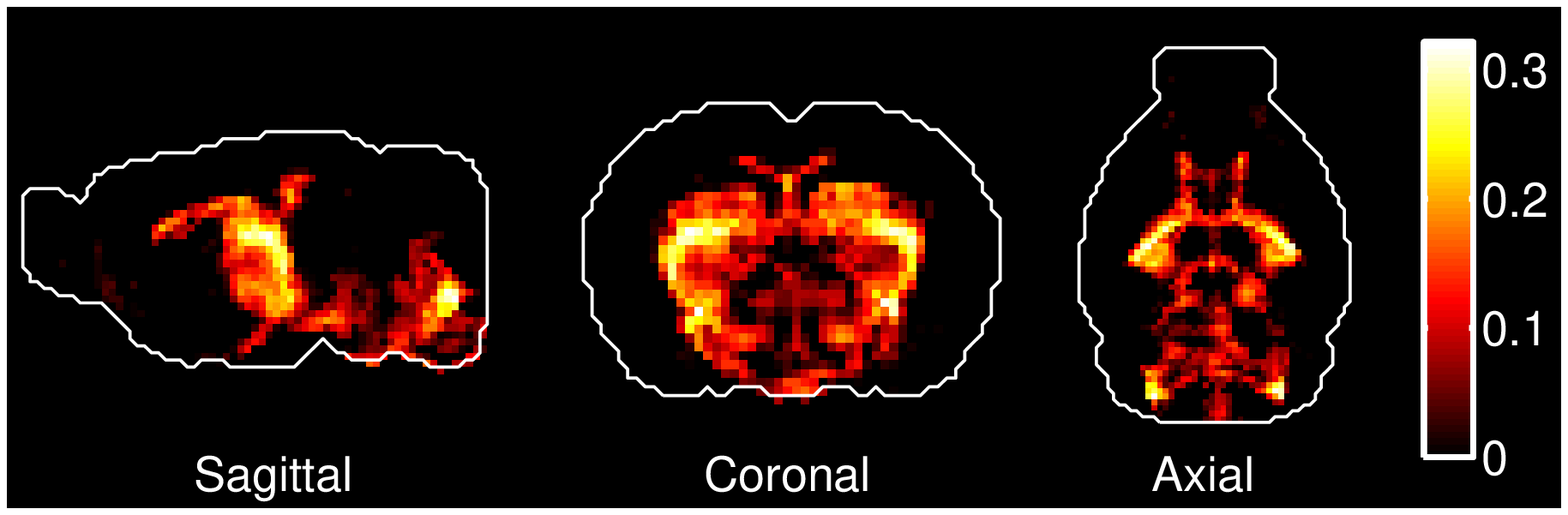}\\\hline
36&\tiny{Oligodendrocytes}&\includegraphics[width=2in,keepaspectratio]{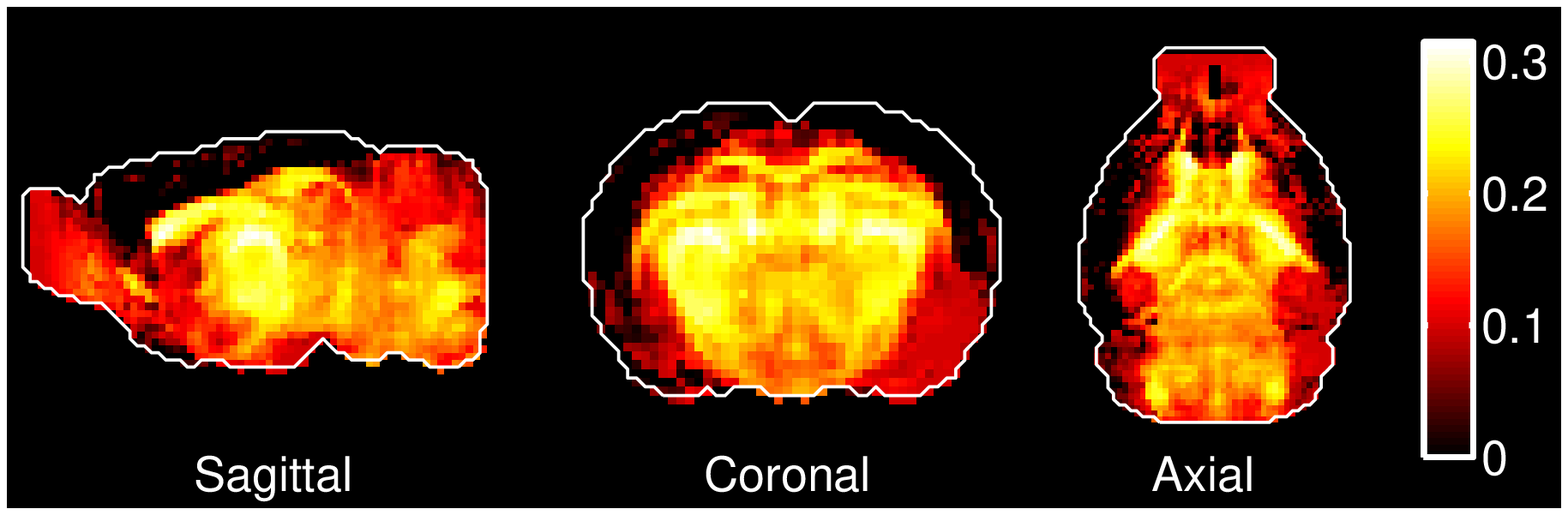}&\includegraphics[width=2in,keepaspectratio]{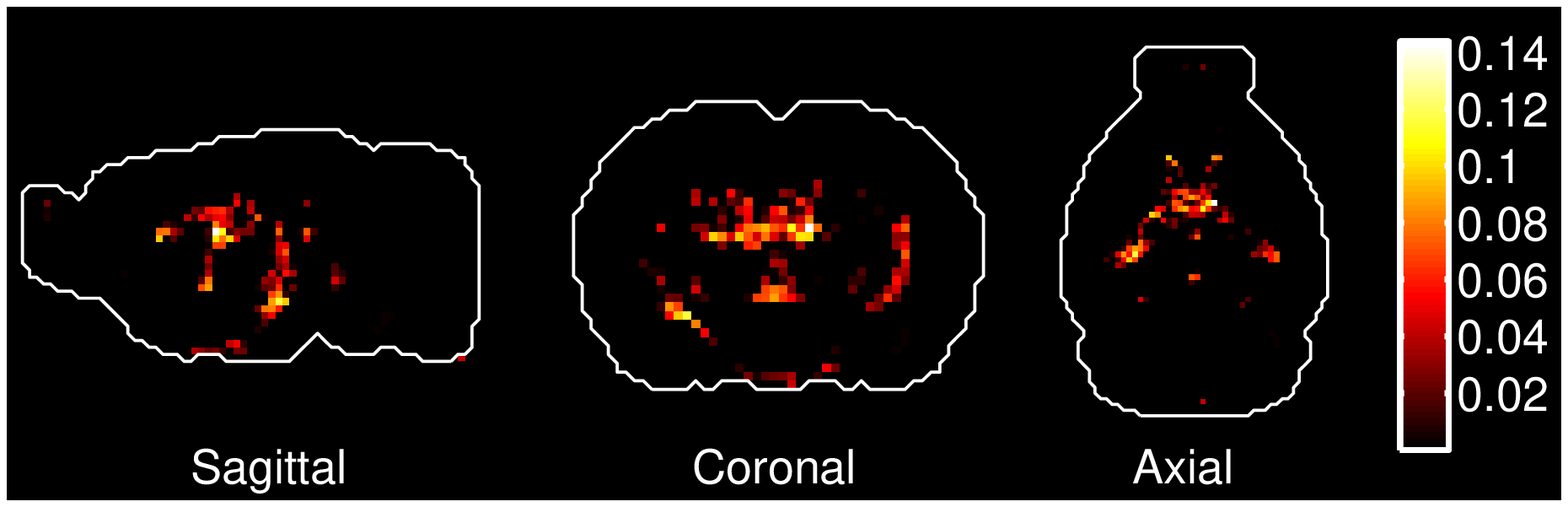}\\\hline
\end{tabular}
\\
\begin{tabular}{|l|l|l|l|}
\hline
\textbf{index}&\textbf{Cell type}&\textbf{Heat map of correlations}&\textbf{Heat map of weight}\\\hline
37&\tiny{Oligodendrocyte Precursors}&\includegraphics[width=2in,keepaspectratio]{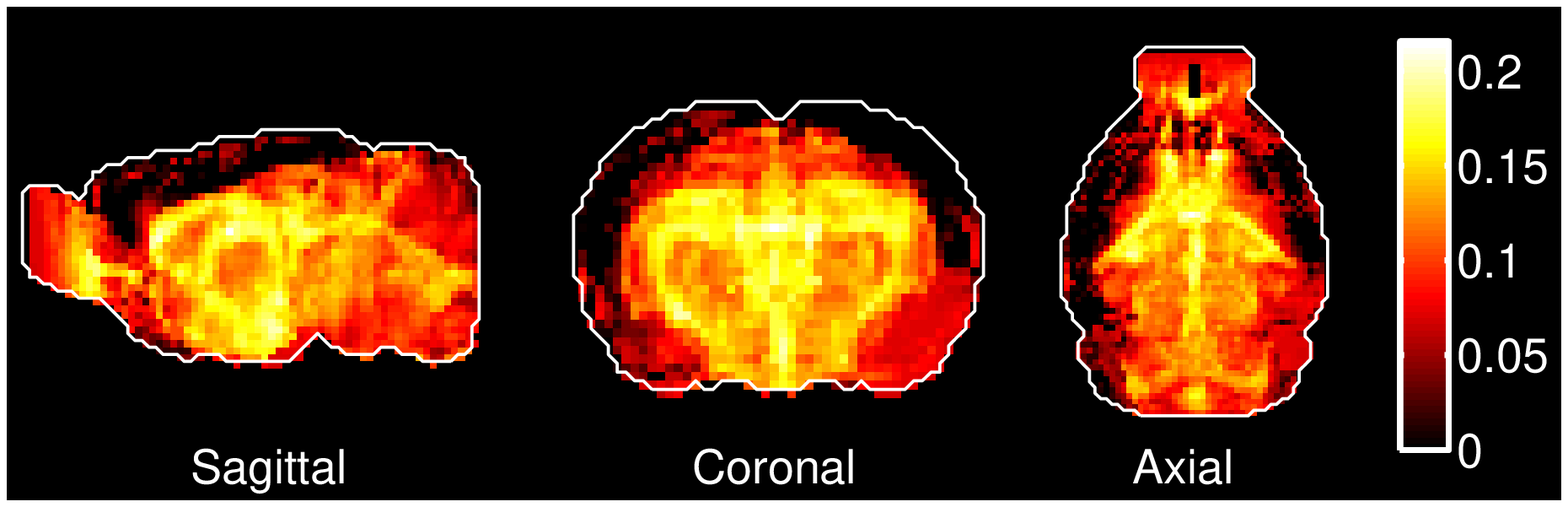}&\includegraphics[width=2in,keepaspectratio]{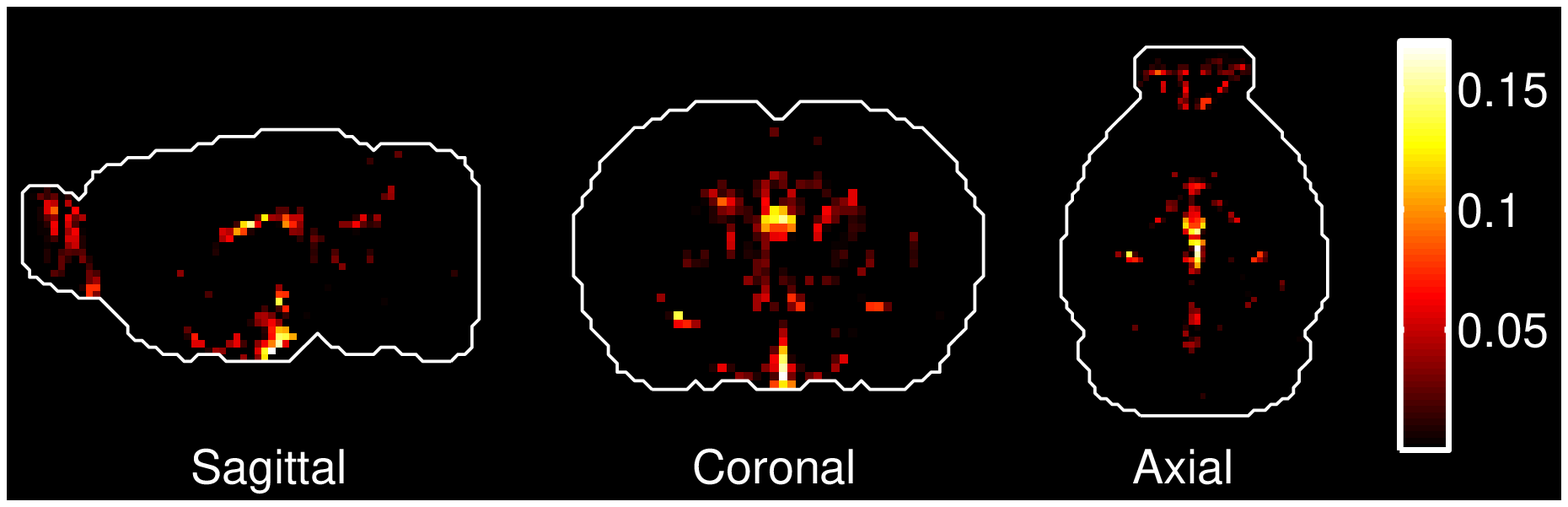}\\\hline
38&\tiny{Pyramidal Neurons, Callosally projecting, P3}&\includegraphics[width=2in,keepaspectratio]{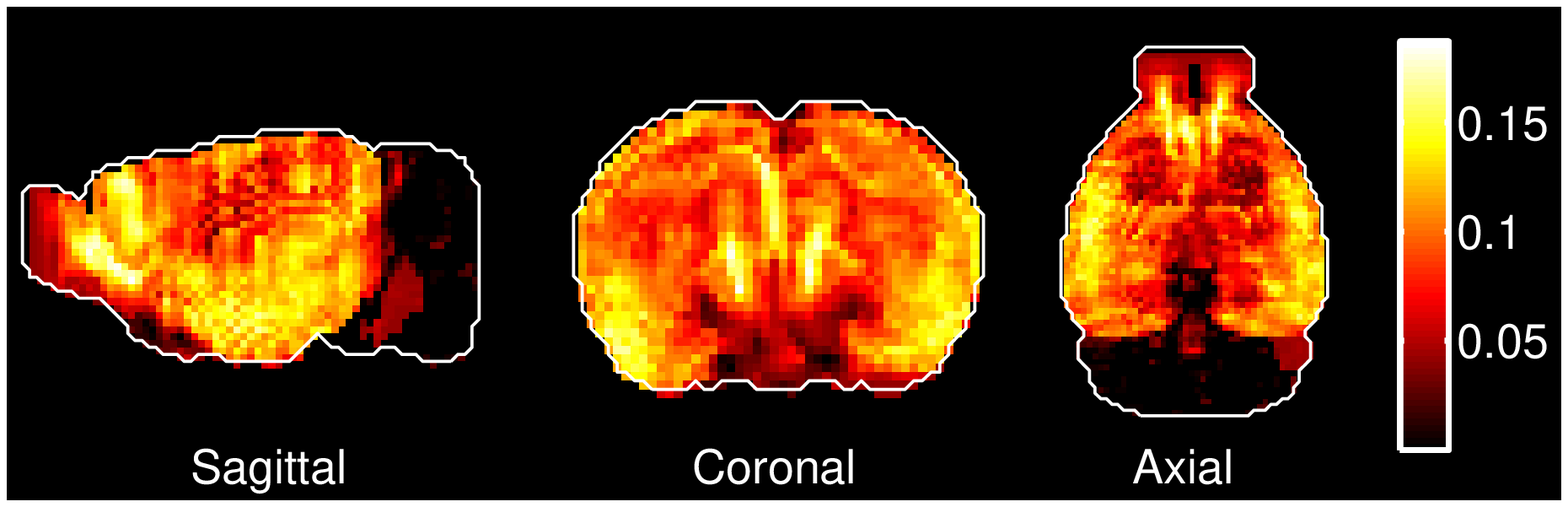}&\includegraphics[width=2in,keepaspectratio]{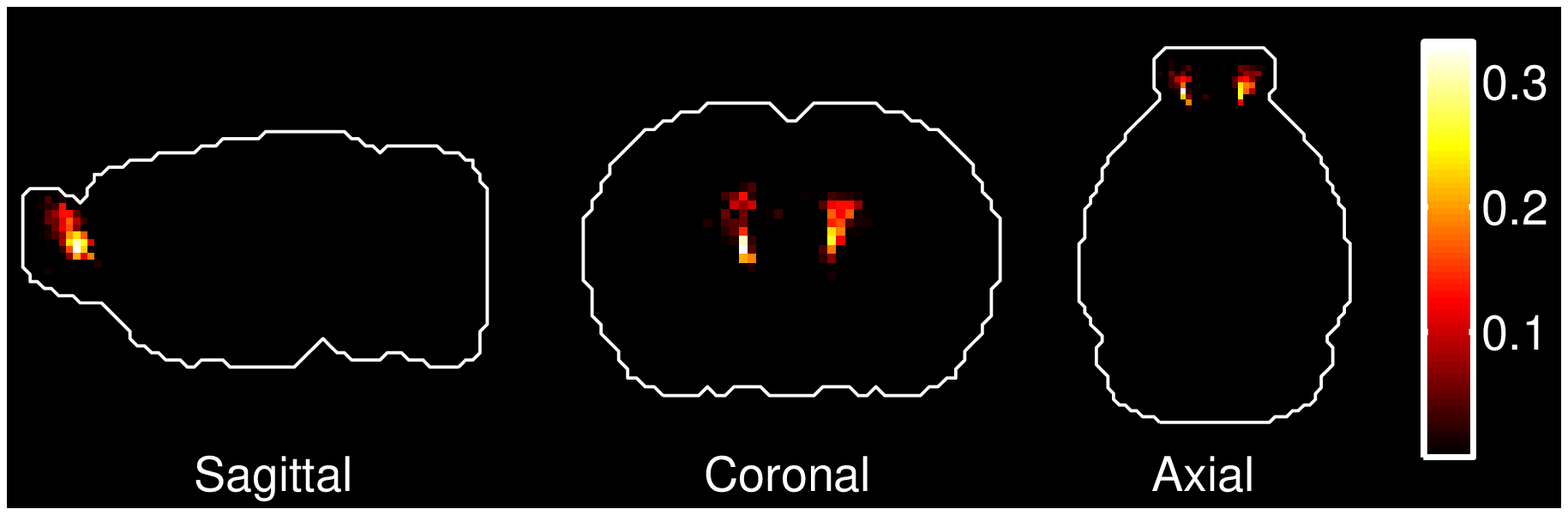}\\\hline
39&\tiny{Pyramidal Neurons, Callosally projecting, P6}&\includegraphics[width=2in,keepaspectratio]{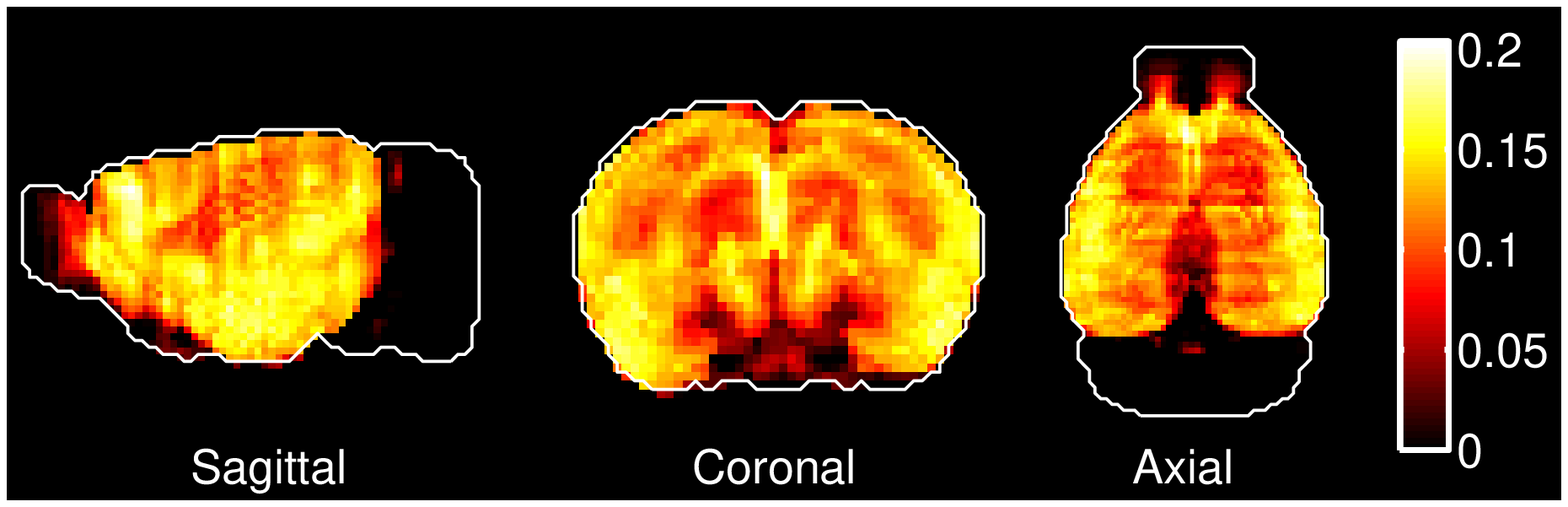}&\includegraphics[width=2in,keepaspectratio]{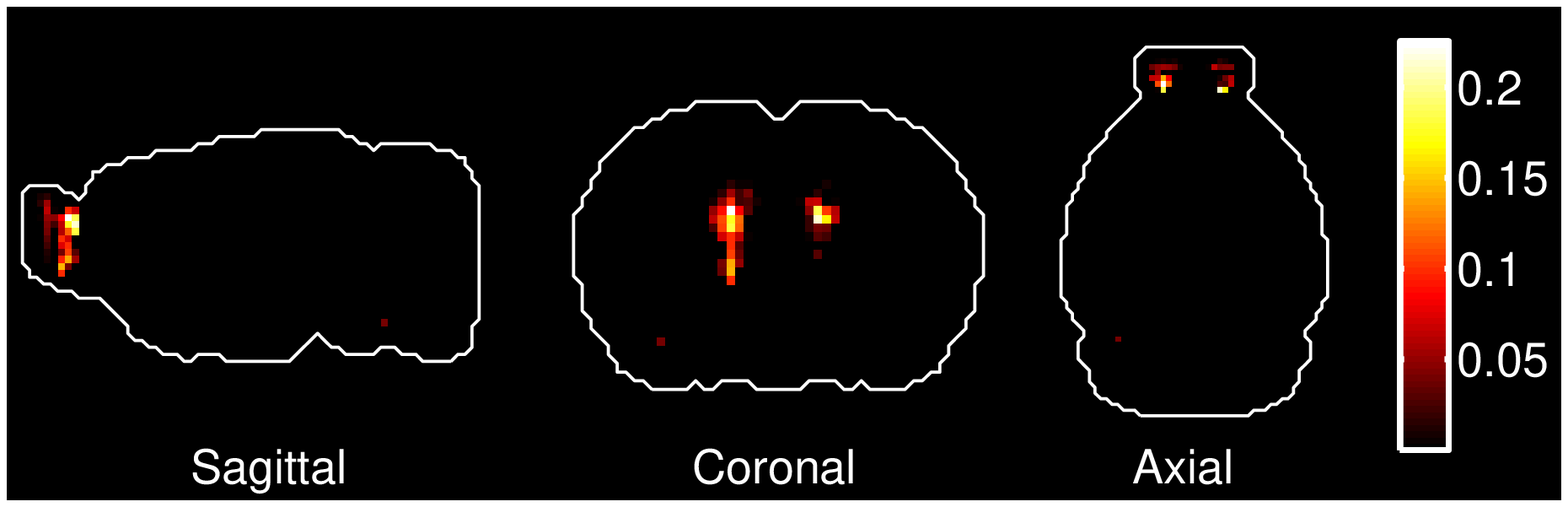}\\\hline
40&\tiny{Pyramidal Neurons, Callosally projecting, P14}&\includegraphics[width=2in,keepaspectratio]{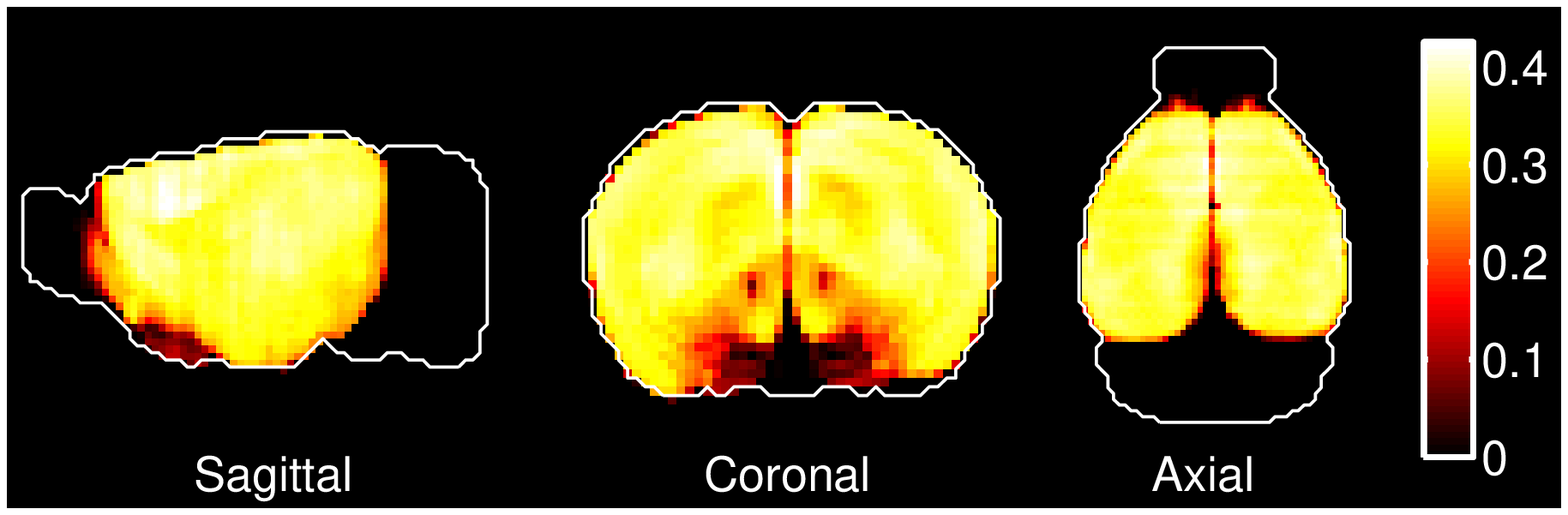}&\includegraphics[width=2in,keepaspectratio]{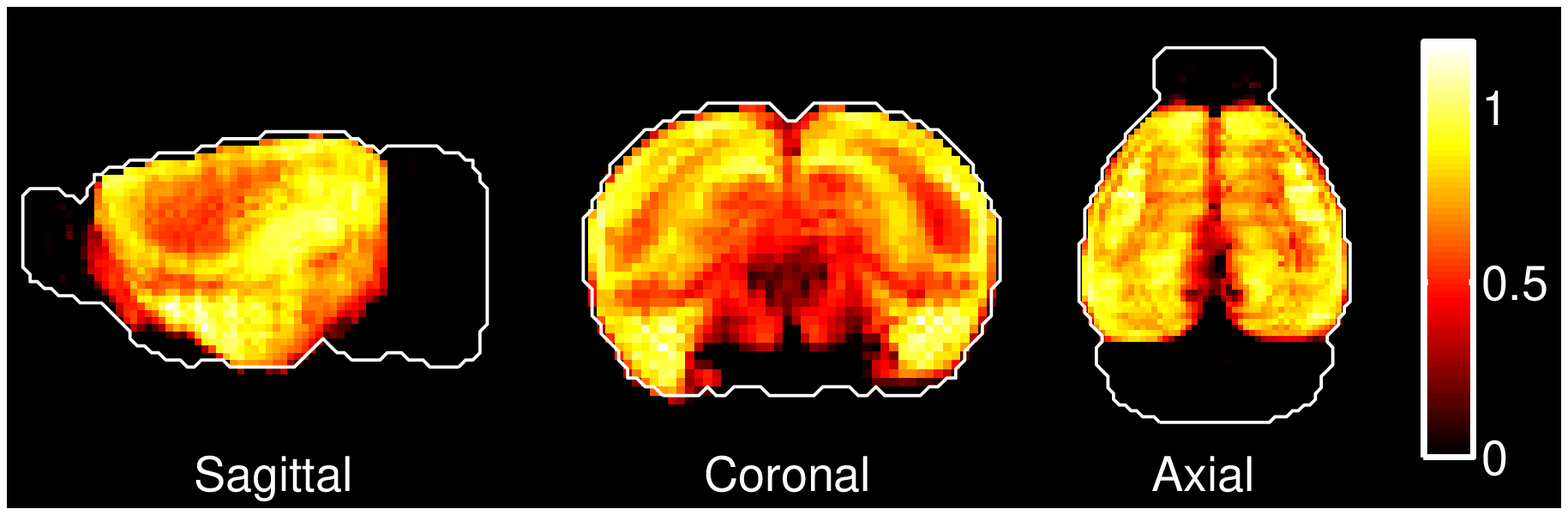}\\\hline
41&\tiny{Pyramidal Neurons, Corticospinal, P3}&\includegraphics[width=2in,keepaspectratio]{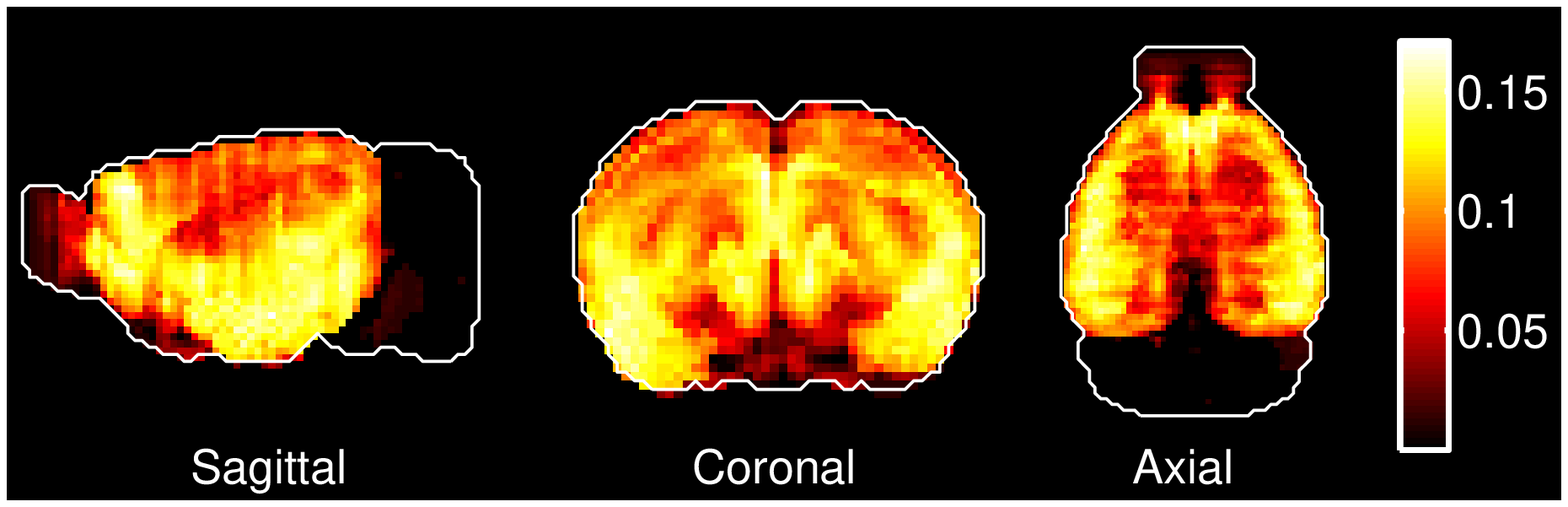}&\includegraphics[width=2in,keepaspectratio]{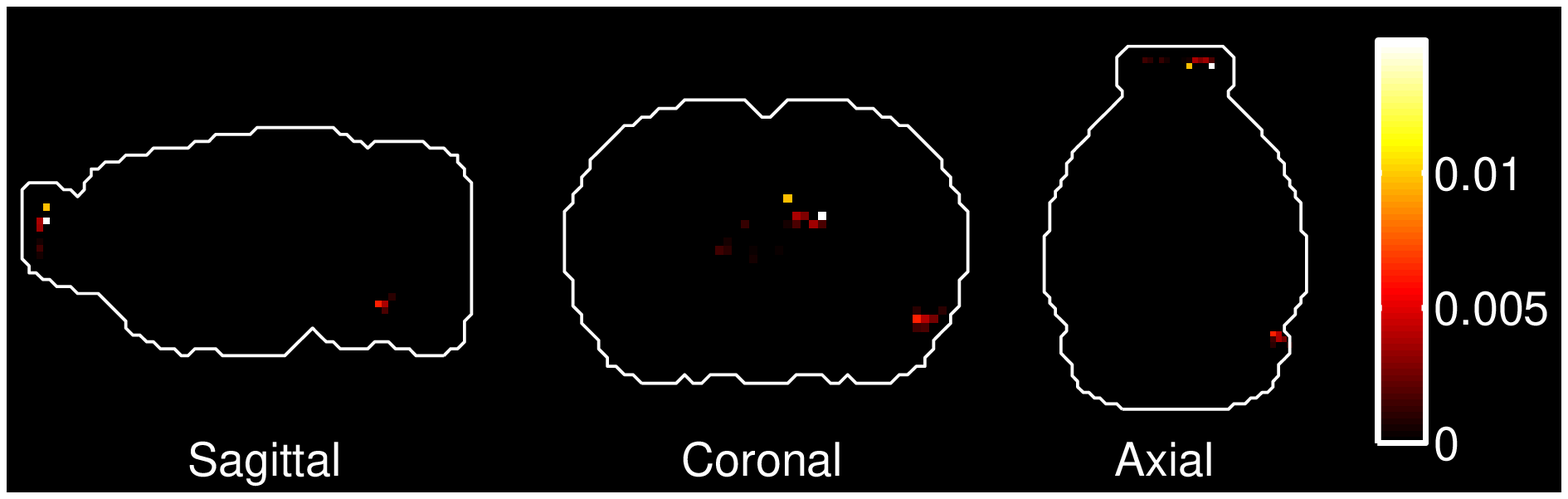}\\\hline
42&\tiny{Pyramidal Neurons, Corticospinal, P6}&\includegraphics[width=2in,keepaspectratio]{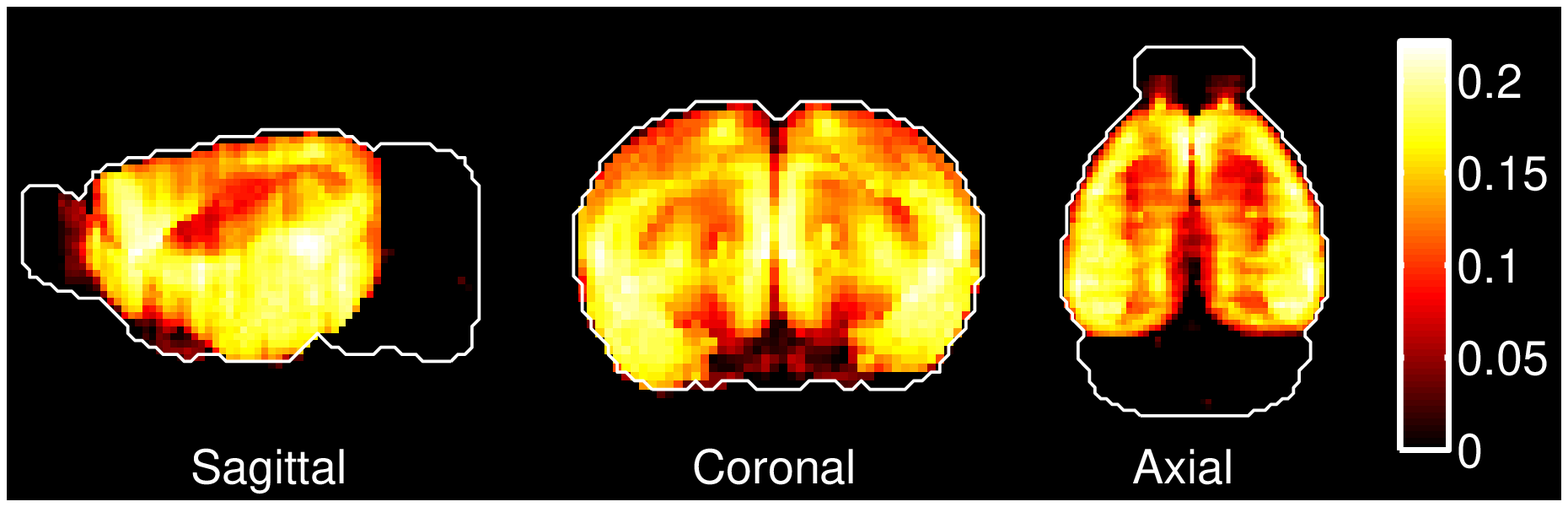}&\includegraphics[width=2in,keepaspectratio]{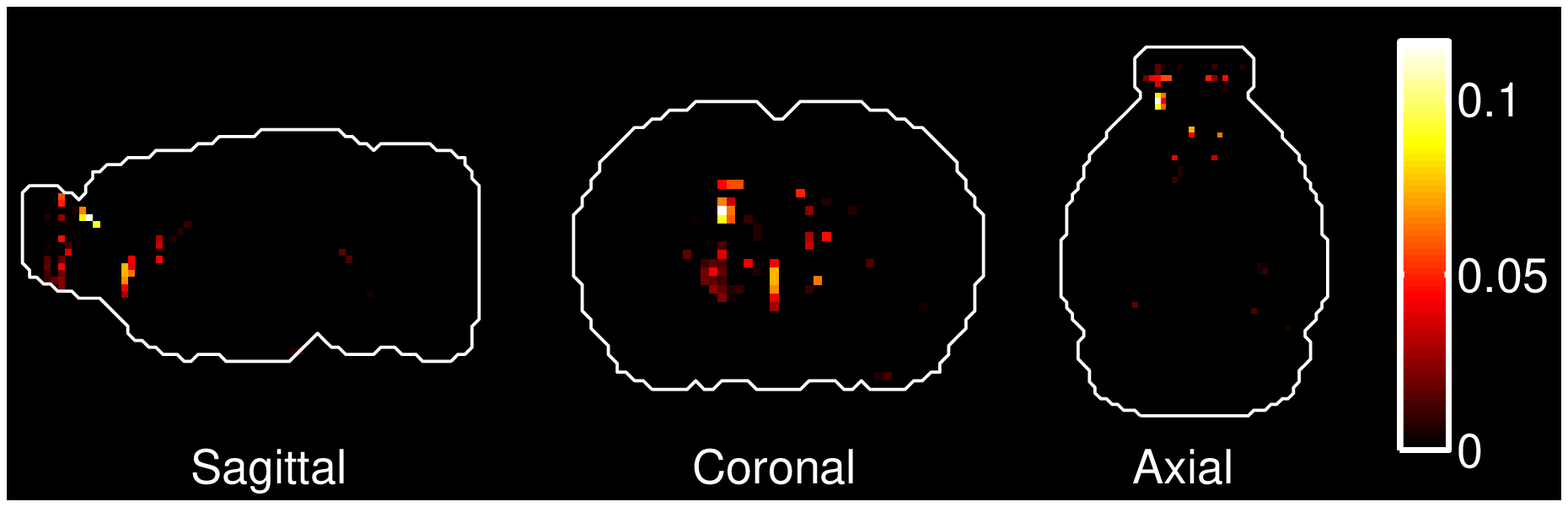}\\\hline
43&\tiny{Pyramidal Neurons, Corticospinal, P14}&\includegraphics[width=2in,keepaspectratio]{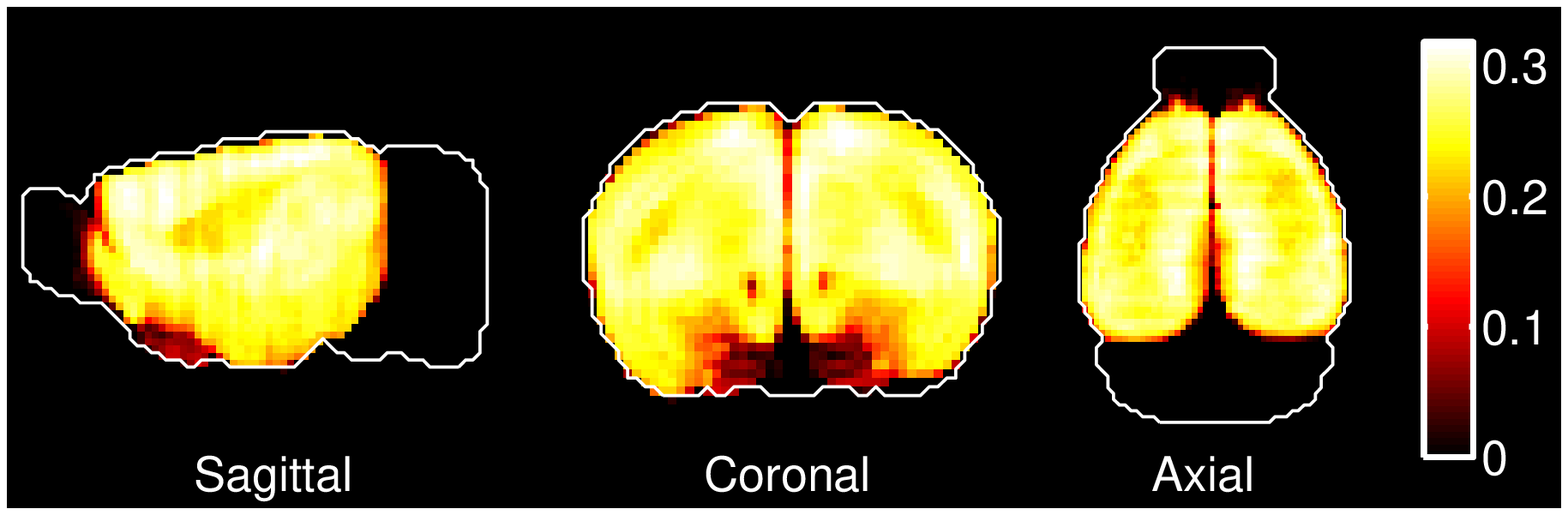}&\includegraphics[width=2in,keepaspectratio]{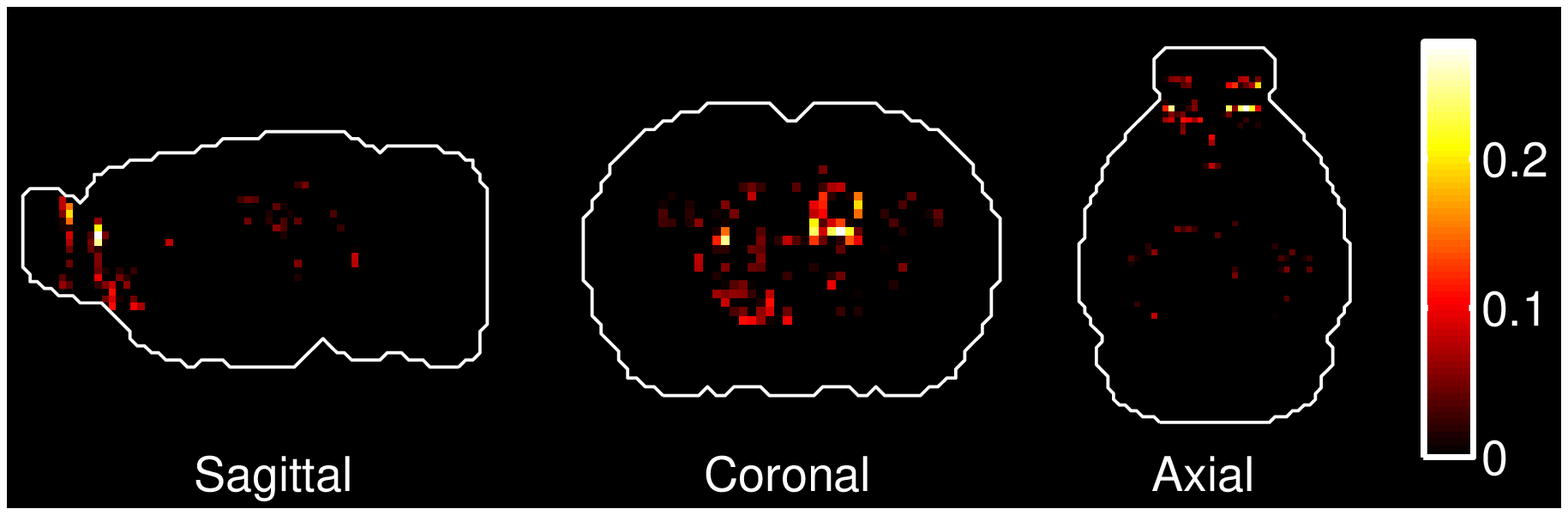}\\\hline
44&\tiny{Pyramidal Neurons, Corticotectal, P14}&\includegraphics[width=2in,keepaspectratio]{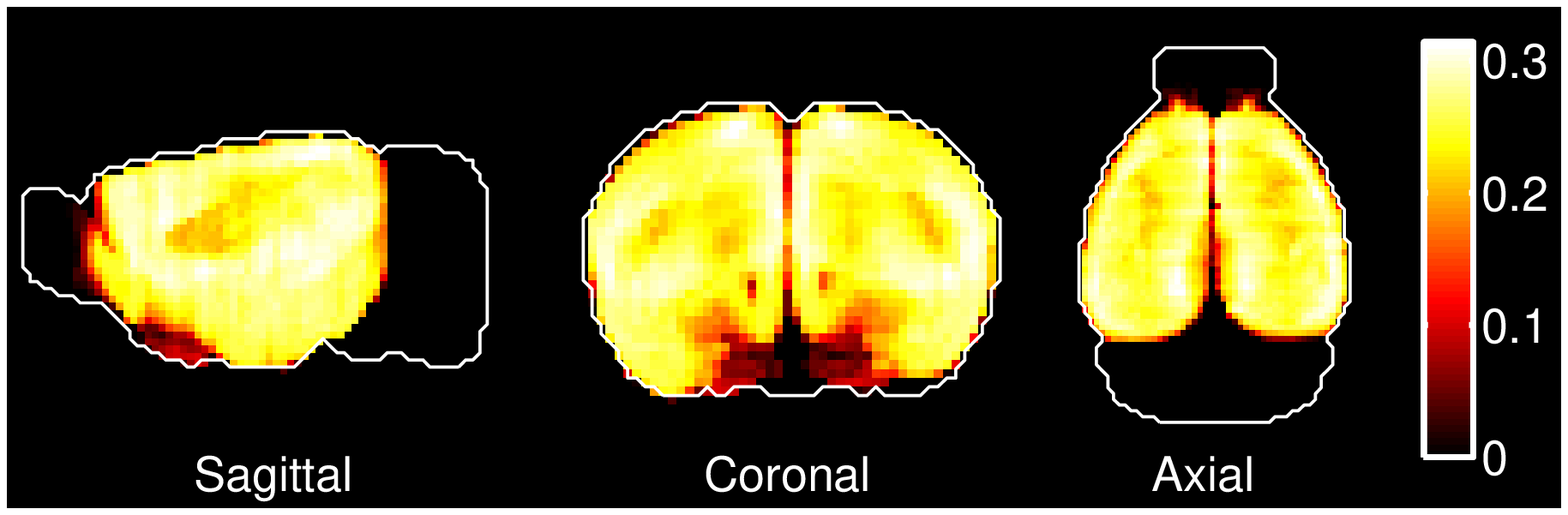}&\includegraphics[width=2in,keepaspectratio]{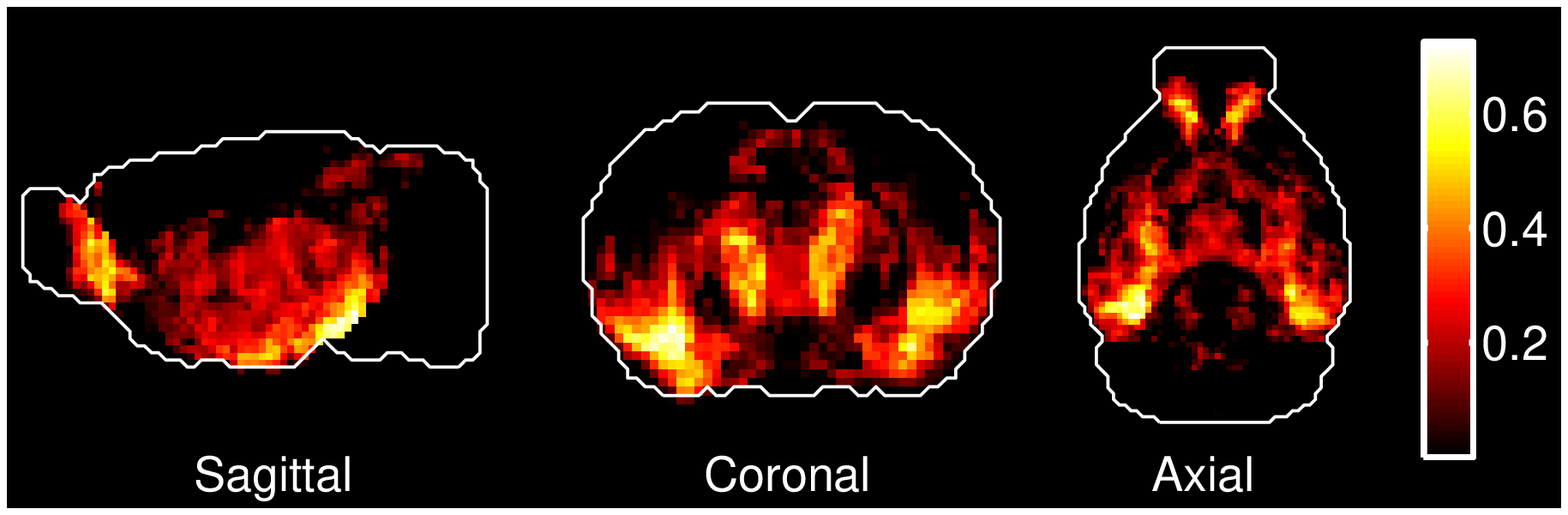}\\\hline
45&\tiny{Pyramidal Neurons}&\includegraphics[width=2in,keepaspectratio]{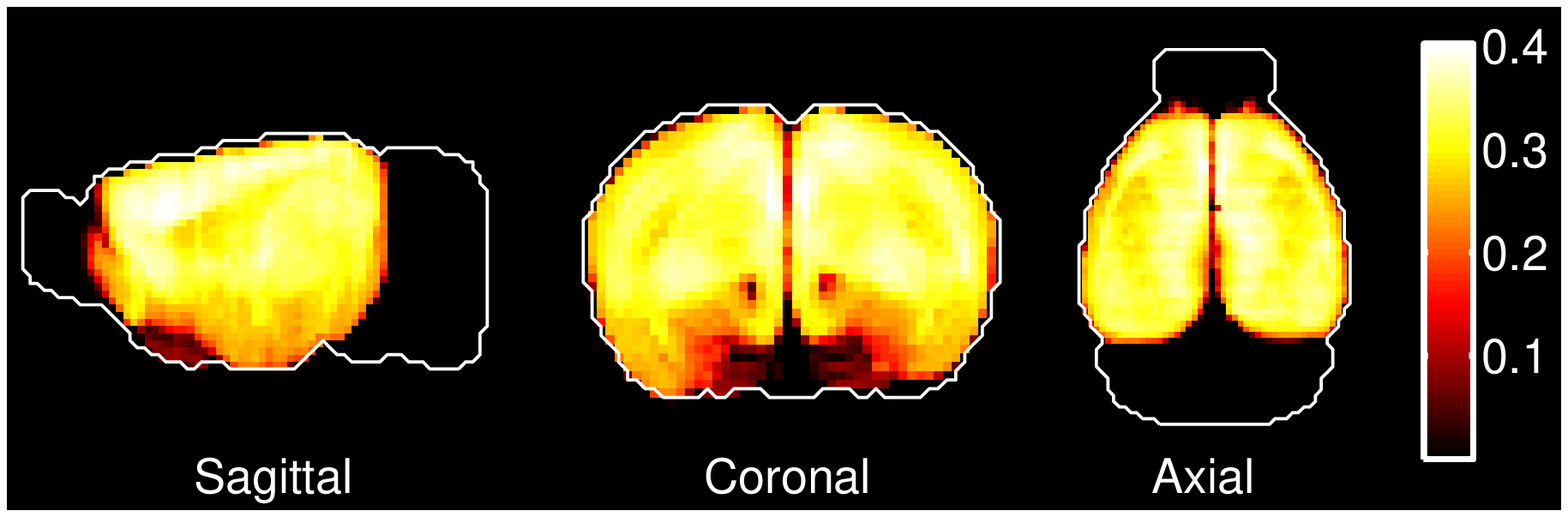}&\includegraphics[width=2in,keepaspectratio]{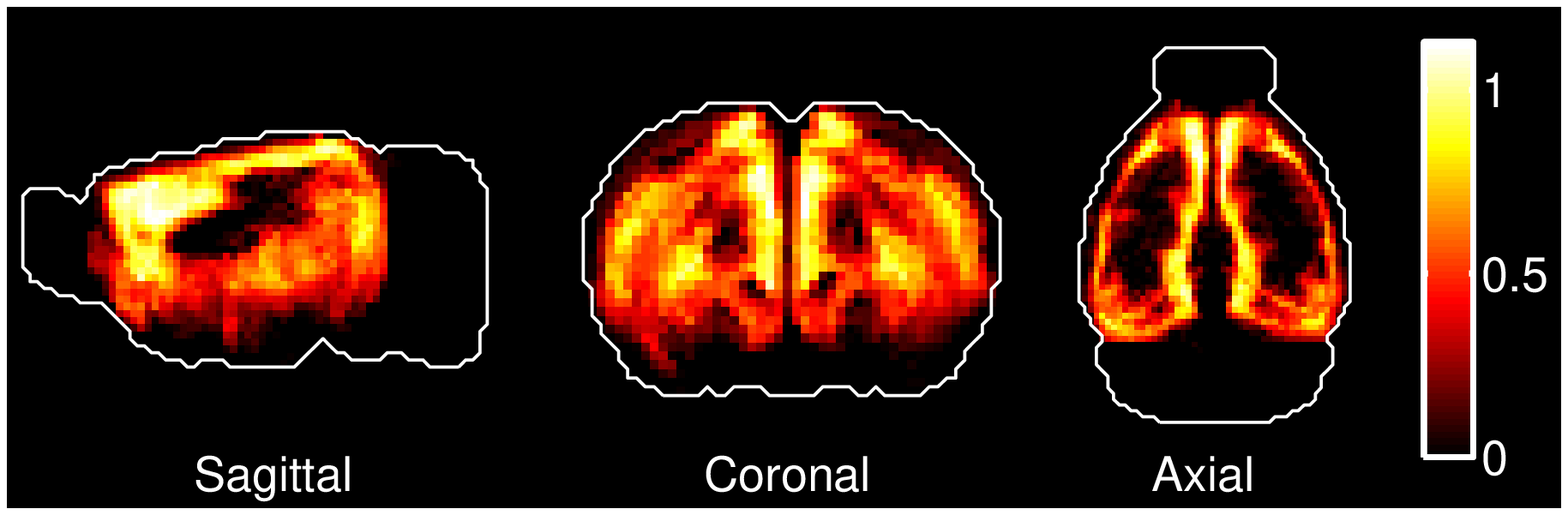}\\\hline
46&\tiny{Pyramidal Neurons}&\includegraphics[width=2in,keepaspectratio]{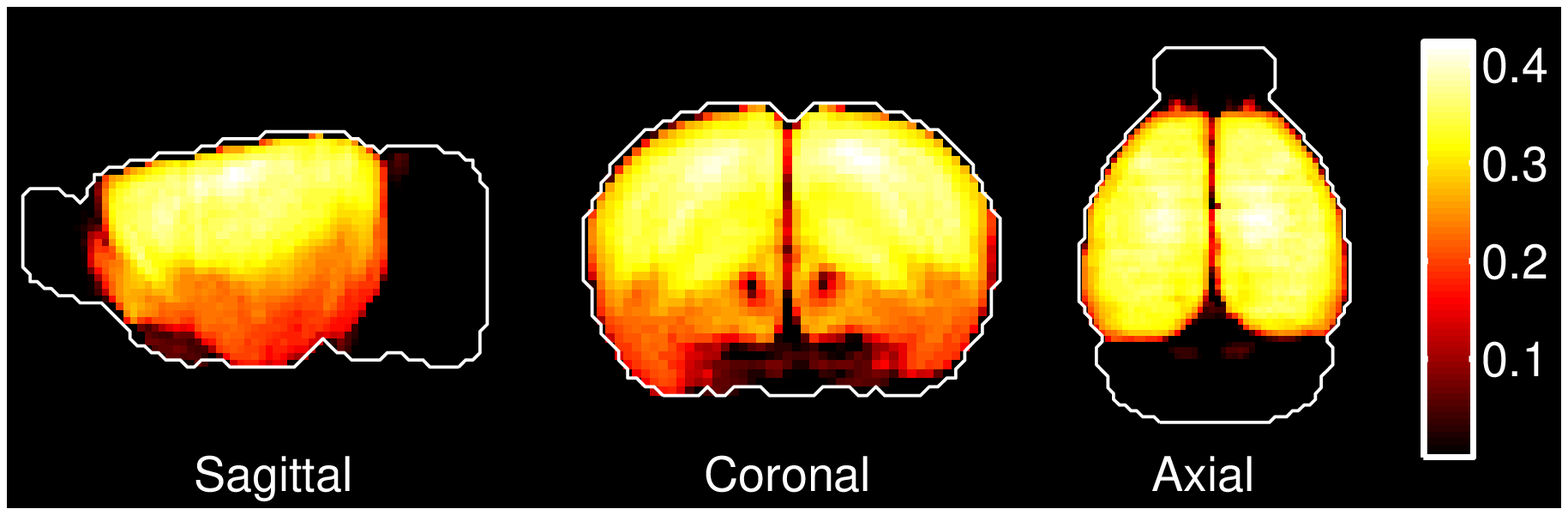}&\includegraphics[width=2in,keepaspectratio]{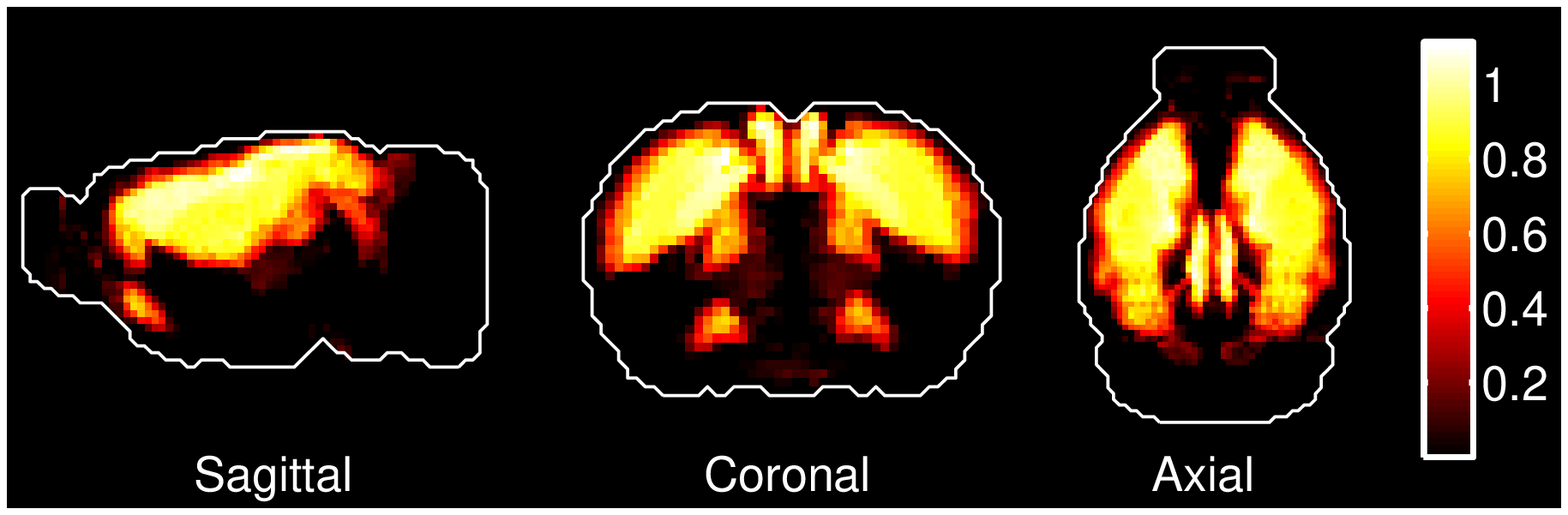}\\\hline
47&\tiny{Pyramidal Neurons}&\includegraphics[width=2in,keepaspectratio]{cellTypeProj47.eps}&\includegraphics[width=2in,keepaspectratio]{cellTypeModelFit47.eps}\\\hline
48&\tiny{Pyramidal Neurons}&\includegraphics[width=2in,keepaspectratio]{cellTypeProj48.eps}&\includegraphics[width=2in,keepaspectratio]{cellTypeModelFit48.eps}\\\hline
\end{tabular}
\\
\begin{tabular}{|l|l|l|l|}
\hline
\textbf{index}&\textbf{Cell type}&\textbf{Heat map of correlations}&\textbf{Heat map of weight}\\\hline
49&\tiny{Pyramidal Neurons}&\includegraphics[width=2in,keepaspectratio]{cellTypeProj49.eps}&\includegraphics[width=2in,keepaspectratio]{cellTypeModelFit49.eps}\\\hline
50&\tiny{Pyramidal Neurons}&\includegraphics[width=2in,keepaspectratio]{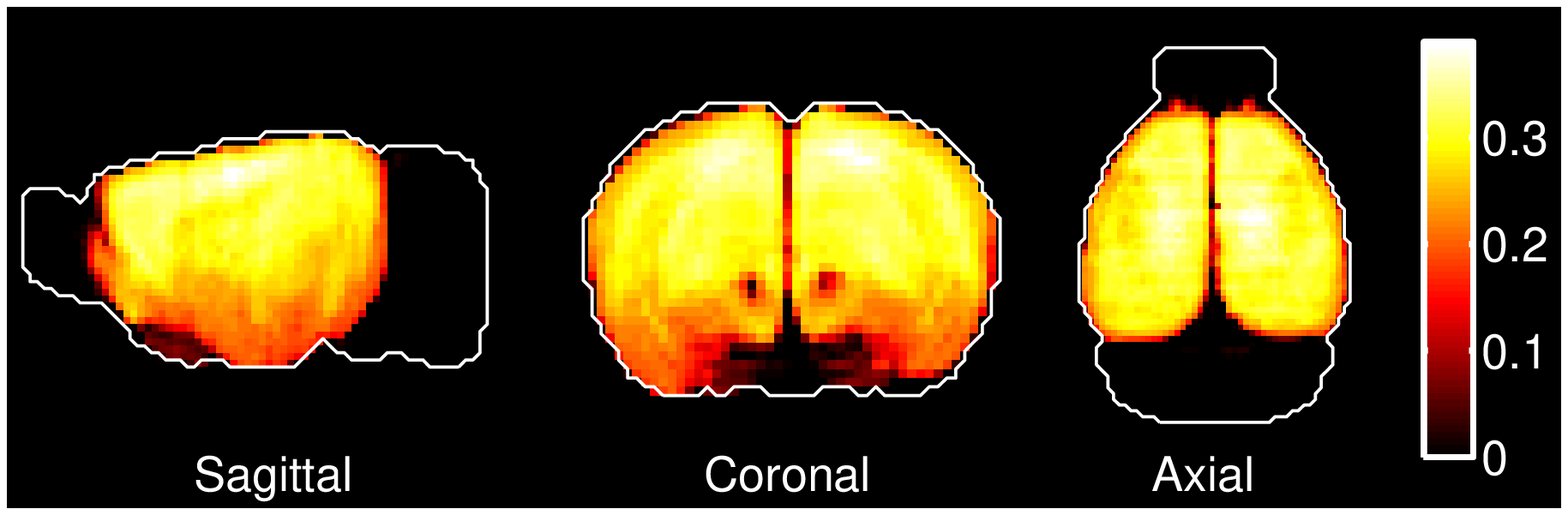}&\includegraphics[width=2in,keepaspectratio]{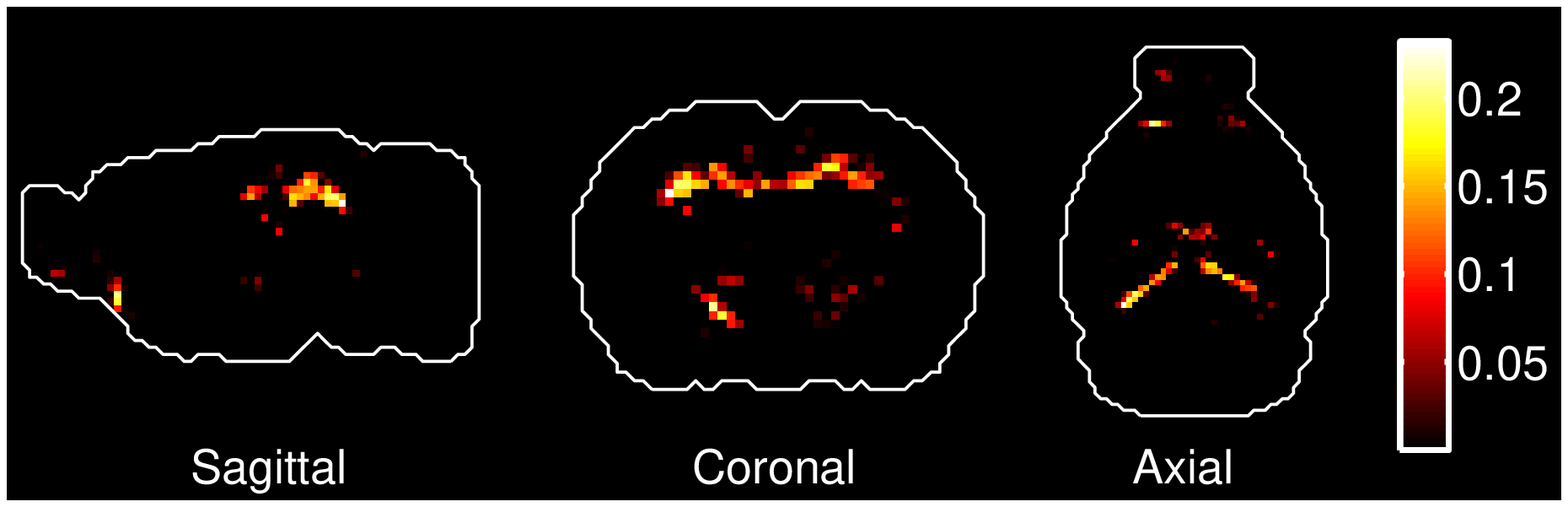}\\\hline
51&\tiny{Tyrosine Hydroxylase Expressing}&\includegraphics[width=2in,keepaspectratio]{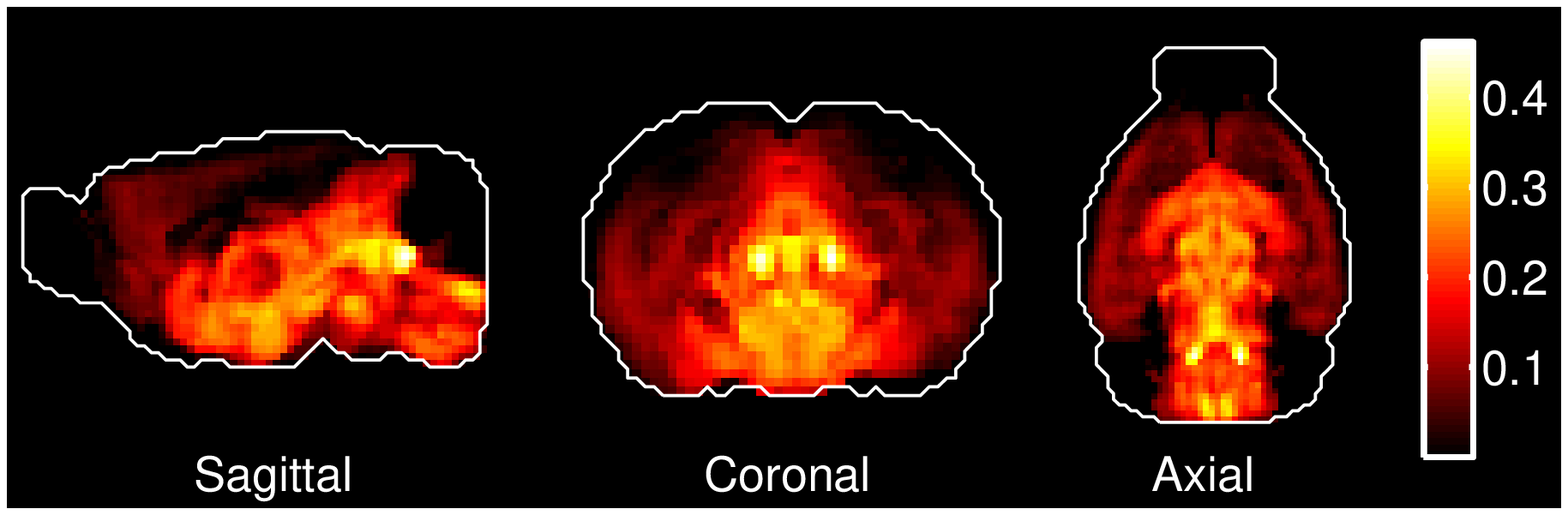}&\includegraphics[width=2in,keepaspectratio]{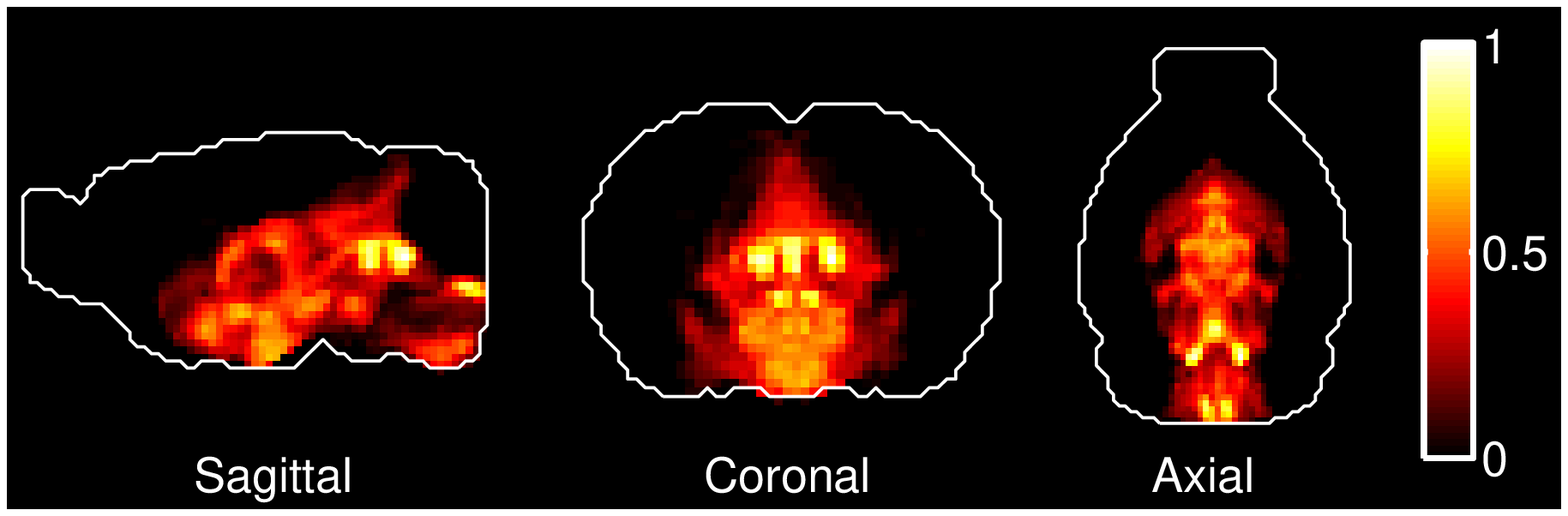}\\\hline
52&\tiny{Purkinje Cells}&\includegraphics[width=2in,keepaspectratio]{cellTypeProj52.eps}&\includegraphics[width=2in,keepaspectratio]{cellTypeModelFit52.eps}\\\hline
53&\tiny{Glutamatergic Neuron (not well defined)}&\includegraphics[width=2in,keepaspectratio]{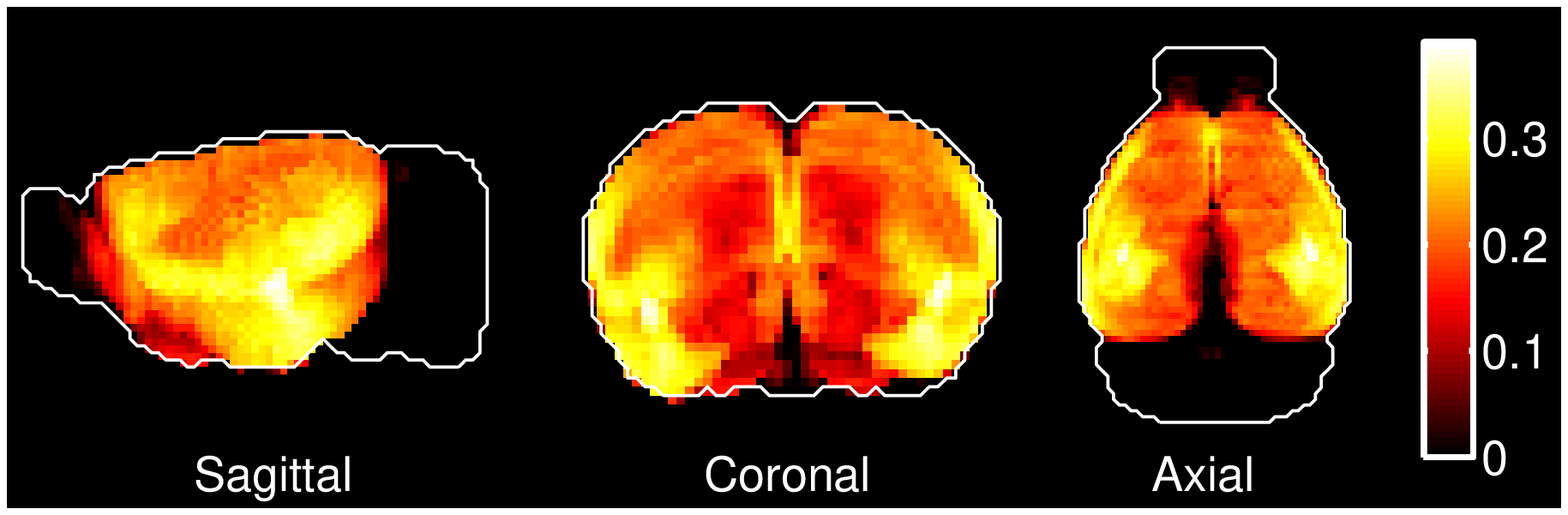}&\includegraphics[width=2in,keepaspectratio]{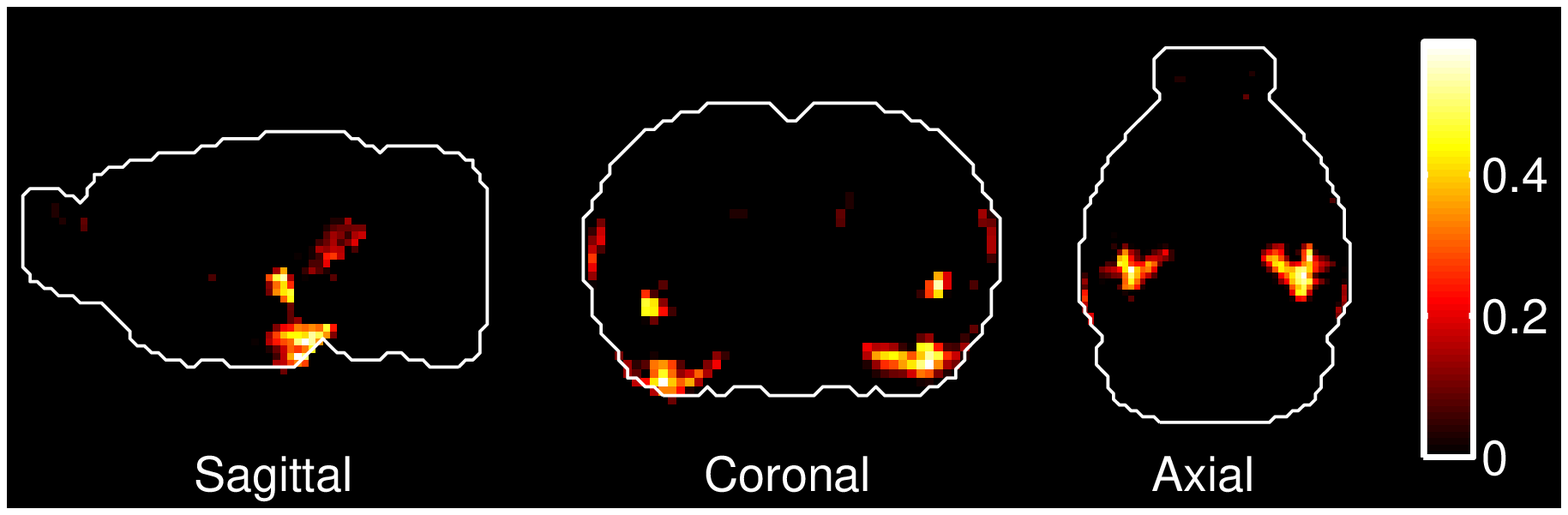}\\\hline
54&\tiny{GABAergic Interneurons, VIP+}&\includegraphics[width=2in,keepaspectratio]{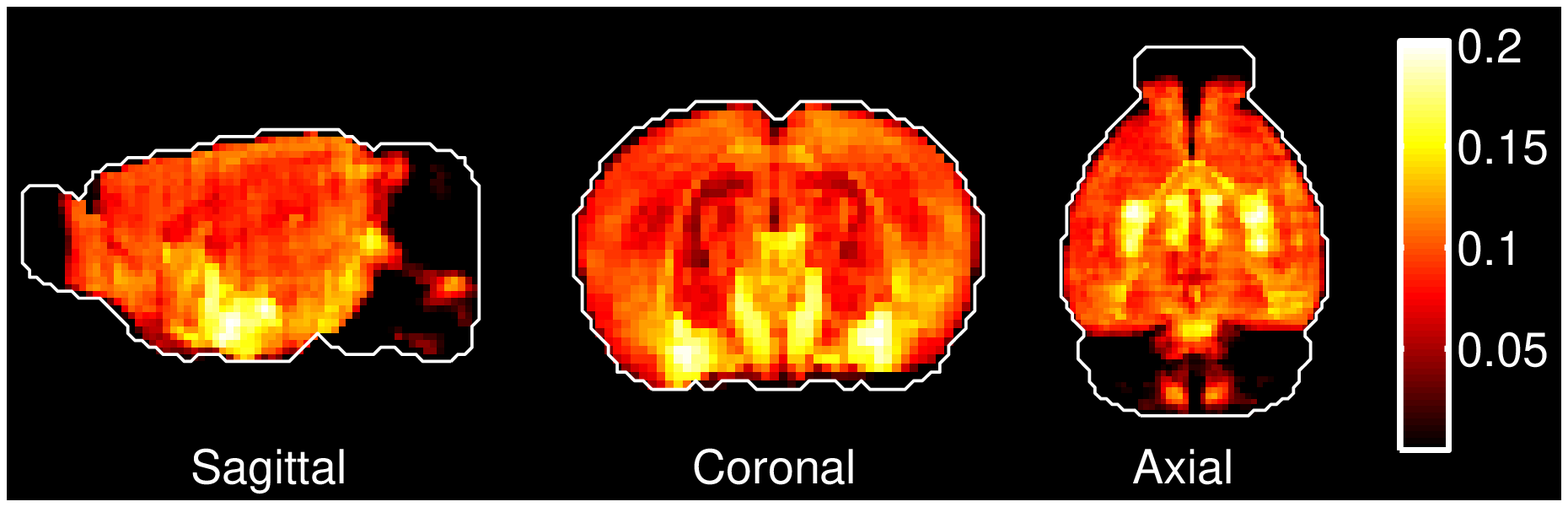}&\includegraphics[width=2in,keepaspectratio]{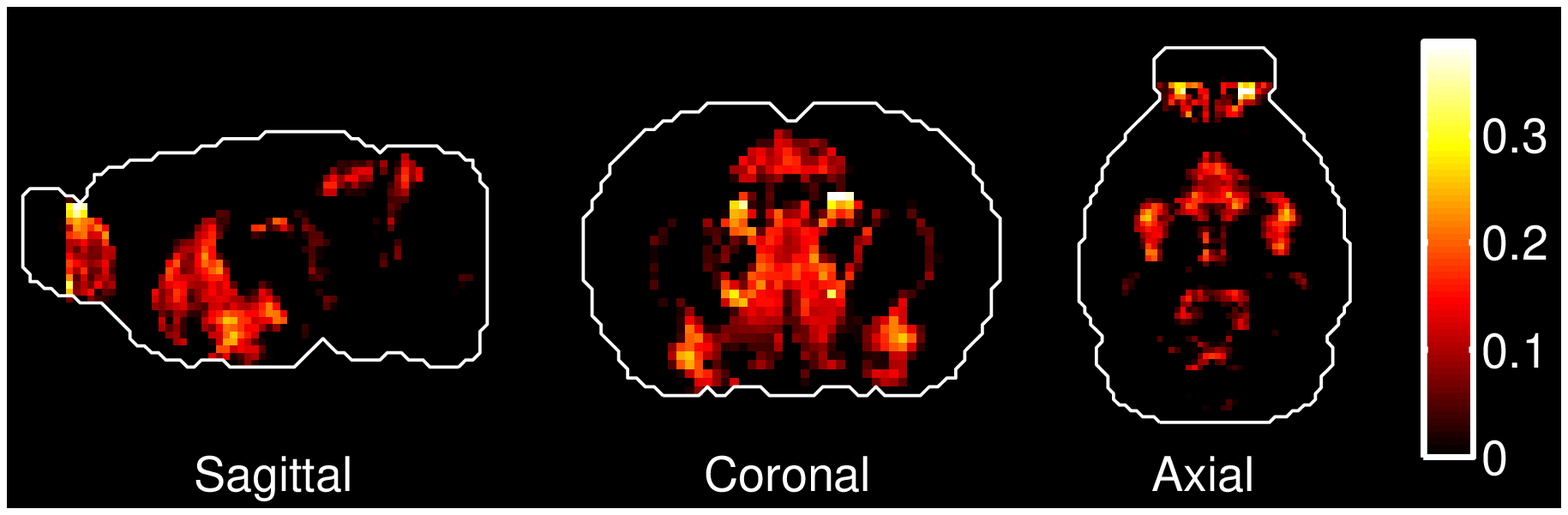}\\\hline
55&\tiny{GABAergic Interneurons, VIP+}&\includegraphics[width=2in,keepaspectratio]{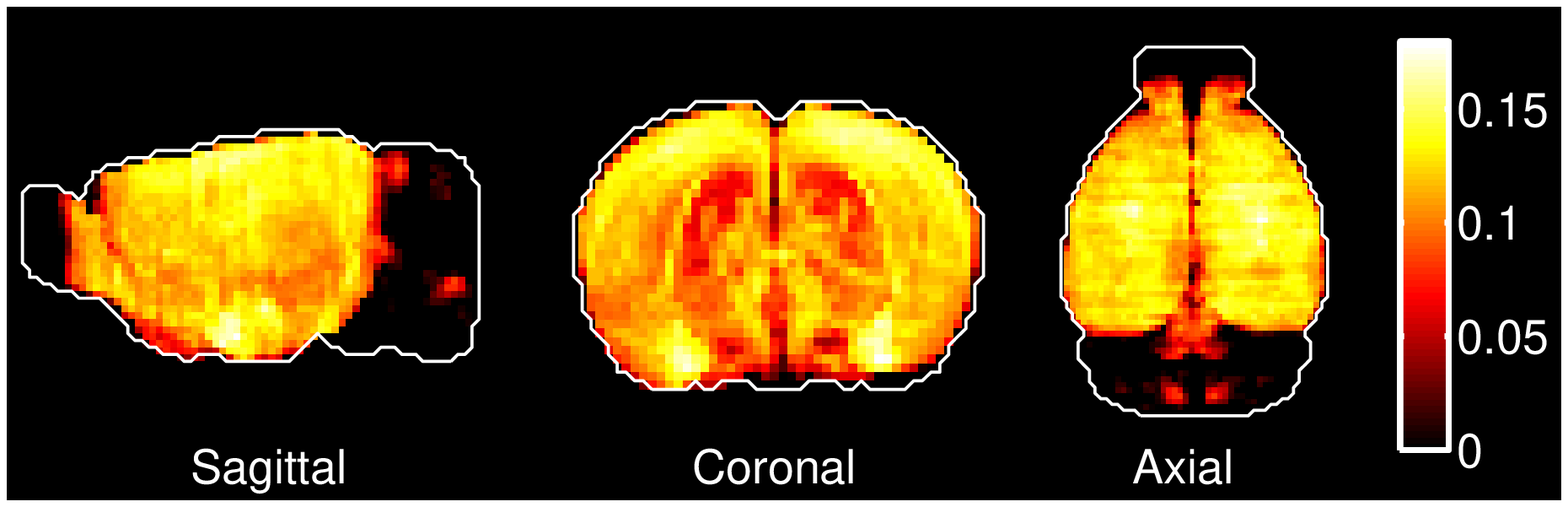}&\includegraphics[width=2in,keepaspectratio]{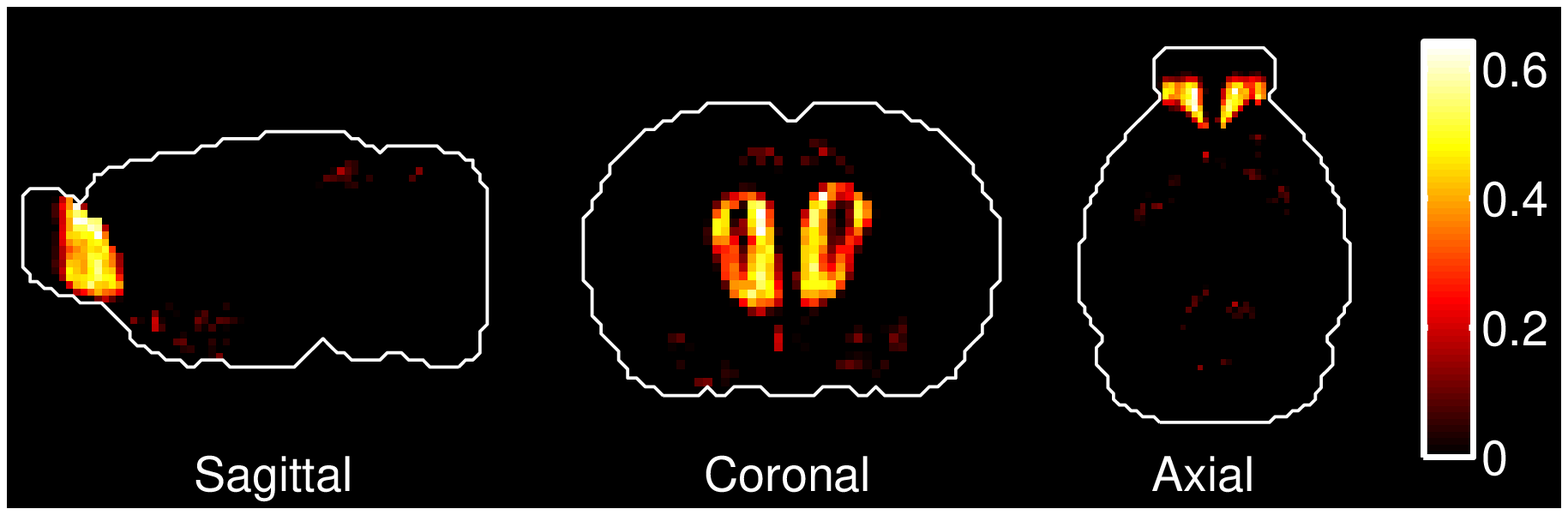}\\\hline
56&\tiny{GABAergic Interneurons, SST+}&\includegraphics[width=2in,keepaspectratio]{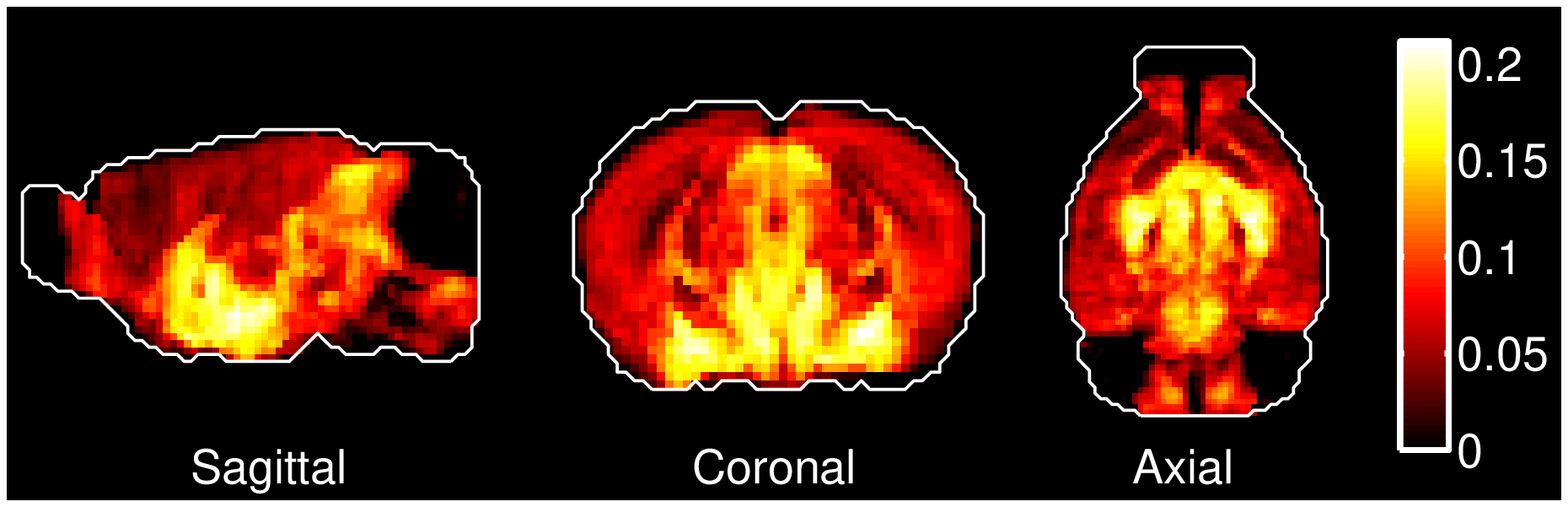}&\includegraphics[width=2in,keepaspectratio]{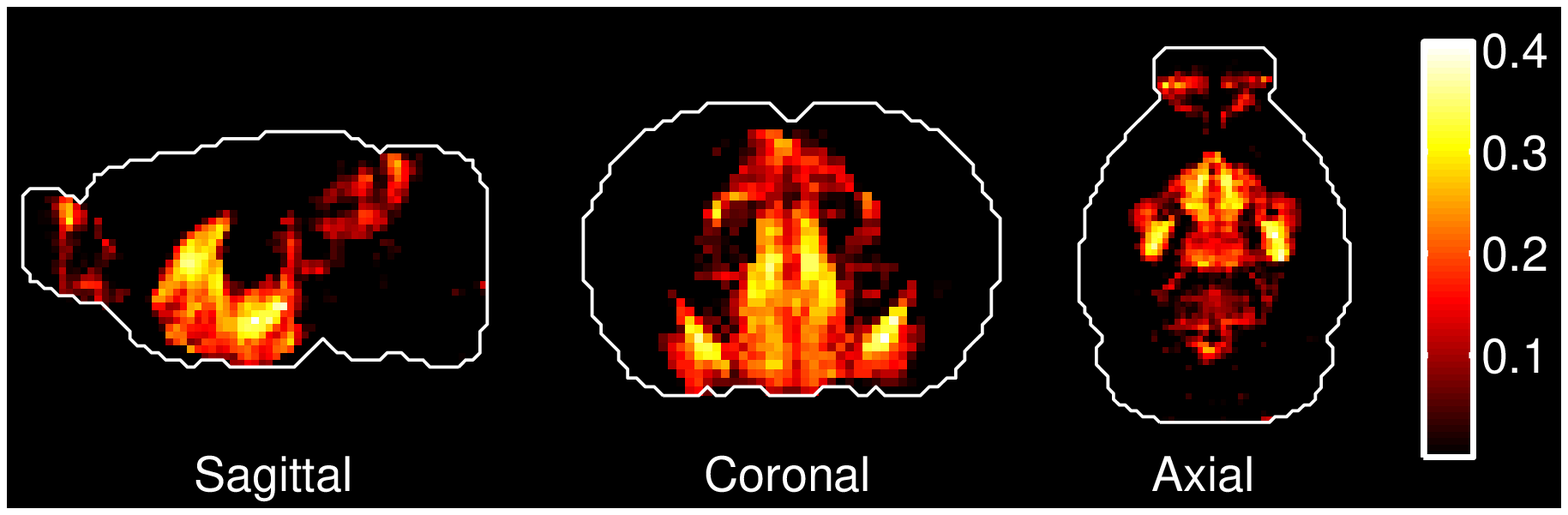}\\\hline
57&\tiny{GABAergic Interneurons, SST+}&\includegraphics[width=2in,keepaspectratio]{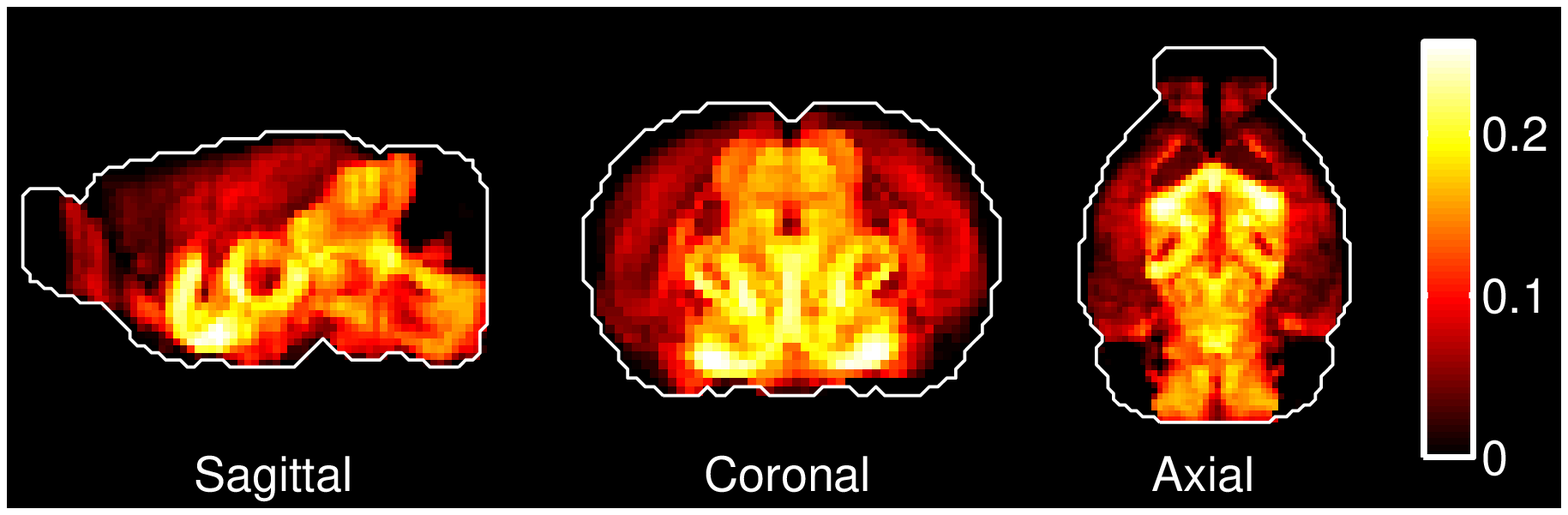}&\includegraphics[width=2in,keepaspectratio]{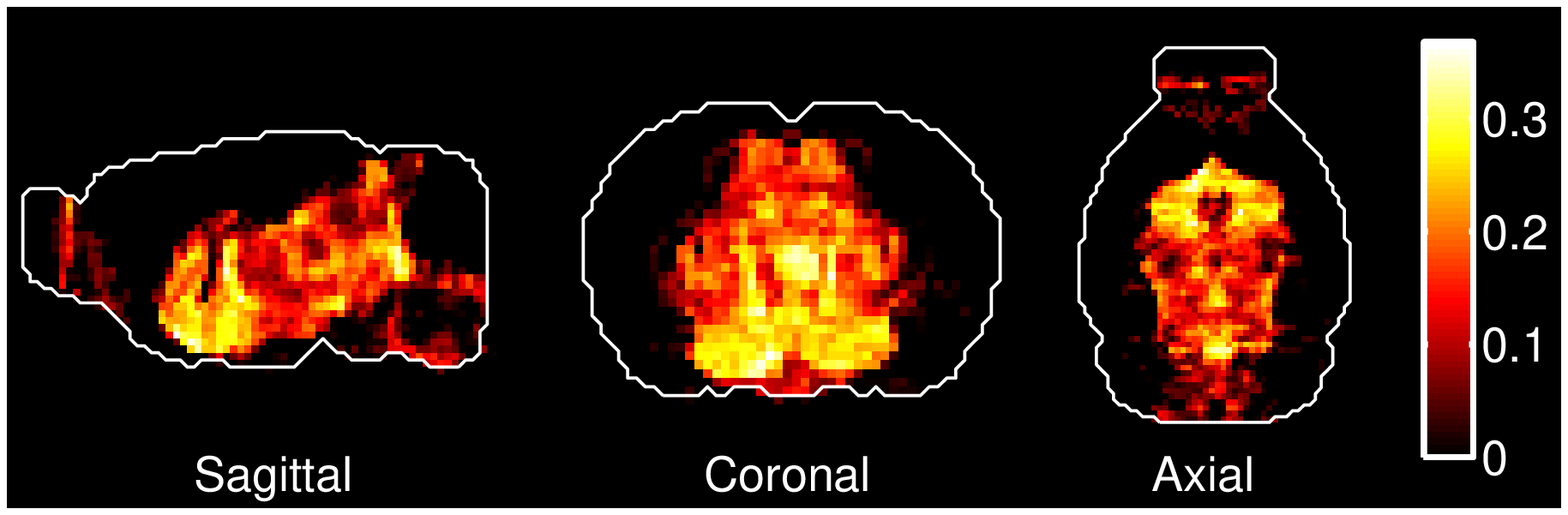}\\\hline
58&\tiny{GABAergic Interneurons, PV+}&\includegraphics[width=2in,keepaspectratio]{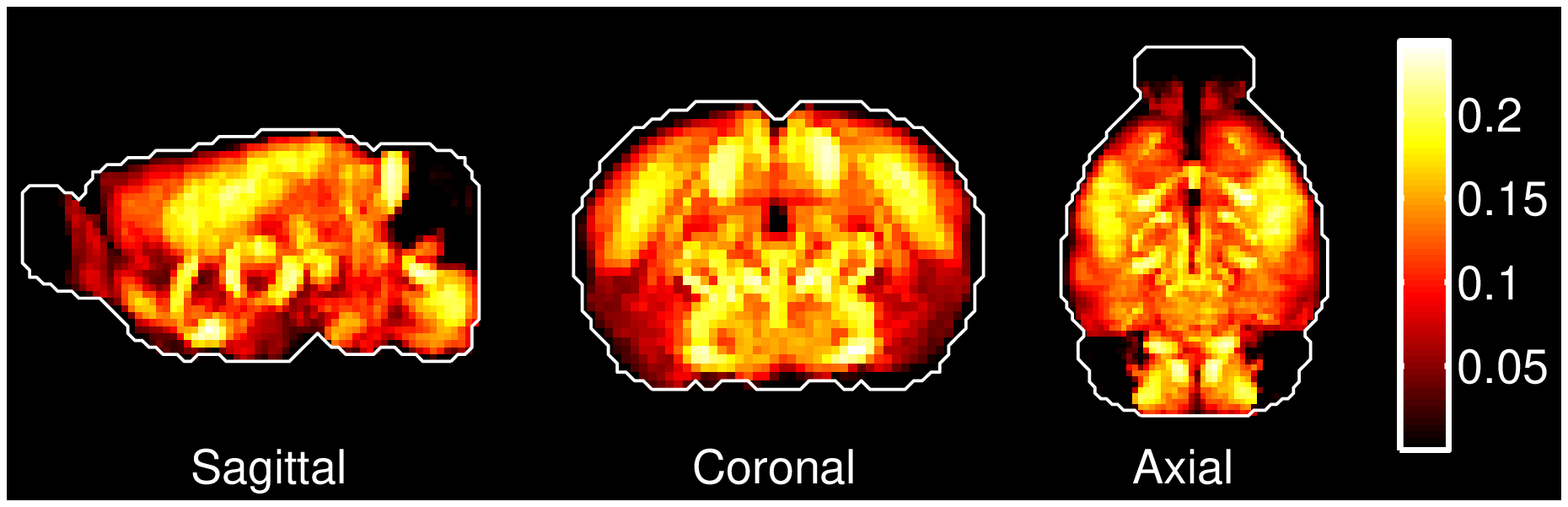}&\includegraphics[width=2in,keepaspectratio]{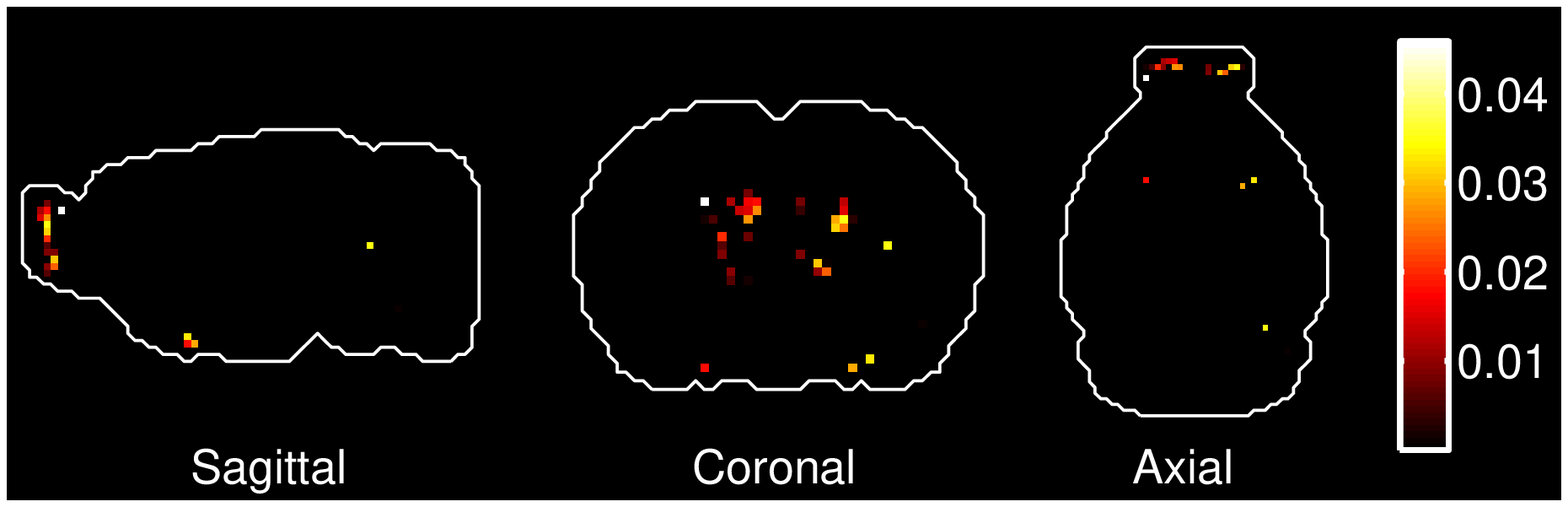}\\\hline
59&\tiny{GABAergic Interneurons, PV+}&\includegraphics[width=2in,keepaspectratio]{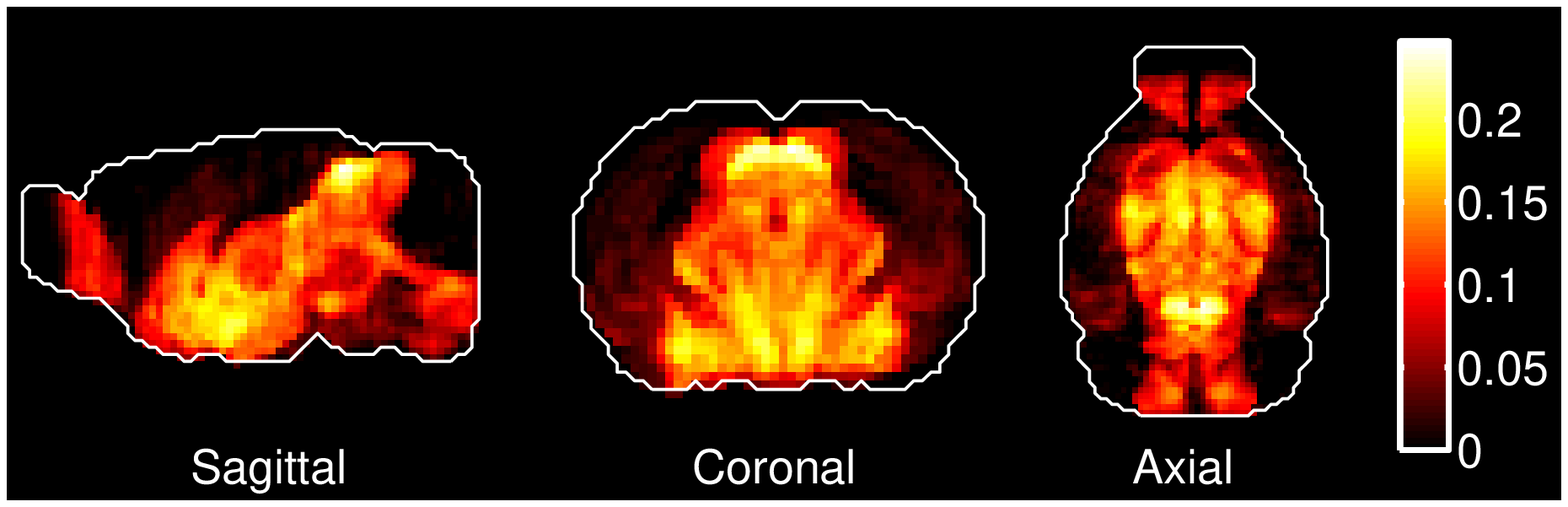}&\includegraphics[width=2in,keepaspectratio]{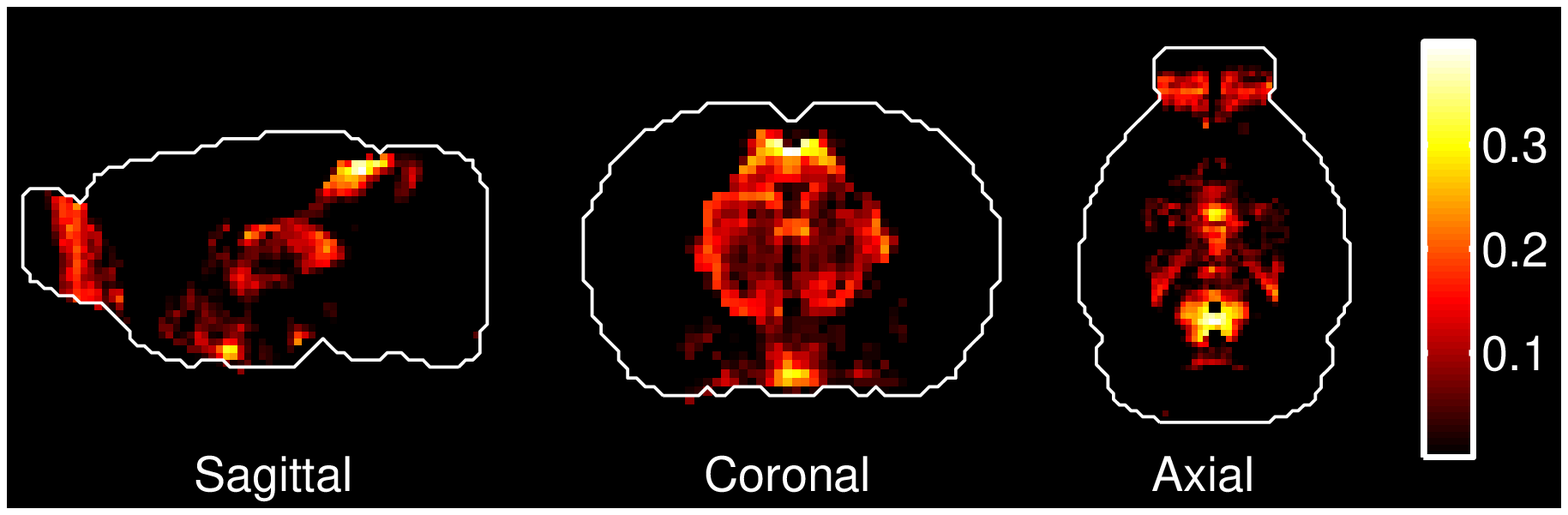}\\\hline
60&\tiny{GABAergic Interneurons, PV+, P7}&\includegraphics[width=2in,keepaspectratio]{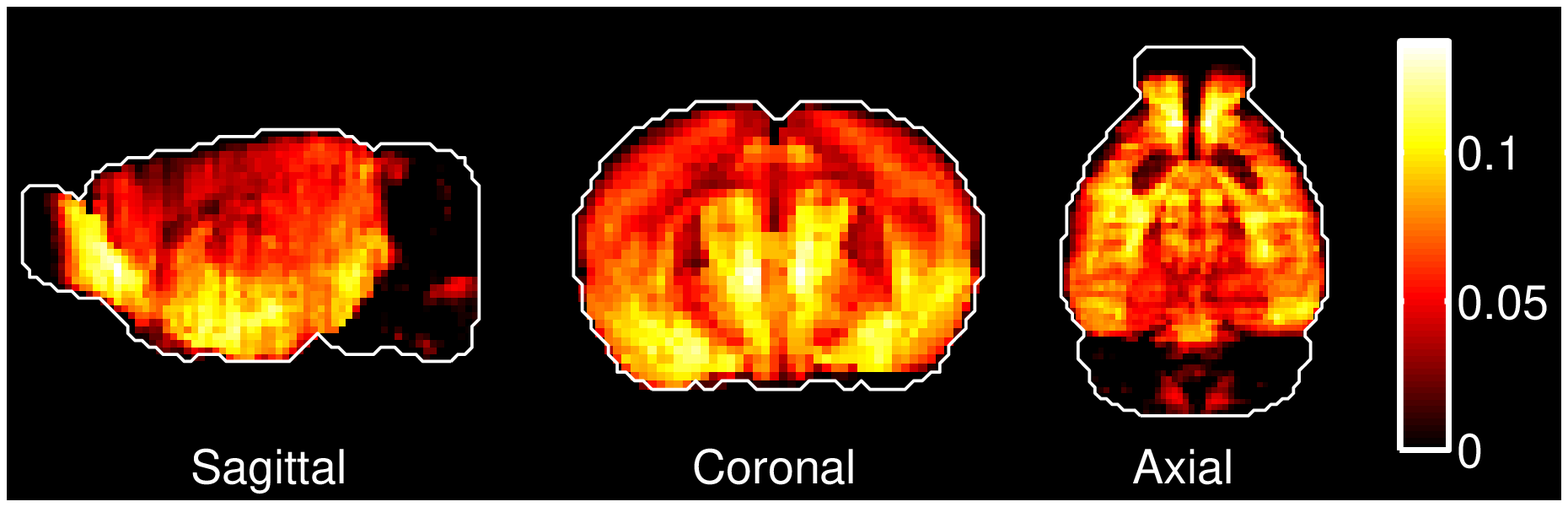}&\includegraphics[width=2in,keepaspectratio]{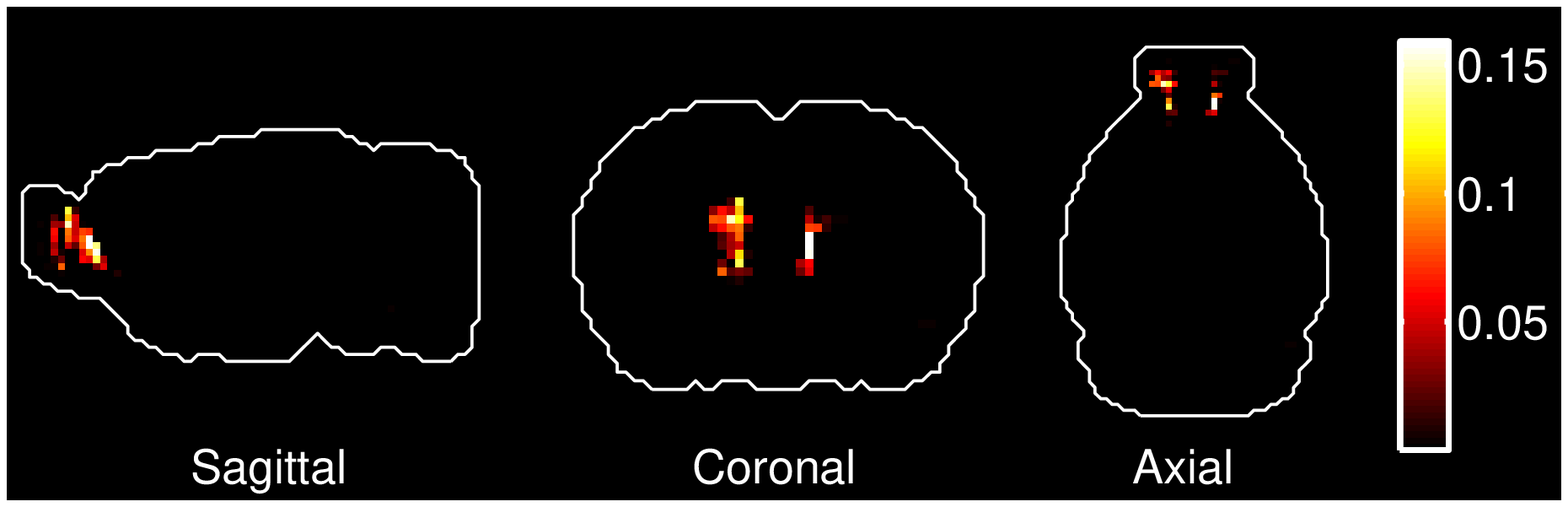}\\\hline
\end{tabular}
\\
\begin{tabular}{|l|l|l|l|}
\hline
\textbf{index}&\textbf{Cell type}&\textbf{Heat map of correlations}&\textbf{Heat map of weight}\\\hline
61&\tiny{GABAergic Interneurons, PV+, P10}&\includegraphics[width=2in,keepaspectratio]{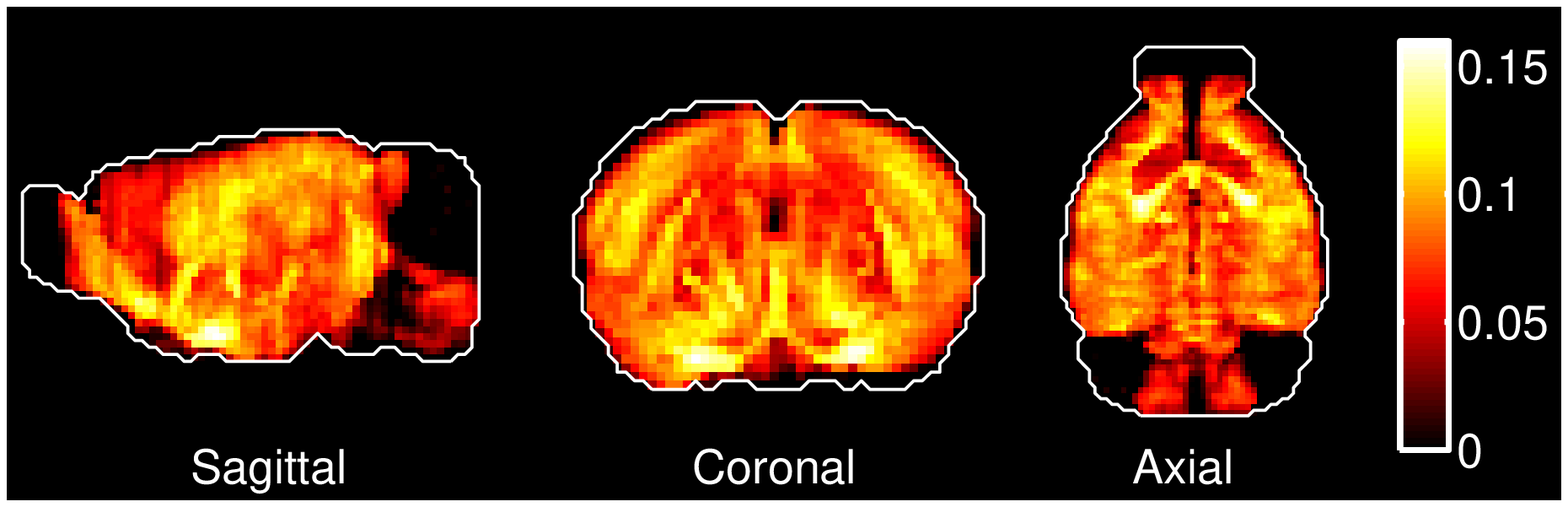}&\includegraphics[width=2in,keepaspectratio]{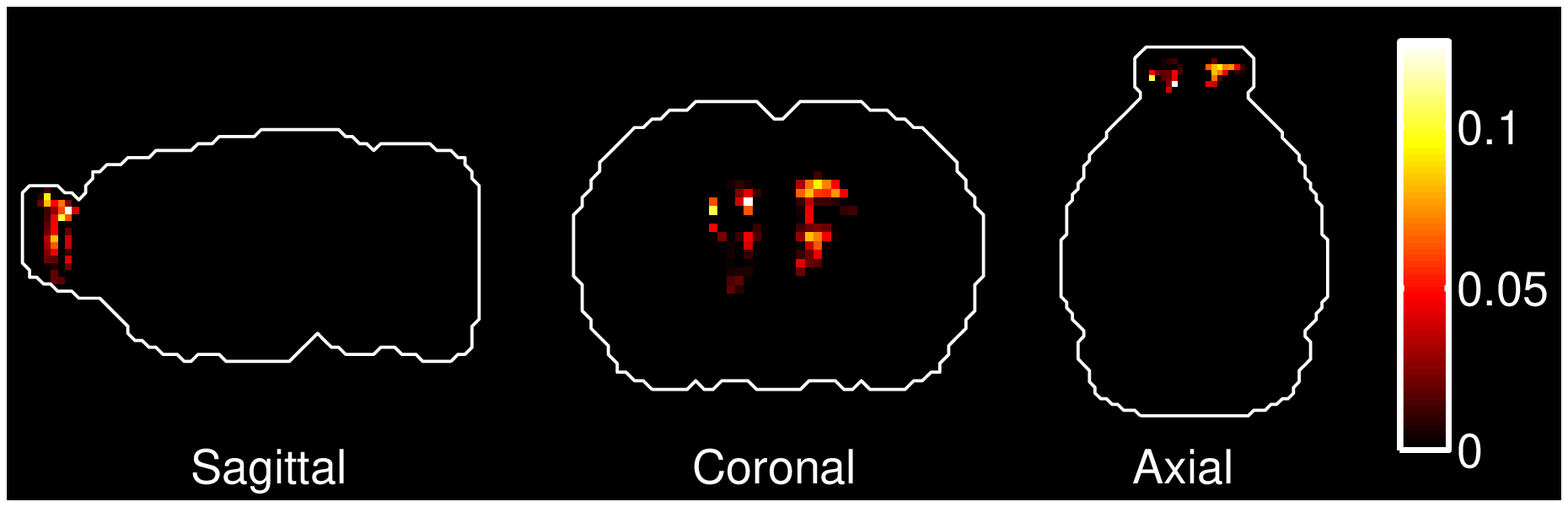}\\\hline
62&\tiny{GABAergic Interneurons, PV+, P13-P15}&\includegraphics[width=2in,keepaspectratio]{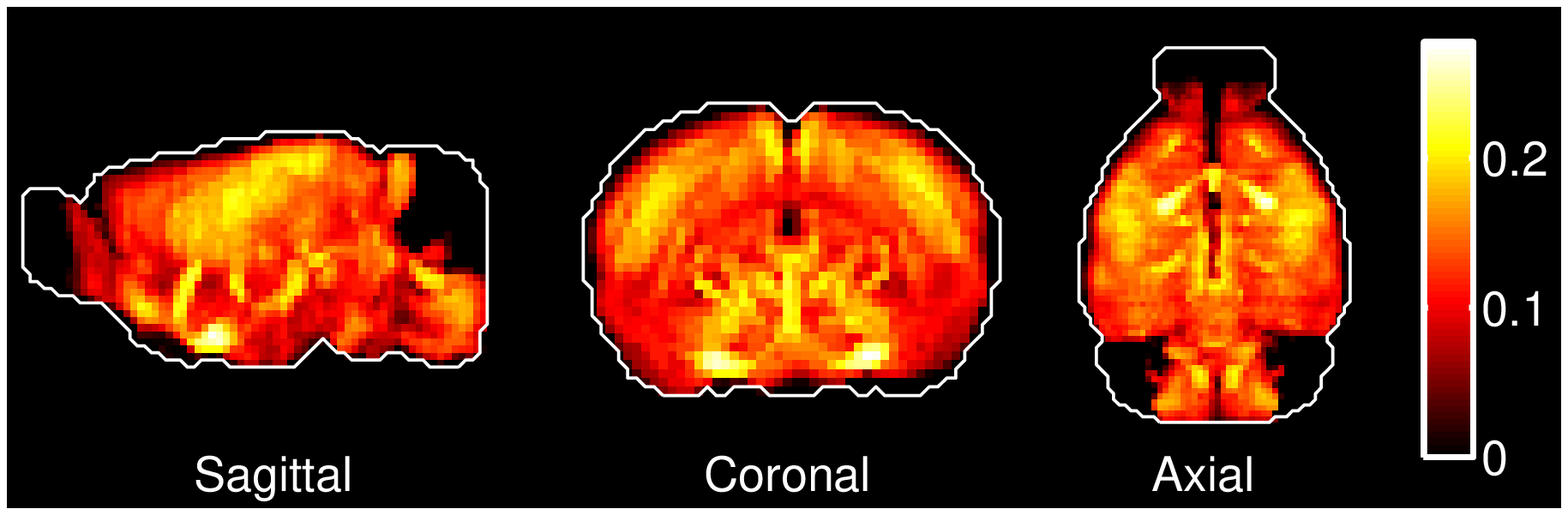}&\includegraphics[width=2in,keepaspectratio]{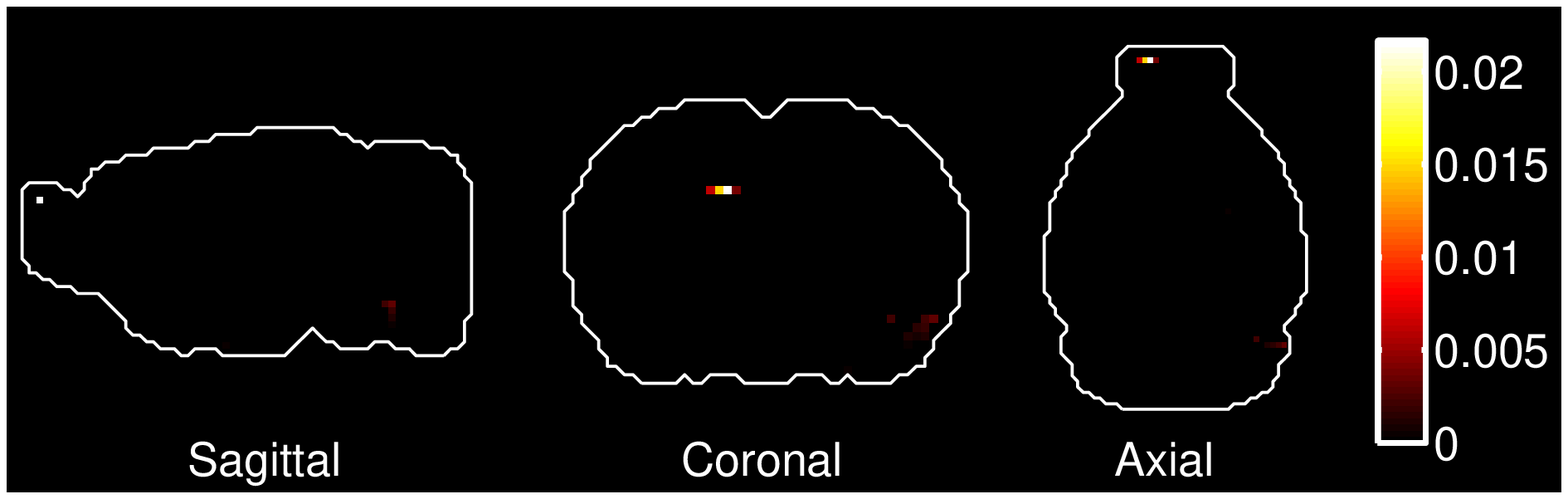}\\\hline
63&\tiny{GABAergic Interneurons, PV+, P25}&\includegraphics[width=2in,keepaspectratio]{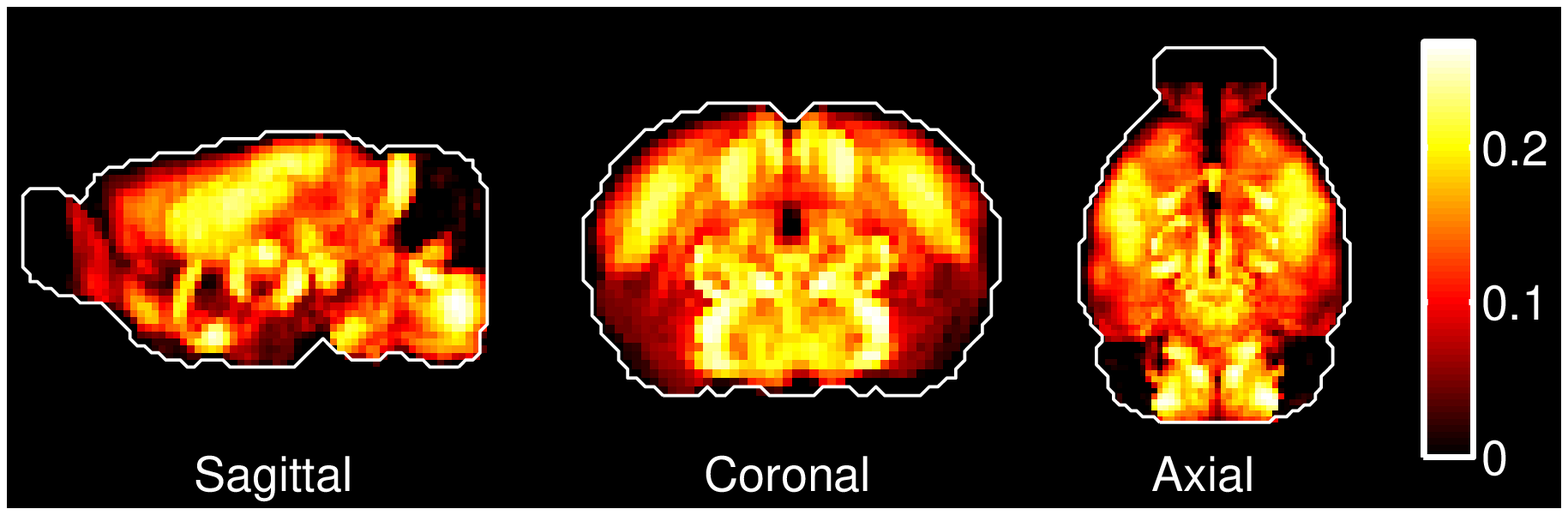}&\includegraphics[width=2in,keepaspectratio]{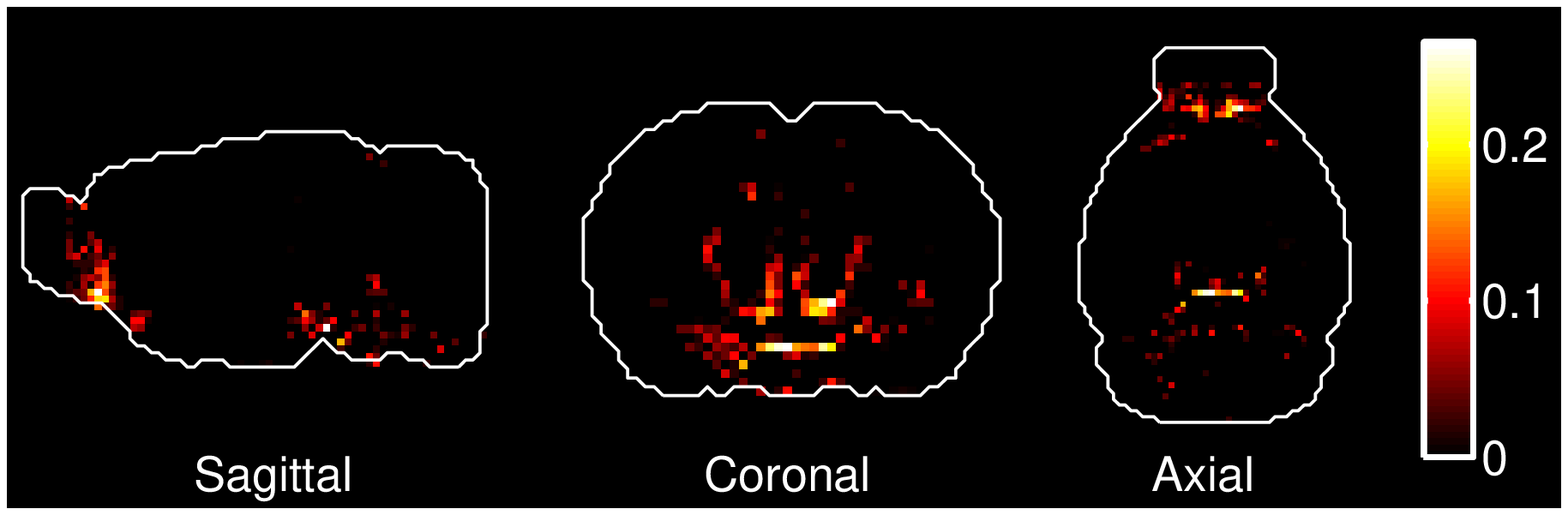}\\\hline
64&\tiny{GABAergic Interneurons, PV+}&\includegraphics[width=2in,keepaspectratio]{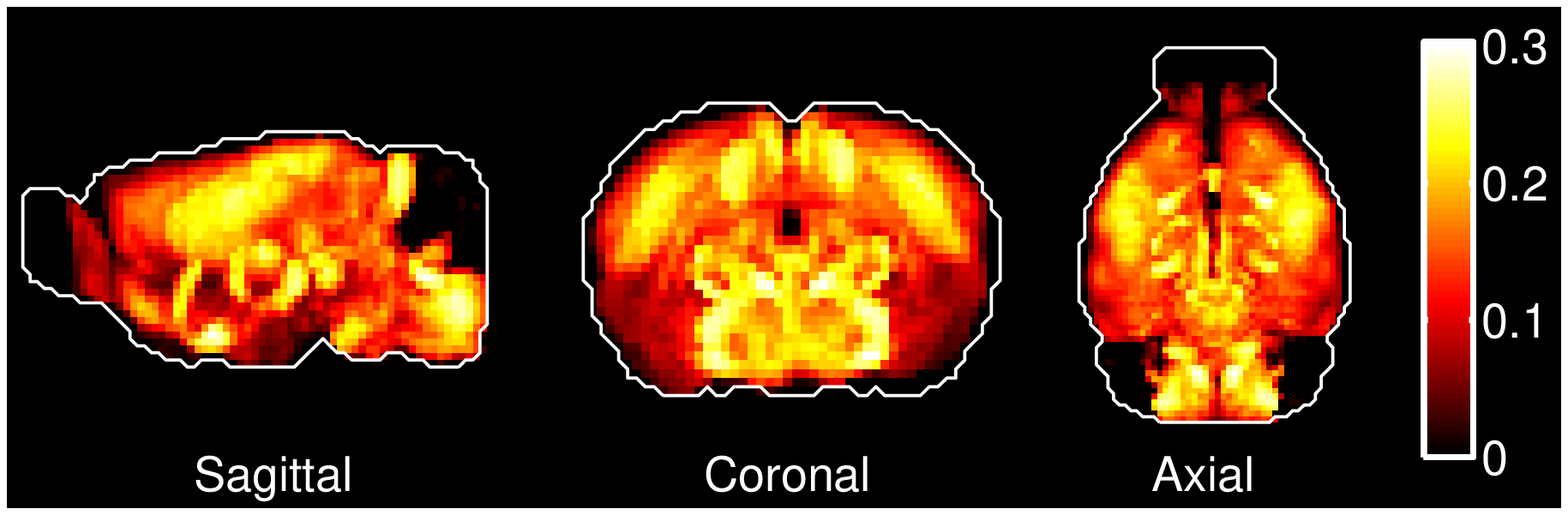}&\includegraphics[width=2in,keepaspectratio]{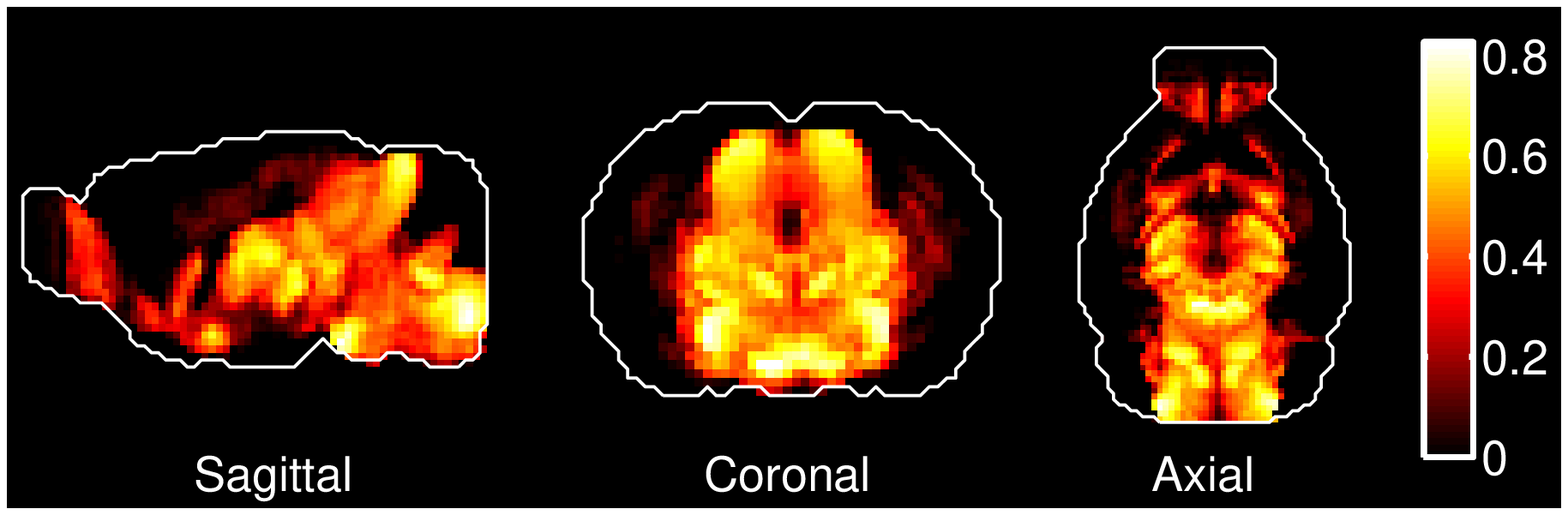}\\\hline
\end{tabular}
\\

\end{document}